# 2023 Astrophotonics Roadmap: pathways to realizing multi-functional integrated astrophotonic instruments

**Nemanja Jovanovic[1,56,57], Pradip Gatkine[1,56,57], Narsireddy Anugu[2], Rodrigo Amezcua-Correa[3], Ritoban Basu Thakur[10,50], Charles Beichman[4], Chad Bender[5], Jean-Philippe Berger[6], Azzurra Bigioli[7], Joss Bland-Hawthorn[8], Guillaume Bourdarot[9], Charles M. Bradford[10], Ronald Broeke[11], Julia Bryant[8], Kevin Bundy[12], Ross Cheriton[13], Nick Cvetojevic[14], Momen Diab[15], Scott A. Diddams[16], Aline N. Dinkelaker[17], Jeroen Duis[18], Stephen Eikenberry[3], Simon Ellis[19], Akira Endo[20], Donald F. Figer[21], Michael Fitzgerald[22], Itandehui Gris-Sanchez[23], Simon Gross[24], Ludovic Grossard[25], Olivier Guyon[5,26,27,28], Sebastiaan Y. Haffert[5], Samuel Halverson[10], Robert J. Harris[29,30], Jinping He[31,32], Tobias Herr[33], Philipp Hottinger[34], Elsa Huby[35], Michael Ireland[36], Rebecca Jenson-Clem[12], Jeffrey Jewell[10], Laurent Jocou[37], Stefan Kraus[38], Lucas Labadie[39], Sylvestre Lacour[35], Romain Laugier[7], Katarzyna Ławniczuk[11], Jonathan Lin[22], Stephanie Leifer[40], Sergio Leon-Saval[8], Guillermo Martin[37], Frantz Martinache[14], Marc-Antoine Martinod[7], Benjamin A. Mazin[41], Stefano Minardi[42], John D. Monnier[43], Reinan Moreira[44], Denis Mourard[14], Abani Shankar Nayak[45], Barnaby Norris[8], Ewelina Obrzud[46], Karine Perraut[37], François Reynaud[25], Steph Sallum[47], David Schiminovich[48], Christian Schwab[49], Eugene Serbayn[10], Sherif Soliman[18], Andreas Stoll[17], Liang Tang[31,32], Peter Tuthill[8], Kerry Vahala[50], Gautam Vasisht[10], Sylvain Veilleux[51], Alexander B. Walter[10], Edward J. Wollack[52], Yinzi Xin[1], Zongyin Yang[53], Stephanos Yerolatsitis[3], Yang Zhang[54] and Chang-Ling Zou[55].**

[1] Department of Astronomy, California Institute of Technology, Pasadena, CA, USA
[2] The CHARA Array of Georgia State University, Mount Wilson Observatory, Mount Wilson, CA 91203, USA
[3] CREOL, The College of Optics and Photonics, University of Central Florida, Orlando, FL, USA
[4] IPAC/NASA Exoplanet Science Institute, Jet Propulsion Laboratory, California Institute of Technology, Pasadena, CA, USA
[5] Steward Observatory, University of Arizona, Tucson, AZ, USA
[6] Institut de Planétologie et d'Astrophysique, UGA, CNRS, France
[7] Katholieke Universiteit Leuven, Belgium
[8] Sydney Institute for Astronomy (SIfA), School of Physics, The University of Sydney, Australia
[9] Max Planck Institute for Extraterrestrial Physics, Garching, Germany
[10] Jet Propulsion Laboratory, California Institute of Technology, Pasadena, CA, USA
[11] Bright Photonics BV, Eindhoven, The Netherlands
[12] Department of Astronomy and Astrophysics, University of California, Santa Cruz, CA, USA
[13] Advanced Electronics and Photonics Research Centre, National Research Council Canada, Ottawa, Canada
[14] Université Côte d'Azur, Observatoire de la Côte d'Azur, CNRS, Laboratoire Lagrange, France
[15] Dunlap Institute for Astronomy and Astrophysics, University of Toronto, Toronto, Canada
[16] Electrical, Computer and Energy Engineering and Department of Physics, University of Colorado Boulder, CO, USA
[17] Leibniz Institute for Astrophysics Potsdam (AIP), Potsdam, Germany

[18] PHIX Photonics Assembly, Enschede, Netherlands
[19] Australian Astronomical Optics, Macquarie University, NSW, Australia
[20] Electrical Engineering, Mathematics and Computer Science, Delft University of Technology, Delft, the Netherlands
[21] Center for Detectors, Rochester Institute of Technology, Rochester, NY, USA
[22] UCLA Physics & Astronomy Department, Los Angeles, CA, USA
[23] ITEAM Research Institute, Universitat Politècnica de València, Valencia, Spain
[24] MQ Photonics Research Centre, Macquarie University, NSW, Australia
[25] XLIM Research Institute, University of Limoges, France
[26] Subaru Telescope, National Astronomical Observatory of Japan, National Institute of Natural Sciences, Hilo, HI, USA
[27] Astrobiology Center of NINS, Osawa, Mitaka, Tokyo, Japan
[28] College of Optical Sciences, University of Arizona, Tucson, AZ, USA
[29] Max-Planck-Institute for Astronomy, Heidelberg, Germany
[30] Department of Physics, Durham University, Durham, UK
[31] National Astronomical Observatories, Nanjing Institute of Astronomical Optics & Technology, Chinese Academy of Sciences, Nanjing, China
[32] Key Laboratory of Astronomical Optics & Technology, Nanjing Institute of Astronomical Optics & Technology, Chinese Academy of Sciences, Nanjing, China
[33] Deutsches Elektronen-Synchrotron DESY, Germany and Universität Hamburg, Germany
[34] Landessternwarte, Zentrum für Astronomie der Universität Heidelberg, Heidelberg, Germany
[35] LESIA, Observatoire de Paris, Université PSL, CNRS, Sorbonne Université, Université Paris Cité, Meudon, France
[36] The Australian National University, Canberra, Australia
[37] Université Grenoble Alpes, CNRS, IPAG, Grenoble, France
[38] Department of Physics and Astronomy, University of Exeter, Exeter, UK
[39] I. Physikalisches Institut, Universität zu Köln, Cologne, Germany
[40] The Aerospace Corporation, El Segundo, CA, USA
[41] Department of Physics, University of California, Santa Barbara, CA, USA
[42] Ams-OSRAM, Jena, Germany
[43] University of Michigan, Department of Astronomy, Ann Arbor, MI, USA
[44] Ultra-Low Loss Technologies, Santa Barbara, CA, USA
[45] Institut für Angewandte Physik, Friedrich-Schiller-Universität Jena, Germany
[46] Centre Suisse d'Electronique et de Microtechnique, Neuchâtel, Switzerland
[47] Department of Physics and Astronomy, University of California, Irvine, CA, USA
[48] Department of Astronomy, Columbia University, NY, USA
[49] School of Mathematical and Physical Sciences, Macquarie University, NSW, Australia
[50] Department of Physics, California Institute of Technology, Pasadena, CA, USA
[51] Department of Astronomy and Joint Space-Science Institute, University of Maryland, College Park, MD, USA
[52] NASA Goddard Space Flight Center, Greenbelt, MD, USA
[53] College of Information Science and Electronic Engineering, State Key Laboratory of Modern Optical Instrumentation, Zhejiang University, Hangzhou, China
[54] Electrical and Computer Engineering Department, University of Maryland College Park, USA
[55] CAS Key Laboratory of Quantum Information, University of Science and Technology of China, Hefei, Anhui, China

[56]Guest editors of the Roadmap.
[57]Author to whom any correspondence should be addressed.

E-mails: nem@caltech.edu, pgatkine@caltech.edu

## Abstract

Photonic technologies offer numerous functionalities that can be used to realize astrophotonic instruments. The most spectacular example to date is the ESO Gravity instrument at the Very Large Telescope in Chile that combines the light-gathering power of four 8 m telescopes through a complex photonic interferometer. Fully integrated astrophotonic devices stand to offer critical advantages for instrument development, including extreme miniaturization when operating at the diffraction-limit, as well as integration, superior thermal and mechanical stabilization owing to the small footprint, and high replicability offering significant cost savings. Numerous astrophotonic technologies have been developed to address shortcomings of conventional instruments to date, including for example the development of photonic lanterns to convert from multimode inputs to single mode outputs, complex aperiodic fiber Bragg gratings to filter OH emission from the atmosphere, complex beam combiners to enable long baseline interferometry with for example, ESO Gravity, and laser frequency combs for high precision spectral calibration of spectrometers. Despite these successes, the facility implementation of photonic solutions in astronomical instrumentation is currently limited because of 1) low throughputs from coupling to fibers, coupling fibers to chips, propagation and bend losses, device losses, etc., 2) difficulties with scaling to large channel count devices needed for large bandwidths and high resolutions, and 3) efficient integration of photonics with detectors, to name a few. In this roadmap, we identify 23 key areas that need further development. We outline the challenges and advances needed across those areas covering design tools, simulation capabilities, fabrication processes, the need for entirely new components, integration and hybridization and the characterization of devices. To realize these advances the astrophotonics community will have to work cooperatively with industrial partners who have more advanced manufacturing capabilities. With the advances described herein, multi-functional integrated instruments will be realized leading to novel observing capabilities for both ground and space based platforms, enabling new scientific studies and discoveries.

## Contents

# 1 | Introduction

Nemanja Jovanovic and Pradip Gatkine
**Department of Astronomy, California Institute of Technology, Pasadena, CA, USA**

**Status**

Astrophotonics is simply defined as the application of photonic technologies to astronomy. Like many fields of technology-driven science, astrophotonics directly benefits from the multi-decade and multi-billion dollar investments in photonics by industry, especially by telecommunications. Photonic technologies are appealing because they can provide many avenues for controlling and manipulating light, including spectral dispersion, spectral, spatial, and polarization filtering, phase and amplitude modulation, light generation, frequency shifting, and light detection to name a few. In addition, operating at the diffraction limit of the input telescope means that the instrument can have the smallest possible footprint for a given set of specifications, reducing volume, mass and cost [1,2]. Guiding the light in fibers or waveguides in a photonic integrated circuit (PIC) allows further reductions in instrument size by enabling flexible integration of many functions optimized for a specific science case and the elimination of bulk optics in most cases leading to extreme miniaturization, which can more readily be thermally and mechanically stabilized. Unlike bulk optics, photonic technologies can be highly replicable, offering dramatic cost savings once produced in volume. A highly replicable diffraction-limited instrument can be readily deployed to numerous observatories and achieve the same performance, potentially providing cost-savings by eliminating the prototypical nature of current astronomical instruments.

With such favorable properties, photonic technologies stand to impact many areas of astronomy. Leveraging the replicability for example, means that low cost massively multiplexed spectroscopic surveys could be enabled in the future. These would allow the measurement of large samples of spectroscopic redshifts, constrain galaxy evolution through tracers of star formation, outflows, etc, and enable detailed compositional study of stellar populations to name a few. On the other hand, the diffraction-limited nature of photonics makes them directly applicable to point-source-like targets that typically rely on adaptive optics (AO) observations. These include the study of exoplanets and disks, young stellar objects, evolved stars, active galactic nuclei, and nuclear star cluster kinematics for example.

Despite its origin in the late 1970s, and pioneering overviews about the prospects of photonics in the 90's [59], the field of astrophotonics was first recognized as a sub-field of astronomical instrumentation in 2009, when the editors of Optics Express [3] solicited a special issue on the topic for the first time. Although then already a thriving field, it has grown over the last 13 years as shown in Figure 1, which highlights some of the various technical developments resulting in two further special issues being developed in 2017 and 2021 respectively [4,5].

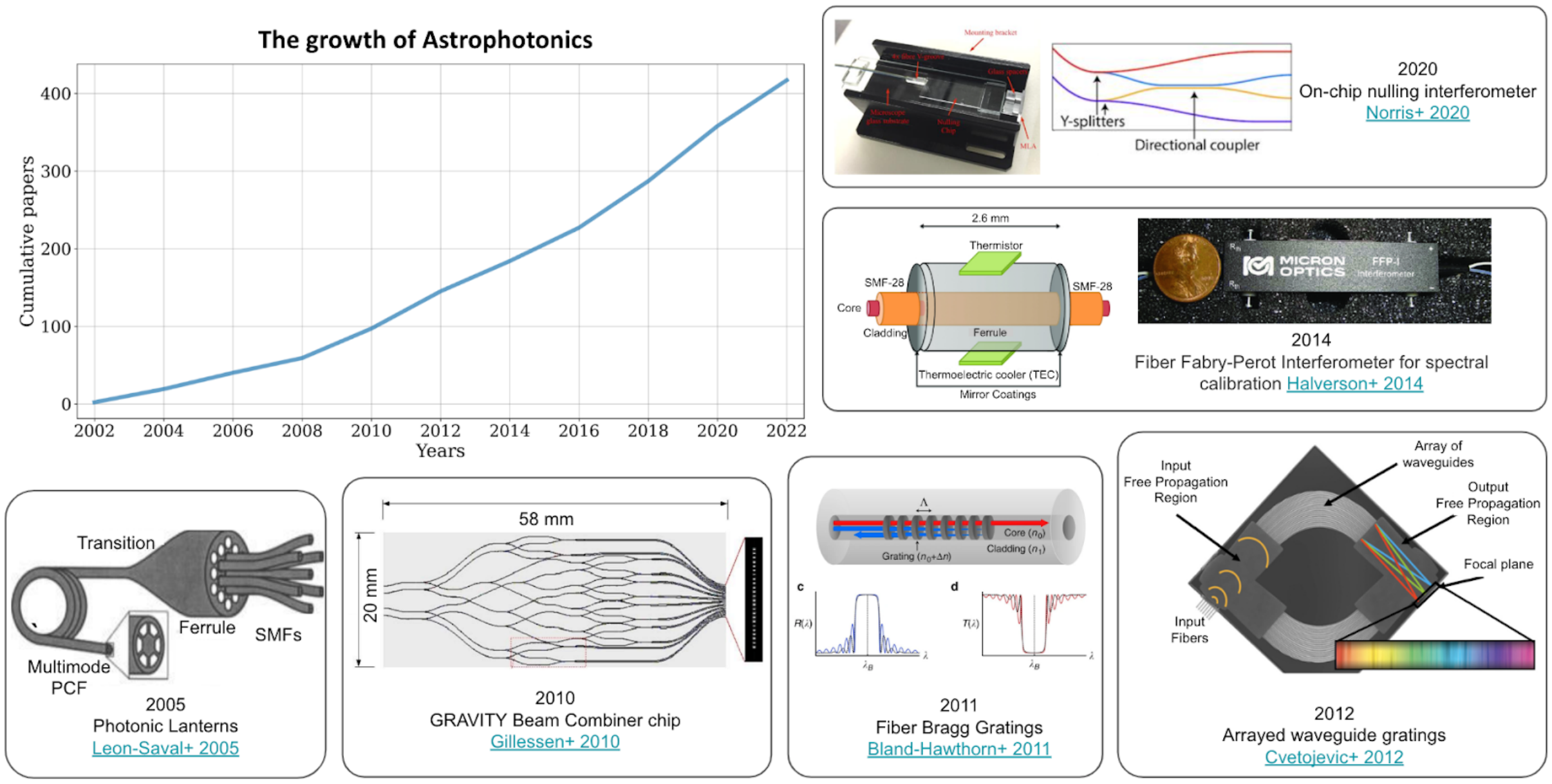

Figure 1: Top left: A plot showing the growing trajectory of research in the field of Astrophotonics, compiled using NASA ADS search engine. The surrounding insets show examples of various technologies representing the sub-fields in astrophotonics, as they matured for on-sky demonstration. These examples include: photonic lanterns [6], photonic beam combiners for GRAVITY interferometry [7], fiber Bragg gratings for atmospheric OH emission suppression [8], arrayed waveguide gratings for spectroscopy [9], fiber-based Fabry-Perot interferometer for spectral calibration [10], and an on-chip nulling interferometer [11]. This is just a small subset of a large family of astrophotonic technologies that have been developed and demonstrated over the last two decades.

Numerous technologies have been advanced over the past decade. Here we briefly highlight a few success stories. The photonic lantern was invented to exploit the coupling advantages of multimode and the stability of single-mode fibers (SMFs) for use in astronomy [6]. This mode converting device has since undergone extensive technical development [12-20] and now a device that converts a multimode input into 19 single-mode outputs can routinely do this with a throughput >95% at 1550 nm. The lantern device was leveraged to realize a complex spectral filter that could suppress the telluric OH emission [8, 21]. This was achieved by inscribing complex fiber Bragg gratings (FBGs) and attaching them to each of the single-mode outputs of a lantern, before recombining the filtered signal into a multimode fiber and feeding a downstream spectrograph [22-24]. This spectral filtering capability enabled accurate sky-subtraction in the NIR, key to advancing faint observations in that region. Regarding calibration, laser frequency combs (LFCs) provide an ultra-stable set of highly dense comb lines across large bandwidths [25,26], making them the perfect tool for calibrating a spectrograph [27,28]. These tools have been optimized for astronomy for the past decade [29,30] and are now becoming commonplace for use with precision radial velocity spectrographs [31]. Finally, in the field of interferometry, the GRAVITY instrument is a long-baseline interferometer that can combine the four 8-m telescopes at the VLT in the near-infrared K-band centered at 2.2 μm. It does this using a photonic beam combiner chip [32] and provides milli arcsec spatial resolution, microarcsecond astrometric precision, and high throughput thanks to the large telescopes and AO systems, which have enabled many transformative discoveries [33-36].

With these successes and a long list of possible technologies and functions to explore, astrophotonics has a huge potential to impact astronomy. This potential was recently outlined in several whitepapers [37, 38] solicited by the National Academy of Sciences for the US Decadal survey and later explicitly called out in the survey itself [39]. Specifically the survey stated "The possibility of obtaining extremely high-precision radial velocities, of the order of a 10 cm/s or better, as well as direct imaging of exoplanets may largely rely on the maturity of single-mode fibers and on-chip nulling interferometers."

Importantly, astronomy driven technology developments have broader applicability which could improve funding for such developments. For example, the all-photonic flattener on a chip [40,41] may find application beyond astronomy for gain flattening [42,43], temporal pulse shaping [44, 45] as well as for targeted excitation of particular molecular species [46]. In addition, photonic lanterns are being considered for spatial division multiplexing in telecommunications [47] as well for high efficiency free-space optical communications [48]: the latter is critical to ensuring high data rates for future astronomy and planetary science missions and may also find application in microscopy. These are just a few examples of the potential of astrophotonics to impact society more generally.

Despite these successes, astrophotonic technologies require further developments in nearly all cases to advance them to facility-class instrument science readiness and/or expand capabilities. With this roadmap, we aim to outline the specific areas development should be focused on to advance the field and allow for truly integrated, multi-functional instruments to be realized and advance scientific investigation.

**Roadmap organization and goals:**

The roadmap explores the status of currently used technologies and outlines other promising choices that should be investigated. The aim is to highlight the potential astronomical applications of a range of technologies and provide guidance on the specific developmental path to realize that promise including, developing more advanced design tools, prototyping, fabricating, characterizing, packaging, integrating, field testing, and science demonstration. The roadmap is broken up into 5 key thematic areas which span all of astrophotonics as follows:

***Light injection, wavefront control, and light transport*** - The efficient coupling of light into photonic technologies is the first step in being able to utilize them for astronomical instruments. Given photonic devices generally operate at or near the diffraction limit, the ability to couple light from a telescope into such a device relies on either an AO system to correct for the wavefront in the optical/infrared [49] (common to most large 5-m+ telescopes) or a mode-matching solution, be it a with a photonic lantern [17, 19], an integral field unit (IFU), or a series of small telescopes more closely matched to the Fried parameter, $r_0$ [50]. In addition, with advances in AO systems that can now generate reasonable correction across large fields (MCAO; ~25% Strehl over ~2 arcminute fields-of-view [51]) and references therein, when combined with low-mode-count photonic lanterns, efficient multi-object photonic instruments can be realized.

Photonic technologies can also be used to produce signals critical to driving the AO system, and offer numerous advantages including eliminating non-common path and chromatic errors. Next-generation instrument architectures will merge photonic wavefront sensing and science instruments and result in integrated instruments with superior performance.

Once light is coupled to a photonic system, light transport is the next critical aspect to ensure the light can be efficiently routed to the science instrument. Optical fibers are ideally suited to this and here we explore a range of new silica fiber architectures that enable photonic applications to currently un-explored wavelength ranges with advantageous properties (dispersion, polarization, bandwidth, few-mode vs single-mode, etc.).

***Spectroscopy and spectral filtering*** - Spectroscopy is a key analytical tool in an astronomers toolbox. It can be used in several ways: to disperse the light for scientific measurement as well as for spectral filtering applications. Photonic dispersing elements in the form of Arrayed Waveguide Gratings (AWGs) have been tested for their suitability for application to astronomy, but need further developments to optimize throughput, bandwidth, polarization response as well as to scale to higher resolution before wide astronomical application. Apart from AWGs, other photonic technologies provide a plethora of novel approaches to the dispersion problem including dispersed Fourier transform interferometers consisting of Mach-Zehnder Interferometers (MZIs) and AWGs, ring-resonator enhanced spectrometers and integrated serpentine grating spectrometers to name a few. These technologies need to be explored to ascertain their scientific potential and demonstrate the technical advantages and to assess challenges to implementation.

Critical to exploiting photonic spectroscopy is being able to realize the devices in wavebands other than the NIR, where telecommunications has already invested heavily. Photonics is evolving in the visible, so we outline photonic efforts to develop platforms, both PICs and fiber based, that are promising at more extreme wavelengths like the UV (<400 nm) and in the MIR (>2.5 µm). AWGs for example have already been built and demonstrated efficiently at sub-mm wavelengths, and in the roadmap we look at further advances needed to optimize these devices, as well as a pathway to large field-of-view instruments.

Spectral filtering is a critical capability. Technologies like FBGs are very mature, but require further development to be applicable to wavebands other than the NIR. Micro-ring resonators, which form tiny resonant cavities and waveguide Bragg gratings, are less mature and need more significant advancements, including utilizing smaller feature sizes, enabling complex filters and reducing cladding mode coupling, but offer the potential to integrate them with other components on PICs.

The applications of filtering devices are growing as well. Optical cross-correlation for example looks at introducing a paradigm shift in how instruments are built, eliminating the need for extremely costly high pixel count arrays, if efficient and flexible filters can be realized to process the light optically. This would be transformative for instruments designed for extremely large telescopes.

***Spectral calibration*** - Key to extracting physical quantities via spectroscopy is the ability to calibrate the spectrograph. Photonics offers the ultimate wavelength calibration tool: the Laser Frequency Comb (LFC). An LFC can be stabilized to extreme levels over decades [25,26], meeting even the most demanding science needs in astronomy (e.g. extreme precision radial velocity). Although LFCs are becoming common at many observatories, they are still complex and costly, and lack reliable spectral coverage in the blue, as needed for the most precise radial velocity measurements of Sun-like stars. In addition, LFCs offer the potential for non-spectroscopic calibration applications, including extreme time keeping for very long baseline interferometers as well as metrology.

To maximize the benefit to spectroscopy, the spectral profile of the comb lines in an LFC spectrum must be flattened, or be made more uniform to fit within the spectrometer's dynamic range. On-chip flattener technology has only recently been explored for astronomy and requires further development. In particular, investigation is needed into the optimal device design given the LFCs optical properties, the number of channels that can be realized on a single chip, how to build devices that span several astronomical bands, and the prospects for devices in the visible region.

An alternative calibration technology that offers a compact, portable, cost effective solution to wavelength calibration is the Fabry-Perot etalons. In recent years, these devices have been commissioned at several observatories with precision radial velocity instruments (e.g. ESPRESSO, HARPS, NEID, HPF, SPIROU, Maroon-X). Their spectrum can be readily engineered to provide very broad wavelength coverage including the UV range, and linewidth and spacing precisely matched to the spectrograph. While etalons have demonstrated very reliable operation, and excellent short term stability, they exhibit long term drifts that are wavelength dependent, likely due to aging effects in the mirror coatings, necessitating the use of other, absolute calibrations sources to periodically recalibrate the etalon spectrum and track the long term drift. The mechanism responsible for the chromatic drifts needs to be identified and the coating effects mitigated before Etalons can become stand-alone calibrators.

***Interferometry -*** Interferometry is a powerful approach to reach angular resolutions beyond that of single telescopes so one can study the environment in the immediate vicinity of the host target. While GRAVITY has led to groundbreaking results in the NIR, several technical advancements can be undertaken to extend the scientific reach including improving the throughput in several spectral ranges, increasing the Fourier coverage by combining more telescopes, and integrating active fringe tracking capabilities onto the beam combining chips. In addition, discrete beam combiner technologies present an alternative approach that may simplify beam combination circuits when >4 telescopes need to be combined due to their implementation of straight single-mode waveguides in a lattice. But, the technology also requires their operational bandwidths to be extended and at the same time, realize a smart algorithm for finding the best input configurations and the geometry of the lattice when the number of input telescopes increases.

In parallel to advancing beam combiner technologies, developing the components to allow kilometric baselines by using optical fibers is also critical. This requires the development of low loss fibers across the NIR and MIR regions (1-17 μm), a careful understanding of the dispersion properties of fibers and how to compensate for delays with fibers as well as FBGs, and to determine if photonic fringe tracking, via piezo stretched fibers or thermal or electro-optic phase shifters on-a-chip will be sufficient to replace bulk optic delay lines. Birefringence and diattenuation will also need to be closely studied.

For high-contrast applications, nulling interferometry is the desired approach and the roadmap outlines the components (tri-couplers, achromatic phase shifters, etc) needed to realize high performance circuits in future at both NIR and MIR wavelengths. To eliminate any stellar leakage the circuit architecture also needs further development, with approaches like kernel nulling showing promise, but careful evaluation is needed. Fiber and photonic lantern nulling rely on a fiber or a mode-selective photonic lantern to be positioned directly in the telescope focal plane. Although they are simple to implement, the contrast will be limited without further investigations into 1) the ability of the cladding to suppress the rejected stellar light and 2) the limitations of cross-coupling in mode-selective photonic lanterns. A PIC-based beam combiner is a promising solution to calibrate the leaked light of a photonic lantern nuller to enhance the null, but needs to be studied in detail to determine the optimum architecture of the circuit.

Finally we can consider other techniques used in radio astronomy, including heterodyning and frequency conversion. The latter allows the thermal IR photons to be converted to the visible or the near infrared where detectors are abundant and thermal background is no longer the limiting factor. Moreover the combination of the heterodyne or frequency conversion and the use of photonic technologies borrowed from the telecommunication world offers the possibility of linking the telescopes of an array to the central interferometric correlator without the need for a costly and hard-to-maintain infrastructure. Interestingly, the combination of a heterodyne approach and single-photon quantum technologies may allow highly efficient interferometric schemes. However, to realize functioning instruments in the MIR one needs broader bandwidths using faster mid-infrared detectors, phase-stable mid-IR laser frequency combs and broadband nonlinear conversion crystals.

***Realizing efficient, multi-functional instruments*** - The ultimate goal is to route and process the light collected from the telescope completely in photonics to the detectors (see Figure 2 for an example of an integrated photonic instrument concept). Photonics offers many useful functions to achieve this, each optimized across a broad range of disparate materials and platforms. To advance towards realizing instruments on-a-chip, efficient hybridization will be necessary. Mastering repeatable high-efficiency packaging to deliver ultra-low loss coupling from fiber arrays to PICs, and between disparate PIC platforms optimized for various functions (Silica-on-silicon, ion-exchange waveguides in glass, Silicon Nitride, Silicon on insulator, etc) will be necessary. In addition, fan-out devices, needed to go from the output of multi-core fibers (MCFs) to PICs will need to be further developed to reduce losses. Active circuits will also require electrical integration and possibly laser sources for calibration and/or metrology.

Repeatable and robust packaging - integration of various technologies to make a single functioning device will be critical. Fabrication and packaging of most photonic devices is still a niche industry or done on individual basis, we need to better utilize the expertise of the broader integrated photonics industry and for their capabilities to evolve and mature (eg: multi-project wafer (MPW) runs, detector integration, fiber to chip to free space/detector packaging). Standardization within the photonic industry will dramatically reduce costs for future devices.

Detector integration with PIC or fiber devices will be key to realizing an integrated instrument. Semiconductor detectors operating at temperatures closer to where photonics are typically operated could be integrated via single pixel photo-diodes on the chip, on an active chip which is flip chipped onto the passive device, via linear arrays which are edge coupled to the PIC or 2D arrays that image the beams ejected out of the top surface of the PICs, via grating or vertical couplers. These options need to be studied in more detail to understand the relative pros and cons and for example at which pixel count to transition from single pixel photodiodes to arrays. Losses of the various outcoupling options (edge, vertical, grating couplers) need to be minimized over broadbands as well. The role of micro-optics needs to be better understood.

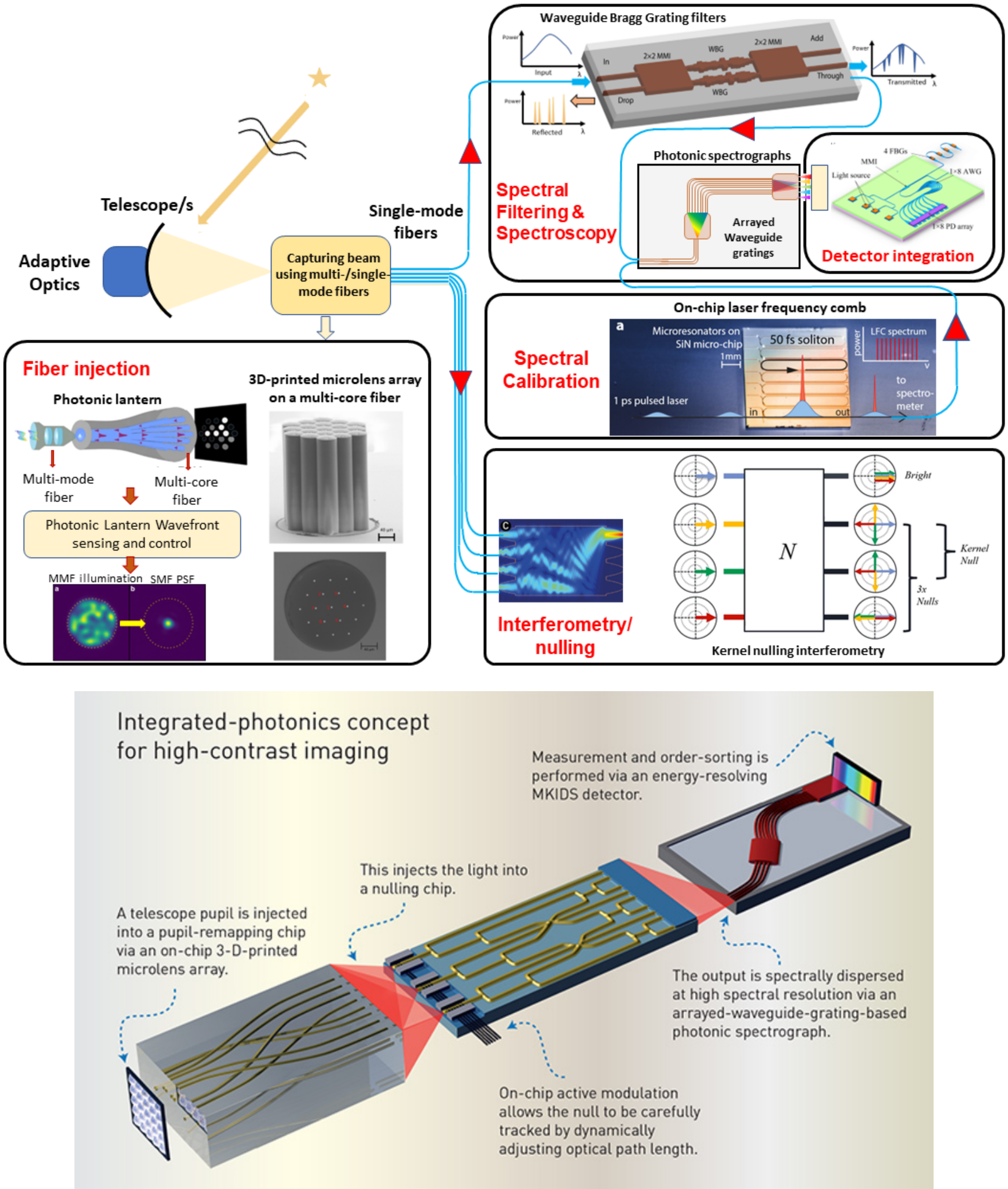

Figure 2: (Top) The five main subcategories of the roadmap are shown with example technologies that can be used in each, as well as an example potential implementation. Some of the technologies shown here include: fiber injection (with photonic lanterns [19], lantern-based wavefront sensing [11], and 3D-printed microlens arrays on multi-core fibers [52]), spectral filtering and spectroscopy (using complex waveguide Bragg gratings [53] and arrayed waveguide gratings [54]), spectral calibration (with on-chip laser frequency combs [55]), interferometry (using Kernel-phase nulling [56]), and detection integration [57]. (Bottom) A future vision for an integrated photonic high-contrast imaging instrument. Micro-lenses segment the pupil plane, a pupil remapper routes the light to a nulling beam combiner chip which has active control for fringe tracking, before light is injected into an arrayed waveguide grating (AWG) to disperse the light and then onto a detector (image taken from [58])

To push the sensitivity of detection, photonics could be integrated with superconducting detectors as well. Both MKIDs and SNSPDs run at much lower temperatures (<4K), but provide extreme sensitivity photon counting capabilities, free from dark current and read noise. The mismatch in materials between those used for PICs and detectors means further studies are needed to determine the optimum route to integrate them. In addition, superconducting detectors and circuit structures should be optimized for IFUs in the THz as well.

***Community development*** - In addition to the technical developments outlined above, the community needs development as well. Firstly, there is a lack of astrophotonics experts primarily because universities nurture astronomers or photonics technologists. Although specialists are needed, the field also requires interdisciplinary experts that can understand the astrophysics science cases, formulate requirements and then design and realize photonic instruments. New bridges between workforce training centers and new pipelines to bring people into the field are needed. In addition, multidisciplinary centers such as innoFSPEC (Germany) fuel innovation in astrophotonics by bringing together expertise from different fields.

Secondly, the community is not representative of society at large as is the case of the parent field - astronomical instrumentation. As a first step towards addressing the latter point, the community should leverage the initiatives developed by professional societies like SPIE, Optica and AAS, who frequently have discussions, training and workshops on various topics about diversity, equity and inclusivity at conferences, to name a few. At the local level, revising admissions and hiring, retention and advancement practices to eliminate biases and increase diversity and equity are critical.

Realizing the potential of astrophotonics is contingent upon developing a talented, robust and diverse workforce. This work force will not be developed without significant and directed attention and effort by the community.

## Concluding Remarks

The series of papers that follow elaborate on these 5 technical themes, each outlining the status, current and future challenges, and advances in the science and technology to meet the challenges for 23 key areas of the astrophotonics field. The technical developments specified in the roadmap will require development of the workforce, adding new interdisciplinary specialists and creating a more diverse and a robust community. Through close collaboration between academia, which is better positioned to advance concepts, designs and qualify devices, and industry partners who are better positioned to provide fabrication and packaging, the technical developments outlined can most efficiently be realized. While astrophotonics has great science potential, currently the field is not funded at a level that can sustain large infrastructure upgrades for PIC, fiber, and packaging advancements. Instead, the field will benefit from continued investments by industries such as telecommunications, AR/VR headsets, photonics for satellites, and LIDAR for autonomous vehicles. Conversely, those fields will also benefit from astrophotonic driven developments. Despite this, the astronomical community must stay engaged with industry partners to provide them with guidance on the needs of the community as well as to be inspired by advances in vendor capability.

**Acknowledgements**

*This work was supported by the National Science Foundation under Grant No. 2109232.*

# 2 | Symbiosis Between Adaptive Optics and Photonic Components: the Path to Fully Integrated Instruments

Philipp Hottinger[1], Olivier Guyon[2,3,4,5] and Rebecca Jenson-Clem[6]
**[1] Landessternwarte, Zentrum für Astronomie der Universität Heidelberg, Heidelberg, Germany**
**[2] Subaru Telescope, National Astronomical Observatory of Japan, National Institute of Natural Sciences, Hilo, HI, USA**
**[3] Astrobiology Center of NINS, Osawa, Mitaka, Tokyo, Japan**
**[4] Steward Observatory, University of Arizona, Tucson, AZ, USA**
**[5] College of Optical Sciences, University of Arizona, Tucson, AZ, USA**
**[6] Department of Astronomy and Astrophysics, University of California, Santa Cruz, CA, USA**

## Status

Most 8-10 meter class telescopes are equipped with adaptive optics (AO) to compensate for atmospheric turbulence. All AO systems rely on wavefront sensor(s) (WFS) to measure optical aberrations, deformable mirror(s) (DM) for optical correction, and computational methods for linking WFS measurements to DM commands. Yet their design details vary considerably depending on the intended science application.

Laser guide stars (LGS) assisted wavefront sensing enables AO-corrected observations of fields with no bright star, such as the galactic center (Ghez et al., 2008; Gillessen et al., 2009). The AO-corrected field of view (FoV) can be extended by using multiple DMs and WFSs (Rigaut & Neichel, 2020) to support, for example, crowded-field astrometry or multi-object spectroscopy (e.g. Gemini's GEMS and GNAO, VLT's MAVIS (Rigaut et al., 2020) and Subaru's ULTIMATE (Minowa et al., 2020)).
Extreme-AO systems (ExAO), on the other hand, are optimized for performance over a small field-of-view around bright natural guide stars (NGS), integrating high actuator count DMs with advanced wavefront control methods. Several such systems have been deployed for imaging exoplanets, including Gemini/GPI (Macintosh et al., 2014), VLT/SPHERE (Beuzit et al., 2019), Subaru/SCExAO (Lozi et al., 2018), and Magellan/MagAO-X (Males et al., 2020). Focal plane WFSs can address remaining wavefront distortions such as non-common path aberrations (NCPA) and thermally-induced phase discontinuities (Milli et al., 2018), using a wide range of algorithms and approaches (Skaf et al., 2022; Vievard et al., 2019).

Thanks to excellent AO correction over a small FoV of ~<50'' on large telescopes, starlight can now efficiently be coupled into single-mode fibers (SMF) for high angular and spectral resolution spectroscopy, with an efficiency closely linked to Strehl ratio (SR). Jovanovic et al. (2017) achieved coupling efficiencies of over 50% with SRs of 60% in H-Band at Subaru/ScExAO, Crass et al. (2017) of more than 35% in Y- and J-Band with LBTI/iLocater, and Delorme et al. (2021) aim to reach coupling efficiencies of 60% in K- and L-Band with Keck2/KPIC. Photonic single-mode components extend SMF use to multiple telescopes (GRAVITY Collaboration et al., 2017), and compact integral-field spectroscopy with approaches including hexabundles (Bland-Hawthorn et al., 2011), SCAR (Por & Haffert, 2018; Haffert et al., 2020), and 3D-M3 (Anagnos et al., 2021).

With the intermediate SRs of 10-20% delivered by wide FoV (~>100'') AO systems, coupling into a single SMF becomes less efficient. Modal conversion with photonic lanterns (PL) (Leon-Saval et al., 2005) can couple light to multiple SMFs to feed downstream SMF-base instrumentation such as a fiber-bragg gratings (FBG) for airglow suppression (Horton et al., 2012). Alternatively, few-mode fibers can make use of this partial AO correction like NIRPS (Cabral et al., 2022) but these are prone to the negative impact of modal noise.

AO subsystems could individually be replaced by maturing photonic technologies, offering identical or enhanced functionalities in a miniaturized and integrated footprint. The manufacturing processes often allow in-situ alignment (Dietrich et al., 2018) that reduces operational complexity and increases optical stability. One of the most promising applications is the use of PL as focal plane WFS with the potential to supplement well-established pupil plane WFSs (see Chapter 3 for more details). Large systems would benefit from the reduced complexity and smaller footprint of these sensors, enabling better scalability for Multi-Object-AO and Multi-Conjugate-AO systems requiring multiple WFSs. Goodwin et al. (2014) introduced the concept of a miniaturized Shack-Hartmann WFS with similar benefits.

To optimally exploit photonic technologies, they should not simply replace individual conventional components but rather aim to be integrated as part of the science instrument and the AO system. This type of hybridization will make telescope optics more resource efficient as it reduces optical and mechanical footprint and complexity. One partially integrated approach has been proposed by Dietrich et al. (2017) for reconstructing tip-tilt with a multi-core SMF equipped with an 3D-printed lenslet array and tested on-sky with a refined design utilizing a micro-lens ring tip-tilt sensor (MLR-TT sensor) by Hottinger et al. (2021). There, wavefront sensing is integrated into a vital part of the science instrument, in this case with simultaneous SMF coupling. While tip-tilt sensing is only a limited functionality, it shows the advantages such an integrated approach can have as it reduces complexity by replacing multiple bulk optic components while almost completely eliminating non-common path aberrations (NCPA).

This interplay between AO and instrument has led to demand and existence of test environments that allow transition of the development from laboratory to on-sky performance in order to mature existing concepts. These are core objectives of the SCExAO testbench at Subaru (Lozi et al., 2018) and Canary at WHT (Gendron et al., 2011).

## Current and Future Challenges

**Efficient injection of astronomical light into photonic devices** requires exquisite wavefront quality and stability, beyond what is achieved with AO on large telescopes. Reliable, efficient and stable coupling will require further improvements in AO performance, with particular attention to vibration control. Proper optical matching of the telescope beam to SMF is also important; Phased induced amplitude apodization (PIAA) has already been shown to have the potential to significantly increase single mode fiber coupling (Jovanovic et al., 2017).

**Modern AO systems' demanding wavefront quality requirements** on large telescopes drive them to employ fast, sensitive and accurate wavefront sensors driving high-count deformable mirrors, with higher optical complexity and use of advanced wavefront control algorithms. Precise calibration of system components (WFS, DM) and control of residual non-common path aberrations is becoming increasingly critical. This is especially essential in HCI applications requiring starlight suppression by nulling or coronagraphy to support the direct imaging and spectroscopic characterization of exoplanets. Photonic technologies can provide the high performance interferometric wavefront sensing solutions to meet this challenge, combining high sensitivity with large dynamical range, and providing wavelength diversity. These benefits have recently been demonstrated on-sky with the PL (Norris et. al 2022) and the GLINT instrument (Guyon et al. 2022, Norris et al., 2022). To fully realize this potential, the interferometric WFS output signals will need to be used for real-time active wavefront control, either by DM actuation or in-chip phase modulation. One challenge will be to accurately interpret the interferometric signals for closed-loop wavefront sensing.

**Manufacturing capabilities** of photonic components are often still insufficient for astronomy applications, with most new developments driven by the much more broadly funded telecommunication industry which usually has less demanding requirements on optical throughput, pathlength control, and broadband performance. The astronomical community has the chance to contribute in these new areas, expanding the range of applications of photonic devices. When astrophotonics aims to integrate or replace full-scale AO systems, scaling of these devices also becomes an issue as the number of coherent elements scales with $\sim(r_0/D)^2$ and so will the required number of individual sensor readout and pathlength control units. In the near future, photonics-based AO systems will be deployed as small units downstream of AO systems using bulk optics, while full-scale photonics AO systems on large telescopes will require more development.

**Practical challenges** of photonic technologies need to be considered as these are often a new class of manufactured devices that need special attention. 3D-printed microlenses using two-photon lithography allow high flexibility in free-form shapes and in-situ printing and thus integrated alignment (Dietrich et al., 2018). But this also means that care needs to be taken for handling these devices during transportation and employment as their structural integrity and the bond joint between components is vital to their performance. Single-mode fibers feature a mode-field diameter (MFD) that varies slightly between manufacturing batches and needs to be matched for best throughput between individual fiber segments. Ultra-fast laser inscription (Thomson, Kar & Allington-Smith, 2009) is continuing to mature but many manufacturing challenges such as reliable waveguide printing, limited refractive index contrast, and high attenuation still remain. In particular, these considerations will also be essential when astrophotonic devices are to be employed at space based observatories as the environmental impact regarding vibrations, temperature, pressures and background radiation pose an additional challenge.

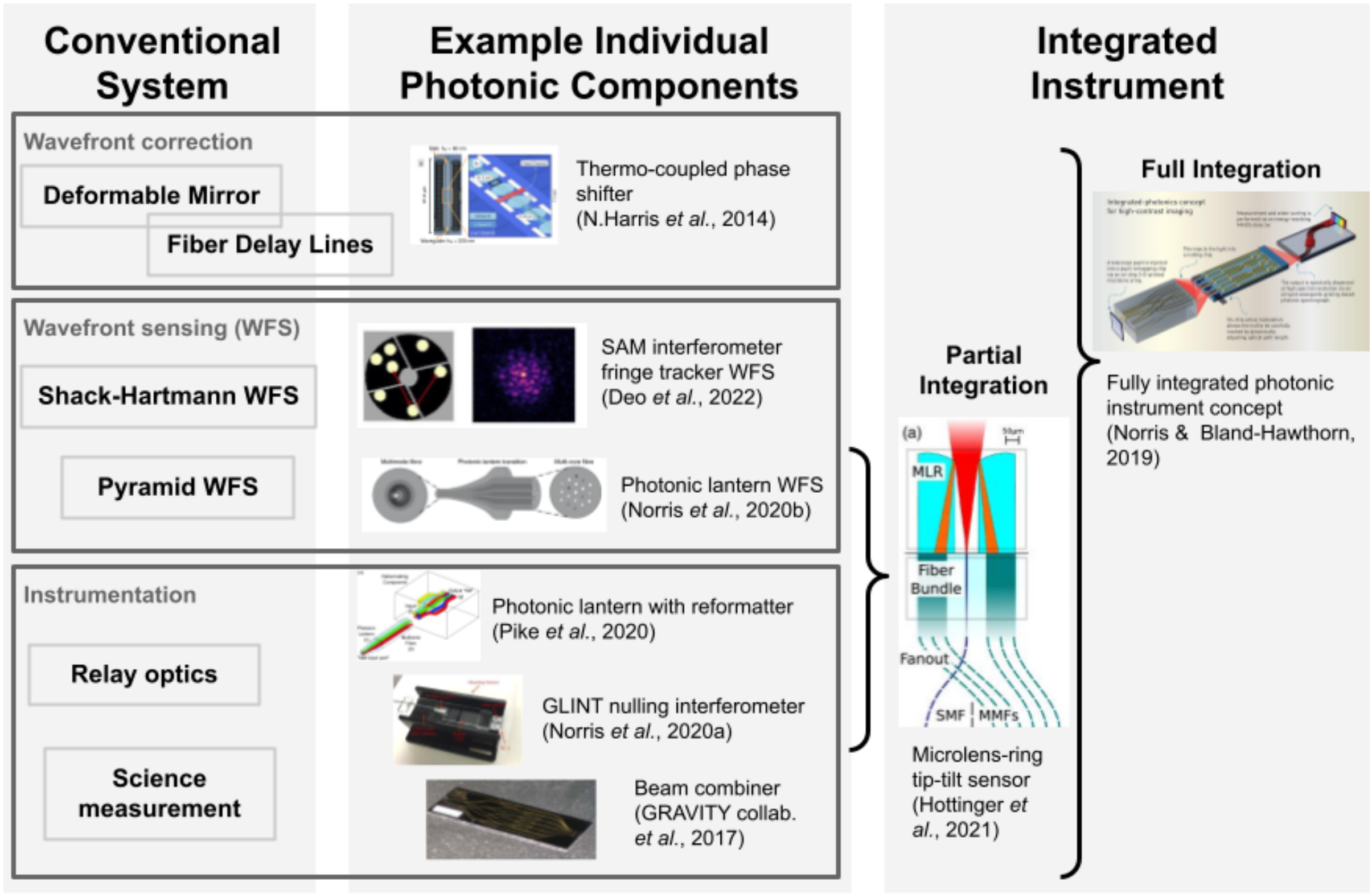

***Figure 1*** *- Overview of conventional adaptive optic (AO) systems, potential photonic counterparts and additions, as well as integrated instrumental concepts.* ***(Permission request pending).*** *Figures reprinted: Thermo-coupled phase shifter (N.Harris et al., 2014, Fig.1), sparse aperture masking (SAM) interferometry fringe tracker WFS (Deo et al., 2022, Fig.3), photonic lantern WFS (Norris et al., 2020b, Fig.1), photonic lantern with reformatter (Pike et al., 2020, Fig.2), GLINT nulling interferometer (Norris et al., 2020a, Fig.4), Gravity beam combiner (GRAVITY Collaboration et al., 2017, Fig.5), microlens-ring tip-tilt (MLR-TT) sensor (Hottinger et al., 2021, Fig.), integrated photonic instrument illustration by Phil Saunders (Norris & Bland-Hawthorn, 2019).*

## Advances in Science and Technology to Meet Challenges

There are two main roads towards an increased symbiosis between astrophotonic instruments and adaptive optics. Firstly, individual photonic components can be employed to augment conventional systems, for example by providing a second stage WFS, or by performing a science measurement well suited for photonic approaches downstream of high performance AO correction. In the long term, they can potentially also replace parts of the conventional systems, for example as the main WFS. Secondly, the photonic components will be tightly integrated together as a system performing a wide range of functions, leading to more compact, less complex instrument systems that enable high automation both in manufacturing as well as in operation. Some concepts enable new optical approaches potentially surpassing the performance and capability of conventional systems. Figure 1 brings together these points, showing the existing individual photonic components that correspond to or can supplement conventional systems (center column) as well as integrated concepts where photonic components are part of a larger instrument (right column).

The deployment of the photonic beam combiner at GRAVITY (GRAVITY Collaboration et al., 2017) has indeed shown that photonics can already individually replace conventional components and uniquely enable high precision astrophysical measurements. In general, evolution for these components is slow as these concepts need to prove their optical performance and optical efficiency before consideration for large scale investments.

As outlined in the previous sections, astrophotonic technologies are already bringing advances in wavefront sensing, although they currently support a moderate number of wavefront modes. Two main approaches are on the horizon. Firstly, interferometric wavefront sensing between multiple telescopes for facilities like VLTI/GRAVITY and CHARA or within the same large telescope aperture as demonstrated in FIRST (Huby et al., 2012) and GLINT (Norris et al., 2020), providing high sensitivity as well as wavelength diversity. Secondly, PLs and multi-core fibers can reduce optical elements and thus provide a simple high-throughput alternative to bulk optics slit formatting and light transport, etc. In these applications, the photonic device only performs low-order wavefront sensing of a few modes. The photonic device then operates as a final-stage wavefront sensor downstream of the high-order adaptive optics system. In the near future, these few-mode devices will be integrated with the kHz-speed adaptive optics correction by feeding back into the upstream AO system.

On the control side, communication between the different subsystems will be established. The use of machine learning (ML) algorithms is explored in a number of projects and shows potential to solve some of the challenges associated with analysis of complex sensor data (eg. Norris et al., 2020). The use of these algorithms is already widespread in many real time applications outside astronomy, and will need to be accommodated in observatories as well. The astronomical context will require operation in the kHz regime with a wide variety of input data from both optical and environmental sensors. The computational infrastructure will be expanded to fulfill this need. This includes the corresponding telemetry data filter and storage which is needed for both analysis and future algorithm training.

An important component in fully integrating adaptive functionalities into a single system is wavefront control. While current system use mechanically driven fiber delay lines, photonics provides several solutions that can potentially offer control authority into the kHz regime, including thermoacoustic devices (Lewoczko-Adamczyk et al., 2009), thermo-coupled (N.Harris et al., 2014), electro-optic phase shifters (Errando-Herranz et al., 2020) and piezo-actuated phase shifters (Dong et al., 2022). Additionally to classical wavefront correction, fast photonic-based pathlength modulation can reveal incoherent exoplanet light within the coherent starlight residual when performing HCI.

Integration of AO and science acquisition in a single compact photonic device will allow for new self-calibrating and self-tuning capabilities to be a core feature of the instrument design. This architecture is particularly attractive for HCI, where a photonic nulling chip could iteratively reconfigure its internal phase delays (Miller, 2013) or drive external deformable mirror(s) to maintain optimal broadband null depths. By integrating science measurement and WFS functions within the same photonic chips, the stable relationship between WFS and science outputs can be measured and leveraged for self-calibration of science measurements down to the photon noise residual (Guyon et al. 2022) .

## Concluding Remarks

Given the rapid pace of development of photonics applications to adaptive optics and astronomical instrumentation over the last decade, and the growing number of successful on-sky demonstrations, photonic technologies and devices are poised to play an increasing role in astronomy. Progress in this area is driven by advances in underlying core technologies, maturation of instrument designs incorporating emerging photonics solutions, and access to increasingly powerful AO systems providing the suitable wavefront quality on large astronomical telescopes.

Integration of adaptive optics and astronomical measurement functions in a single photonics-based system will be particularly enabling, providing in a compact format a rich set of functionalities with high stability and accurate calibration of science measurements. In order to realize this potential, further developments will need to include both lab prototyping activities and on-sky prototyping demonstrations. In this paper we have summarized the vast variety of different components involved, hinting at the challenges that lie ahead when all these systems are aimed to be combined. Ongoing collaborative efforts providing phononics experts with access to on-sky development testbenches such as Subaru/ScExAO and WHT/Canary have been, and will continue to be particularly instrumental in maturing photonics systems for astronomy.

## Acknowledgements

*We thank R.J.Harris for the fruitful discussions on the topic. P.H. is supported by the Deutsche Forschungsgemeinschaft (DFG) through project 326946494, 'Novel Astronomical Instrumentation through Photonic Reformatting'. O.G. acknowledges support from NASA grant 80NSSC19K0336 and JSPS grant 21H04998 supporting development of advanced WFS techniques including photonic devices. O.G. and R.J-C. acknowledge support from the Heising-Simons foundation.*

# 3 | Photonic Lantern Wavefront Sensing and Control

Barnaby Norris[1], Jonathan Lin[2], Robert J. Harris[3,4], Aline N. Dinkelaker[5] and Momen Diab[6]
**[1] Sydney Institute for Astronomy, School of Physics, The University of Sydney, Australia**
**[2] UCLA Physics & Astronomy Department, Los Angeles, CA, USA**
**[3] Max-Planck-Institute for Astronomy, Heidelberg, Germany**
**[4] Department of Physics, Durham University, Durham, UK**
**[5] Leibniz Institute for Astrophysics Potsdam (AIP), Potsdam, Germany**
**[6] Dunlap Institute for Astronomy and Astrophysics, University of Toronto, Canada**

### Status

Astronomy has been revolutionized by the adoption of adaptive optics (AO) [1], where the wavefront of light, corrupted by Earth's turbulent atmosphere, is sensed, analyzed and corrected. This allows diffraction-limited, high-contrast imaging of targets such as exoplanets, and is key to enabling the full potential of the upcoming Extremely Large Telescopes.

Existing AO systems generally measure the wavefront error at the pupil plane, using wavefront sensors (WFSs) such as a Shack-Hartmann or pyramid sensor [2]. These sensors are blind to highly problematic modes such as petaling (phase shear at the pupil plane around the telescope spiders [3]), and suffer from non-common-path aberrations (NCPA) with respect to the science focal plane. In a standard imaging system, the focal plane image alone cannot be used to sense the wavefront since it contains only intensity (not phase) information about the PSF.

Placing a photonic lantern (PL) at the focal plane, allows the *complex* amplitude of the PSF, and hence the wavefront, to be directly measured (see Figure 1 for examples). A PL is a passive device that consists of sets of waveguides with different numbers of modes, between which light is transferred [4, 5], see also Chapter 4. Of particular interest is when the multimode (MM) waveguide at the input tapers into several single-mode (SM) waveguides at the output, the number of which matches or exceeds the number of spatial modes at the input. The excitation of modes in the MM end of the PL is a direct function of the spatially-dependent complex amplitude of the injected PSF. If the transfer function of the PL is known, the wavefront can be reconstructed from the PL's SM outputs. But this transfer function is only known by measurement post-fabrication, since the fabrication process is not deterministic.

The PL-WFS offers several advantages: it is sensitive to any mode which affects the PSF, including modes to which pupil-plane WFSs are blind, it makes optimal use of detector pixels (one pixel per mode) minimizing detector noise, and the entire device fits within a standard fiber connector making it ideal for multi-object (MO) systems. The output can also be spectrally dispersed via a prism or grating (see Chapters 6, 7), to achieve wavelength-resolved wavefront sensing, useful in breaking phase-wrapping degeneracy (such as seen in petaling) and to measure atmospheric scintillation. It can also be used for PSF reconstruction. Perhaps most significantly, a fully photonic device has high stability and is ideal for optimally injecting light into a SM fiber (for subsequent spectroscopy [6], interferometry [7], etc.) wherein it offers a truly zero NCPA WFS (see *hybrid MSPL* below).

The relationship between low-order wavefront modes (such as Zernike aberrations) and PL outputs has been shown [8] as well as laboratory and on-sky demonstrations of wavefront reconstruction. These include a 19-mode PL-WFS and a neural network reconstructor using a multi-core-fiber based PL [9] and a pupil plane low-order WFS [10].

Research is underway to evaluate the efficacy of different PL types and algorithms for WFSing. Both optical-fiber-based [4] and Ultrafast Laser Inscribed (ULI) lanterns [11] are being investigated, as well as mode-selective photonic lanterns (MSPL) [12, 13]. While in a standard PL each SM output is a complex linear combination of the complex input mode amplitudes, in a MSPL a one-to-one mapping between the modes excited at the MM port and the SMFs can be engineered [14]. This can ideally match the modes coupled to those in the PSF [15]. However, the number of modes that MSPLs can multiplex is limited by device length and core diameter, with the length of the required taper proportional to the square of the number of modes [16], and are not expected to be effective in WFSing applications [17].

A so-called *hybrid* MSPL can circumvent this limitation; here, all light propagating in the fundamental ($LP_{01}$) mode is routed to a single output, and higher order modes are used for WFSing. In this case, the higher order outputs drive the AO system to maximize coupling in the $LP_{01}$ mode, which is then routed to the SM science instrument. Alternative modes could be made selective instead if desired. Importantly, this selectivity is maintained over a broad bandwidth [18].

With ongoing developments and diverse configurations, PLs have the potential to provide solutions for a broad range of WFS applications (see Figure 2 for an example of PL WFS integration).

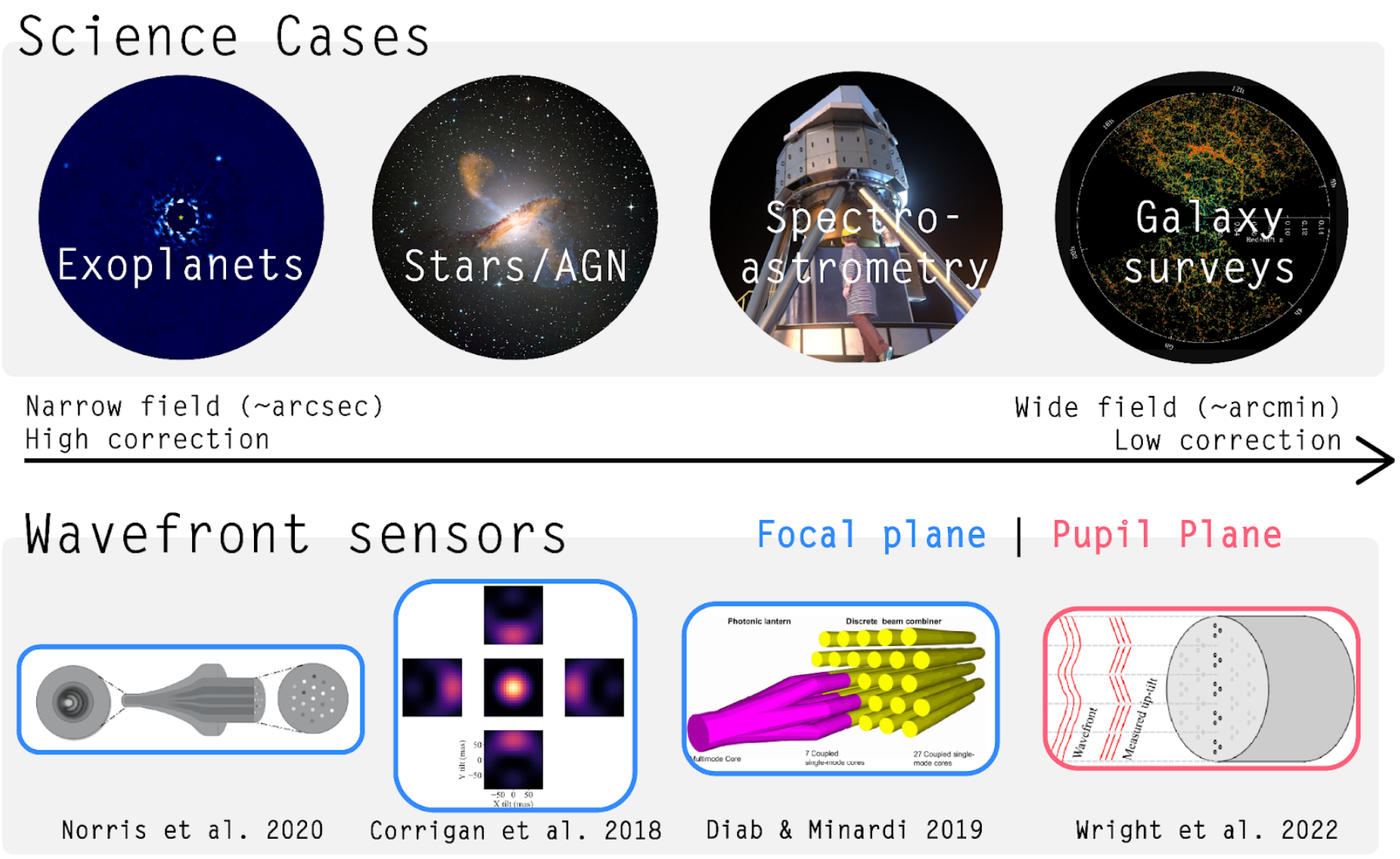

***Figure 1*** *- Different photonic lantern wavefront sensor designs (bottom) and their potential applications (top): the highest level of correction/narrow field-of-view on the left, and lower level of correction/wide field-of-view on the right. Blue indicates focal plane and red indicates pupil plane WFS.*

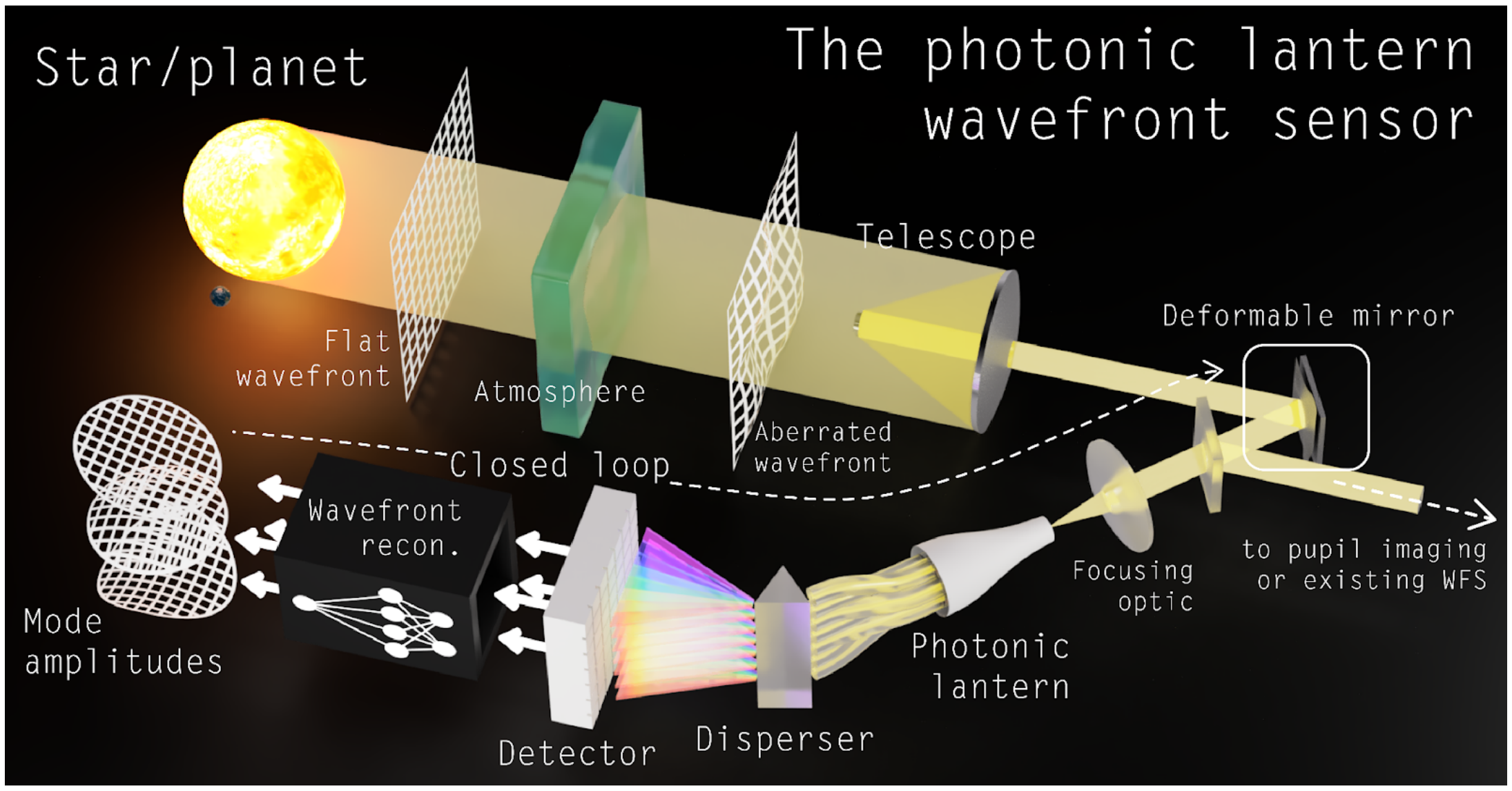

***Figure 2*** *- Overview of a PL-WFS system. Light arrives at the telescope distorted from the atmosphere. After the telescope mirrors, the light is guided to a photonic lantern, where it is distributed between different channels and exits at multiple outputs. Here, it can be spectrally dispersed and is then detected. The data can be used directly for scientific analysis as well as wavefront reconstruction (e.g. using a neural network), with the results feeding back to a deformable mirror earlier in the optics chain for wavefront correction.*

### Current and Future Challenges

One difference to overcome between the PL-WFS and traditional WFSs is that PL-WFSs are non-linear. Since the measured quantity is the SM output intensity (the square of the complex amplitude), the transfer function mapping wavefront phase to output flux is non-linear (and non-monotonic for large phase errors [9]).

For wavefront reconstruction (e.g. for closed-loop AO) the standard linear reconstruction algorithms only work over small wavefront error ranges, where the transfer function is locally linear. One option is to use a higher order (e.g. quadratic) reconstruction algorithm [17] or to use a fully non-linear inference model (e.g. neural network [9]). An alternative approach is to instead optimize for PL linearity, either altering the PL itself or through additional optics such as phase masks and beam recombiners.

While measurements of on-sky wavefronts have been demonstrated [19], the next step is to deploy a PL-WFS in a closed loop configuration on-sky. The current challenge is deploying a suitable (e.g. non-linear) wavefront reconstruction algorithm that can operate with sub-millisecond latencies and integrating it into existing AO software, while running in parallel with existing high-order WFSs. Additionally, while the compact size of a PL-WFS makes it ideal for MO WFSing, algorithms and instruments to exploit this are yet to be developed.

To date, only low-order (<20 modes) PL-WFSs have been demonstrated. For high-order AO applications such as speckle suppression many more modes would be advantageous. Photonic lanterns with many 100s of modes have been produced (e.g. using a cascaded design [20]), but not yet routinely. The difficulty in implementing algorithms for 100 or even 1000 mode PL WFsing is yet to be determined. An alternative could be to use microlens arrays to reduce the number of modes in each lantern, such as the WFS proposed by [10].

A high mode-count PL becomes especially important for imaging, wherein each mode corresponds to one spatial element of the image. This is an important goal, as obtaining a full complex image of the science target would allow atmospheric speckles and the science object to be optimally disambiguated - currently the greatest challenge in high contrast imaging. A suitable architecture (and algorithm) to measure this, along with the coherence properties of the source, is an active area of research. One possibility is to interferometrically recombine the outputs of the PL to measure their complex amplitudes [21, 22]. Alternatively, the interferometry could take place within the PL itself, with the number of PL outputs being made greater than the number of input modes, thus oversampling them.

While the design parameters of fiber-based PLs can be broadly specified, the tapering-based fabrication process means that the exact transfer function cannot be specified, and tapering parameters are set empirically [23]. A precise, deterministic fabrication method would enable more specialized and optimized designs. Deterministic fabrication would address the mismatch between the “ideal” PLs assumed in simulations, and real PLs, which feature slightly mispositioned cores, non-circular claddings, and other defects. The simulation process also faces other challenges, such as the high computational cost and approximations of standard beam propagation algorithms (such as the beam propagation method (BPM)), reducing the accuracy of future high mode-count models.

One promising application is the use of the PL-WFS (possibly with hybrid MSPL fiber injection) on arrays of small telescopes. Their small apertures offer high Strehl ratios with only low-order correction such as that provided by a MSPL. Light from multiple telescopes could then be efficiently injected into single-mode fibers and (in)coherently combined for diffraction-limited spectroscopy, or interferometry. A challenge, however, is the overall lower flux collected by these smaller telescopes, requiring the PL-WFS system to be very sensitive.

**Advances in Science and Technology to Meet Challenges**

More accurate, deterministic control of PL parameters during fabrication would enable devices to be manufactured whose properties match those in the simulated design. Most mature photonic fabrication techniques such as photolithography are limited to two-dimensional structures, making them unsuitable for PLs. Current fiber-based PLs are produced via tapering on fusion splicing glass processing machines, making reliable and repeatable fabrication difficult. Increased demand for PLs in industry (e.g. telecommunications) may lead to PL-oriented glass processing machines. ULI is a promising technology, allowing the use of more arbitrary structures to construct the PL. This technology is still developing, and generally has lower refractive index contrast and worse control than fiber-tapering methods, but the field is evolving rapidly [24, 25].

For any fabrication technique, accurate simulations require a better understanding of how errors in the manufacturing process propagate to actual WFS behavior. For end-to-end simulations, the source of wavefront error represented should be refined to incorporate errors from a wider range of on-sky and instrumental sources. Advances in numerical simulation algorithms are also needed, both for increases in accuracy (e.g. adopting a fully vectorial BPM technique or finite-difference time-domain method) and speed. Fully differential numerical models would also be extremely useful to enhance the optimization processes. Together these will allow production of devices with precisely optimized imaging / wavefront sampling.

Advancements in AO algorithms are required, including optimization of non-linear algorithms (e.g. neural networks) at low-latency kilohertz rates for closed-loop operation. The performance of these algorithms under a range of conditions must be carefully characterized to achieve the reliable, routine operation demonstrated by standard AO algorithms. Beyond WFSing, performing full coherent imaging – simultaneously measuring the image and wavefront, and its spatial coherence properties – will require innovation in analysis and image reconstruction.

Due to their stability, compact form factor and high sensitivity due to optimal detector usage, PL WFSing may become invaluable for space-based telescopes. Since they operate in the low wavefront error regime, a high mode-count PL could be the only WFS (both for speckle control and mirror phasing), relaxing the need for separate pupil-plane WFS and optics. As highlighted by the Roman space telescope coronagraph technology demonstration [26] the success of future high resolution space telescopes will be contingent on WFSing and control. WFSing and control is also critical for optical communications between space and ground. Before PLs can be deployed in these contexts, these components will have to be space qualified by means of environmental testing and technology demonstrator missions using small satellites.

**Concluding Remarks**

The use of photonic lanterns for wavefront sensing is a novel application and has already seen a surge of interest. Advantages include the ability to perform direct measurements of both phase and amplitude at the focal or pupil plane, inherent compactness and stability, optimal use of detector pixels, and the ability to perform optimal injection of starlight into one single-mode fiber (acting as a zero non-common-path wavefront sensor). While the concept has been demonstrated in laboratory and in open loop on-sky experiments, further development is required to enable widespread adoption. This includes implementing low-latency non-linear control algorithms and developing deterministic, accurate fabrication methods coupled with precise simulations. Full coherent imaging (wavefront and image measurement) is also on the horizon. There is a strong research effort by multiple groups worldwide to investigate these aspects, and programs to test photonic lanterns on major telescopes are already underway.

**Acknowledgements**

*B.R.M.N. acknowledges support from the Australian Research Council, Discovery Early Career Researcher Award (DE210100953).*

*R.J.H. would like to thank the Deutsche Forschungsgemeinschaft (DFG) for their support through project 326946494, `Novel Astronomical Instrumentation through photonic Reformatting'*

*A.N.D. acknowledges support by the Bundesministerium für Bildung und Forschung (BMBF) through the project 03Z22AN11.*

*This material is based upon work supported by the National Science Foundation Graduate Research Fellowship Program under Grant No. DGE-2034835. Any opinions, findings, and conclusions or recommendations expressed in this material are those of the author(s) and do not necessarily reflect the views of the National Science Foundation. This work was also supported by the National Science Foundation under Grant No. 2109232.*

# 4 | Spectroscopic Applications Enabled By Photonic Lanterns

Sergio-Leon Saval[1], Steph Sallum[2], Stephen Eikenberry[3] and Kevin Bundy[4]
**[1] Sydney Institute for Astronomy, School of Physics, The University of Sydney, Australia**
**[2] Department of Physics and Astronomy, University of California, Irvine, CA, USA**
**[3] CREOL, The College of Optics and Photonics, University of Central Florida, USA**
**[4] Department of Astronomy and Astrophysics, University of California, Santa Cruz, USA**

## Status

Conventional astronomical fiber-fed spectrographs use multimode (MM) optical fibers to feed telescope light to a disperser. These fibers can be sized to match the width of the point-spread function (PSF), providing more efficient light capture compared to their telecom counterpart, single-mode fibers (Ellis 2021). However, these spectrographs are larger and limited in spectral resolution compared to diffraction-limited platforms fed with single-mode (SM) fibers. These have recently been demonstrated for precision radial velocity measurements (Crepp 2016, Gibson 2019, Mawet 2022) for example, but require high performance adaptive optics (AO) systems to prevent significant fiber coupling losses (Bechter 2020, Jovanovic 2017). Photonic lanterns (PLs) offer a powerful solution for efficient diffraction-limited spectroscopy by converting MM inputs to SM outputs (Figure 1). This reduces AO performance requirements, enabling near-diffraction-limited spectroscopy at the outputs (Leon-Saval 2012, Betters 2016, Schwab 2012). As a result, PLs have recently emerged as a viable technology for the following applications: spectro-astrometric imaging; efficient combination of light from small telescope arrays; and massively-multiplexed spectroscopy for wide fields of view.

Spectroastrometry (SA; Bailey 1998) involves using spectral centroid shifts to infer the presence of wavelength-dependent morphology within the diffraction limit. Traditionally carried out with slit spectrographs both with and without AO (e.g. Whelan et al. 2008), conventional SA's astrometric precision is set by the PSF width and signal-to-noise ratio, with precision < 1 mas for the best AO experiments (Pontoppidan 2011). Even AO-corrected SA can suffer from PSF variability (e.g. Brannigan et al. 2006), and photonic lanterns have recently been demonstrated in simulation as an avenue for recovering SA signals in the presence of variable correction (Kim et al 2022). For example, hydrogen line signals from accreting planets can be recovered for achievable tip-tilt jitters of 0.1 $\lambda$/D. Furthermore, since SM output intensities encode pupil plane phase variations, PL spectro-astrometry does not suffer from position angle degeneracies present in conventional approaches, making it a promising technique for complex scenes such as resolved circumstellar disks and outflows.

A second PL application involves combining light from arrays of incoherently-linked smaller telescopes. Such arrays have recently emerged as a reduced-cost (by ~10x per unit area) method of increasing telescope collecting area for spectroscopy of faint sources (Swift et al., 2015; Eikenberry et al., 2019; Roth et al., 2022; Bender et al., 2022). In the PolyOculus implementation (Figure 2), packs of semi-autonomous, small (25-40 cm), off-the-shelf telescopes comprise the array (Eikenberry et al., 2019). Photonic lanterns that incoherently combine several multi-/few-mode fibers into a single MM fiber offer particular advantages for combining the light, because they preserve etendue without oversized downstream pupils (Eikenberry et al., 2019). PolyOculus has demonstrated “7-pack” PL light combiners with >90% efficiency across the visible (Moraitis et al., 2021), producing a collecting area equivalent to a ~0.9 m telescope. Furthermore, using multiple 7-pack arrays, the PL outputs can be spliced to optical fibers which then feed into another PL combiner. This approach can be extended, providing a hierarchical scaling to build collecting areas equivalent to the 30-40 m diameter mirrors of upcoming Extremely Large Telescopes.

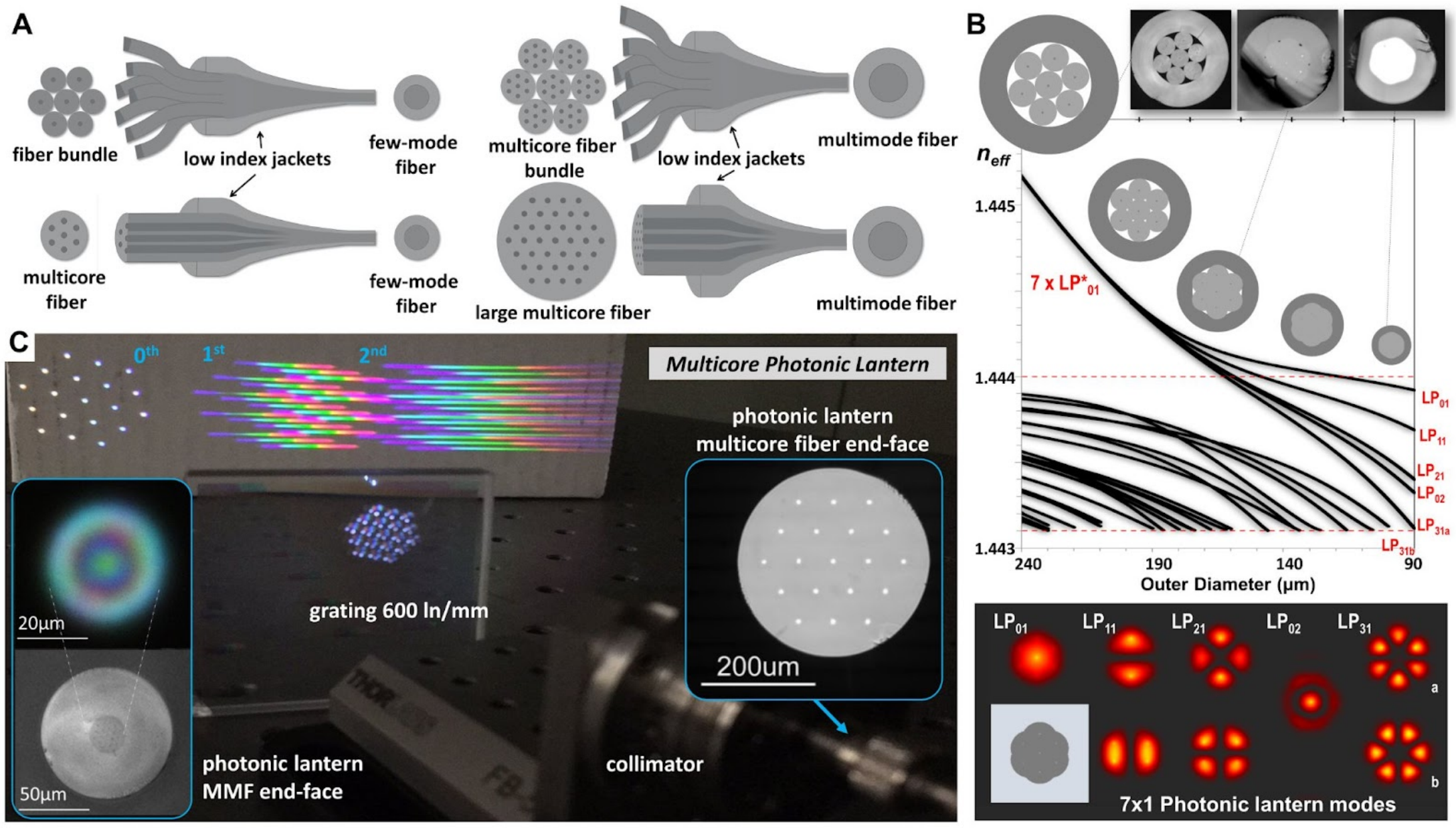

***Figure 1*** *- (A) Schematics of all-fiber photonic lanterns, including: few-moded photonic lanterns conventionally fabricated using single-mode fibers or a single multicore (left), and highly-multimode photonic lanterns fabricated from one or several multicore fibers (right). (B) Simulation of the modal evolution as a function of the waveguide tapering factor, including cross-sectional images of a 7x1 photonic lantern along the transition and the supported simulated modes. (C) Actual images of a multicore photonic lantern device “converting” multimode light in the visible (450-750 nm) to an array of single-mode outputs wavelength-dispersed by a volume holographic grating in the lab.*

Finally the ability of astronomical spectrometers to measure multiple target spectra simultaneously (i.e., the multiplex; e.g. SDSS, DESI, SAMI; Croom et al. 2012, Bundy et al. 2015, DESI Collaboration 2016), has enabled major progress in cosmology, galaxy formation, and stellar astrophysics. The state-of-the-art is the DESI facility, which collects 5000 simultaneous fiber spectra across an 8 square degree field on a 4-m telescope. Future instruments with 4-10 times greater multiplex on larger telescopes are already under consideration. With a conventional design, such an instrument would be impractical, exceeding the cost of the telescope itself. Mass-produced photonic spectrometers, however, offer a path toward greatly reducing the “cost per spectrum” (e.g., a “spectrometer-on-a-chip”; Chapters 6 and 7). Because these devices typically demand single- or few-mode input, photonic lanterns will play an important role by extracting the modal components of potentially thousands of on-sky sources.

## Current and Future Challenges

Currently, photonic lanterns have achieved high performance with transmission efficiency of >90%. However, lanterns have been constrained in terms of the number of inputs and outputs. PLs with up to 19 output single-modes are fairly standard, but larger port number devices would be desirable. They would enable more efficient combiners for telescope arrays, more versatile reformatting for integral field and multiplexed spectroscopy, and more detailed recovery of asymmetries via spectro-astrometry.

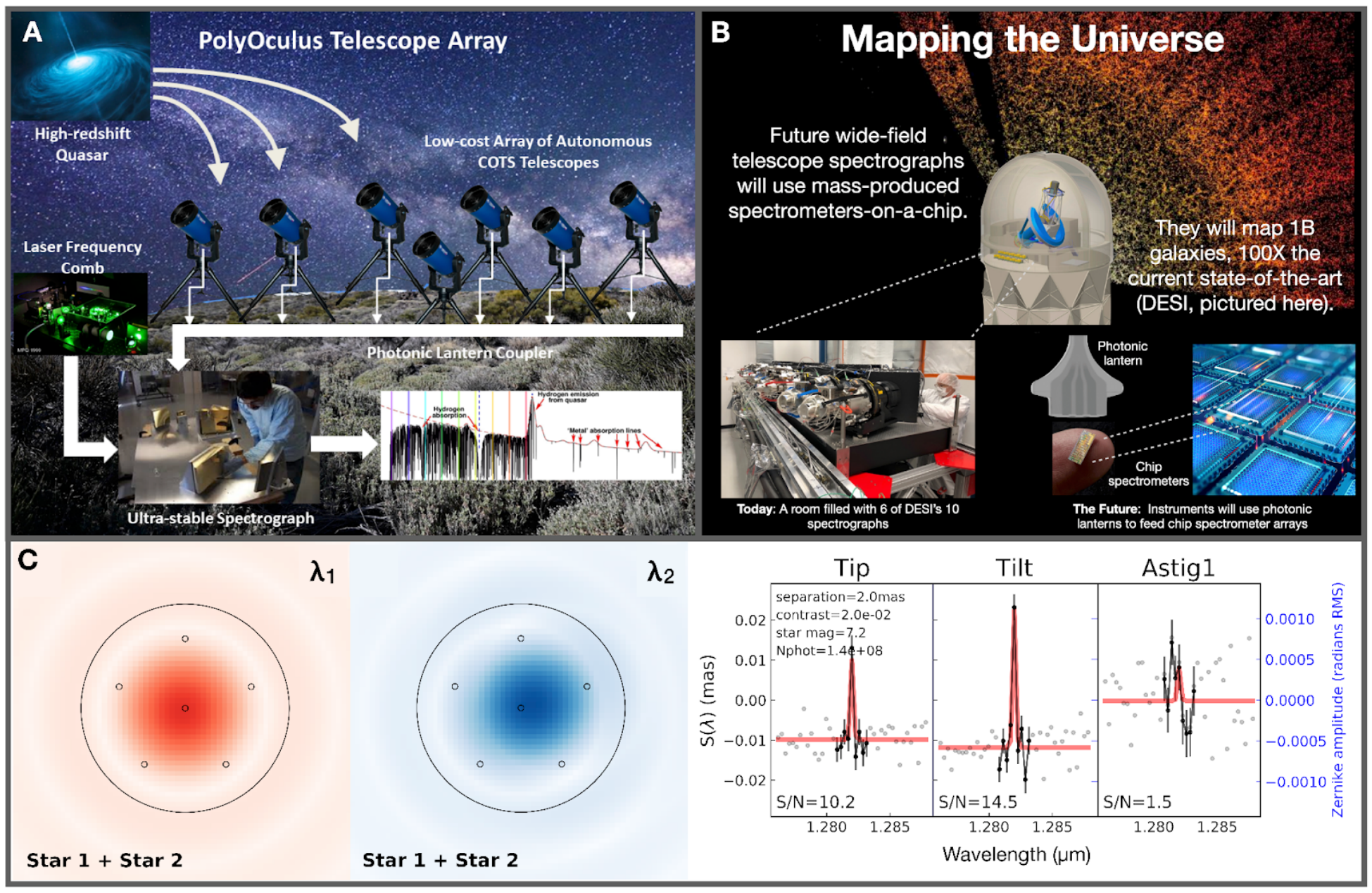

***Figure 2*** *- Spectroscopic photonic lantern applications. (A) Schematic layout of the PolyOculus telescope array concept. An array of small autonomous telescopes point to the same target on the sky, with fiber optic feeds coupled via high-efficiency photonic lanterns into a single fiber, which feeds into a scientific detector. (B) Illustration of the state of the art in massively-multiplexed spectrographs, future instruments enabled by photonic lanterns and chip spectrometer arrays, and their anticipated science outcomes. (C) Simulation of photonic lantern spectroastrometry. Companions with excess flux relative to the primary star at specific wavelengths (left) can be detected within the classical diffraction limit by measuring spectral centroid shifts. These signals can be modeled as Zernike modes, illustrated in the right panels showing an injected and recovered protoplanet with Pa-β excess emission (adapted from Kim et al. 2022).*

Increasing the number of PL modes relies on increasing the number of fibers or cores in a multicore (Figure 1A). Scaling to a large number of fibers is challenging due to manufacturing limitations related to handling hundreds of individual fibers and limitations on the maximum diameter handled by glass processing equipment (e.g. splicing, tapering). However, multi-core fibers reduce some manufacturing limitations and could offer hundreds to thousands of cores. To date, the devices with the highest port counts constructed from individual single-mode fibers are 1×61 (Noordegraaf 2010), and 1×88 with reduced single-mode fiber cladding diameters (Birks 2015). The largest single multi-core photonic lantern with single-mode cores have 121 cores (Chandrasekharan 2017). A 600-μm-diameter, 511-core multi-core lantern was reported in Birks 2015. However, this photonic lantern, intended for fiber scrambling, contained a mix of single-mode and few-mode cores.

The need for high-performance, wide-field adaptive optics presents an additional challenge for high-multiplex instruments with large fields-of-view [e.g. > 1° diameter; Croom et al. 2012, Bundy et al. 2015, DESI Collaboration 2016). Diffraction-limited performance on large ground-based telescopes across several arcminutes is currently infeasible, although modest corrections are possible with ground-layer adaptive optics (GLAO, 0.3'' FWHM over 8 arcminute fields; Chun et al. 2018) and multi-conjugate adaptive optics (MCAO; ~25% Strehl over ~2 arcminute fields-of-view; Rigaut & Neichel 2018 and references therein). For instruments with fields-of-view comparable to MCAO correction, the architectural goal would be to optimize coupling efficiency, number of modes, and field-of-view, to achieve acceptable coupling efficiency without prohibitively high requirements on the number of lanterns, outputs, and instruments. While the AO system can reduce the number of modes, high-performance low-mode-count (<19) lanterns will be needed to take full advantage. Although such devices have been demonstrated, reliable reproduction and connectorization of such lanterns remains a challenge. For wider fields - i.e. where Strehl remains low - the challenge lies in identifying and filtering the most valuable single modes from the low-Strehl, multi-mode light. For example, a PL "speckle spectrograph" (where individual speckles are spatially directed to a spectrograph; Sheinis 2006) may reduce sky noise and boost signal-to-noise, compensating for ignoring many fainter, more spatially-distributed modes.

**Advances in Science and Technology to Meet Challenges**

One solution to increase the desired number of modes (i.e. inputs/outputs) in photonic lanterns systems may be to implement "hierarchical" PLs (Leon-Saval 2017). In this concept, a large field-of-view photonic lantern efficiently separates the light into a large (but manufacturable and manageable) number of multi-mode output fibers. Each multi-mode output feeds into another photonic lantern that further subdivides the light. This cascade can be repeated until each output reaches the desired modal content - be it single-, few-, or (smaller) multi-mode. With this approach, lanterns with relatively modest output numbers (~19) could approach ten-thousand-mode outputs with only three hierarchical levels.

Generally, fabricating complex lanterns as described above is possible, with limitations imposed by the system's scientific and/or technical applications. For example, limitations to realizing practical, high-order lanterns are imposed by the need to potentially share light between a science instrument and a wavefront sensor, in which case some cores would physically be split off from others. These requirements will impact both the manufacturability and functionality of the lantern. Additionally, manufacturing and engineering challenges will need to be overcome to achieve integration and connectorization of these systems for rugged use in observatory environments.

Hybrid photonic lantern devices may soon allow for increasing the number of modes while improving adaptive optics performance (Norris 2022). Hybrid PLs are made with dissimilar fibers, and can divide light from a multimode fiber into separate science and wavefront sensing fiber arrays. These could potentially bring the best of both worlds: single- or few-mode fiber coupling to a science instrument, and simultaneous wavefront sensing without non-common path errors (as discussed in Chapter 3). This would, for example, improve spectroastrometric precision by providing a better, more stable point spread function, and by allowing for selection of science fibers that optimally recover asymmetries in the target. Furthermore, deploying many of these devices could improve adaptive optics correction for large fields-of-view, highly-multiplexed instruments, enabling more efficient spectroscopy of faint, distant objects.

**Concluding Remarks**

Photonic lanterns present an opportunity for efficient, diffraction-limited spectroscopy using both single-object and wide-field, highly-multiplexed instruments. While these devices have been demonstrated to provide high coupling efficiency for a small number of inputs/outputs, their versatility for astronomical observations would be improved by: (1) manufacturing devices with larger numbers of inputs/outputs, increasing the number of PL modes, (2) developing techniques to extract the most valuable modes from low-Strehl light, and (3) manufacturing devices where outputs can be physically separated for simultaneous applications (i.e. science and wavefront sensing, which would improve AO performance for both narrow and wide fields-of-view). Future applications of these enhanced photonic lanterns will lead to new observational constraints for a wide range of sub-fields, from cosmology, to galactic astrophysics, to stellar and planetary formation and evolution.

**Acknowledgements**

*This work was also supported by the National Science Foundation under Grant No. 2109232.*

# 5 | New Optical Fibers for Astrophotonics

Stephanos Yerolatsitis[1], Rodrigo Amezcua-Correa[1], Itan Gris-Sanchez[2] and Julia Bryant[3]
**[1] CREOL, The College of Optics and Photonics, University of Central Florida, USA**
**[2] Institute of Telecommunications and Multimedia, Universitat Politècnica de València, Spain**
**[3] Sydney Institute for Astronomy, School of Physics, The University of Sydney, Australia**

## Status

Fiber optics have been used in several diverse areas of astronomy since the early 1980s [Angel et al, 1977, Hill et al, 1980, Gray et al, 1984, Vanderriest et al, 1984] and are nowadays commonly deployed at a large number of telescopes. They have been routinely used to transport light from the telescope's focal plane to a spectrograph and to optically multiplex telescopes, linking remote instruments to enable interferometric observations [Hill et al, 1980, Hubbard  et al, 1979, Gies et al 2019]. In this regard, fiber optics has become a powerful tool used to simultaneously observe thousands of astronomical objects, allowing the realization of massive spectroscopic surveys for the first time [2dFGRS: Colless et al, 2001, SDSS: York et al 2000] and integral field spectroscopic (IFS) surveys [SAMI: Croom et al, 2012, MANGA: Bundy et al, 2015] Most of these developments in astronomical instrumentation have so far relied almost exclusively on using optical fibers as efficient light pipes. However, processing light in an optical fiber system can provide unprecedented opportunities in many areas of astronomical instrumentation where current approaches are still in need of a solution. In practice, fibers are already providing some degree of signal processing in telescope settings, for example, they are used as spatial [Coude du Foresto et al, 1992, Mawet et al, 2017]  and spectral [Trinh et al, 2013] filters, efficient mode scramblers [Birks et al, 2012], to feed light into photonic integrated circuits (PICs) [Cvetojevic et al, 2012, Pfuhl et al, 2014] and in the development of photonic lanterns (PL) to mention a few [Leon-Saval et al, 2010, Birks et al, 2015]. The difficulty of implementing new fiber technologies and PICs in general for astronomical instruments arises from the multimode nature of the light being collected at the telescope's focal plane. This demands fibers and fiber devices supporting 1000s of spatial modes with low loss, broadband operation, and for some applications without significant mode mixing. High performance adaptive optic systems can help reduce the number of modes that need to be supported, but are limited to narrow fields of view (1-2 arcminutes) [Rigaut & Neichel 2018].

In the last decade, the demand to increase capacity in optical communication systems ensued a flurry of activities on space division multiplexing that utilizes spatial modes in multimode and multicore fibers in addition to conventional multiplexing techniques. These research activities not only led to new classes of multimode and multicore optical fibers but also new components to tailor and control multimode light {Leon-Saval et al, 2017, Richardson et al 2013, Velazquez-Benitez et al. 2018, Alvarado-Zacarias et al. 2019}. These optical fiber-technology advances greatly overlap with and can drive the development of specialty tools for astronomical instrumentation.

This paper focuses on silica fibers, but we note that fibers for mid-IR wavelengths based on other glass materials have evolved tremendously and are also important for astrophotonics instrumentation. Here, we outline opportunities with new fiber designs and describe their challenges and necessary advances for their successful employment in current and future astronomical instruments.

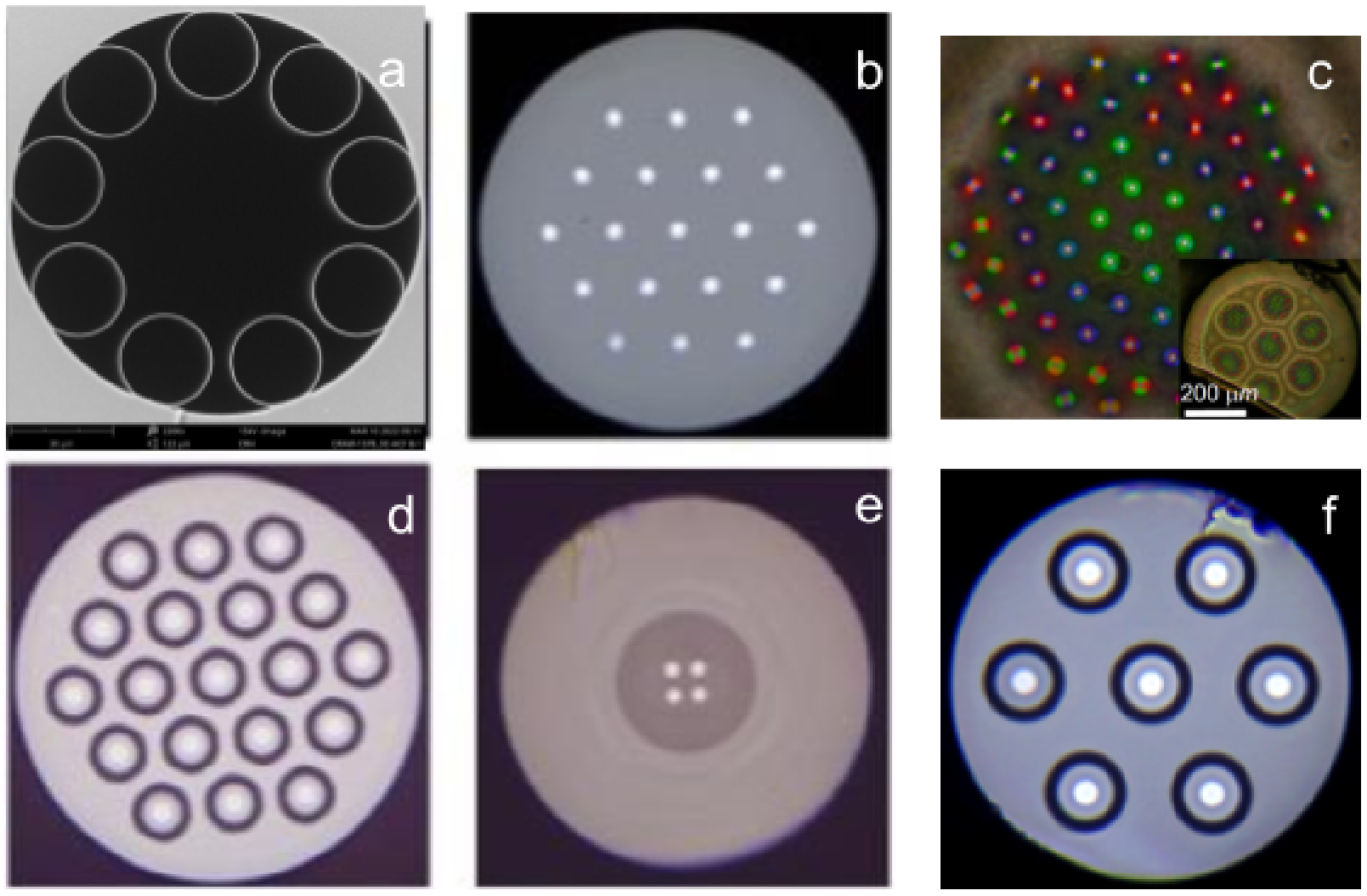

Figure 1: a) Scanning electron image of an AR-HCF. The hollow-core is surrounded by an antiresonant cladding structure which consists of nine thin silica tubes. b-f) Micrographs of multicore fibers featuring different index profiles and wavelength operation range.

**Current and Future Challenges**

**Few Mode, Multimode, and Multicore Fibers** are any fiber supporting more than 1 spatial mode, including graded-index, flat-top, and fibers with unique refractive index profiles.

Graded index fibers exhibit less mode coupling compared to step-index fibers, due to their refractive index profiles. Therefore, graded-index fibers or fibers with specially tailored refractive index profiles can be used to reduce focal ratio degradation (FRD). On the other hand, flat-top fibers are specially designed to scramble the mode content propagating in their core, by tailoring the mode power distribution. They can thus achieve a complete mixing between the modes of the fiber, enabling strong mode-scrambling of the output pattern of the fiber. These fibers can be used to mitigate modal noise which can cause inaccuracies in the measured spectra for radial velocity observations Gris-Sánchez et al, 2018. Multicore fibers with graded index cores or flat-top cores can also be realized. Different types of multicore fibers have been extensively investigated in telecommunication applications as a way to increase transmission capacity. We show some examples in Figure 1. It is important to highlight complementary technologies such as fan-ins [Alvarado-Zacarias et al, 2019] and 3D-printed microlenses (Dietrich et al, 2017, Haffert et al, 2020, Jovanovic et al, 2020) to couple light to each core/channel and to increase coupling efficiency and decrease the gaps in the IFU have seen tremendous advances. Nevertheless, developing broadband low-loss complementary technologies with a large-number of spatial channels remains a challenge and an active area of research.

**Fibers with unique index profile:** By tailoring the refractive index profile of solid core fibers, fibers with unique performance capabilities can be realized. Here, we discuss two such examples. The first is an Airy fiber: a fiber featuring a complex refractive index profile (shown in Figure 2) that allows the efficient coupling of an Airy pattern mode, increasing the coupling efficiency from the telescope to the fiber up to 93.7%. A significant improvement over the theoretical limit (80%) of coupling a beam from a circular aperture to the fundamental mode (gaussian-like) of an ordinary step-index single-mode fiber. This was achieved by inversely designing the fiber to support an approximate Airy pattern [Gris-Sánchez et al. 2016.]. Nevertheless, due to the complex refractive index profile, the fiber guides more than one mode, while the airy mode is not the lowest order mode of the structure.

The second example is a fiber featuring a logarithmic index profile [Harrington et al, 2017]. In many applications including astronomical instrumentation, it is desirable to effectively combine light from different sources. This is usually done by fusing and tapering gradually many fibers together. In this process, the main limiting factor on the number of fibers that can be combined is the fulfillment of the adiabatic criterion which in turn dictates the requirement of a minimum transition length. In contrast, an optical fiber with a logarithmic refractive index profile can be adiabatically tapered over any length, however short, enabling the development of a new class of fiber components, including couplers and photonic lanterns. The fiber's mode field distribution is independent of the fiber's size and therefore does not change as the fiber is tapered down (i.e. it remains the same along the transition). Nevertheless, developing low-loss photonic lanterns using such fibers that support a large number of modes (>1000) still remains a significant challenge that needs to be addressed. The required transition length for adiabatic propagation increases with increasing number of supported modes and can become the limiting factor for realistic implementations.

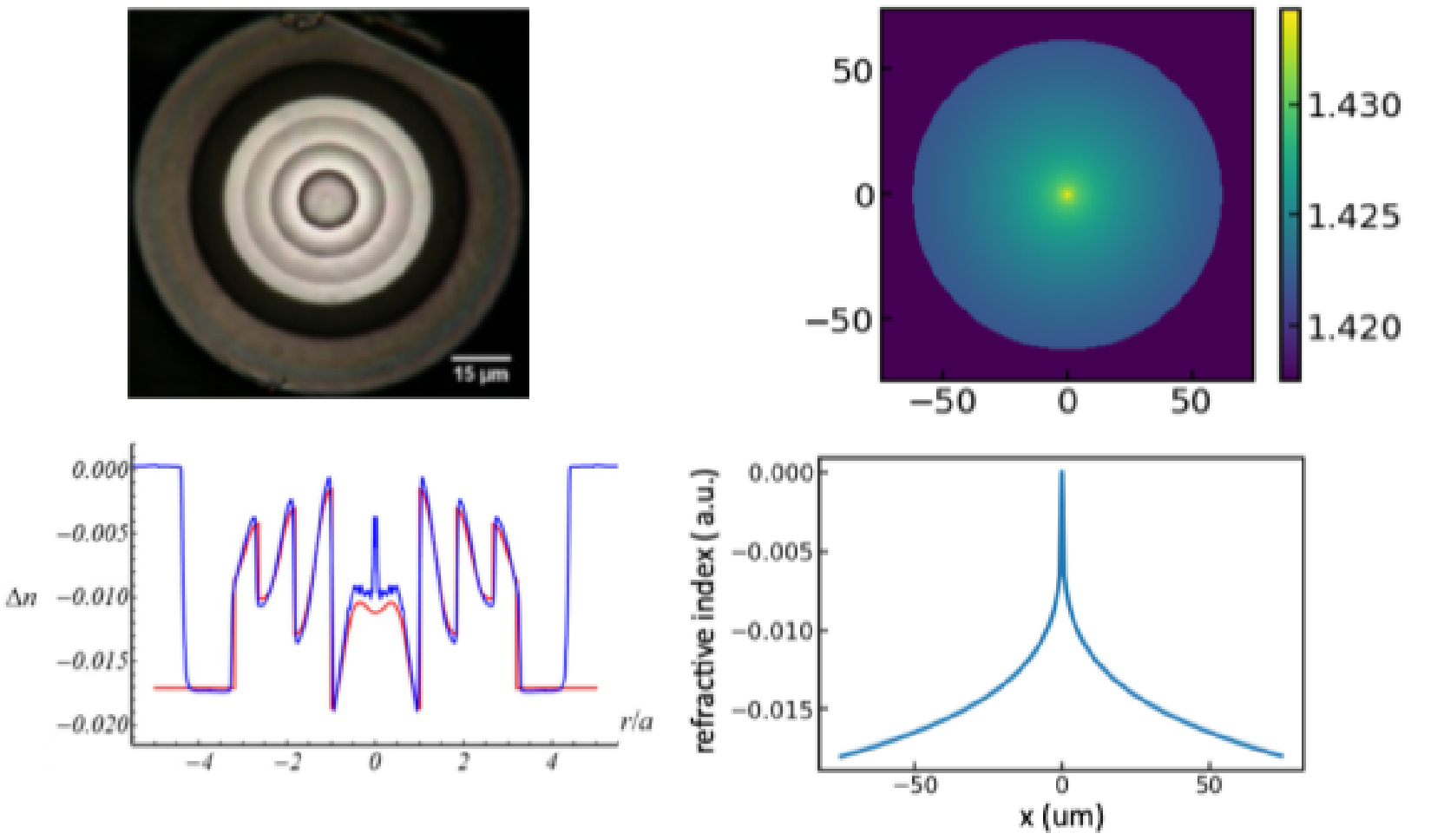

Figure 2: (top left) Optical micrograph of the Airy fiber (bottom left) Physical refractive index profiles of the designed Airy (red), and of that of the PCVD preform (blue). (top right) 2D theoretical refractive index profile of a fiber with logarithmic profile.(bottom right) 1D refractive index profile of the same fiber (relative to undoped silica).

**Hollow core fibers (HCFs):** Recent advances in the design of HCFs led to the development of a new class of antiresonant HCFs with remarkably low losses [Jasion et al, 2022, W. Ding 2020]. In their simplest form, these fibers consist of a hollow core with a ring of anti-resonant cladding tubes surrounding the core as shown in Figure 1. A unique feature of HCFs is that most of the light (>99.99%) can be guided in the central air-core with only a tiny fraction of light overlapping with the surrounding glass structure, hence increasing the optical damage threshold, and reducing material absorption significantly. In addition, these fibers are simpler to manufacture, omitting the complicated structure of previous HCFs. By precisely designing the structure of this fiber to match specific conditions (such as the core-to-cladding tubes ratio), single-mode behavior can be achieved. Due to their remarkable transmission properties, recent interest led to the first experimental realization of AR-HCFs with extremely low loss (0.174 dB/km at 1550 nm which is on the same scale as single mode fibers) and improved guidance [Jasion et al, 2022]. This recent spark of interest opens up possibilities in developing novel HCFs for guidance in wavelengths that silica fibers cannot operate such as UV (even extreme-UV) and mid-IR. In turn, these wavelengths can result in spectroscopic measurements that are currently prohibited due to the use of conventional fibers. For example, the UV spectrum can be utilized to interrogate the planetary atmospheres for specific molecular species. In addition, UV laser guide stars are interesting alternatives to Rayleigh guide star adaptive optics applications [Dekany et al, 2022 & Baranec et al, 2012]. HCFs offer the prospect of being able to mount the laser in a stable environment off-telescope and transport the light to the launch telescope via fibers, improving safety at the observatory as compared to mirror relays as well as providing improved stability for the launched laser. In the mid-IR (>2.0 microns), there is currently only one fiber fed spectrograph in operation (Delorme et al, 2021) and it utilizes ZBLAN fibers. These fibers are significantly more lossy than HCFs and therefore HCFs could enable a wave of fiber-fed spectrographs operating at the diffraction-limit in the mid-IR (Fitzgerald et al, 2022). Such spectrographs could be utilized for radial velocity measurements, the direct characterization of exoplanets and brown dwarfs (Wang et al, 2021) and for studying highly redshifted light of distant galaxies, newly forming stars, and faintly visible comets. Finally, HCF fibers could also be used to realize very long baseline (km scale) MIR interferometers, which are not possible with current technologies (see Chapter 17 for details). Processes leading to the routine connectorization and termination of AR-HCFs need also to be investigated. This in turn will enable the successful incorporation of AR-HCFs to existing photonic technologies.

.

AR-HCFs are inherently multimode although they suffer from mode-dependent loss; higher-order modes tend to attenuate quicker compared to the fundamental mode. Nevertheless, AR-HCFs can be designed such that the higher-order modes have comparable losses to the fundamental mode. In order to achieve this and maintain low-loss guidance, additional cladding elements are utilized (Winter et al. 2019, Shere et al 2021, Petry et al. 2022). Similar to standard solid-core fiber, we can increase the number of guided modes by increasing the HCF's core size. Multimode AR-HCFs can significantly increase the collection efficiency of any system at wavelengths (such as visible/UV and mid-IR) where the performance of silica is hindered due to photodarkening/solarization and where other multicomponent glass fibers haven't reached the required performance levels. Current fabrication processes prohibit the realization of continuous long lengths of AR-HCFs, especially for mid-IR applications where the core size and the surrounding structure of the HCF need to be significantly larger to achieve the desired low-loss guidance compared to near-IR/visible wavelengths. It is important to mention technologies such as mode adapters/reformatters, fiber combiners and splitters also need to be developed to effectively incorporate these fibers with current technologies (Sulsov et al. 2021).

**Fiber materials other than silica:** Alternative materials such as ZBLAN and Chalcogenide are also used to construct optical fibers and have been used in astronomy (Delorme et al, 2021), and silicon core fibers with low-loss optical windows upto 3.3 um (Ren et al, 2019) However, the former technologies are extremely brittle. This not only makes handling them difficult, but also limits the maximum length of a single piece to the order of 70-100 m. Although there are efforts being made to improve losses, these fibers still have a long way to go to reach the maturity and applicability of silica fibers.

## Advances in Science and Technology to Meet Challenges

In addition to the fiber technologies outlined above, complementary technologies like fiber components need to advance for these fibers to be applicable to astronomy. For future applications, we envision the development of a range of in-line fiber technologies based on these novel fibers. These can include broadband fiber splitters, polarization beam splitters, large-channel fan outs, etc. New functionalities can be incorporated to the fibers, for example, separating the cores from a MCF to perform different measurements. These devices will require new fabrication techniques and may need other materials.

Hollow core fibers with polarization-maintaining properties and HCF devices such as beam-splitters could have a major impact on future astronomical instruments. Nevertheless, as mentioned above, to achieve HCFs with improved performance and low-loss guidance in the mid-IR and UV region new fabrication techniques need to be exploited. At these wavelengths, any non-uniformities at the final structure of the fiber will have significant implications in their optical performance. Precise control of the drawing process of these fibers needs to be exploited.

A significant challenge that needs to be addressed is the scalability of the technology for the next-generation of fibers that scramble or prevent scrambling of the modes; i.e. to achieve the desired performance with an increased number of modes. To accomplish this, designs with complex refractive index profiles need to be exploited as increasing the number of modes (by increasing the core size) of the current fiber designs will result in degraded performance. Artificial intelligence (AI) techniques can be utilized for this purpose, enabling for the first time the realization of fibers with unique refractive index profiles designed specifically for astronomical applications.

Current MCF technologies have enabled significant advances in various areas of astronomy. Nevertheless, there is a constant desire to increase the number of cores as well as the packing efficiency (>90%) of these MCFs. Minimizing the gaps between cores enables more efficient collection of astronomical light and eliminates the need of lens arrays at the input to these fibers. Nevertheless, the core-to-core coupling increases significantly with decreasing separation. A significant challenge that needs to be overcome is minimizing the core-to-core coupling and therefore maintaining the uncoupled nature of the fiber. One way to increase the packing efficiency while minimizing the coupling between cores is by using cores with different refractive-index profiles including complex profiles. This is again an area where AI can be of significant value. In addition, it is important to mention that developing fan-outs for MCFs with many cores is not straightforward. Nevertheless, complementary technologies such as ultra-fast laser inscription can be utilized for this to develop high-precision fan-out systems [Thomson et al, 2007].

## Concluding Remarks

In recent years, fiber optic technology has seen tremendous advances which resulted in the development of unique fiber designs. These advances can feed the development of a new class of astronomical instruments. Fibers with unique profiles can be used either to effectively scramble the light or to minimize FRD. Novel fiber designs such as the log fiber and Airy fiber can be utilized in the fabrication of fiber devices for feeding light to the next generation of spectrographs. Furthermore, hollow-core fibers hold great promise in advancing the field of astronomical instrumentation. Low-loss hollow-core fibers for visible/UV and mid-IR guidance can be employed for new astronomical surveys, collecting light at wavelengths where previously we couldn't imagine that it would have been possible.

## 6 | Getting on-chip Arrayed Waveguide Grating Spectrographs Ready for Astronomy

Pradip Gatkine[1], Andreas Stoll[2] and Yang Zhang[3]
**[1] Department of Astronomy, California Institute of Technology, Pasadena, CA, USA**
**[2] Leibniz Institute for Astrophysics Potsdam (AIP), Potsdam, Germany**
**[3] Electrical and Computer Engineering Department, University of Maryland College Park, USA**

### Status

The volume, mass, cost, and complexity of seeing-limited spectrographs grow as the square of the telescope diameter. Building on-chip astrophotonic spectrographs is a promising approach for both ground- and space-based telescopes in order to achieve compactness (given the single-mode fiber feed), flexibility (with the ability to manipulate the amplitude, phase, and polarization in waveguides), thermo-mechanical stability (given the small size and no moving parts), and cost-effective replicability. They are ideally suited for capturing the adaptive-optics-corrected light and realizing various science cases, such as characterizing exoplanet atmospheres, precision radial velocity measurements, stellar chemistry/kinematics, planetary missions, and remote sensing. Multi-object spectroscopy can also be achieved by leveraging the stackability and replicability of the chips. The benefits of astrophotonic spectrographs go well beyond astronomy including biomedical sensors, industrial chemical detection, etc [1].

Arrayed waveguide gratings (AWGs) and photonic echelle gratings (PEGs) are well-known and mature technologies used in photonic wavelength-division-multiplexing [2]. However, astronomical spectroscopy poses unique and stringent specifications including high efficiency, broadband (eg: J and H bands), high resolving powers ($\lambda/\Delta\lambda$ > 10,000), and polarization independence [3,4]. While PEGs can deliver high resolving powers, achieving smooth on-chip reflecting surfaces is challenging due to fabrication limitations [5]. AWGs, on the other hand, require simpler fabrication schemes and can deliver high efficiency (> 75%), which makes them more suitable for astrophotonic spectroscopy [6]. Hence, we focus on improvements needed in the AWG architecture to make it science-ready. Cascaded dispersive architectures (two or more on-chip dispersers) are also widely discussed offering various benefits, such as on-chip order separation, adaptable component design, and potential tunability (eg: micro-ring resonator + AWG [7,8] or cascaded AWG [9,10]). Since AWGs require long path delays with bends, low-loss material platforms are desirable, with the most promising being Silica and Silicon Nitride.

**Silica:** Silica-based passive integrated photonic device technology is an attractive choice for astrophotonics due to the high coupling efficiency to fibers and excellent transmission characteristics from the visible to the near-infrared (J, H, and K-bands). Astrophotonic AWGs on a 2% refractive index contrast silica platform with R ~ 10,000 – 36,000 and efficiencies of up to 72% in the H-band have been recently demonstrated experimentally, using three-stigmatic-point geometry for focal field flattening [11].

**SiN and high-contrast AWGs*:*** AWGs need curved waveguides to introduce the path differences. Smaller radii of curvatures can help achieve smaller device footprints, and hence lower susceptibility to material/process variation (and thus, the effective index variation) across the chip. The higher contrast ratio offered by the silicon nitride ($Si_3N_4$) material platform allows low-loss radii of curvature as small as 100 micron, depending on the thickness of $Si_3N_4$ [12]. Further, ultra-low propagation loss has been demonstrated in Si3N4 over a broad band (400-2400 nm) [13], making it suitable for Astronomy [14]. This has encouraged several implementations of $Si_3N_4$ AWGs [15-17].

## Current and Future Challenges

For astrophotonic spectrographs, the primary losses occur at the interfaces between the telescope focal plane and optical fibers, between fibers and photonic chips due to imperfect mode matching, and within the photonic integrated circuits due to propagation losses and scattering at discontinuities of the effective refractive index. Efficient coupling of starlight to single-mode fibers becomes increasingly challenging with growing telescope diameter and focal spot size. However, coupling to SMFs with adaptive optics on large telescopes (> 4m) is becoming more common now [18] and is discussed in chapter 2. The current challenges in making the photonic spectrographs science-ready are discussed below.

**AWG loss:** The biggest source of insertion loss in an AWG is the waveguide-slab interface [16]. Interface losses occur mostly during the two transitions between the waveguide array and the free propagation slabs. Hence, achieving the highest efficiency across these transitions is crucial. Further, coupling of light into and out of the chip is a key contributor to the loss due to mode mismatch and Fresnel reflection, particularly for high-contrast integrated platforms such as SiN. Recent developments in low-loss on-chip spot-size converters/tapers [19] and off-chip interposers have demonstrated high fiber-chip coupling efficiency [20].

**Phase errors:** High-resolution AWGs (R > 20,000) require a larger number and longer length of arrayed waveguides, leading to a significant increase in the footprint of the device. Phase errors arise due to material/process variations across the chip, causing systematic and random variations in the effective refractive indices of individual waveguides. They become more significant as AWG size increases [21, 22]. Phase errors degrade the AWG performance by decreasing power in the main lobe and increasing the cross-talk. For instance, an AWG at R $\gtrsim$ 24,000 requires precise control of the optical path lengths of the order of ~10 ppm to ensure sharp constructive-interference peaks with minimal crosstalk [23, 24]. Fabrication of such high-R AWGs becomes increasingly difficult with a larger (N>300) number of arrayed waveguides due to path length accumulation of several centimeters. Correcting the phase errors using electro-optic, thermo-optic effect or piezoelectric modulation is possible [25]. However, incorporating a phase shifter (~ 1mm x 1mm size) for every waveguide (N ~ hundreds) can prohibitively increase the chip size and cost (see Chapter 22 for more details).

**Cross-dispersion:** The free spectral range (FSR) of high-resolution AWG devices varies between 10 – 20 nm which necessitates spectral order separation by cross-dispersion to cover a range of several hundred nanometers or an astronomical band. Semi-integrated spectrographs achieve order separation using bulk-optics cross-dispersion, which limits the potential for miniaturization [26, 27]. Truly integrated spectrographs require full integration of imaging optics and detectors on the chip [28, 29]. Alternatively, on-chip order separation using tandem/cascaded AWG architectures [**30**] will help achieve direct edge-coupling with off-chip detector arrays (Fig. 1). Implementation of ultra-low-loss single-stage AWGs and minimizing the inter-channel spectral dropout in the coarse AWG stage are essential to enable low-loss tandem AWGs.

**AWG focal plane:** The AWG focal plane is curved (Rowland circle). In conventional AWGs, the output spectrum is discretely sampled by the waveguides in discrete spectral channels, leading to inter-channel gaps. However, in Astronomy, a continuous, gapless spectrum is desired. To achieve that, the curved focal plane of the AWG needs to be imaged, which poses challenges in cleaving along the curved surface and imaging a curved plane on a flat detector [30]. The three-stigmatic-point technique can help achieve a flat focal plane (Fig. 1b), thus alleviating these challenges [31-34]. The flat focal plane minimizes the defocusing aberration of off-centered channels and allows a direct bonding of the detector to the focal plane of the chip. Accurately etching/cleaving/polishing the chip-edge to expose the exact focal plane is important to minimize defocusing.

**Polarization dependence:** Most photonic waveguides are birefringent (i.e. different effective indices for TE and TM modes) due to imperfect symmetry of the waveguide cross-section shape. At high resolving powers (R > 20,000), waveguide birefringence causes significant relative separation in the TE- and TM-mode spectral channels, resulting in broadening of the composite channels and reduced spectral resolution in the unpolarized regime. High-resolution AWGs are therefore restricted to fully polarized light unless waveguide birefringence can be adequately compensated for. One solution is to split the polarization using low-loss free-space or fiber-based or on-chip polarization splitters (eg: [35]), and then rotate the TM-polarized output (eg: [36, 37]) to then feed two copies of the TE-optimized AWG. Alternatively, birefringence-compensating designs such as waveguide cross-section engineering [38], varying waveguide width within the array [39], or slanted FPR-waveguide array interface [40] can be used.

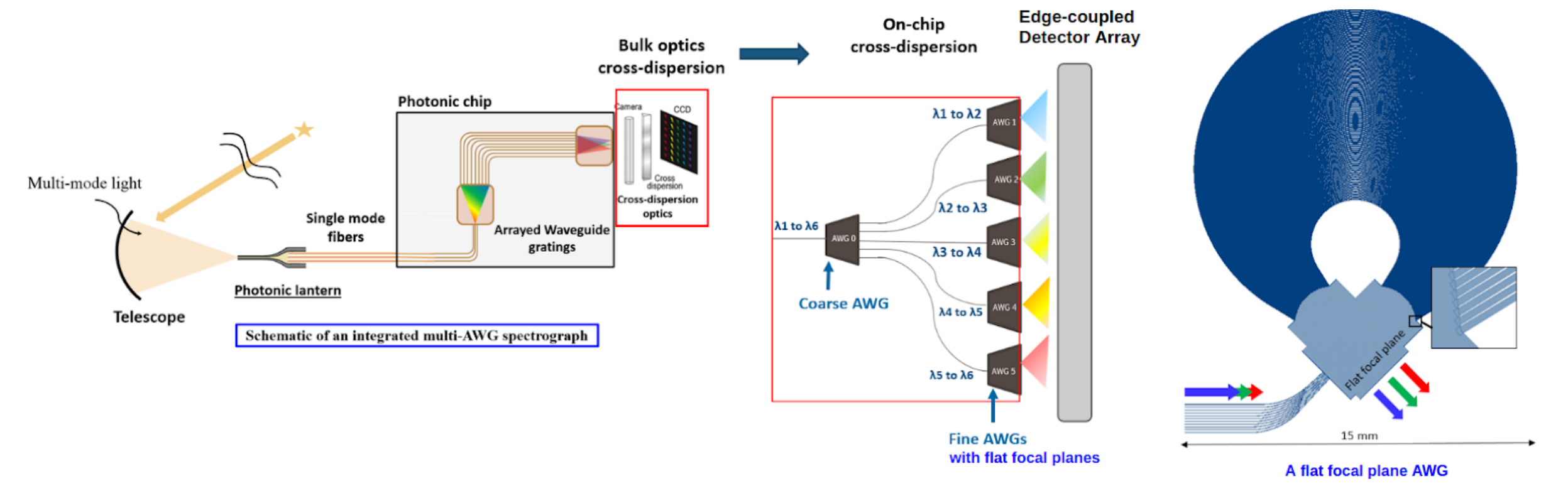

***Figure 1** - **Left:** The schematic of a conventional AWG spectrograph from telescope to detector (with bulk optics cross-dispersion), and that of a cascaded AWG architecture with on-chip order separation and an edge-coupled detector array. **Right:** 3-stigmatic-point AWG architecture in a folded configuration*

## Advances in Science and Technology to Meet Challenges

To achieve the vision of compact, fully integrated, high-resolution/broadband, and massively replicable astrophotonic spectrographs/IFUs for ground- and space-based telescopes, the following advances are critical, addressing the challenges of throughput, phase errors, and imaging of the dispersed spectrum.

1. **Innovative AWG component designs for low on-chip loss:** Tapers have proven effective in minimizing FPR-waveguide interface loss. Specific taper geometries can achieve an adiabatic transition. Sub-wavelength structures [41, 42], MMIs, and inverse-design tapers [43] can also be used to eliminate interface losses. UV-lithography limits the closest separation between the tapers (due to minimum feature size ~200 nm), thus creating gaps, which lead to losses. E-beam lithography is well-suited to fill these gaps (minimum feature size ~5 nm), however, it is not scalable. Hence, novel designs (eg: with thinner waveguide cores) are needed that can offer ultra-low loss with the scalable stepper-lithography process [44, 45].

2. **Phase-error correction:** Current state-of-the-art AWGs use electro-optic, piezoelectric [46], or thermo-optic phase shifters integrated on the arrayed waveguides for post-fabrication phase-error correction [47, 48]. However, the elimination of phase errors at the source using tighter fabrication process control and minimal-phase-error designs [49] is the ideal solution for achieving high-throughput, high-resolution AWGs. Phase error control techniques for silica AWG devices require post-fabrication phase error measurements and treatment by UV trimming [50], which is not feasible for mass production unless automated post-processing facilities are implemented. An alternative novel method of phase control can be implemented via programmable photonics [51], allowing in-situ phase adjustments on individual optical paths, which opens the possibility of mass-produced self-correcting AWGs.

   A novel architecture called Reusable-Delay Line AWG (RDL-AWG) utilizes an optimized array of directional couplers to distribute the input signal from the delay line into the free propagation region (Fig 2.(4)). It is a potential solution to reduce the impact of fabrication imperfections and non-uniformity for ultra-high resolution (R>50,000) AWGs [52]. This architecture can be ~100 times more compact than the traditional AWGs.

3. **On-chip order separation:** Tandem AWGs can be designed to separate the spectral orders to cover broad bands uniformly (Fig. 1). For the micro-ring resonators + AWG concept (Fig. 2), the quality factor of the resonators is the source of the high spectral resolving power. It is challenging to choose the radius of the rings such that the set of peaks is shifted very slightly to achieve uniform coverage. However, the peaks can be tuned after fabrication using thermo-optic modulators.

4. **Efficient coupling of light into and out of the chip:** The light can be efficiently injected from SMFs to waveguides using lensed fibers and/or adiabatic inverted tapers (~95%) [18]. For certain material platforms, inverse designs [53,54] or low-loss interposers are more appropriate [55]. Getting the light out of the chips efficiently is challenging for high-index-contrast materials (SiN, SOI). Currently, high-NA microscope objectives are used to image the AWG focal plane onto a detector. However, they have a small field of view (~ 1 mm). With th goal of moving away from bulk optics, one could consider solutions such as NA-matched microlens arrays with 100% fill factor [56-58] or 3D-printed lenses on chip [59, 60] need to be explored for better integration.

5. **Detector integration:** Several approaches are being tried to achieve true integration of photonic spectrographs + detectors, with no moving parts. Integration of energy-resolving MKIDs as cross-dispersing detectors is a promising approach [See chapter 24]. Alternatively, tandem AWGs perform on-chip cross-dispersion and allow direct integration of detectors via heterogeneous material platforms (eg: Si+Ge or Si+InP) [61-63]. Small-pitch linear detector arrays need to be developed for directly butt-coupling with the output FPR of cascaded AWGs (Fig. 1a). Other approaches include sending the light upwards from the chip using efficient grating couplers/metamaterial optimized for different wavelengths [64] (for discrete AWGs) or an inclined reflective edge or micro-mirror arrays (for both discrete and continuous AWGs) [Chapters 23, 24]. It is much simpler to image the discrete waveguide outputs as opposed to a continuous AWG focal plane onto a detector. New design efforts are needed in FPR-waveguide tapers (at output) to achieve a discrete-output AWG that can still sample the complete spectrum gaplessly.

6. **Beyond near-IR:** The scientific scope of photonic spectrographs can be dramatically increased by extending to wavelengths ranging from near-ultraviolet to mid-infrared. Further developments are needed in new materials such as AlN-on-Sapphire for UV [66] and chalcogenide glass or Si/Ge for developing low-loss AWGs in the mid-IR. The specific challenges and advances are discussed in Chapter 8.

7. **Space readiness:** A stable, durable, and crack-free fiber-chip bond is of particular importance for applications that require operation at cryogenic temperatures, as well as space applications in hard vacuum. Refractive index matching gap fillers/gels/oils are not suitable for these conditions due to solidification at low temperatures and potential outgassing. The impacts of radiation-induced degradation and temperature-induced core-cladding stresses need to be assessed on the overall throughput.

8. **AWG spectrograph throughput outlook:** The main throughputs in a photonic spectrograph are SMF-to-chip, on-chip, and chip-to-detector throughputs. We present the current state-of-the-art and optimized throughputs in the future to give a near-term throughput outlook.

| **Component** | **Current throughput** | **Optimized throughput** |
|---|---|---|
| SMF-to-chip efficiency | 95% | 95% |
| AWG throughput | 70% (Silica AWG) | 85% (optimized tapers) |
| AWG to free-space | 80% | 95% (with AR coating + output facet optimization) |
| Relay optics + cross dispersion | 50% (prism + lenses) | 70% (optimized cascaded AWG) |
| **Total throughput** | **~ 26%** | **~53%** |

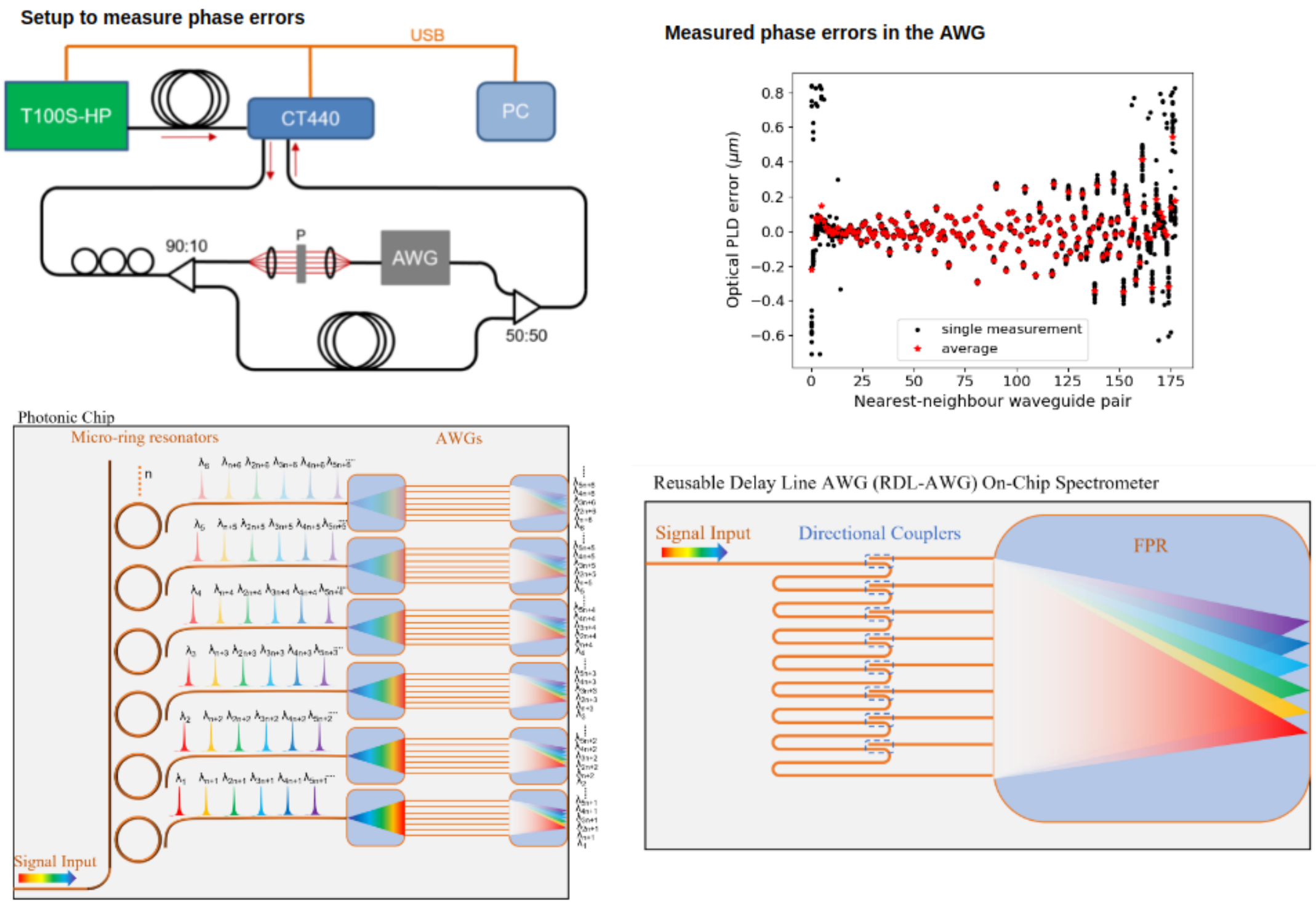

***Figure 2** - **Top-left:** Fiber-based frequency domain interferometer setup to measure AWG path-length errors; **Top-right:** measured optical path length error between adjacent waveguides in a silica AWG [67]; **Bottom-left:** An*

*integrated photonic chip with a cascaded array of ring resonators (thermally tunable) and AWGs.* ***Bottom-right:*** *A schematic of the arrayed waveguide grating with reusable delay lines (RDL-AWG) [52].*

## Concluding Remarks

Astrophotonic spectrographs offer significant advantages for both ground- and space-based telescopes, particularly at high spectral resolution. Multi-pronged developmental work is currently being pursued with several novel photonic technologies to achieve the vision of compact, high-throughput, fully integrated (with fibers and detectors), and highly replicable photonic spectrographs to augment both ground- and space-based astronomy.

## Acknowledgments

*Support for P. Gatkine was provided by NASA through the NASA Hubble Fellowship grant HST-HF2-51478.001-A awarded by the Space Telescope Science Institute, which is operated by the Association of Universities for Research in Astronomy, Incorporated, under NASA contract NAS5-26555. Y. Zhang is supported by National Science Foundation (NSF) under grant 1711377 and the National Aeronautics and Space Administration (NASA) under grant 16-APRA 16-0064.*

## 7 | New Directions in Photonic Spectrograph Architectures

Pradip Gatkine[1], Chang-Ling Zou[2], Zongyin Yang[3], Jinping He[4,5] and Ross Cheriton[6]
**[1] Department of Astronomy, California Institute of Technology, Pasadena, CA, USA**
**[2] CAS Key Laboratory of Quantum Information, University of Science and Technology of China, Hefei, Anhui, China**
**[3] College of Information Science and Electronic Engineering, State Key Laboratory of Modern Optical Instrumentation, Zhejiang University, Hangzhou, China**
**[4] National Astronomical Observatories, Nanjing Institute of Astronomical Optics & Technology, Chinese Academy of Sciences, Nanjing,China**
**[5] Key Laboratory of Astronomical Optics & Technology, Nanjing Institute of Astronomical Optics & Technology, Chinese Academy of Sciences, Nanjing, China**
**[6] Advanced Electronics and Photonics Research Centre, National Research Council Canada, Ottawa, Canada**

### Status

Bulk-optic, seeing-limited astronomical spectrographs have limitations like high complexity, large size, mass, and high cost, which scale quadratically with the telescope diameter. Integrated photonic spectrographs can address these concerns by enabling high-stability, high-precision dispersion spectral processing for many science cases. Astronomical spectrographs require high efficiency, large spectral bandwidth (e.g., J-H bands: 1200-1700 nm), low or high resolving powers ($R=\lambda/\Delta\lambda \sim$ 1,000-100,000), and in most cases, polarization insensitivity, which each present a different set of challenges. Some of the potential use cases include: **(a)** high-precision radial velocity (RV) measurements of stars to constrain exoplanet masses (R > 80,000, RV precision ~10 cm/s), **(b)** high-resolution spectroscopy to characterize exoplanet atmospheres composition, measuring stellar composition and kinematics (R ~ 30,000), **(c)** diffraction-limited multi-object spectroscopy of crowded fields to measure stellar properties in extreme environments (R > 10,000), **(d)** low-cost spectrographs for rapid spectroscopy of a large number of transients on small telescopes (R ~ 3000), and **(e)** ultra-compact spectrographs for space-based telescopes and planetary missions.

Currently, arrayed waveguide gratings (AWGs) are the most prominent candidate architecture for astrophotonic spectroscopy (Cvetojevic+2012, Jovanovic+2016, Stoll+2021). Conventional high-resolution AWG spectrographs require bulk-optic cross-dispersion to achieve broadband coverage. However, we envision fully integrated on-chip spectrographs that would lead to ultra-stable monolithic instruments. Challenges with the AWGs and potential solutions are discussed in Chapter 6. While astrophotonic AWG spectrographs have been demonstrated over the last decade, the phase errors in the AWG chips pose a significant barrier for R > 10,000 (Stoll+2020, Gatkine+2021). This paper therefore focuses on emerging photonic architectures to expand to higher resolving powers (R>10,000).

Several novel photonic spectrograph architectures (Blind+2017, Gatkine+2019) are on the horizon with the potential to overcome the limitations of AWGs and enable improved integration with detectors. In this paper, we present a non-exhaustive list of promising new spectrograph architectures, their unique advantages, and the developments needed to make them science-ready.

**Current and future challenges**

Here, we discuss the specific architectures, the advantages they offer, and the challenges that need to be addressed in the future to make them suitable for astronomy.

**Ring resonator-enhanced spectrographs:**

Achieving a truly integrated broadband photonic spectrograph requires an on-chip separation of the overlapping spectral orders of the AWG. One solution is to use an AWG with high-resolution and low free spectral range (FSR) (Eg: R~30,000 FSR~10 nm) followed by an array of microring resonators (MRRs) as filters to separate the spectral orders [Hu+2021, Fig. 1a]. MRRs can also be placed after a coarse AWG to achieve high spectral resolution within each AWG output channel. Cascaded AWG-MRR spectrographs have also been produced with detectors on a single chip [Zheng+2019, Zhu+2020], creating a fully integrated spectrograph.

The resonance wavelengths of MRRs are highly sensitive to fabrication errors and temperature fluctuations without active thermal tuning. In addition, to achieve a broadband, high-resolution AWG+MRR spectrograph (Fig 1a), we need:

Total # of MRRs = (# of AWG channels) x (# of AWG spectral orders).

This implementation can quickly result in thousands of MRRs and tuning elements, presenting a serious integration challenge. Advanced routing and packaging strategies, such as the use of flip chip bonding with vias, may be necessary to integrate these components.

**Fourier-transform spectrographs:**

Fourier-transform spectrographs (FTSs) are an architecture that can offer ultra-high-resolution spectroscopy for bright sources (Ridgway & Brault 1984) by deducing the optical spectrum from an interference pattern generated as a function of the optical path length difference. This method enables the use of a single-channel detector that can be directly integrated into a photonic FTS chip. However, broadband photonic FTSs are limited by photon noise and waveguide dispersion, which degrades the fringe quality, and the throughput limit (Souza+2018). Stationary wave integrated FTS (SWIFTS), which reconstructs the spectrum by sampling the standing wave created by forward and reflected waves within a waveguide-mirror configuration, faces a similar narrow-band challenge due to the undersampling of the standing waves (Le Coarer+2007, Bonneville+2013). However, a dispersed FTS architecture (e.g. FTS + AWG) could resolve these issues (Hajian+2007).

An FTS+AWG concept (Fig. 1c) effectively multiplies the resolving power of low-resolution, broadband AWGs (e.g. FSR = 100 nm, R = 5,000) using an upstream FTS with switch-selectable path delays. Kita+2018 experimentally demonstrated the switchable FTS concept with a 64-channel FTS (Fig. 1c) and a single photodiode. This concept can be adapted to multiply the resolving power of the AWG by a factor of $2^k - 1$, where $k$ is the number of phase-shift units incorporated in the circuit. The switchable FTS modulates the spectrum as we switch through the path delays. This allows us to reconstruct the spectrum at a much finer wavelength step within each AWG channel, effectively multiplying the AWG resolving power, while maintaining a high FSR.

Various challenges need to be addressed to make FTS-AWG architectures suitable for astronomy. It requires high-precision measurement of the output power and detailed high-resolution calibration to reliably retrieve the spectrum at high resolving power. Hence, a stable, low-noise detector array is highly desirable. Implementation of integrated photodetectors will enable a highly compact, and stable high-resolution photonic spectrograph. Furthermore, this architecture requires broadband 50-50 splitters and thermally-controlled photonic switches (to switch between path delays) that offer ultra-low loss and minimal crosstalk to enable easier spectral reconstruction. Achieving ultra-low propagation losses (<0.1 dB/cm) is critical to minimize the impact of the long path delays.

**Serpentine integrated grating spectrograph**

Current high-resolution AWGs require cross-dispersion to separate the spectral orders and prevent spectral overlap in each channel. While cascaded AWGs can help resolve this issue, they can be too large for fabrication on a single chip and would require very long (>1000 pixels) linear detector arrays to sample the spectral elements. The serpentine integrated grating (SIG, Fig. 1b) can help alleviate this problem by naturally allowing a 2D sampling of the spectrum. This architecture uses a long folded delay line with grating couplers to create a large optical delay path along two dimensions (high-dispersion along the column and 'order separation' along the serpentine row) in a compact integrated device footprint (Dostart+2020). The grating couplers send the light upward, thus achieving two goals - high spectral resolution and easy imaging of the 2D spectrum from above the chip. This method requires high index-contrast waveguides, such silicon-on-insulator (SOI) and silicon-nitride (SiN) platforms, which allow sharp low-loss bends (with radii as small as ~ 10 um), to enable several centimeters of path delay in a small footprint (~ few mm^2). An SIG spectrograph with R~100,000 and ~6750 spectral bins over the 1540-1650 nm range has recently been demonstrated on the SOI platform (Brand+2021).

High losses of the device in the SIG demonstration (-36 dB) need to be addressed to make it viable for astronomical spectroscopy. The main contributors to the loss were: straight-to-bend waveguide tapers (8.4 dB = 0.07dB per taper x 120 tapers), input fiber-waveguide edge couplers (7 dB due to mode-mismatch), and the on-chip grating couplers (16 dB). Optimized transition tapers can offer ultra-low-loss U-bends (Shen+2010, 0.0022 dB per taper). Similarly, high-efficiency broadband edge-couplers need to be engineered (eg: Cheben+2015, Papes+2016: >90% efficiency). High-efficiency grating couplers are challenging to realize. However, apodized designs (Marchetti+2017) and dimension reduction techniques can reduce the design parameter-space and device optimization time (Melati+2019). Alternatively, low-loss 3D circuit fabrication will allow easier integration of the device (Shang+2015).

**Microcomb-based spectrometer**

The ultimate resolution of an on-chip disperser is limited by the total optical path difference in the dispersive elements. Further spectral resolution increases could be obtained by converting the optical frequency to electronic RF frequency by measuring the heterodyne beating signal between the signal light and a local oscillator. The microcomb source, which is generated by dissipative Kerr solitons in a laser-driven integrated microresonator [Kippenberg+2018], provides an excellent local oscillator for an integrated spectrometer (for bright objects), as it delivers a series of comb lines with a precise and fixed free spectral range. Innovations such as battery-operated integrated comb generators [Stern+2018], octave-spanning microcombs [Obrzud+2018, Rao+2021], and frequency stabilization [Del'Haye+2008, Liu+2020] have recently been demonstrated experimentally.

By mixing the input signal with the microcomb, the optical signal frequency can be converted to an RF signal due to the beating between the input light and a nearby comb line. Using an integrated silica microresonator to generate a microcomb with a 22 GHz repetition rate (i.e., frequency spacing between comb lines), a spectrometer has been demonstrated with a frequency uncertainty of around 4 MHz [Yang+2019].

This spectrometer architecture faces various challenges: i) The resolution of the frequency measurement is limited by the frequency instability (~MHz) of the comb lines of an unlocked microcomb. The frequency instability mainly comes from two sources, the instability of the drive laser, and the thermal stability of the microring resonator. (ii) There is ambiguity in determining the frequency and power of the heterodyne signal due to ambiguity in determining the closest comb line. This ambiguity could be resolved by simultaneously beating the signal with two independent comb sources with different repetition rates, and the optical signal frequency could be determined by comparing the two beating frequencies (dual-comb spectroscopy). Alternatively, the two comb sources could be provided by a single microresonator by either employing different mode families [Yang+2019] or switching the repetition rate in real time [Niu+2022].

**Computational spectroscopy**

A significant departure from the traditional dispersive approaches is the use of computational techniques to decode spectral information from precalibrated encoded information to produce an enhanced spectrum beyond the limitations of gratings/filters. So far, complex spectral-to-spatial (Hadibrata+2013, Redding+2013) and multiplexed spectral response [Yang+2019, Bao+2015] strategies have been proposed for encoding spectral information (Fig. 2a). In the complex spectral-to-spatial encoding process, incident light is dispersed by random media, such as disordered materials or multimode fibers to form an overlay of encoding patterns created by individual wavelengths. The multiplexed spectral response encoding is an alternative to spatially dispersed wavebands. This encoding can be achieved in the filters or detectors themselves. A unique precalibrated wavelength-responsivity library is then used for decoding the incident spectrum.

Both encoding strategies can offer high throughput compared to dispersive spectrometers. Moreover, a higher spectral resolution can be achieved in the same device area compared to traditional photonic spectrographs, since the scattering pattern is not restricted by the effective index or path lengths as in traditional architectures. Emerging strategies use inverse design to specify the desired output spectrum, and iterative steps to generate non-intuitive, disordered geometries for computational spectroscopy that can meet the requirements of the output spectrum (Hadibrata+2021). The compactness of the geometry means shorter propagation lengths and smaller chips, leading to lower phase noise and higher thermal stability.

Another computational architecture photonic cross-correlation [Cheriton+2020] uses tunable spectral filters to process the entire spectrum for a set of specific spectral lines, enabling higher throughputs and the use of a single detector channel (detailed discussion in Chapter 11).

However, computational spectrometers face certain challenges in reconstructing robust results. The results are accurate with sparse spectra, but if the input is a broadband spectrum, the result is vulnerable to distortion. Furthermore, significant computational resources are required for both the spectrograph design generation and spectrum-interpretation algorithms to achieve high performance. Another challenge is their limited spectral bandwidths. This can be mitigated by appending compact computational/inverse-designed spectrographs to AWG output channels to perform broadband, high-resolution spectroscopy.

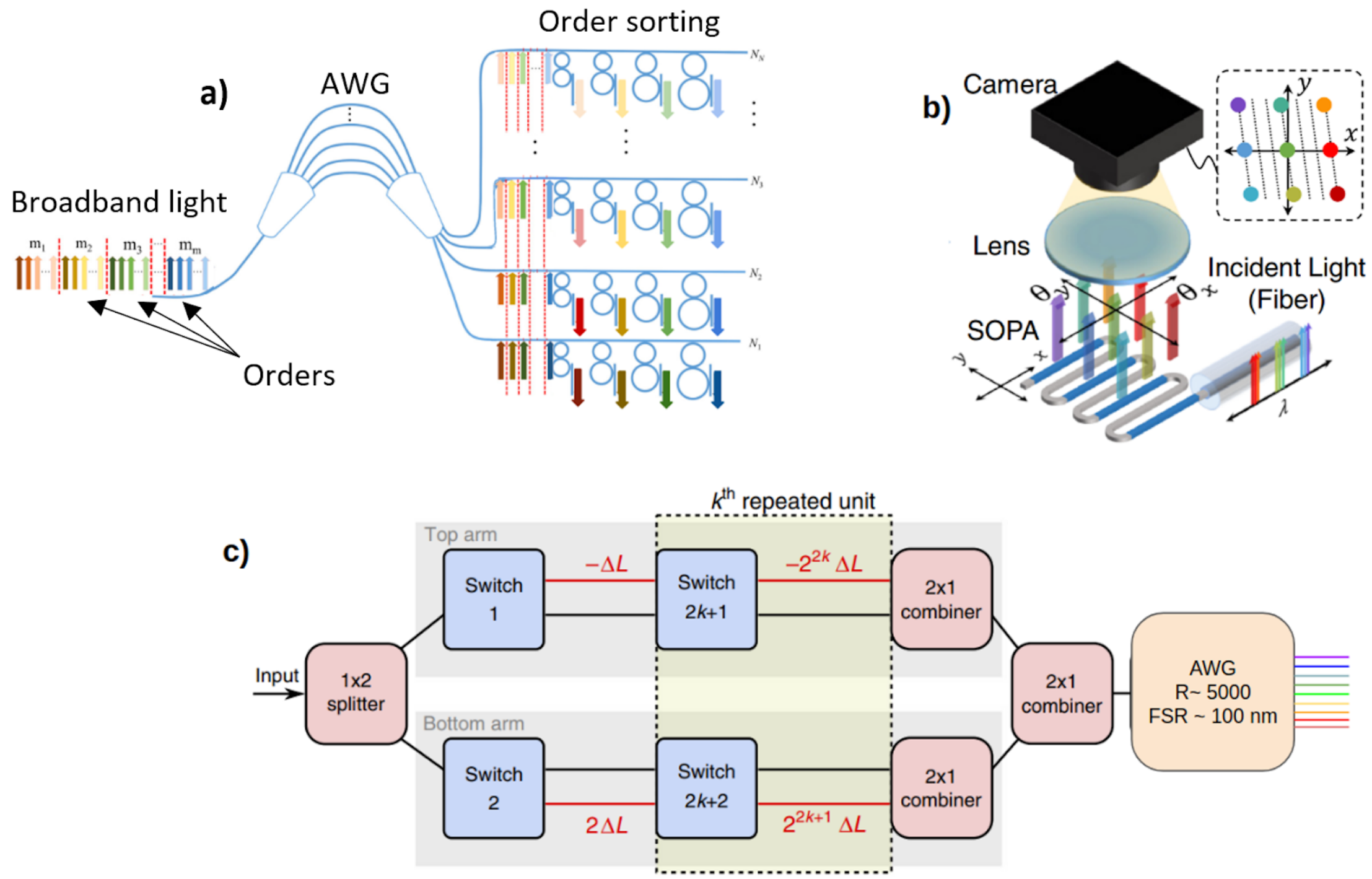

***Figure 1*** *- a) A schematic showing the concept of high-resolution AWG with MRRs at the output channels to separate the overlapping spectral orders. b) The concept of a serpentine integrated grating using a serpentine optical phased array (SOPA), providing high-resolution dispersion along the y-axis (with long cumulative phases), and low-resolution separation along the x-axis (from* [*Brand+2021*]*). c) A schematic showing the concept of switchable path delays in an FTS and the AWG dispersing the FTS fringes to effectively create a dispersed FTS architecture (adopted from* [*Kita+2018*]*).*

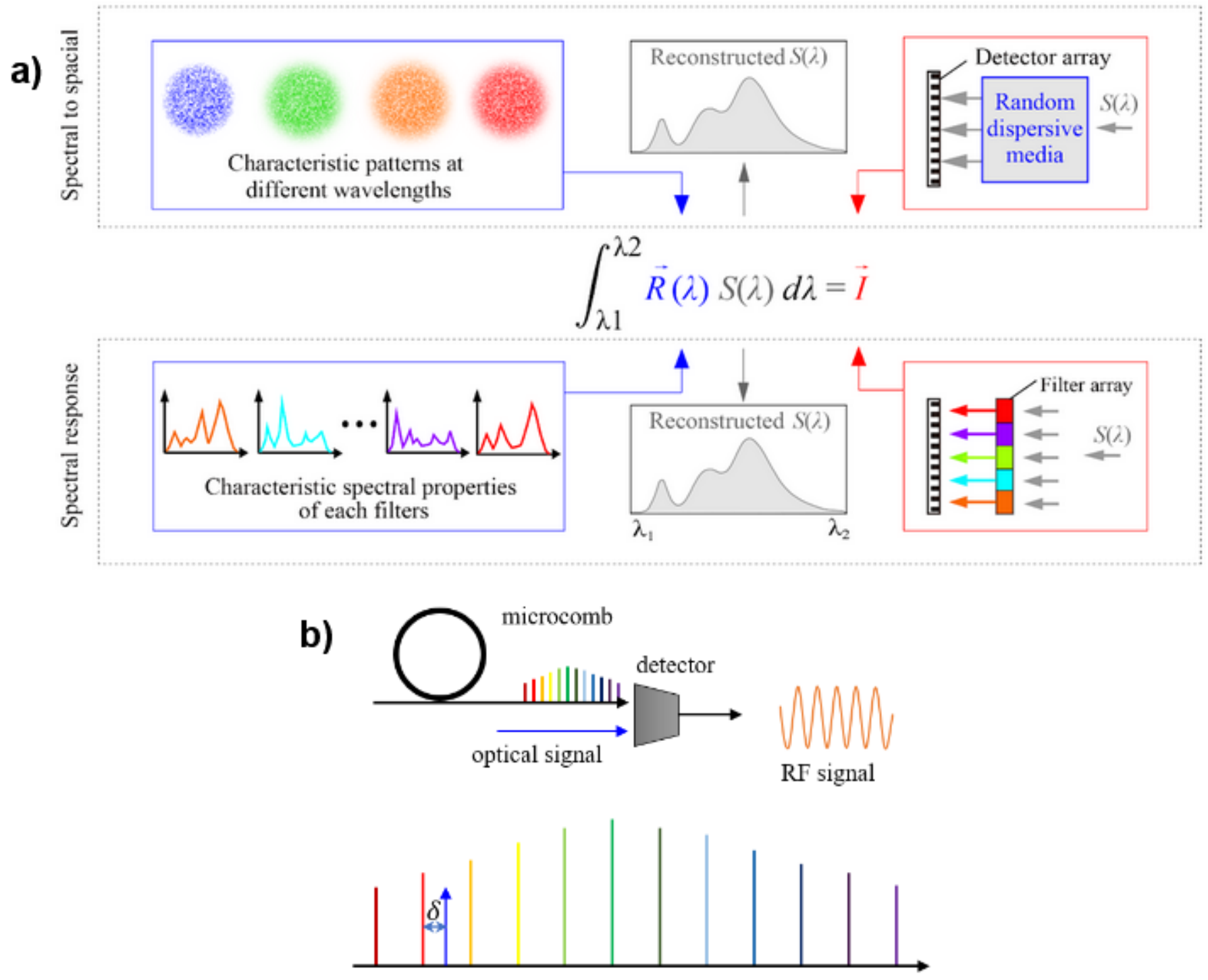

***Figure 2 -*** *a) Two encoding schemes of computational spectrographs are shown here. The top part shows the pattern-to-spectrum reconstruction for random dispersive media, and the bottom part shows the reconstruction from a calibrated spectral response library. Both require advanced reconstruction algorithms for robust reconstruction. b) The concept of an optical signal mixed with a microcomb to create an RF beat signal for high-resolution spectral reconstruction.*

## Advances in Science and Technology to Meet Challenges

Significant efforts are needed in several directions including computational reconstruction, heterogeneous material platforms and fabrication, inverse design, precision calibration, light routing schemes, and integration of photonics, electronics, and detectors, to achieve fully integrated high-resolution, broadband photonic spectrographs for ground- and space-based telescopes. These advances are summarized in Table 1. We can classify key advances in 3 categories:

a. **Design and fabrication of low-loss components:** Fully integrated cascaded spectrograph concepts are viable only with low-loss, broadband individual stages. Major losses occur at interfaces with mismatched refractive indices. Highly efficient taper designs are required for waveguide to SMF, waveguide to free space, single-mode to multi-mode, and straight-to-bend waveguide transitions. Similarly, ultra-low loss and broadband splitters are essential for all interferometric architectures (Eg: FTS+AWG). For SIGs, a significant effort is required in developing high-efficiency vertical grating couplers for launching the dispersed light perpendicularly for simpler 2D detector integration (Hong+2019).
b. **Computational advances:** Computational spectroscopy and dispersed FTS implementations will greatly benefit from high-precision spectral calibration, and state-of-the-art spectral reconstruction techniques (eg: ML-based elastic-D1 regularization, Kita+2018). Further, novel algorithms are needed for forward modeling and inverse-design optimizations (bandwidth, throughput) on large photonic components in various spectrograph architectures (Molesky+2018, So+2020). Fast approximate solvers for scattered fields are critical to optimizing large surfaces that can be thousands of wavelengths in diameter (Pestoruie+2018).
c. **Integration with detectors and electronics:** Future implementations should aim to integrate the detection array monolithically into the photonic chip itself or couple the light onto a nearby external detector with minimal optics for low cost, compactness, and improved throughput. Efforts are needed to develop reliable fabrication recipes in heterogeneous material platforms (eg: InGaAs-SiN, Piels+2014, NbN/WSi/MoSi-SiN, Kahl+2017) allowing a large number (thousands) of on-chip pixels. Ongoing efforts in kilopixel SNSPD arrays (Wollman+2019) are promising with improvements expected in efficiency and dark current. Industrial foundries are well-placed to offer these to the broader photonics community (See Chapters 23 and 24).

***Table 1 -*** *Spectrograph architectures, challenges, and future advances*

| | New architectures (with main advantages) | Challenges | Advancements required |
|---|---|---|---|
| 1 | Ring resonator-enhanced (high resolution and bandwidth) | Thermal cross-talk,<br>Complex electrical integration | High-throughput AWG designs<br>Thermal stability when driving MRRs<br>Scalable thermo-optic control of large arrays of MRRs |
| 2 | FTS+AWG hybrid (multiplies the resolving power of the AWG) | high-precision measurement of output power,<br>Additional loss in the path delays,<br>broadband splitters | Integrated detector arrays,<br>Ultra-low propagation loss,<br>Ultra-broadband splitters, photonic switches (sub-wavelength inverse-design can help) |
| 3 | Serpentine integrated spectrograph (easy integration with detectors and R ~ 100,000) | Throughput,<br>Straight-to-bend taper losses,<br>Fiber-waveguide losses,<br>Grating coupler losses, | Cross-section optimization,<br>Inverse design,<br>Broadband grating couplers, optimized for the specific waveband of the spectrograph |
| 4 | Microcomb-based (ultra-high R > 100,000) | Comb line stability, ambiguity in determining the relevant comb frequency | Self-referenced comb<br>Advanced on-chip resonator stabilization<br>Fast and accurate repetition rate switching |
| 5 | Disorder modes (no FSR overlap, compactness) | Spectral bandwidth,<br>Fabrication quality to enable higher resolution and throughput | Large device areas to enable greater performance<br>Greater computing resources to support larger 3D structure simulations<br>Novel metasurface realizations |
| 6 | Computational spectrometers | Low stability,<br>Noise sensitivity,<br>Reliability over larger bandwidths, | Accurate and fine coding strategies,<br>Powerful reconstruction algorithm,<br>Incorporation of computational techniques to complement traditional dispersive spectrometers such as AWGs |

## Conclusion

The new directions in astrophotonic spectrographs have the potential to overcome many of the limitations with the prevalent AWG architectures. Further developments in photonic fabrication, computation, complexity and complete integration need to be leveraged for building compact, efficient, and broadband astrophotonic spectrographs.

## Acknowledgements

*Support for P. Gatkine was provided by NASA through the NASA Hubble Fellowship grant HST-HF2-51478.001-A awarded by the Space Telescope Science Institute, which is operated by the Association of Universities for Research in Astronomy, Incorporated, under NASA contract NAS5-26555.*

# 8 | Advances in UV and Mid-IR Astrophotonics

Lucas Labadie[1], Reinan Moreira[2], Simon Gross[3] and David Schiminovich[4]
**[1] University of Cologne, Germany**
**[2] Ultra-Low Loss Technologies, Santa Barbara, CA, USA**
**[3] MQ Photonics Research Centre, Macquarie University, NSW, Australia**
**[4] Department of Astronomy, Columbia University, NY, USA**

## Status

Common photonic integrated circuits (PICs) are designed for the near-infrared (NIR) wavelengths of 1.3 µm and 1.5 µm, utilizing silicon and its oxide form, $SiO_2$, such as silica-on-silicon (SOS) and Si/SiN-on-insulator (SOI) technologies [Takahashi 2003, Siew 2021, Xiang 2022]. At 2.2µm, the GRAVITY instrument [Eisenhauer 2017] is an excellent example of the astrophysical yield of photonics in the NIR (cf. Chapter 15). Lithium niobate, as an alternative, is employed for signal modulation in data transmission and integrated quantum photonics [Boes 2018, Li 2020]. In parallel, three-dimensional material structuring by the means of ultrafast laser inscription (ULI) offers new opportunities for manufacturing photonic devices [Osellame 2012, Gross 2015].

From an astrophysical perspective that needs panchromatic information, there is great interest in extending integrated photonics to more "extreme" frequencies towards the mid-infrared (MIR) range beyond ~2.5 µm [Labadie 2009], or towards the ultraviolet (UV) regime (e.g., λ<380 nm).

At MIR wavelengths, the fields of young stellar objects, the spectral characterization of nearby exoplanets, or the study of the circumnuclear regions of active galaxies fully justify the extension of PICs into the 3-20 µm wavelength range. Similarly, accessing - in part from space - the near- to far-UV range at high-spectral resolution is of critical importance for the study of the galactic and extragalactic Interstellar Medium (ISM) phases and the mass reservoir of cold molecular hydrogen.

To date, UV/MIR photonics have yet to demonstrate the "on-sky" maturity achieved by their NIR counterparts. In the MIR, the TISIS experiment at IOTA using a fiber X-coupler in the L-band (~3.8 µm) is the only "on-sky" astrophotonic experiment undertaken thus far [Mennesson 1999]. Most MIR integrated photonics and fiber demonstrations are laboratory-based and focused on the characterization of new devices and platforms. Originally motivated by the DARWIN/TPF mission, the effort has moved from simple waveguides and fibers [Ho 2006, Vigreux-Bercovici 2007, Ksendzov 2007, Ksendzov 2008] to more elaborate optical functions [Labadie 2011, Rodenas 2012, Arriola 2014, Tepper 2017, Diener 2017] for ground-based astronomy. In the near future, the NOTT instrument at the VLTI [Defrère 2018] will eventually reach a new stage in on-sky use of MIR astrophotonics. In the UV domain, the field of astrophotonics and the related experimental characterization of devices is emerging. Recently, novel material platforms have been developed that enable the operation at shorter wavelengths close to the visible, at visible and at UV wavelengths driven by life-science and quantum optics applications, sensing, spectroscopy and optical clocks [Blumenthal 2020].

## Current and Future Challenges

*Identifying priorities in the spectral bands of operation of MIR astrophotonics:* Most astronomers interested in photonics tend to broadly define the MIR as the 3 to 20 µm range, which is difficult to cover with a single photonic technology/platform. Furthermore, PICs usually require single-mode behavior, which limits the usable wavelength bandwidth to less than half an octave before waveguides have either become multimoded or guide the light poorly. One could isolate sub-bands of operation, such as matching the typical atmospheric windows. In recent years, the 3-5 μm spectral range has seen the largest progress, for instance, with significant advances in beam combination capabilities for interferometry using ULI [Labadie 2018]. In the 8-12 μm spectral range, a proof-of-concept of astronomical functions has been achieved [Rodenas 2012, Butcher 2018]. Recently, a silicon-germanium integrated Fourier Transform spectrometer [Montesinos-Ballester 2019] was demonstrated at 7.7 μm using thermo-optic modulation. To date, the only attempt to cover the ultrabroad MIR range 6-20 μm with a single technology used telluride(Te)-rich chalcogenide single-mode waveguides, albeit at the cost of high propagation losses [Vigreux 2015].

*A future challenge resides in dividing the astrophotonic MIR range into several narrower spectral regions over which the overall performance can be maximized.*

*Photonic technologies for the UV domain:* With photon energies exceeding the Si bandgap, ultraviolet and optical photonics pose distinct materials, design and fabrication challenges as compared with the NIR. Issues include limited material, dispersion and photon source selection; increased absorption and scattering losses; decreased performance due to material impurities and defects; and operational degradation, especially at higher power and in strong fields. Some of these challenges and issues may ultimately be exploited for the design of sensitive probes and sensors, and engineered quantum emitters e.g., for room temperature on-chip applications. Astrophotonic applications have lagged the broader photonic research and development effort, increasingly so at shorter wavelengths (deep UV). At the same time, space-based UV instruments stand to substantially benefit from new technologies that reduce volume, mass and power.

*Material choices and fabrication platforms:* Highly transparent materials in the MIR (chalcogenides, heavy metal oxides and fluorides, …) are often more difficult to handle than the well-established Si/SiN/$SiO_2$ systems. MIR materials are generally less compatible with established fabrication techniques. For lithography, highly pure and defect-free thin-films are required and etch processes with minimal sidewall roughness. Suitable cladding materials are required to protect the waveguides and a careful choice of carrier materials is also needed to avoid delamination due to thermal expansion coefficient differences. In contrast, ULI can directly modify highly pure bulk material, but generally cannot reach the high index contrasts of lithographic platforms. With several platforms/materials now tested (e.g., SiN/Si, SOI, Ti:LiNbO3, Si/$Al_2O_3$, GLS, Ge/Si..), the current status suggests that ULI in chalcogenide glasses (e.g. GLS, $As_2Se_3$, $Ge_{33}As_{12}Se_{55}$) is a promising combination for the spectral extension beyond 3.5 μm of interferometric beam combiners, owing to the relative simplicity and versatility of the manufacturing process [Tepper 2017b, Butcher et 2018]. The photolithography and etching of chalcogenide films, already proposed for channel waveguides [Vigreux-Bercovici 2007], is being studied to develop multimode interference couplers [Kenchington Goldsmith 2017]. Silicon- and silicon/germanium-based integrated photonics [Lin 2017, Marris-Morini 2018] are attractive for the field, but have not been investigated in the context of astronomical requirements yet.

At UV and visible wavelengths, Blumenthal (2020) identifies materials for PIC and wafer-scale blue and UV photonic applications, highlighting silicon nitride $Si_3N_4$ (>400 nm), tantalum pentoxide $Ta_2O_3$ (>300 nm), amorphous Aluminum oxide $Al_2O_3$ (>250 nm) and aluminum nitride AlN (>200 nm) as leading material platforms. These will be supplemented by other 3D and 2D wide bandgap materials (e.g. III-N, AlGaN, hexagonal boron nitride (hBN), silicon carbide, diamond), metamaterials, transmissive step-index fiber (UV grade silica or fluoride/ZBLAN fibers, Hoang 2021) and solarization-free anti-resonant hollow core fiber technologies (Gao 2018, Li 2020). Astronomical applications have included UV-transmitting fibers in the FIREBall experiment (Tuttle 2010) and design concepts incorporating hollow core fibers in the deep UV (Gilliam 2021).

*A future challenge for UV/MIR astrophotonics is therefore to clearly highlight a dominant and optimal platform (or platforms) in terms of device quality and broadband performance for identified sub-ranges. Furthermore, the availability of glasses with excellent purity and batch-to-batch repeatability is a key aspect for manufacturing MIR PICs using different platforms.*

*Transparency/Losses (including Fresnel losses):* Low loss PICs are a major requirement due to the intrinsic weakness of the distant sources being observed. The propagation losses for MIR single-mode waveguides are in the 0.1 dB/cm to 3 dB/cm range (or 98% to 50% throughput over 1 cm) [Lin 2017, Gretzinger 2019, Madden 2017], an order of magnitude larger than for NIR silica waveguides (~ 0.01-0.05 dB/cm, or down to 99% throughput per cm).

MIR materials with indices of ~2.5 exhibit ~15% Fresnel reflection losses per facet, and silicon or germanium even >30%. This requires broadband anti-reflection coatings on the small-sized input/output facets. Alternatively, materials with n~1.5 have also been investigated, such as ZBLAN glass [Gross 2015b], though the transparency is limited to <~4 µm.

*Processing high purity and defect-free MIR materials to produce waveguides with <1db/cm losses, as well as being able to deposit broadband AR coatings on the chip/waveguide facets are important future challenges for the progress of MIR astrophotonics.*

*Coupling and Interfacing with fibers:* Pigtailing PICs components for astronomical applications is an important aspect at all wavelengths, including in the MIR. Typically, Mode Field Diameters (MFD) of MIR single-mode devices range from a few microns in high index contrast platforms to a few tens of microns for low field-confinement waveguides. The low-loss interfacing of MIR integrated photonics to commercial MIR fibers (e.g., fluoride, chalcogenide, silver halide or hollow core fibers) requires close matching in MFDs, as well as a good level of index matching. At MIR wavelengths, the diversity in the properties of the used materials (e.g., low-index fluoride fibers and high-index chalcogenide PIC) or in the design of the fiber (solid-core, hollow core, microstructured fibers; see Chapter 5) can be a stronger challenge than at NIR wavelengths where both interfaced ends are often made of silica glass. Furthermore, polarization becomes important in high index contrast systems, which exhibit a strong waveguide birefringence and polarization dependence in mode-field profile.

*For the development of pigtailed MIR PICs, a critical challenge is to optimize the mode matching between fibers and the integrated optic waveguides, at least for specific spectral sub-bands. One additional challenge is to identify a suitable type of IR-transparent glue, and characterize its thermal behavior, in order to connect the fibers to the chips.*

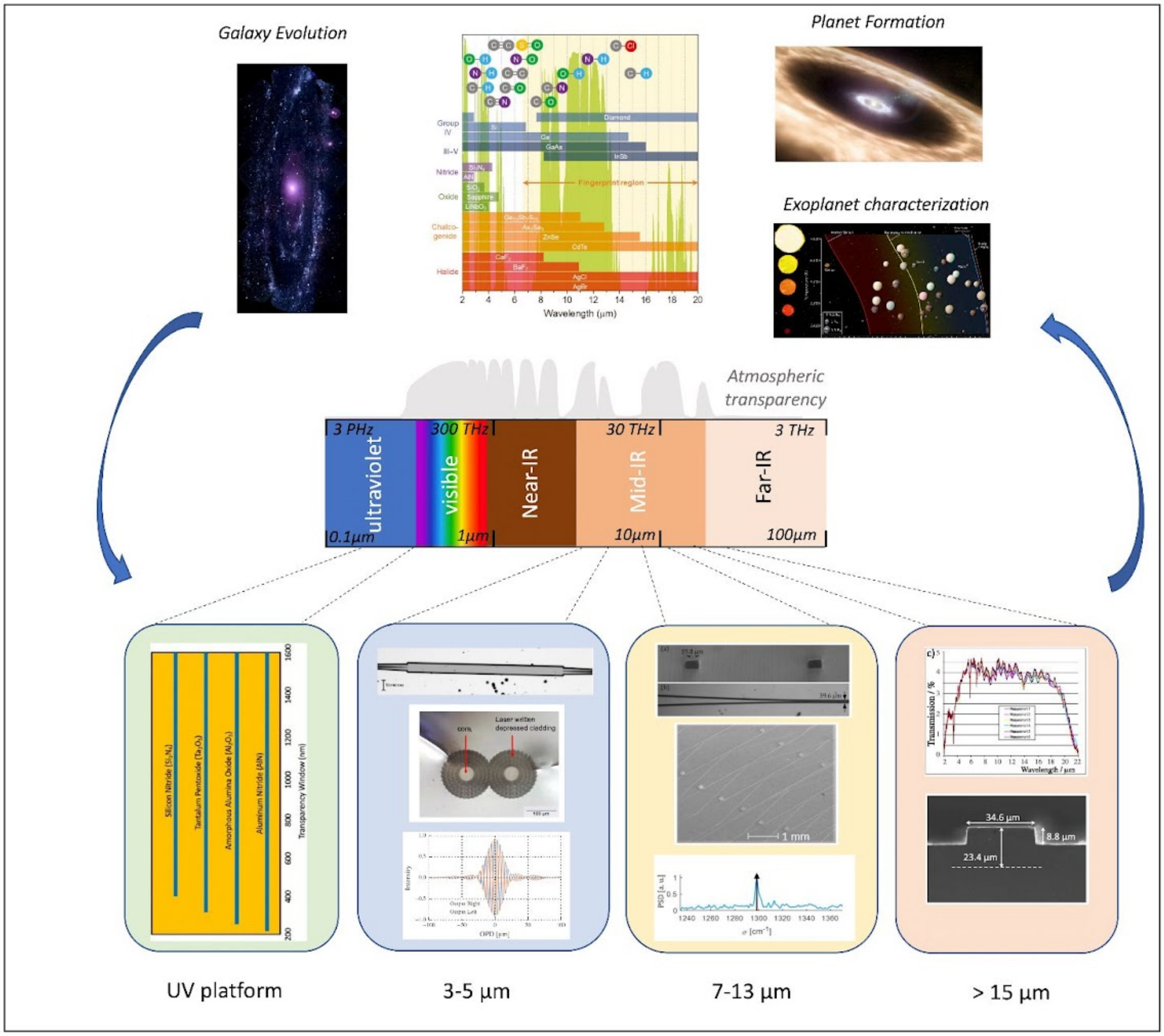

Figure 1: Overview image of the current landscape of astrophotonics in the ultraviolet and mid-infrared range. The central panel illustrates the wavelength ranges of interest, coupled to the atmospheric transparency. The top panel illustrates possible science themes associated with the technological development discussed in this paper: (Left) Image of M31 in the ultraviolet; credit NASA. - (Center) overview of the fingerprint region in the mid-infrared with overplotted the associated potential material platforms [Lin 2017]. - (Right) Artist view of a planet forming disc; credit ESO / Kepler population of terrestrial exoplanet [Kane 2016]. The bottom panel illustrates the current yield of astrophotonics. UV panel: table illustrating the possible technologies for UV astrophotonics for future first prototypes [Blumenthal 2020]; Panel 3-5 µm: *Top:* Micrograph of a chalcogenide-GLS multimode interferometer fabricated by lithographic techniques [Kenchington Goldsmith 2017]; *Center*: Image of a ZBLAN coupler's end face [Gross 2015]; *Bottom*: MIR interferogram obtained with a ULI coupler in GLS glass delivering high interferometric contrast [Tepper 2017]; Panel 7-13 µm: *Top*: view of the front-face and coupling region of a ULI coupler in GeAsSe chalcogenide glass [Butcher 2018]; *Center*: SEM picture of an MZI array using the Si:Ge platform for integrated Fourier Transform spectroscopy at 8 µm. Bottom: reconstructed spectrum with a resolution R~100 at 1300 cm-1 [Montesinos-Ballester 2019]; Panel >15µm: SEM picture (bottom) of a chalcogenide rib waveguide showing continuous transparency from 4 to 20 µm [Vigreux 2011].

## Advances in Science and Technology to Meet Challenges

MIR integrated photonics have focused on wavelengths around or below 10 µm. The 10-20 µm range, of interest for space missions, remains largely unexplored. The propagation losses are expected to be larger than a few dB/cm, at least in the single-mode regime [Vigreux 2015]. While single-mode Hollow Core Waveguides (HWG) were proposed to circumvent the limitation of dielectric materials, they were found to strongly underperform [Abel-Tibérini 2007] as has the alternative InGaAs/InP technology [Jung 2019]. It has been estimated that a Germanium-Tin-Silicon platform could cover the 3-19 µm range with ~2 dB/cm losses [Soref 2006]. However, this remains to be proven experimentally.

*The optimization of a 10-20 µm platform for astronomy could be expensive and would need to be supported by a long-term technology program justified by a major astrophysical initiative.*

A critical aspect for MIR astrophotonics is the testing of operation at low or cryogenic temperatures, required to minimize thermal background. The GRAVITY experience has proven the operation of pigtailed germanium-doped silica beam combiners at -80˚C [Perraut 2018], but MIR photonics may have to operate at lower temperature as, for instance, for the NOTT instrument.

UV PICs development is likely to include further progress towards robust, low loss, higher-Q devices. With Rayleigh scattering increasing into the blue/UV, improvements such as those seen in CMOS foundry-based devices with low-roughness waveguides patterned on $Si_3N_4$ (Morin 2021) are needed. Recent developments include the fabrication of UV ring resonators (Shin 2021), UV arrayed waveguide gratings (Hu 2021) with high bandgap photonic materials coupled with UV/VIS transmitting substrates.

Finally, radiation hardening will become an increasingly important challenge in the mid-term future for space applications of astrophotonics, regardless of the frequency range. This is also relevant for other fields such as satellite-based quantum technologies where efforts in this direction have already been made [Piacentini 2021], targeting predominantly the VIS/NIR domain. In the future, since radiation hardness strongly depends on the material, specific tests for MIR integrated photonics will have to be implemented.

## Concluding Remarks

We have discussed the opportunities to extend the operation range of astrophotonics beyond the traditional VIS/NIR range. At the wavelength extremes of UV and the MIR, the choice of material and compatible waveguide fabrication technologies is limited. Generally, MIR waveguides exhibit larger losses and have to operate at extreme temperatures, which adds to the challenges. However, MIR photonics (waveguide and advanced devices such as on-chip spectrometers) have progressed significantly over the past 10 years, driven by other applications such as sensing, which the astrophotonics community can benefit from.

## Acknowledgements

*L.L. acknowledges support from the German Ministry of Education and Research (BMBF, ALSI project 05A14PK2) and the DFG (NAIR project, grant 326946494) in the field of mid-infrared astrophotoncis.*

# 9 | Integral Field Units for THz Astrophysics

Akira Endo[1], Charles M. Bradford[2], Ritoban Basu Thakur[2,3] and Edward J. Wollack[4]
**[1] Electrical Engineering, Mathematics and Computer Science, Delft University of Technology, Delft, the Netherlands**
**[2] Jet Propulsion Laboratory, California Institute of Technology, Pasadena, CA, USA**
**[3] Department of Physics, California Institute of Technology, Pasadena, CA, USA**
**[4] NASA Goddard Space Flight Center, Greenbelt, MD, USA**

## Status

Superconducting astrophotonics will revolutionize submillimeter / millimeter astronomy (f~0.1–1 THz, λ~0.3–3 mm) by enabling massive integral field units (IFUs), that can simultaneously measure the wideband spectrum of all points in a 2-D image. A wide-field and wideband spectrometer in the far-infrared offers a sensitive and unique 3-dimensional approach to topics which bridge cosmology and astrophysics, in particular the earliest galaxies as they are born in the Reionization epoch (Chluba et al. 2019 , Chlube and Yacine 2016). Mm/submm wave-IFUs have yet to be realized because they are out of reach of existing technology such as heterodyne receivers, direct-detection cameras, and quasioptical spectrometers (Stacey 2011). The physical wavelength scale makes it impractical to design quasioptical IFUs with many spatial pixels (spaxels) and spectral channels (voxels). The photonic architecture addresses this challenge by shrinking the optical path into equivalent guided waves along single-mode superconducting transmission lines and metamaterial structures (Mirzaei et al. 2020). The advantages of superconducting astrophotonic instruments (compared to their quasioptical equivalents (Stacey 2011)) are mechanism-free, compact, lightweight, and ultimately scalable designs. The utility of the superconducting photonic devices goes hand-in-hand with the rapid advances in far-IR/submm/mm detector array format and sensitivity, particularly for kinetic inductance detectors (KIDs) (Day et al. 2003). A full discussion of KIDs is beyond the scope here, but we refer the interested reader to articles by Zmuidzinas, 2012, Baselmans et al. 2017, and references therein. KID-based instruments with total pixel counts of $10^5$ or more are now conceivable, and photonic techniques are required to make optimal use of these large formats. For related developments at optical wavelength see Section 24 (Mazin, Walter and Zou).

The development of superconducting astrophotonics to date has been largely focused on ultra-wideband filterbank spectrometers (Figure 1a) or pseudo-grating spectrometers (Figure 1b), which sort the light from a single sky mode into an array of spectral channels for detection. Examples of these implementations can be found in Endo et al. 2019 and Cataldo et al. 2014, respectively. These dispersive approaches offer high sensitivity to individual sources, and are being used for detecting [CII] and CO lines in dusty star-forming galaxies at high redshift (Bigiel et al. 2008). These spectrometers typically use a guiding structure created with patterned superconducting films (often niobium or NbTiN) and a thin dielectric (typically silicon nitride or silicon). This architecture allows operation up to the Nb (or NbTiN) gap frequency of 690 GHz (1100 GHz), with spectral resolution generally limited by the loss in the dielectric (currently $R = f/\Delta f$ ~ 300-500). Examples of on-chip filterbank spectrometers are Deep Spectroscopic High-redshift Mapper (DESHIMA) (Endo et al. 2019) that has seen astronomical first light on the ASTE (Atacama Submillimeter Telescope Experiment) 10-m, and SuperSpec (Redford et al. 2018  Karkare et al. 2020) which will soon be deployed on the 50-m Large Millimeter Telescope (LMT).  Both use microstrip lines, with half-wave filters to define channel bandapsses.  The Micro-Spec device employs a phased-array approach in which interference creates spectral channels via a lithographically patterned 2-D multi-mode region (Cataldo et al. 2014, Switzer et al. 2021).

Another complementary approach which is emerging is an on-chip Fourier-transform spectrometer (FTS), called Superconducting On-chip Fourier Transform Spectrometers (SOFTS), which employs an electrically tunable wave speed to modulate the phase delay (Basu Thakur et al. 2020). See Figure 1c for a schematic representation and operational principle. Each spatial mode of the spectrum is encoded in time and detected by a pair of detectors, enabling improved spectral recovery (Basu Thakur et al. 2021). This allows spatial multiplexing at the expense of sensitivity in each spectral channel (due to the full band photon load on the detector).  Another virtue of the SOFTS is that, like a classical FTS, the working resolving power is selectable in operation by simply adjusting the amount of phase delay used.  SOFTS prototypes have been demonstrated at 10 GHz, with expectations for adaptation to higher frequencies.

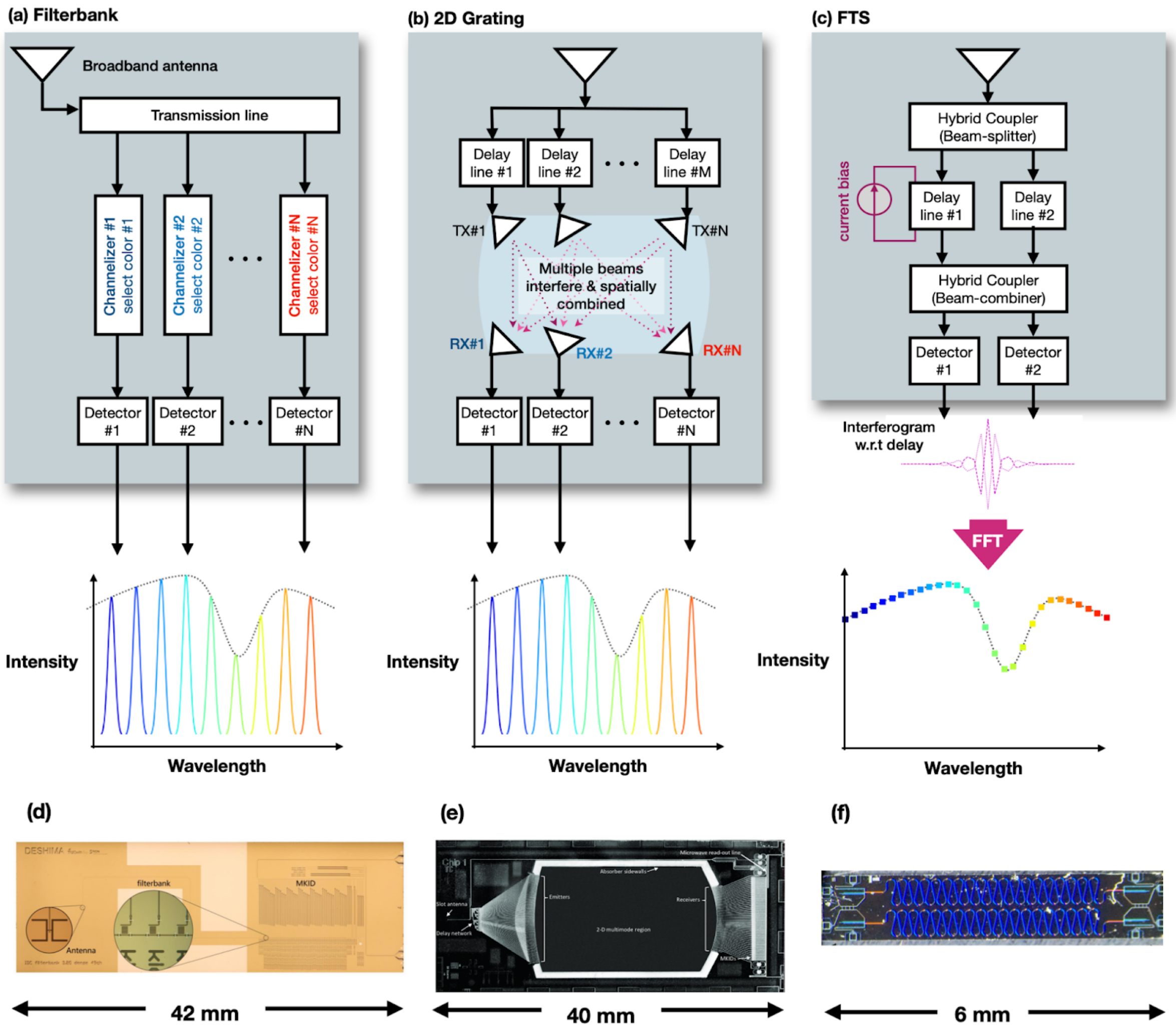

***Figure 1 -*** *The top row shows the working principles of the superconducting photonics circuits of the (a) filterbank, (b) quasi-grating, and (c) FTS spectrometers. The bottom row shows photographs of examples of corresponding on-chip implementations: (d) Endo et al. 2019, (e) Catalodo et al. 2014, (f) Basu Thakur et al. 2021.*

## Current and Future Challenges

The first generation of on-chip spectrometers (DESHIMA, SuperSpec) are offering photon-noise-limited performance, and others are coming soon (e.g., EXperiment for Cryogenic Large-Aperture Intensity Mapping (EXCLAIM) (Switzer et al. 2021), South Pole Telescope Summertime Line Intensity Mapper (SPT-SLIM) (Karkare et al. 2021), SOFTS (Basu Thakur et al. 2021), Cambridge Emission Line Surveyor (CAMELS) (Thomas et al. 2014)). These demonstrations pave the way for larger, more powerful instruments. In considering approaches to the next generation instruments for wide-field science, a first-order figure of merit is simply the number of detectors that can be fielded (see Figure 2). Whether they are multiplexed primarily spectrally as in the dispersive filter banks and pseudo-gratings, or spatially as in the SOFTS, the net spatial-spectral survey speed scales as the number of detectors, provided the detectors are photon-noise limited at their respective backgrounds (Stacey 2011). While the legacy submm telescopes have modest field-of-view (e.g., Caltech Submillimeter Observatory (CSO) /Leighton Chajnantor Telescope (LCT) (Vial et al., 2020) with 15 arcmin giving 170 feeds at 220 GHz), new wide-field telescopes under development offer ample etendu (e.g., Fred Young Submillimeter Telescope (FYST) which could feed some 10,000 devices at 220 GHz). As a result, in the near term the SOFTS and dispersive approaches could offer comparable survey speed. Assuming continued development in KID format to $10^5$ pixels, the dispersive approach offers the ability to field more total detectors, and has the potential to become the preferred solution.

In any case, scaling to larger and dual-polarization array formats presents packaging challenges. For full spatial packing, ultimately the spectrometers must be either miniaturized greatly (so that each spectrometer unit cell is only a few wavelengths on a side and they can be positioned between the antennas in focal plane), or they need to be arrayed on chips which extend behind the telescope focal plane. The latter approach will require new antenna / coupling approaches.

## Advances in Science and Technology to Meet Challenges

We observe that there are compelling scientific opportunities to be realized by pushing astrophotonic spectrometers to higher resolving power ($R > 10^3$), higher frequency (> 1 THz) and total sensitivity. See Figure 2 for a graphical summary of the achieved total spatial pixels (spaxels) and spectral channels (voxels) number. Expanding upon these implementations will require careful attention to dielectric material loss, potentially higher-gap superconductor materials, and either improving lithographic precision (down to tens of nanometers: Thoen et al. 2022) or development of new designs. The challenges in the photonic elements (filters, delay lines, etc.) reside in the direction of higher frequencies and higher spectral resolution, combined with a high optical efficiency on a system level. Because the resolving power scales with pathlength in the medium, higher frequency resolution requires low-loss transmission lines in the mm/submm frequency range. Finally, combining the ultra-wideband (octave or more) bandwidth with many spaxels that fill a wide field-of-view poses a severe challenge for the optics design.

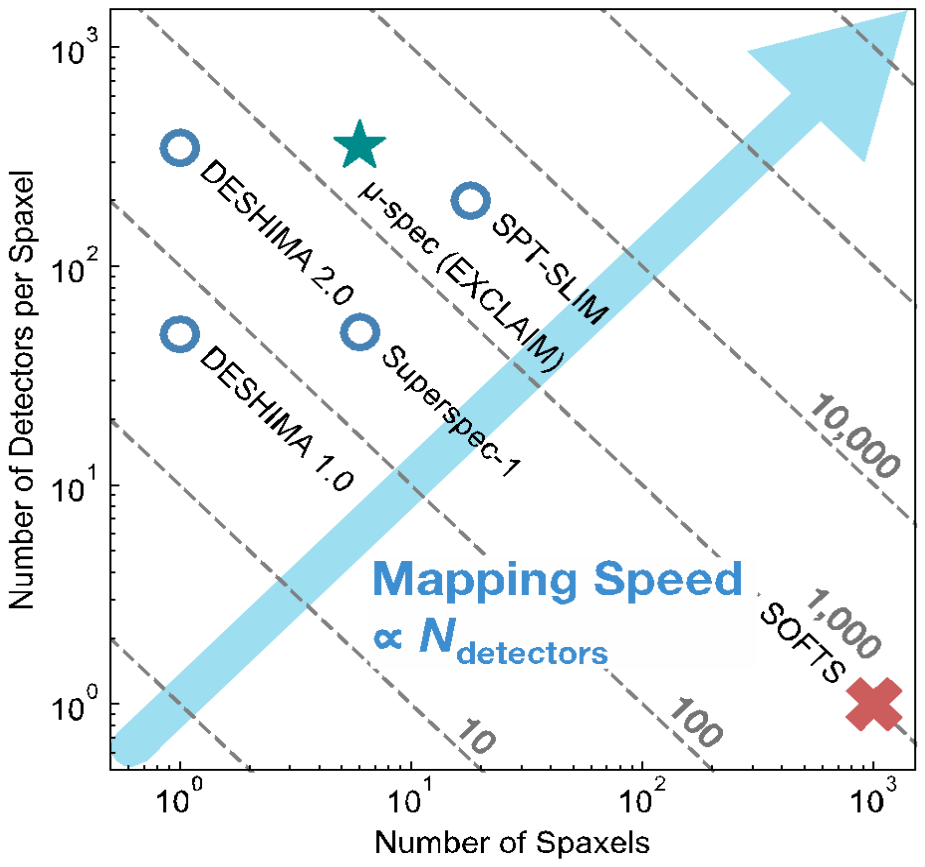

***Figure 2 -*** *Current astrophotonic (imaging) spectrometers. The dashed lines represent constant total number of detectors $N_{detectors}$, which is the product of the number of spaxels (horizontal axis) and the number of detectors per spaxel (vertical axis). The mapping speed is proportional to $N_{detectors}$, regardless of whether the photonic device is a filterbank (○), an FTS (✖), or a 2D grating (★).*

To improve the frequency resolution and optical efficiency of the photonic elements, dielectric materials with low losses are being investigated. Because coplanar waveguides suffer from large radiation losses, microstrip lines are a preferred option at mm/submm wavelengths. One possibility is to take a crystalline Si membrane from a SOI wafer and process on both sides (Mirzaei et al. 2020). Another approach is to deposit a dielectric film, which is more convenient for processing but has the disadvantage that amorphous dielectrics (e.g., SiNx with $Q_{loss\sim}1\times10^3$ at 235 GHz (Hailey-Dunsheath et al. 2016), Si with $Q_{loss\sim}3\times10^4$ at 350 GHz (Hähnle et al. 2021), and SiCx with $Q_{loss\sim}1\times10^4$ at 270 GHz (Buijtendorp et al. 2022), at cryogenic temperatures) exhibit larger losses than crystals. It is notable that the physical loss mechanisms of amorphous and poly-crystalline dielectrics in the mm/submm are not well understood at cryogenic temperatures (McRae et al. 2020) and present an opportunity for potential improved performance.

To push beyond the gap frequency of NbTiN (1.1 THz), higher gap superconductors (e.g., $MgB_2$) must be used. Dielectric losses also increase with frequency (i.e., due to the presence of IR absorption bands wings) and may represent the next limiting factor (Cataldo and Wollack 2016). In scaling to larger superconducting integrated photonic device formats control over transmission line losses, inter-element crosstalk (Noroozaian et al. 2012) and unintended modal coupling (Yates et al. 2017) will become more pressing design considerations. These challenges are seen as manageable, however, will ultimately influence the device topology, layout, and packaging strategies.

An exciting platform for superconducting astrophotonics is space observatories. Here it is encouraging that KIDs have recently shown excellent Noise Equivalent Power (NEP) down to $3\times10^{-20}$ W/Hz$^{0.5}$(Baselmans et al. 2022). The next challenge is to integrate this type of KID in an astrophotonic spectrometer. For space applications, mitigation of cosmic rays is also important (Karatsu et al. 2019). Recent advances in wideband planar lens-antennas can be leveraged to meet the need for wideband optical coupling. Using these large-format IFUs strategies to achieve wide- band and field-of-view optical systems represents an additional challenge.

### Concluding Remarks

Similar to astrophotonic implementations in the optical, superconducting IFUs employ low-loss guided-wave structures, but also utilize the unique properties of superconductors to enable a variety of functions. It is a young and growing field of research, with multiple architectures being explored in parallel and new astronomical applications being identified (e.g., line intensity mapping (Switzer et al. 2021, Karkare et al. 2021), recombination studies (Chluba et al. 2019), etc.). The aim is to pioneer 3D-volumic astronomy in the mm/submm band, uncovering the large-scale structure of the Universe in the first few billion years using multiple tracers. We encourage the participation of experts from the related fields, including astronomy, solid-state physics, electrical engineering, and data science.

**Acknowledgements**

*A portion of this research was carried out at the Jet Propulsion Laboratory, California Institute of Technology, under a contract with the National Aeronautics and Space Administration (80NM0018D0004). A.E. was supported by the European Union (ERC Consolidator Grant No. 101043486 TIFUUN). Views and opinions expressed are however those of the author(s) only and do not necessarily reflect those of the European Union or the European Research Council Executive Agency. Neither the European Union nor the granting authority can be held responsible for them.*

# 10 | Astrophotonic Spectral Filtering

Simon Ellis[1], Sylvain Veilleux[2] and Joss Bland-Hawthorn[3]
**[1] Australian Astronomical Optics, Macquarie University, NSW, Australia**
**[2] Department of Astronomy and Joint Space-Science Institute, University of Maryland, College Park, MD, USA**
**[3] Sydney Institute for Astronomy (SIfA), School of Physics, The University of Sydney, Australia**

## Status

The frontiers of astrophysical knowledge inevitably lie at the limits of detection of the most advanced telescopes and instruments. To advance beyond these limits requires increasing the signal-to-noise of detections, via (i) larger telescopes, (ii) more efficient instruments, (iii) isolating the signal and reducing the sources of noise, or (iv) probing new properties of light, either different wavelengths or higher order moments (e.g. polarization). Astrophotonic filters are aimed at (iii), either by selectively filtering and measuring the signal of interest, or by removing the sources of noise.

Filters (absorptive, interference, holographic, etc.) have long been a part of observational astronomy, but astrophotonic filters (i.e. filters embedded into waveguides) offer much more complex filtering with the ability to isolate multiple (>100) specific signals in a single waveguide. The first astrophotonic filters were fiber Bragg gratings (FBGs), first developed in 2004 [1] to selectively filter the emission from atmospheric OH lines to reduce the background noise in near-infrared spectroscopy. FBGs have their origins in telecommunications, where typically one fiber can filter a single wavelength. A breakthrough in their application for astronomy was the ability to print *aperiodic* FBGs which can filter >100 wavelengths in a single fiber [2,3]. Since FBGs require single-mode behavior, the photonic lantern (see chapters 3 and 4) was developed in order to incorporate FBGs into the multimode fibers necessary for efficient coupling with seeing-limited astronomical telescopes. Further development and refinement of FBGs led to on-sky experiments, GNOSIS [4,5] and PRAXIS [6], which have demonstrated their efficacy, suppressing 103 OH doublets from 1.47 - 1.7 μm by factors of up to 40 dB, and resulting in a reduction of the integrated background by a factor 9. PRAXIS achieved an end-to-end throughput of ≈ 18 % [6].

To date, single mode FBGs in photonic lanterns are the only astrophotonic filters to have been incorporated into an astronomical instrument, but there has been considerable attention given to other platforms and other applications. Table 1 lists the different platforms currently under development, as well as an estimate of their TRL (using NASA's TRL scale of 1-9) and comments on their advantages and challenges. These will be discussed more fully in the next section.

A different use for astrophotonic filters is to isolate the desired signal, for example using the reflected light in an FBG [7] or the drop port of an add-drop ring-resonator [8,9] or cascaded multi-mode interferometers (MMI) [10]. Not only can a specific signal be selected this way, but multiple features can be selected and combined *photonically* in the same waveguide, with no associated electronic noise penalty. For example, exoplanets could be detected by selecting and combining multiple stellar spectral lines (see chapter 11), and monitoring the rise and fall in intensity (after photometric calibration using e.g. an 90/10 splitter to monitor the variable coupling due to changes of seeing etc.) as the spectral features are Doppler shifted into and out of the filter bandpass. This technique could potentially dispense with the need for Doppler-spectroscopy, and the subsequent loss of signal-to-noise, as well as the requirement to observe parts of the spectrum not useful to the analysis, thereby reducing the need for expensive large 2D detector arrays.

***Table 1 -*** *Techniques of astrophotonic filters, along with an estimate of their Technology Readiness Level (TRL) and comments on their advantages (+) and challenges (-).*

| Technique | TRL | Comments |
|---|---|---|
| **Fibers** | | |
| FBGs | 8 | + on-sky demonstration [4,5,6]<br>+ complex phase masks for routine manufacture<br>- single-mode usage only<br>- backward cladding-mode coupling<br>- inexpedient to scale to large numbers of modes |
| Direct write FBGs | 4 | + does not require phase masks and photosensitized fibers like standard FBGs [11]<br>- single-mode usage only<br>- not yet demonstrated for complex filtering (e.g. 100 notches) |
| MCFBGs | 3 | + multi-mode usage<br>- uniformity across the multiple cores [12] |
| Ultra-deep ultra-narrow filters (UDUN) | 3 | + very deep (> 100 dB) notches<br>+ very narrow (< 100 MHz)<br>- currently single notch only |
| **Direct write waveguides** | | |
| Direct write Bragg gratings | 3 | + integrated structure reduces # of interfaces [12,13]<br>- depth and precision of notch profiles<br>- not yet demonstrated for complex filtering (e.g. 100 notches) |
| **Integrated photonic circuits** | | |
| Complex waveguide Bragg gratings (CWBGs) | 4 | + compact design reduces stitching errors [14]<br>+ end-to-end throughput comparable to that of FBGs when used in combination with fiber-waveguide couplers with adiabatic tapers [14,16]<br>+ ease of integration with other photonic components on a chip<br>+ thermo-mechanical stability for precision applications<br>+ lithographic printing enables replication<br>+ can be incorporated with MMIs to provide add-drop filtering [9] |
| | | - single-mode usage only<br>- backward cladding-mode coupling [14, 17]<br>- polarization dependence |
| Ring resonators | 3 | + optimized for single notch filtering [8,9]<br>+ tunable using on-chip micro-heaters or stress-optic actuators<br>+ lithographic printing enables replication<br>- scalability to complex filtering is not yet demonstrated<br>- small FSR limits bandpass<br>- single-mode usage only<br>- polarization dependence |

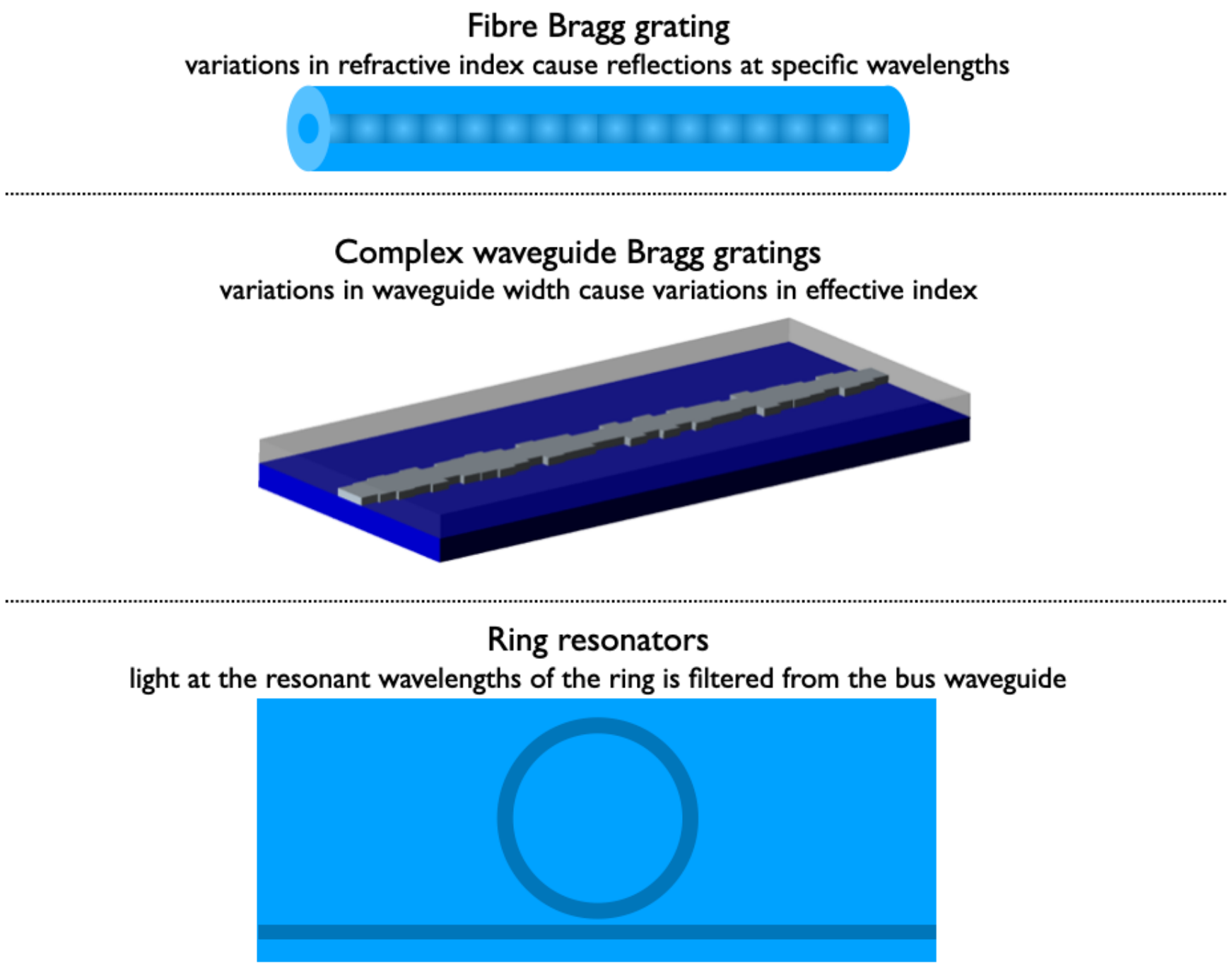

***Figure 1*** *- Sketch of the different methods of astrophotonic filtering. The top panel shows a fiber Bragg grating, in which there is a variation in the refractive index of the core, causing light of specific wavelengths to be rejected. The middle panel shows a complex waveguide Bragg grating, which works on the same principle, but the variations in effective index are achieved by varying the waveguide width. The bottom panel shows a ring resonator, in which light with a wavelength which is a unit fraction of the optical path length of the ring is filtered from the input bus waveguide. These same techniques can be applied to different platforms, e.g. fibers, planar waveguides, direct write ultrafast laser inscription etc., with different manufacturing techniques.*

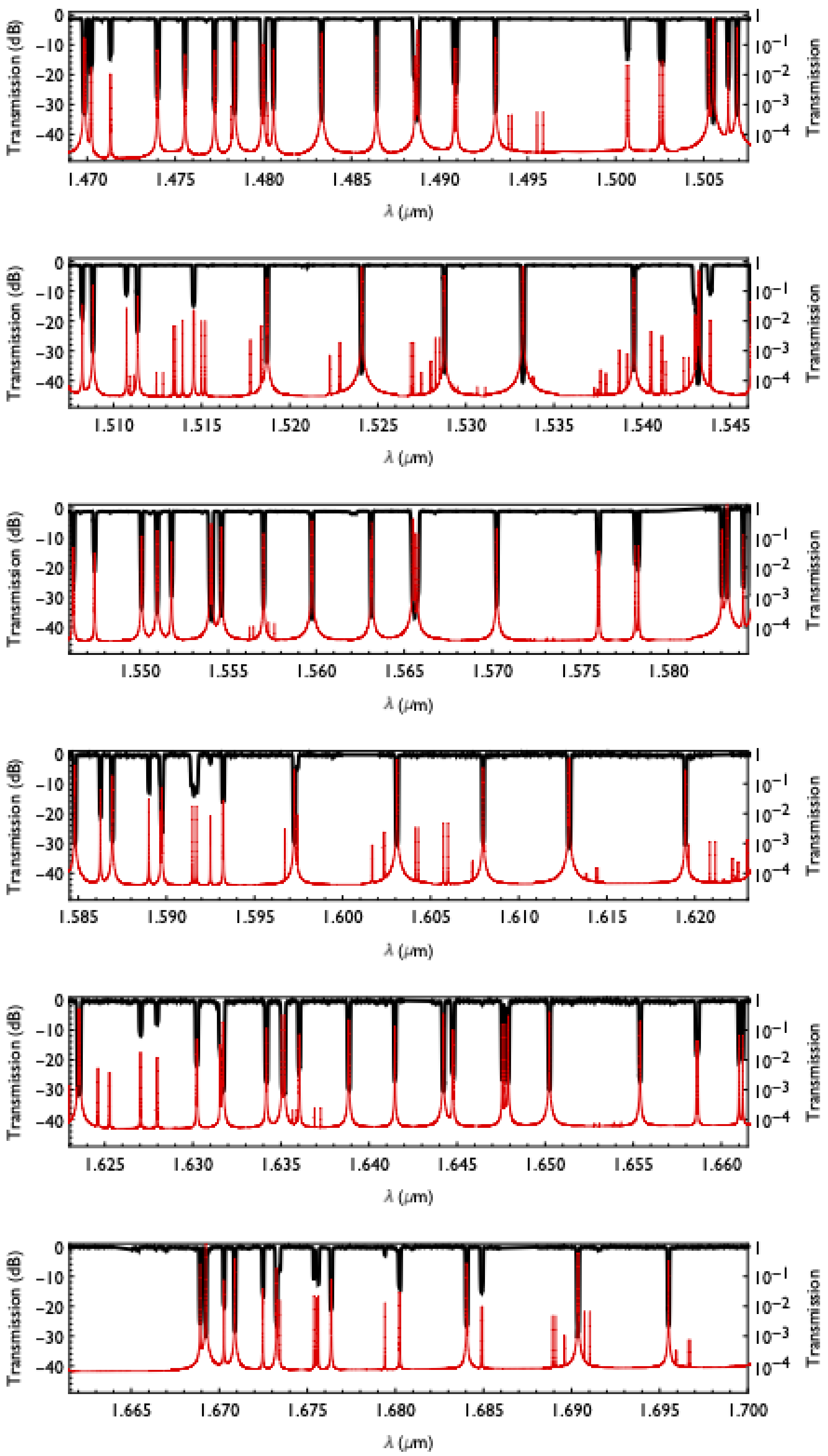

***Figure 2 -*** *The measured transmission (black) of the FBGs used in the GNOSIS and PRAXIS experiments [4], which consists of two FBGs in series covering 1.47 - 1.58 μm and 1.58 - 1.7 μm respectively and filtering 103 OH doublets in total. Also shown is a model of the night sky OH spectrum (red). Note that the FBG response matches the wavelengths, line strength, and OH doublet separation perfectly. This remains the state-of-the-art in astrophotonic filtering, although significant progress is being made with other technologies, see* ***Table 1****.*

## Current and Future Challenges

The main challenges for astrophotonic filtering in general are (i) scaling the technologies to increase the number of modes, and thereby accessible field-of-view and multiplex, (ii) extending the wavelength range of the filters, (iii) integrating the technologies with other photonic components, and with astronomical instruments. Astrophotonic filtering with asymmetric waveguides has the added challenge (iv) of being polarization (TE/TM) dependent. Specific solutions also have their own particular challenges. We discuss these challenges in turn.

(i) *Multi-mode filtering.* Photonic filtering is fundamentally a single-mode process, since it requires precise control of the phase of the light propagating in the waveguide to produce interference effects, whereas the modes of a multimode waveguide will necessarily propagate with different phase velocities making this control impossible. In the diffraction-limited case, e.g. when using adaptive optics with high performance correction, it is possible to feed single mode waveguides directly [21]. However, the input beam in seeing-limited astronomy is multi-mode and therefore has to be split into single modes without incurring significant losses. Furthermore, this requires high numbers of replicated devices, all of which must achieve nearly identical performance (requiring high yield fabrication, or low cost). These must then be packaged into the astronomical instrument. For multiplexed observations typical of fiber optic spectroscopy this problem is likewise multiplied.

(ii) *Broad-band filtering.* To date, most astrophotonic filters have been centered on the telecom C band (1530 - 1565 nm), in order to exploit the R&D and fabrication processes already existing. A key challenge is to extend these processes to other wavelength ranges - ideally across the full UV-visible-NIR spectrum available for use with ground-based optical telescopes. The particular challenges in achieving this are very platform specific. However, some common difficulties are the single-mode cut-off limiting the bandwidth of a specific waveguide; the transmission of different types of waveguides (0.3 μm < $SiO_2$ < 1.8 μm; SOI > 1 μm) requiring the use of more exotic or doped glasses; the higher mode count at shorter wavelengths requiring narrower waveguides; the size of required structures scales with wavelength range (e.g. longer Bragg gratings, higher numbers of ring resonators); limitations in the free spectral range (FSR) of periodic structures; precision of UV photolithography for UV/visible filtering; scattering errors from waveguide sidewall irregularities; stitching errors between e-beam fields. If accessing the reflected light, as for bandpass filters, then any optical circulators or coupling to drop-port waveguides also need to be broad-band and low-loss.

(iii) *Integration.* Integration has two equivalent challenges, coupling light from the telescope into the waveguides, and coupling light from the waveguides into the instrument. In the first place, there is the difficulty of coupling light into single mode fibers already discussed in (i); see also chapter 2. This problem is exacerbated when using high-index contrast lithographic waveguides. For example, SOI may have a ~300 x 400 nm profile, leading to a mode-mismatch with SMF-28 causing ~13 dB of loss. For this reason, other lower contrast waveguides such as $Si_3N_4$ or silica-on-silicon are often preferred. It is usually preferable to minimize the number of channels imaged onto the detector to avoid spreading the light over an unnecessarily large number of pixels increasing the electronic noise. If a MMF feed is used then it will be necessary to recombine the multiple single-mode waveguides back into a multimode feed. If using an extreme-AO feed this problem can be avoided with direct coupling into SMF at reasonable (50 %) efficiency.

(iv) *Polarization dependence of asymmetric waveguides.* Waveguides on wafers are inherently polarization dependent, even when the cross-section of the waveguides is symmetric (square). Although there are some polarization independent waveguides these can only work over a relatively narrow wavelength range.

**Advances in Science and Technology to Meet Challenges**

We now discuss the possible solutions and necessary advances in technology needed to address the challenges described in the previous section.

*(i) Multi-mode filtering.* Photonic lanterns were invented as a direct solution to this problem [18], allowing the conversion between multimode and single-mode fiber and vice versa. Direct-write waveguide versions of this concept have also been proven [13,14]. Scaling photonic lanterns to very high numbers of modes is difficult, and 'divide-and-conquer' methods [19] may need to be employed. Beyond this, multicore fibers offer an attractive solution, allowing the inscription of multiple Bragg gratings at the same time. Devices with Bragg gratings across 121 cores have been demonstrated, but so far all suffer from a variation in Bragg wavelength as a function of the distance from the center of the core to the center of the fiber, which is not fully understood [20]. Direct-write photonic lanterns may offer a solution to this dilemma since then the Bragg gratings are written independently into each waveguide. A totally different approach is the use of extreme AO correction to feed single mode waveguides at the diffraction limit [21,22], with a theoretical coupling efficiency of ~80%, with the best coupling-efficiency to date of > 40% [21]. This technique will be even more apposite in the future era of ELTs with very high AO correction, and very small diffraction limited PSFs, see for example the MODHIS instrument. [23].

(ii) *Broad-band filtering*. Part of the solution to this challenge rests in adapting current techniques to new materials, e.g. using fluoride fiber to access wavelengths > 1.7 μm, or chalcogenide glasses to access wavelengths up to 20 μm, or UV-fused silica to extend to shorter wavelengths (> 195 nm). New recipes for fabricating filters in these different glasses will need to be perfected. Even with the adoption of new materials, the wavelength range will be limited by mode-count, detector sensitivity and FSR, possibly requiring multiple divisions into different wavebands employing differently optimized filters. To extend lithographic filters to shorter wavelengths will require higher precision next-generation nanolithography (e.g. EUV, X-ray, e-beam, focused ion-beam, nanoimprint lithography) to create nm-scale structures and reduce scattering off sidewall irregularities. Losses at wavelengths > 1 μm due to N-H and O-H absorption features in $Si_3N_4$ waveguides can be reduced by annealing *before* patterning to provide wider passbands (0.4 to 2.4 μm) [15].

(iii) *Integration*. There are several proposed solutions for matching fiber and waveguide modes. Lensed fibers can reduce the mode field diameter to < 3 μm, while inverted taper waveguides can expand the waveguide modes to 3.5 μm. These solutions are now becoming mature, with COTS interposer chips available which provide an efficient interface between the chip and a fiber array (see chapter 22). Moreover, these can work over ultra-broad bands [24], unlike some other telecom devices. Other solutions such as grating couplers are wavelength dependent and therefore operate over limited passbands.

The coupling from the waveguide filters into instruments could be significantly improved with customized detectors. For example, directly bonding photonic devices to detectors with a small pixel pitch that properly samples the output of the photonic device would eliminate the free-space interfaces between the filters and detector and make devices more robust and stable; likewise custom formats of linear arrays or rectangular pixels would allow for more efficient integration in some case; energy resolving sensors (e.g. MKIDS; see chapters 23 and 24) could help to alleviate the low FSR of certain types of filters such as ring-resonators.

*(iv) Polarization dependence of asymmetric waveguides.* It is extremely challenging to make a polarization-independent waveguide that can work over a wide wavelength range. A more promising approach is to exploit the modular and replicable nature of lithographic photonics and use broad-band polarization splitters based on MMIs [25] or directional couplers [26], or to avoid the problem with fiber-based solutions or direct-write ultrafast laser inscription of symmetric waveguides in fused silica.

In addition to the specific challenges described above there is a general need to streamline the fabrication of science-ready astrophotonic devices through dedicated access to facilities and expertise in fabrication, characterization, and packaging [27]. This access is becoming easier due to the availability of commercial foundries, especially those offering multi-project wafer fabrication services (MPW) which significantly reduce the cost for making and testing prototypes.

**Concluding Remarks**

Astrophotonic filters have already been shown to have significant promise for astronomy, most notably for near-infrared spectroscopy using fiber Bragg gratings to filter atmospheric emission lines. So far the science benefits of photonic filters have only been partially realized, and there are many science cases other than OH suppression which can profit from the complex filtering enabled by the devices discussed, including Doppler planet searches, emission line diagnostics, and wavelength calibration. Beyond this, there are several promising technology platforms being investigated, including novel ways to inscribe Bragg gratings in multicore fibers and direct-write waveguides which will help to enable a scaling-up of this technology for larger fields-of-view and higher multiplexing capability. A significant, but challenging, development is the integration of photonic filters on lithographic photonic circuits. This has the potential for extremely miniaturized devices, which are modular and easily integratable with other photonic devices such as arrayed-waveguide gratings (Chapters 6 and 7) and photonic beam combiners (Chaps. 15, 16 and 19) with the ultimate aim of fully photonic instruments.

## 11 | Measure what you need with Optical Correlation Spectroscopy

Sebastiaan Y. Haffert[1] and Peter Tuthill[2]
**[1] Steward Observatory, University of Arizona, Tucson, AZ, USA**
**[2] Sydney Institute for Astronomy (SIfA), School of Physics, The University of Sydney, Australia**

### Status

Technological advances driving ever larger telescope apertures have also increased the size of the instruments[1]. Competitive astrophysics demands ever higher spectral resolution, more simultaneous spatial points and larger bandwidths. In particular, high spectral resolving power is needed for identification of atomic and molecular gasses in astronomical objects as their spectral fingerprints may be distinguished at high resolution. Spectral differentiation between species is now used as a successful post-processing technique to search for exoplanets with direct imaging instruments[2,3]. Applying such techniques to large fields would allow us to map the chemical compositions of many objects simultaneously, like stars in dense stellar clusters, or extended objects like the Orion Nebula. However, the competing requirements are difficult to fulfill for integral field units and multi-object spectrographs that require many simultaneously sampled spatial points. Conventional spectrographs - which must be built with real-world limitations on the size of the optics, number of pixels, and data rates - are therefore forced into compromises between spatial sampling, spectral resolving power and bandwidth. Optical cross-correlation offers a pathway to avoid this compromise based on the realization that acquiring the full spectrum may not be necessary for all science applications. We may require, say, only a measurement of the elemental composition, without details of all the lines present. Distinguishing information at high spectral resolution need not mandate the recovery of every single spectral channel.

Such measurements can be achieved through optical correlation spectroscopy (OCS) in which light passes through a spectral filter that partially mimics the spectrum of the element of interest[4] (see Figure 1). The simplest example of a spectral filter might be a template tailored to match the response of one specific molecule, whose presence is then indicated by measuring the total intensity passing through. Most light passes through if the same lines are present in the analyzed spectrum and the filter. However, a lower intensity is measured if there are different lines present in the filter and the spectrum. This means that a single intensity measurement is sensitive to the presence or absence of specific species. Obtaining a high spectral specificity in a single sensor element breaks the trade-off above between resolution, spectral bandwidth, and density of spatial sampling. While a single intensity measurement is enough in theory, flux variations require two measurements (with and without the spectral filter) to normalize the incoming amount of light.

There are several different ways to implement an OCS. The earliest methods used gas-cells that contained the gas of interest[4]. These may be considered ideal spectral filters as they contain exactly all the lines of interest. A downside to the gas-cell correlators is that not all elements can be contained in a cell at astronomical temperatures and/or pressures. A tunable optical filter that replaces the gas-cell is preferred for OCS in astronomy where elements occur under conditions that cannot be reproduced on Earth, and also noting that astronomical objects are not observed in the same rest frame causing lines to undergo Doppler shifts.

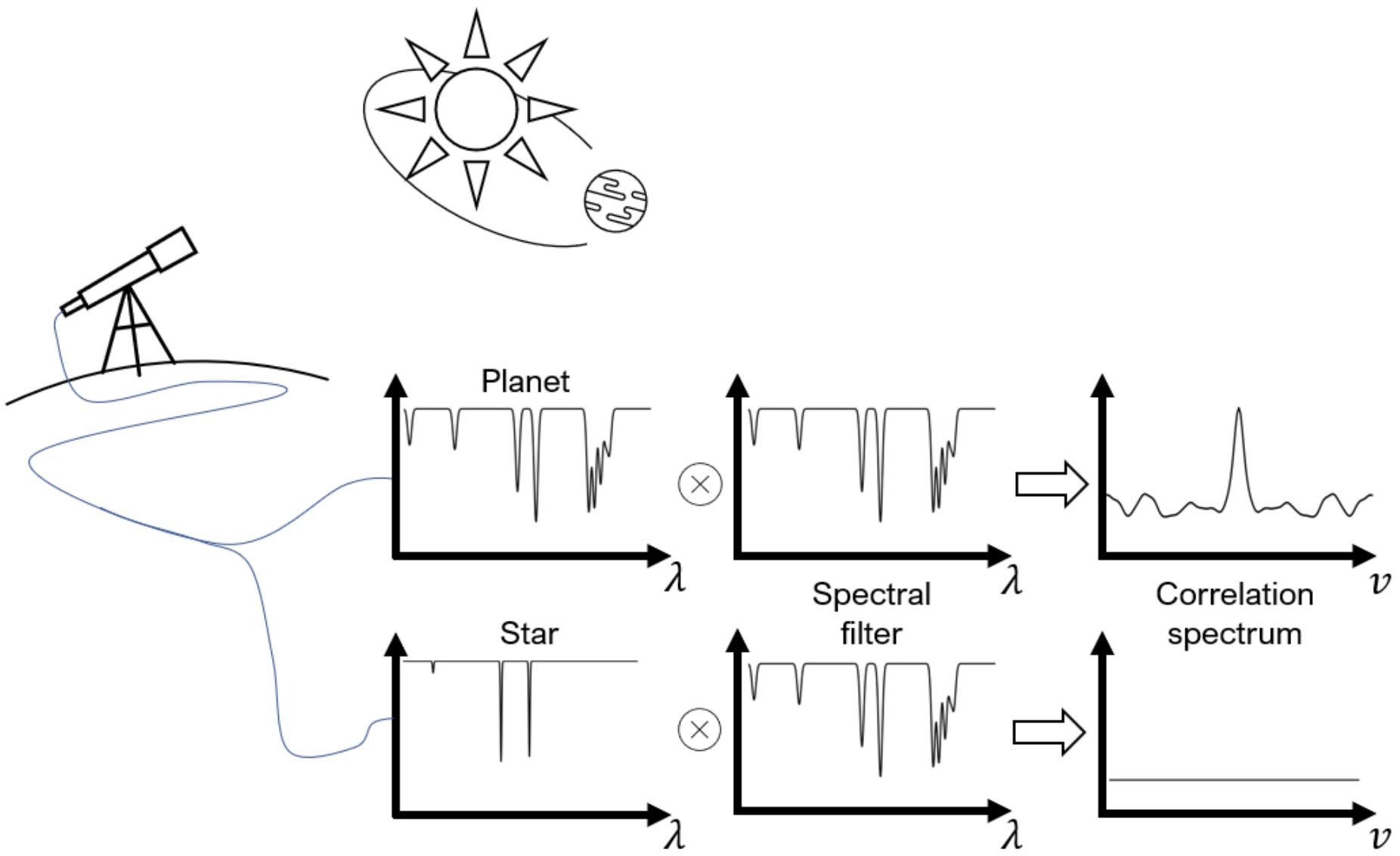

***Figure 1 -*** *Light from two astrophysical sources (a star and a planet in this case) are fed to an optical cross-correlator spectrograph. The OSC passes the light of the objects through a spectral filter that is matched to a particular element in the planet's atmosphere. The planet has a strong correlation, while the star has no correlation signal at all. This allows the OCS to search for particular elements.*

## Current and Future Challenges

OCS recovers a compressed data stream so generally requires simpler input optics, detectors and output signal train. The key challenge for implementation of OCS is realization of the spectral filter. To date this has been accomplished with several different implementations.

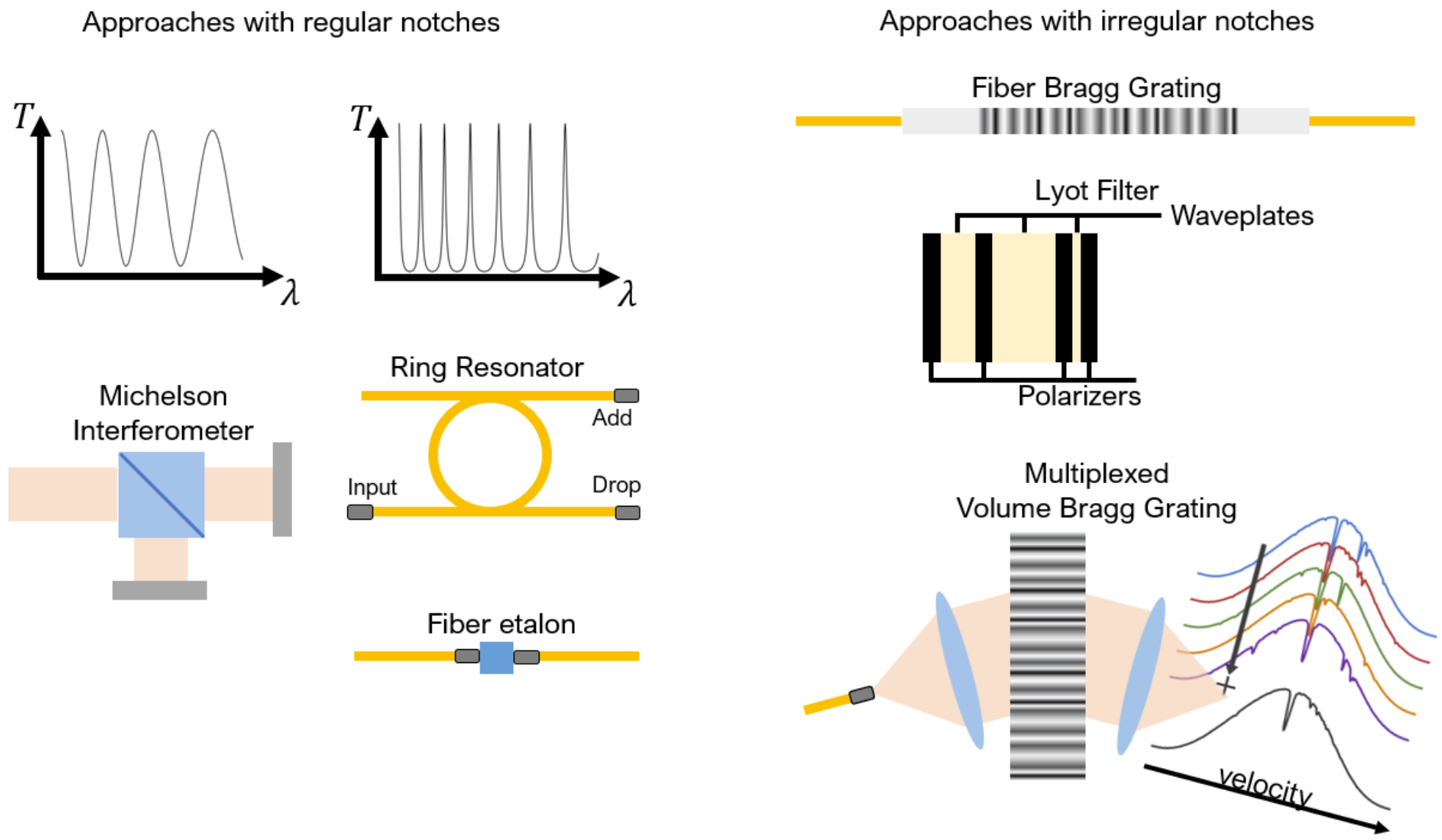

***Figure 2 -*** *An overview of different ways to implement the optical correlation filter. The systems on the left create notches regularly spaced in frequency (diagrams at the top). The systems on the right create irregular notches by multi-step modulation. Figure 1 shows an example of irregular notches.*

### *Interferometric filters*

Interferometric filters use path length differences to create regularly spaced absorption peaks through constructive and destructive interference. This is often done with a Fabry-Perot[7,8] or a Michelson[9] interferometer. More recently, photonic implementations use ring resonators[10] to create a regular pattern. While velocity shifts of the template can be achieved by adjusting the path length, this approach is unable to form a tailored match to spectra of arbitrary line spacing and complexity.

*Fiber Bragg Gratings*

While interferometric methods such as etalons can successfully exploit regularities in the spectral response of many gasses, the use of Fiber Bragg Gratings (FBG) offers dramatically greater flexibility. In particular it is possible with modern aperiodic grating designs to obtain an excellent match even for complicated sets of irregularly spaced atomic or molecular lines. Such bespoke gratings may be tailored to target gas species with high specificity[11] and can be fabricated from the visible through the near-infrared. Such an optic may be integrated into a working sensor system[12] by imposing a variable filter velocity response (potentially by way of a fiber stretcher, or tuning the response with temperature), and ideally facilitates recovery of both the reflected and transmitted beams to optimize capability for accurate calibration. Current FBGs have reached a spectral resolution close to 100,000 which is more than sufficient for scientific observations. Fiber Bragg Gratings require that the light passes through a single-mode fiber (SMF). However, spatially coherent sources (such as stars) need high performing adaptive optics correction (see Chapter 2) for efficient coupling into single-mode fiber[20], which is very challenging towards shorter wavelengths.  Extreme adaptive optics systems deliver much higher performance at wavelengths upwards of 2.0 um, but at those wavelengths the SMF needs different materials as silica is not transmissive anymore. Fiber Bragg gratings have not yet been demonstrated at longer wavelengths.

*Multiplexed Volume Bragg gratings*

Volume Bragg gratings (VBG) are thick transparent gratings that have a periodic modulation of the refractive index. The VBG diffracts only a narrow wavelength range around the Bragg wavelength. Multiple gratings can be created in a VBG at the same time if the refractive index profile is a superposition of the individual grating profiles[13]. Such a grating is called a Multiplexed VBG (MVBG). Each grating then diffracts its specific spectral line in the same direction (angle) where they incoherently combine on the same detector pixel. Effectively creating the OCS signal. An advantage of this method is that it creates a tiny spectrum around the diffracted wavelength. This tiny spectrum is the velocity correlation spectrum. The presence and velocity of the template gas species can be measured at the same time, which is an advantage over most other methods.

Fiber Bragg gratings have recently been made for spectra with very complicated profiles, such as for  OH-suppression filters[14]. The FBGs are inscribed inside the core of silica fibers by using holography and phase masks[15]. This fabrication approach fails for MVBGs due to their thickness: instead a method in which refractive index structures are laser engraved within a substrate (so-called `direct write')[16] is used. However grating-based approaches have a strict tolerance on the written grating period: the relative precision is the inverse of the spectral resolving power. For MVBGs the precision has to be better than 1e-5 to meet a design specification of R=100,000 and unfortunately current direct-write methods are not yet at the level where this is possible over sufficiently large areas (30 x 30 mm) due to aberrations in the writing systems[17].

**Advances in Science and Technology to Meet Challenges**

Most OCS instruments deployed to date have either been based on the gas-cell method or interferometric approaches due to simplicity of the technology and ease of implementation.

They have demonstrated most success for molecules with rovibrational transitions that create lines with relatively regular spacing. However in reality lines are never exactly periodic, imposing a limitation on the interferometric approach to a small part of the spectrum (and consequently to bright objects). Many other species have complex line patterns that are not regular at all. Filters with regular spacings between the notches cannot create the required spectral templates. Complex filters with aperiodic notches are required for most cases. The multiplexed VBG and the aperiodic FBG are therefore the most promising methods for OCS as these provide the most flexibility in template spectra. However, as discussed before several challenges have to be solved.

Integral-field OCS with FBGs present challenges due to the SMF injection step. Single-mode IFUs, often exploiting a micro-lens array to inject into a matched array of fibers, have been successfully demonstrated in the past few years[22]. Nevertheless, the stringent spatial beam requirements for efficient SM injection drive challenges for IFU throughput and efficiency. Few-moded fibers in concert with lanterns may present one pathway for FBG based integral-field OCS to boost efficiency[23]. However such arrays of low mode-count photonic lanterns have not been developed, with further innovations needed to couple them to banks of FBGs (see Chapter 4).

A fundamental issue for the FBG is that at each moment in time only a single correlation match to the template is made, so that time modulation of the response is required to recover velocity information. At the cost of escalating optical complexity, a scheme might be envisaged whereby light from a first-stage FBG is passed to a second FBG tuned to a slightly different velocity. A cascade of such FBGs (where reflected light from one FBG is sent to the next FBG in the cascade and the transmitted light from all the FBGs is measured) can be used to create the velocity spectrum instantaneously with no need for a time sweep. The FBG alters the trade space from spectrograph real estate to telescope observing time and spectral templates. Measurements have to be taken sequentially if several species or Doppler shifts have to be measured. One possible scheme may implement a different FBG filter (tuned for various molecular species) that tiles some region of spatial extent in the image plane. Some scheme of dithering or field rotation would then bring different spatial elements into alignment with different FBG filters, over time sampling the whole scene with templates matched to all species.

The MVBG can provide access to multiple species and velocity channels in a single shot. However, the complexity of the grating increases with every additional information point. This requires very accurate control over the grating pattern which is not available with passive direct write methods. Adaptive optics techniques are deployed to create stable and high precision direct-write setups and have shown promise in improving the write quality[18]. Acousto-optical gratings[13,19], in which sound waves act to modulate the refractive index, are also being explored to provide an alternative. Multiplexed gratings can be realized by sending complex waveforms through the material. With active control of the acoustic synthesizer, the demanding requirements to implement an MVBG can be met. There are two technical challenges to overcome. The high-spectral resolution of R=100,000 requires a large piece of electro-optic crystal with high homogeneity, which is difficult and expensive to source. Furthermore, the required acoustic power scales with $\lambda^4$. Large gratings require injection of significant acoustic power for lines at longer wavelengths which generates waste heat in the crystal that needs to be extracted.

## Concluding Remarks

Optical correlation spectroscopy is a promising technique to map gasses and elements, distilling information at high spectral resolution over significant fields-of-view. This technique allows us to think in a different way about observing. There are many advantages to compressing the information content of the data optically, before the light reaches the sensor, recovering the imprint of the chemical content in the spectrum, rather than the wealth of lines themselves. The method could accomplish our science with fewer targeted measurements, offering relief where constraints from instrumentation or data volume hamper progress. However, it is still a relatively new technique for astronomy and there remain hurdles to be overcome before working systems can be deployed to modern observatories.

## Acknowledgements

*Support for SYH was provided by NASA through the NASA Hubble Fellowship grant #HST-HF2-51436.001-A awarded by the Space Telescope Science Institute, which is operated by the Association of Universities for Research in Astronomy, Incorporated, under NASA contract NAS5-26555.*

# 12 | Laser Frequency Combs

Chad Bender[1], Scott A. Diddams[2], Tobias Herr[3], Stephanie Leifer[4], Ewelina Obrzud[5] and Kerry Vahala[6]
**[1] Steward Observatory, University of Arizona, Tucson, AZ, USA**
**[2] Electrical, Computer and Energy Engineering and Department of Physics, University of Colorado Boulder, CO, USA**
**[3] Deutsches Elektronen-Synchrotron DESY, Germany and Universität Hamburg, Germany**
**[4] The Aerospace Corporation, El Segundo, CA, USA**
**[5] Centre Suisse d'Electronique et de Microtechnique, Neuchâtel, Switzerland**
**[6] Department of Applied Physics, California Institute of Technology, Pasadena, CA, USA**

### Status

Laser frequency combs (LFC), the Nobel prize-winning technology recognized in 2005 (Hänsch 2006, Hall 2006), have revolutionized time and frequency metrology. The LFC or, alternately, optical frequency comb (OFC) is an optical spectrum produced by a laser that consists of an array of delta-function-like frequency modes with perfectly uniform spacing determined by the laser's RF-domain pulse repetition rate, $f_{rep}$. All modes share a common frequency offset, $f_0$, that can be measured and controlled with the technique of *f-2f* self-referencing (Jones 2000). In this way the frequency of each mode of the comb, $f_n$, is given by $f_n = nf_{rep} + f_0$, where $n$ is an integer and $f_0$ and $f_{rep}$ can be referenced to an atomic clock with uncertainty given by the clock itself. For example, when referenced to the clocks in the global navigation satellite system (GNSS), the LFC inherits a typical fractional uncertainty of $10^{-12}$ or $10^{-13}$. It is accordingly a very precise “spectral ruler” for measuring any frequency of light. In precision astronomical spectroscopy, LFCs are ideal calibration sources with performance that surpasses the systematic uncertainty introduced by existing spectrographs and astronomical processes such as stellar activity.  On the other hand, the intrinsic noise in the LFC supports frequency synthesis at and below the $10^{-21}$ level (Yao 2017). As such, it also serves as the clockwork that links RF and optical domains to implement optical atomic clocks with present frequency uncertainty at the $10^{-18}$ level (Beloy 2021), which is two orders-of-magnitude more precise than current time standards. Thus, LFCs can be a critical tool for studying a wide range of challenging questions in physics and astronomy that involve precision frequency measurements.

There are a variety of methods of generating LFCs, and the resultant combs can be characterized by their mode spacing ($f_{rep}$), spectral bandwidth (visible through mid infrared), and optical power.  The most mature comb technology, in terms of commercialization, is the fiber laser comb composed of all-optical-fiber components that form a passively mode-locked laser cavity in sequence with a fiber amplifier, supercontinuum generator, and *f-2f* interferometer (Sinclair 2015). These low repetition rate combs ($f_{rep}$~100 MHz), operating in the near-infrared (NIR, center wavelength of 1550 nm), are used to form an essential component of a new generation of optical atomic clocks. A space-based network of optical atomic clocks that utilize LFCs as the clockwork with $10^{-18}$ accuracy has been proposed for gravitational wave detection (Yu 2017, Loeb 2015), relativistic geodesy, dark matter experiments (Derevianko 2014), opportunities to search for violations of general relativity (Ashby 2009), the standard model, and the Einstein equivalence principle, and to provide the potential to discover new physics (Derevianko 2021). When used for femtosecond-level time transfer or synchronization, LFC technology could enable synchronization of distant sites such as those used for the Event Horizon Telescope (EHT) (Akiyama 2019), a global network of radio telescopes that can provide very high angular resolution imaging, thereby drastically reducing extensive, compute-intensive post processing needed to correctly align the constituent signals.

An LFC referenced to an atomic clock and the SI second has become an essential wavelength calibration tool. As the local oscillator in a heterodyne radiometer (Frederick 2022), they could provide higher resolution Doppler maps of the sun than those produced by NASA's Solar Dynamics Observatory (Scherrer 2012). But in the astronomy community, LFCs have found their greatest application in ground-based Precision Radial Velocity (PRV) measurements, and when specialized for this role, are dubbed “astrocombs” (Murphy 2007, Braje 2008, Li 2008). With absolute stability at the $10^{-12}$ level or better, corresponding to the Doppler shift of <1 cm/s, astrocombs provide sufficient precision in spectral calibrations for PRV spectrographs that aim to measure <10 cm/s Doppler shifts in absorption features of G-type stars, and in so doing, enable characterization of earth-like exoplanets. There are currently more than a dozen astrocombs demonstrated or in operation at major optical and NIR facilities that are pursuing this and related radial velocity measurements (McCracken 2017, Herr 2019, Fischer 2016). They cover a range of wavelengths from ~500 nm to greater than 2 microns, and have been customized to provide spectra with mode spacing appropriate for the unique spectrographs they calibrate (10-30 GHz). This flexibility makes astrocombs attractive calibration sources even for non-PRV spectrographs as a replacement for traditional hollow-cathode lamps that provide inferior calibration spectra due to a nonuniform line distribution over the spectral span, high dynamic range and line blending. Moreover, in some cases, such lamps are difficult to procure due to changing environmental and geopolitical restrictions. Figure 1 depicts comb generation by various methods, as well as various applications in astronomy as a function of radial velocity precision and wavelength.

**Current and Future Challenges**

The promise of astrocombs often belies coupled scientific and technical challenges that limit the widespread application of these highly complex systems. When one considers the details of frequency combs, there are many challenges that must be addressed including comb spectral flatness, low temporal variability, uniform coupling to multimode spectrographs, and polarization effects in single-mode spectrographs. But the most critical issues at this point in time can be essentially described as: (1) overall complexity which is coupled to size, weight, need for maintenance, operation over many years, environmental stability, remote operability, and cost; and (2) full spectral coverage that matches that of stellar spectra, with particular shortcomings on the blue side of the frequency comb spectrum.

Frequency combs are significantly more complicated and costly than traditional calibration sources such as hollow-cathode discharge lamps or gas cells. This complication arises in part from the fact that they are built on active lasers, but it is compounded by the coupled requirements of large mode spacing ($f_{rep}$~10-30 GHz) and broad spectral bandwidth (as large as $\Delta\lambda$=350-2500 nm). To achieve both of these at the same time requires combinations of average power and pulse duration that are not found in any common LFC. The canonical approach, shown in Figure 1a, involves creation of a high repetition rate pulse train (a comb with narrow spectral bandwidth), whose short pulses along with amplification permit spectral broadening through nonlinear optical processes. Finally, the spectral envelope of the broadened spectra is flattened to avoid over- and under-exposure of the detector, and detector effects such as the "brighter-fatter" effect (Blackman 2020) that can induce erroneous radial velocity signatures.

A first challenge is to identify a pulse source that permits operation at high pulse repetition rates. As shown in Figure 1, comb technologies that are being explored include: (a) low repetition rate lasers that are subsequently filtered to provide large mode spacing (Li 2008, Braje 2008, Steinmetz 2008, Ycas 2012), (b) electro-optic frequency combs that intrinsically generate 10 GHz combs (Xi 2016, Metcalf 2019, Obrzud 2018) (c) high-repetition rate active lasers (Bartels 2009, Endo 2015), and (d) microresonator frequency combs, also known as microcombs (Obrzud 2019, Suh 2019). There are tradeoffs that must be made in all of these technological approaches, e.g. wavelength of operation, permissible comb line spacing, size, complexity, cost and ease of operation.

Regardless of the pulse source, a fundamental challenge is that 10 GHz combs have low pulse energy which restricts the nonlinear optics required to generate broad spectral bandwidth, such that maximal radial velocity information can be obtained. The most straightforward way to address this challenge is by increasing the average optical power to the multi-Watt level, but this comes with challenges of thermal management and damage in the LFC system and degradation of components. Such factors further drive the price and complexity, which restricts reliability. Presently, this is most challenging for frequency combs aimed at RV measurements on Sun-like G stars where the desired spectral coverage is ~380 nm to beyond 800 nm, or more than 400 THz and 40,000 comb lines at 10 GHz comb spacing. Nonlinear spectral broadening in silica fibers has been developed to cover this spectral range. However, the reliability of such an approach is limited due to rapid damage of the fiber upon UV exposure (Girard 2019). Thus, reliable operation is only routinely achieved at wavelengths longer than approximately 500 nm, excluding nearly 200 THz of valuable spectral bandwidth. Continued research efforts are needed to provide reliable solutions for continuous nightly spectral LFC calibration down to the shortest blue wavelengths.

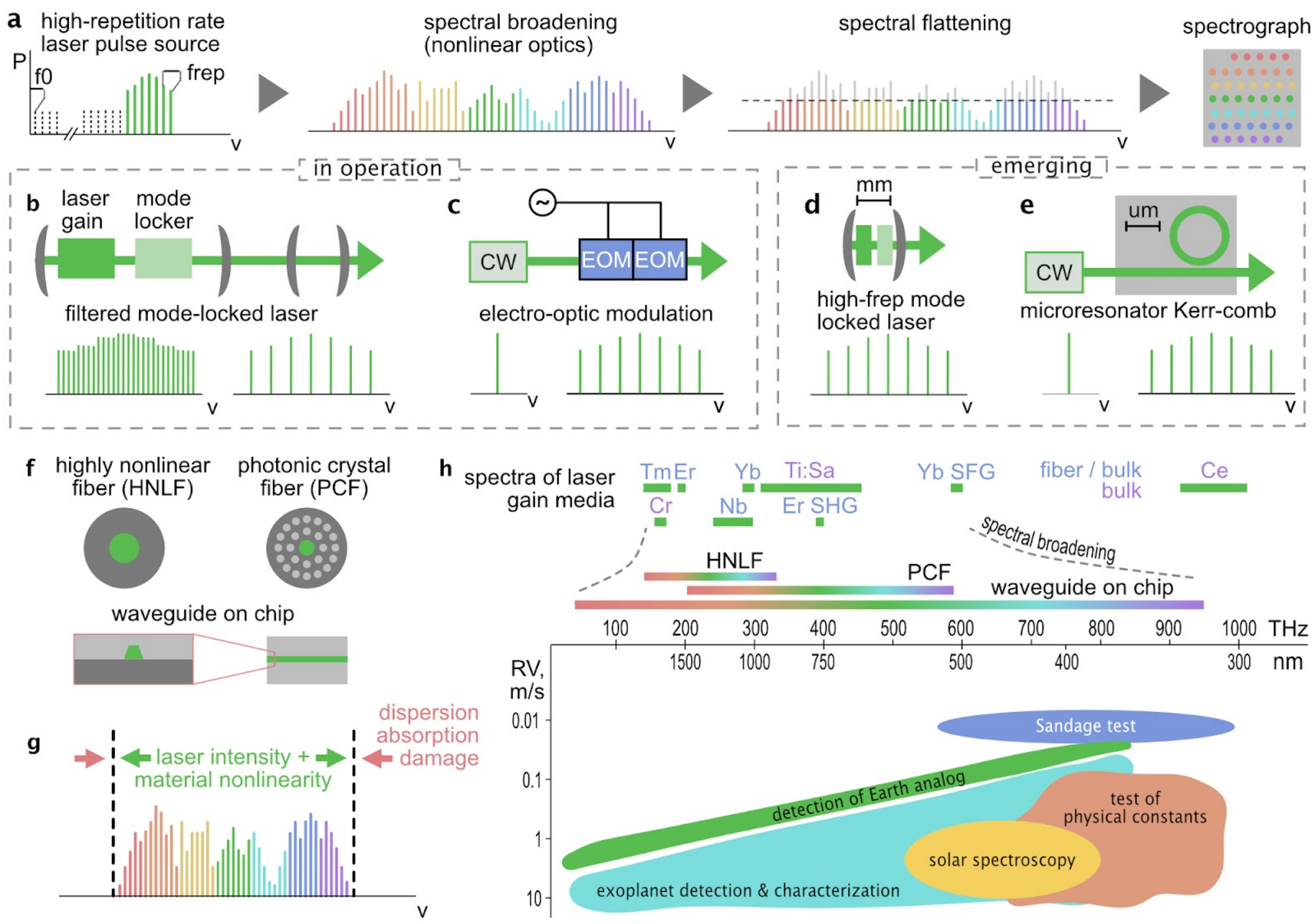

***Figure 1 -*** *(**a**) Steps for astrocomb generation: generating a widely spaced, relatively narrow-band frequency comb from a high repetition rate pulsed laser, spectral broadening into a broadband spectrum with the same line spacing and spectral flattening for equal spectral flux across the detector. The comb spectrum is characterized by a frequency offset (f0) and the comb line spacing, equal to the pulse repetition rate (frep). The initial comb spectrum may be derived from a **(b)** filtered mode-locked laser or an **(c)** electro-optic modulation-based (EOM) comb generator; emerging sources include **(d)** high-repetition rate mode-locked lasers and **(e)** chip-integrated Kerr-nonlinear microresonators. **(f)** Highly-nonlinear fiber (HNLF) and photonic crystal fiber (PCF) are established silica-based fibers for nonlinear spectral broadening. More significant broadening may be observed in chip-integrated waveguides that permit even stronger light confinement and materials with higher optical nonlinearity. **(g)** Broadband spectral broadening relies on high laser intensity and optical material nonlinearity; the achieved bandwidth is usually limited due to chromatic dispersion, absorption or damage of the material. **(h)** Optical spectra of common laser gain media are sparse and narrowband. Spectral broadening by HNLF, PCF or chip-based waveguides can, in principle, generate broadband spectra from infrared to ultraviolet wavelength. Such broadband spectra are required to address important science cases of precision RV measurements; their spectral requirements along with their requirements for RV precision are highlighted.*

## Advances in Science and Technology to Meet Challenges

Major technological advancement can be expected from new materials and their potential to be monolithically integrated on photonic chips for efficient nonlinear optics, compactness and cost reduction. The integration of comb systems requires a range of optical technologies to be co-located on a common silicon substrate. In recent years, the narrow bandwidth comb sources themselves have undergone rapid development based on two technologies: soliton microcombs and thin-film lithium niobate electro-optic modulators (Kippenberg 2018, Zhang 2019). Moreover, integrated nonlinear waveguides have been fabricated showing dramatic spectral broadening and opportunity for spectral tailoring through optimized geometries potentially beyond what is possible in current nonlinear silica fibers (Hickstein 2017, Yu2019, Wu2022, Liu2019, Lamee2020). Each of these platforms has created opportunities for the complete integration of future comb systems.

Soliton microcombs are miniature mode-locked parametric oscillators fabricated from high-Q optical microcavities (Kippenberg 2018). They can be self-referenced and have been used to demonstrate all comb functions including optical clocks and optical synthesizers. While early microcombs were demonstrated as discrete devices, significant progress in optical loss reduction of CMOS-friendly materials such as silicon nitride has made possible microfabrication of more complex microcomb systems. These combine integrated waveguides as well as soliton repetition-rate tuning control. Also, these systems have been heterogeneously integrated with III-V semiconductor lasers for comb pumping. Surprisingly, the pumping of microcombs can be configured to simplify comb activation while also eliminating difficult-to-integrate functions like optical isolators. Compact, butterfly-packaged microcomb systems with optical pump and tuning controls have resulted (Stern 2018, Shen 2020)

While microcombs are transitioning conventional table-top mode-locked combs to a semiconductor chip, there is also a renaissance in the world of electro-optic (EO) combs. Here, the advent of the material system thin-film-$LiNO_3$ on silicon brings the key features of this photonic workhorse material (harmonic generation and electro-optic modulation) into the realm of semiconductor microfabrication. The resulting capability to engineer dispersion and precisely control waveguide dimensions has enhanced the performance of devices like EO-modulators by lowering their V_pi values. It is also dramatically expanding the ways $LiNO_3$ can be fashioned into optical devices and complex systems on-a-chip. Miniature resonant modulators and even electro-optically controlled microcombs are two examples (Zhang 2019, Li 2022).

With this rapid progress in microcomb and EO-comb technology, a pathway exists to compact, integrated astrocomb systems. However, before this can happen, efficient nanophotonic spectral broadening must also be integrated with the comb generator. Again, a challenge here is the low pulse energy that is available from the integrated micro and EO comb generators. Presently off-chip fiber amplifiers must be employed to boost peak power to a sufficient level (~10-100 pJ in 100 fs) to achieve the required spectral broadening. Nonetheless, there is significant progress in the exploration of new materials that offer multi-order-of-magnitude enhancements to optical nonlinearity that are critical for broad spectrum generation. Material platforms of UV transparent media like MgO-doped $LiNO_3$ (Yu2019,Lu2019,Wu2022), AlN (Liu2019,Chen2021) and diamond (Shams-Ansari 2019) show some promise and can be integrated as nanophotonic waveguides, while more research and development is required to achieve high-power handling capability and sufficient flux at short wavelengths.

## Concluding Remarks

Future LFC precision calibrators will need to balance mature and emerging technologies to address conflicting requirements of broadband calibration and high reliability at reduced cost. Integrated photonics will play an important part in this development and hold prospects not only for improved performance but also for low-complexity, robust monolithic systems at lower cost; in case of device failure, such systems could potentially be swapped like a light bulb. On a practical note, the development cost of these miniature systems will be shared with other application areas, including miniaturized precision timing and navigation systems, airborne radar systems, and field-deployable precision chemical sensors. To leverage the full potential of LFCs, dedicated data pipelines and instrumental interfaces (including spectral flatteners - see Chapter 13) will need to be developed. Additional developments in adaptive optics and single-mode fiber-fed spectrographs may bring a significant advance when calibrated with LFCs. And finally, the reduction of size, weight and power consumption creates opportunities for space-based EPRV observations as well as unperturbed studies of planetary atmospheres.

## Acknowledgements

*S. Diddams acknowledges support from NIST, JPL, and NSF grants AST-1310875 and AST-2009982.*

*T. Herr acknowledges support by the Helmholtz YIG VH-NG-1404 and the EU ERC StG 853564.*

# 13 | Spectral Flattening of Laser Frequency Combs On-a-chip

Nemanja Jovanovic[1], Stephanie Leifer[2] and Charles Beichman[3]
**[1] Department of Astronomy, California Institute of Technology, Pasadena, CA, USA**
**[2] Department of Applied Physics, California Institute of Technology, Pasadena, CA, USA**
**[3] IPAC/NASA Exoplanet Science Institute, Jet Propulsion Laboratory, California Institute of Technology, Pasadena, CA, USA**

## Status

The Precision Radial Velocity (PRV) technique is critical to the detection and characterization of exoplanets and relies on the measurement of Doppler shifts of the host star's spectral features [1]. Key to the success of PRV observations is an ultrastable wavelength standard, ultimately requiring sub-cm/s long-term precision and accuracy to enable the detection of terrestrial planets in the habitable zones of solar-type stars, which induce shifts as small as ~9 cm/s over a year [2]. Laser Frequency Combs (LFCs) offer exquisite long term stability making them ideal for this application (See Chapter 12 in the roadmap). However, due to the nonlinear broadening processes that generate broad comb spectra, the amplitude of the comb lines can vary by many orders of magnitude (see Fig. 1). To maximize Doppler precision, all comb lines should be used to derive the wavelength solution. The limited dynamic range of astronomical detectors imposes limits on the amplitude variation that can be tolerated. In addition, the comb line amplitudes should be extremely stable, as changes to the comb profile of the spectrum can masquerade as erroneous Doppler shifts. To meet these requirements, a device called a 'flattener' is often used, downstream of the comb. Commonly, flatteners first collimate the LFC light from an optical fiber, disperse it with a grating, and then use a liquid crystal on silicon spatial light modulator (SLM) to control the amplitude of each spectral channel before recombining the spectrum and injecting it into another fiber [3]. These flatteners are large complex assemblies (see Fig. 1) requiring careful alignment, and the dynamic range afforded by the SLMs (20-30 dB) is often inadequate to flatten the spectrum produced by an LFC. Further, they are tailored for specific spectral bands and require developmental effort to move into new wavebands.

A photonic solution offers the possibility for a compact, portable, replicable, and inherently stable package that may even enable space-based applications. One such solution is an all-photonic spectral flattener based on SiN waveguides consisting of an Arrayed Waveguide Grating (AWG) which disperses the light, Mach-Zehnder interferometers (MZIs) to actively control the amplitude in each channel, thermo-optic phase modulators (TOPMs) to re-phase the channels, and a second AWG used in reverse to recombine the spectrum. Our group has recently demonstrated such a device with 20 channels providing ~40 dB of dynamic amplitude modulation range for a linearly polarized input source [4,5]. It was capable of flattening a temporally incoherent source to a residual amplitude variation of 3 dB, and an LFC spectrum to ~5 dB where both sources started with >25 dB of variation over 250 nm (1400-1650 nm). The device has a millisecond response time (kHz) to make corrections and can handle optical power levels of hundreds of milliWatts. The current device operates in the near-IR, suitable for instruments designed to study planets orbiting cool, mid- to late M stars. Future devices could provide spectral flattening at optical wavelengths appropriate for instrumentation studying warmer, solar-type stars.

Another method for producing broad, relatively flat comb spectra is to accomplish it through engineering the appropriate combination of dispersion and geometry in the nonlinear photonic waveguides used for supercontinuum generation [6]. This would eliminate or reduce the need for an external flattener on-a-chip. Very broad spectra have been demonstrated in, for example, SiN [7], magnesium oxide doped lithium niobate thin films on silicon [8,9], tantala [10,11], and aluminum nitride structures [12].

Beyond LFC flattening, spectral shaping devices can find applications in other fields including add/drop multiplexing [13], gain flattening [14,15], temporal pulse shaping [16, 17] as well as for targeted excitation of particular molecular species [18].

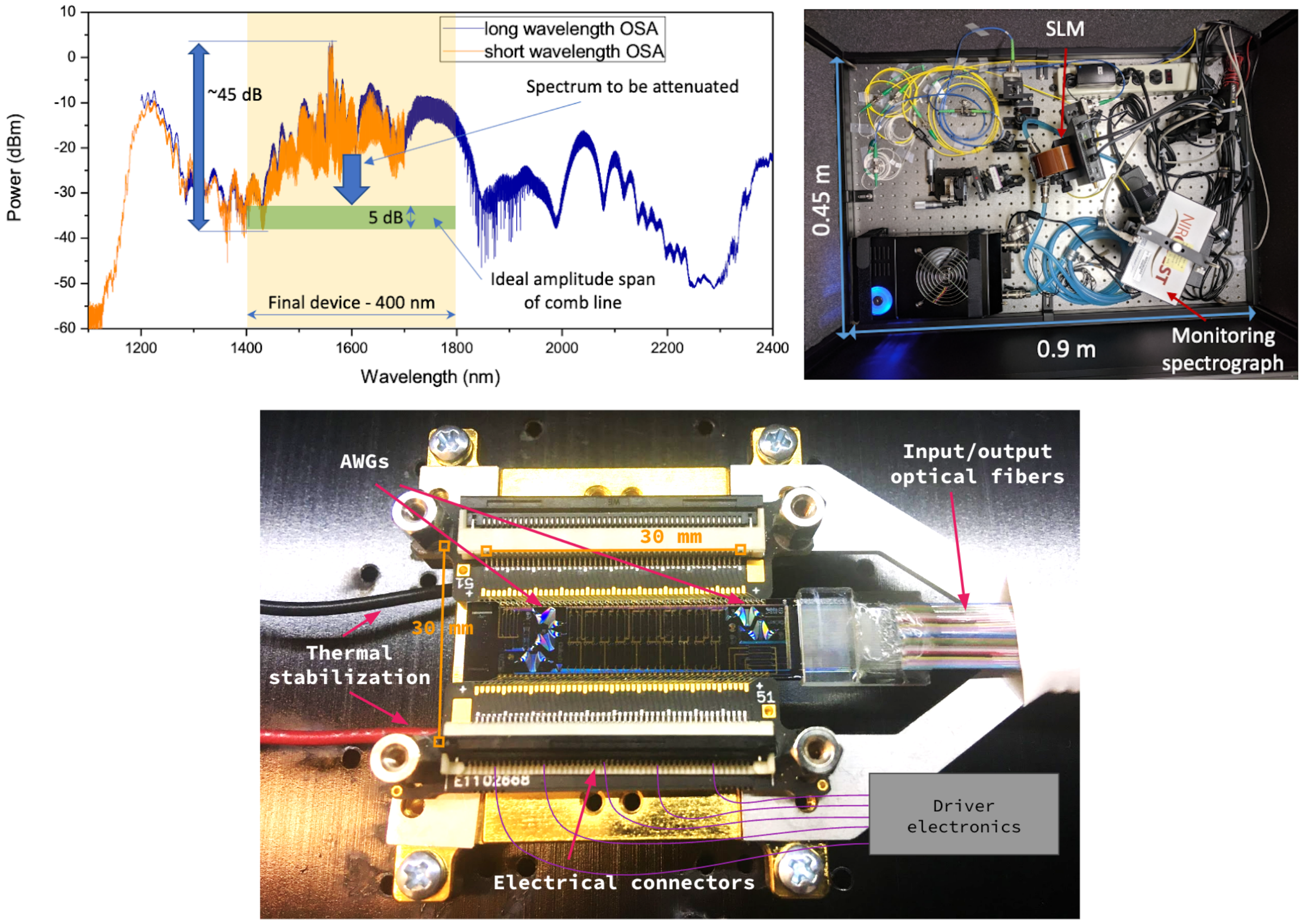

***Figure 1 -*** *(Top) (Left) a spectrum of the Palomar Radial Velocity Instrument (PARVI) LFC showing the strong intensity modulations across the spectrum. (Right) PARVI's bulk optic flattener. The flattener occupies a volume of 1 x 0.5 x 0.5 m*$^3$*. (Bottom) An image of the generation one all-photonic flattener device tested in the laboratory. (Images were reproduced from [4,5]).*

### Current and Future Challenges

To advance all-photonic flattening technologies, requirements for specific science cases must be set with an understanding of how Doppler precision relates to the flatness and power stability of the comb spectrum. This has not been rigorously explored to date and requires simulations to understand how variations in the comb spectrum propagate to Doppler errors. Once the flatness requirement is determined for a specific science case, the device parameters (resolution, number of channels, overall bandwidth etc) can be established. Furthermore, other photonic devices, such as fiber Bragg gratings (FBGs) could be used to compensate for static, large amplitude power variations such as those around the pump region of the comb. The division of flattening capability between these photonic solutions needs to be determined through careful characterization of the specific comb to be used.

Technical challenges include sufficient AWG channel spacing to correct for narrowband features around the pump region. Devices with higher channel counts will be necessary, requiring the addition of more MZIs and TOPMs, and their associated electrodes and driver electronics. These devices are non-negligible in size (2-3 x 1 mm using SiN) and will eventually push the circuit size beyond a single reticle. Devices can be made across multiple reticles, but there could be stitching errors, although these are improving over time with advances in manufacturing processes. In addition, narrower channels may be desirable in some regions of the spectrum and could be accomplished with a cascaded approach [19], where a low resolution device is used over a broad band and narrower devices are used as a secondary stage (see Figure 2 for an example of this concept). Adding more AWGs to the chip or increasing the resolving power of an AWG with a fixed bandwidth both increase the chip size and is limited by either reticle size (limit for a single device) or total usable area on the wafer. In addition, making high resolution, large free spectral range AWGs is plagued with many issues in itself (see Chapters 6 and 7 in this roadmap). Although the number of channels can be increased from the first generation device demonstrated above, it is clear that with current technology and lithographic approaches it would be challenging to scale devices to 1000s of channels, which is the pixel count of commercially available SLMs.

Another challenge is the bandwidth of the device. LFCs are currently used across several astronomical bands ranging from 350 nm to 2.5 microns [20,21]. Photonic devices have a limited range over which they operate with a single guided mode with acceptable bend losses. The prototype outlined above spans only a single astronomical band (H). The two primary options for broader spectral coverage are: 1) making a series of parallel devices optimized for different wavebands, or 2) making broadband devices in a single platform. This technology could have the greatest impact in the visible band where most PRV spectrographs currently operate. Developing efficient SM circuits with the necessary bandwidth could be challenging. LioniX International offers visible waveguides (400-700 nm) in SiN through their multi-project wafer (MPW) offerings, with very low loss (<0.1 dB/cm) [22]. The lifetime of visible flattening devices would need to be assessed given the presence of UV photons generated by the combs.

Finally, the detection scheme used to drive the flattening algorithm could also be integrated onto the chip. Currently, an external, dedicated spectrograph is used to monitor and adjust the flatness of the spectrum. An alternative is to use the light rejected by the MZIs as probes to drive the flattener. This can be done by sensing the signal in the ports with a linear array detector or integrating individual photodiodes onto the chip (see Chapter 23 for details).

Routing the probe channels to the edge of the chip would incur a large number of waveguide crossings (see Fig 2), which could compromise the performance of the device and/or use of the dropped ports as probes due to losses and crosstalk. Integrating photodiodes in place circumvents this issue. Alternatively, the beams could be ejected vertically upward out of the chip via vertical couplers and/or grating couplers onto a two-dimensional array above or a secondary chip with photodiodes flip-chipped bonded to the top. This option is also challenging as vertical and grating couplers are difficult to manufacture without imperfections and grating couplers are large. The concept may also require the use of 2x3 multimode interference coupler at the output of the MZIs, so two rejected ports can be tracked to disambiguate amplitude changes in a given channel with other sources of drift. It would also require careful calibration of the non-common chromatic losses between the output of the MZIs and the final spectrograph.

**Advances in Science and Technology to Meet Challenges**

To advance the technology, all-photonic flattening devices should be designed, fabricated and characterized in various wavebands ranging from 350-2400 nm. Although SiN is transparent and applicable to nearly this entire region, other material platforms like silica-on-silicon or ion-exchanged waveguides [23] could be considered. While the lower index contrast of these platforms would result in larger devices, they offer better mode matching to fibers and in the case of ion-exchanged waveguides offer better transparency beyond 2.2 microns, where SiN starts to become opaque.

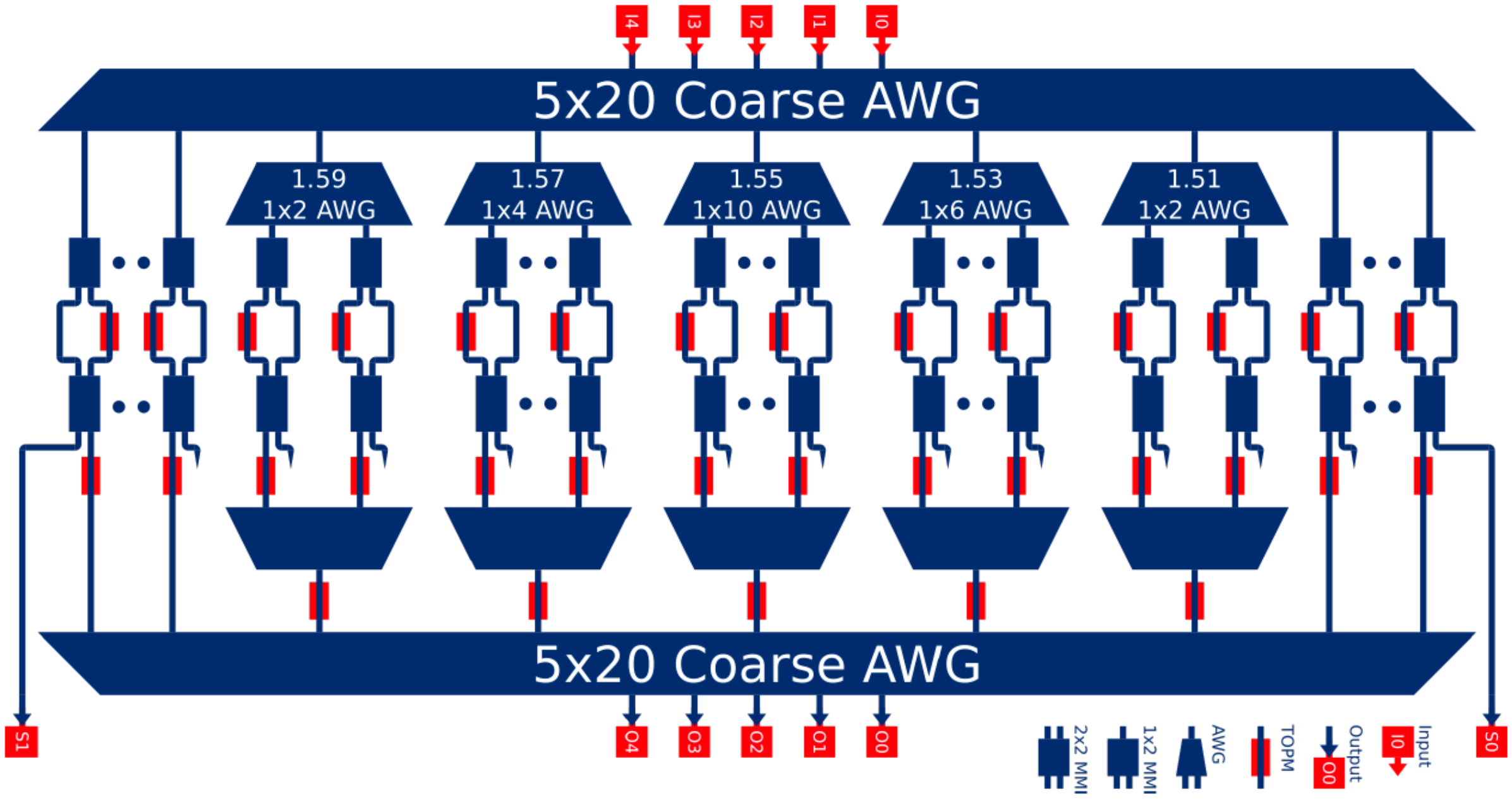

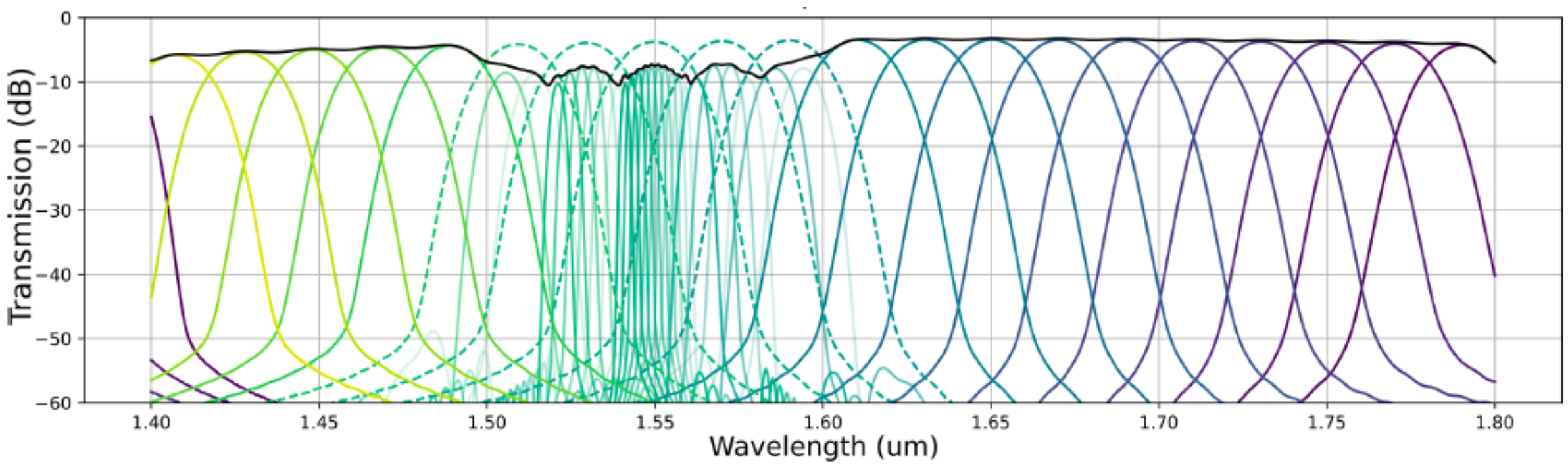

***Figure 2 -*** *(Top) a schematic of an all-photonic flattener that uses cascaded AWGs to enable narrow linewidth channels in some regions of the spectrum. (Bottom) A simulated spectrum for such a device based on SiN waveguides (schematic and simulations carried out by BRIGHT Photonics Inc.).*

Simulations and detailed design work will help to reveal the limitations on properties such as bandwidth, channel count, and losses across large bands, and channel-to-channel crosstalk as well as the potential of different material platforms. In addition, they will provide insight into how large and complex a device can be realized within the limitations of the reticle and/or wafer size. Fabrication and testing will reveal limitations imposed by the manufacturing process including the impact of stitching errors at reticle boundaries.

With the maximum bandwidth of a single device established, several devices could be combined in parallel to span the full wavelength range of the comb. These solutions would require broadband beamsplitters, ideally based on photonics or highly miniaturized technologies Such devices should be developed and tested in conjunction with photonic flatteners to demonstrate large spectral coverage.

Closed-loop control of these devices is a critical step to evaluating the technology and could soon be demonstrated in the field with the PARVI instrument [24], by utilizing a separate monitoring spectrometer. Ultimately, to make the flattener independent of system-specific architectures and portable, approaches that explore integrating the sensing with the chip should be tested as outlined in the section above.

Finally, other approaches or technologies should also be considered for all-photonic flattening as they may offer advantages in terms of channel count scaling. Chapter 7 of the roadmap presents several innovative approaches to dispersing the light that might be suitable for this application.

## Concluding Remarks

All-photonic spectral shapers provide an avenue to extreme miniaturization and simplification of flattening devices for intensity modulation across the output spectrum of an LFC. To become useful across a broad range of applications, these devices will need to undergo further developments to understand bandwidth limitations, demonstrate operability in other wavebands, understand channel number and channel bandwidth scaling limitations, as well as closed loop demonstrations and studies into the optimum architectures for closed loop control as well as field testing.

**Acknowledgements**

The authors acknowledge the Keck Institute for Space Studies for funding and BRIGHT Photonics for their design and simulation efforts of the prototype devices presented herein. The research was carried out in part at the Jet Propulsion Laboratory, California Institute of Technology, under a contract with the National Aeronautics and Space Administration (80NM0018D0004). We would like to acknowledge technical contributions by Pradip Gatkine, Boquang Shen, Maodong Gao, Nick Cvetojevic, Jeffery Jewell and Gautam Vasisht during the development of the first generation all-photonic flattener described above.

# 14 | Fabry-Perot Etalons as Precision Wavelength Calibrators

Samuel Halverson[1], Liang Tang[2,3] and Christian Schwab[4]
**[1] Jet Propulsion Laboratory, California Institute of Technology, Pasadena, CA, USA**
**[2] National Astronomical Observatories, Nanjing Institute of Astronomical Optics & Technology, Chinese Academy of Sciences, Nanjing, China**
**[3] Key Laboratory of Astronomical Optics & Technology, Nanjing Institute of Astronomical Optics & Technology, Chinese Academy of Sciences, Nanjing, China**
**[4] School of Mathematical and Physical Sciences, Macquarie University, NSW, Australia**

## Status

Fabry-Perot Etalons (FPEs) have a rich history in the field of optical physics [1]. At its core, an etalon can be as simple as a pair of planar mirrors, which form an interferometric cavity that acts as a spectral filter. The cavity transmits a discrete set of modes, with wavelengths and line widths dictated by basic properties of the cavity such as mirror spacing, mirror reflectivity, and refractive index of the cavity medium. It is these transmitted modes that can serve as calibration features for a variety of spectroscopic applications, ranging from laser stabilization [2] to stellar spectroscopy [3][4][5]. In astronomy, FPEs have most recently been used as highly precise wavelength calibration references for high resolution, broadband Echelle spectrometers, as they are capable of producing sharp spectral features across wide wavelength ranges.

One of the most demanding spectroscopic applications in astronomy is exoplanet detection via high precision radial velocity (RV) measurements, which inherently relies on exquisite calibration of high resolution spectrometers to measure the minute Doppler reflex motion in stellar spectra due to orbiting planets. The field of *extreme* precision RV measurements (EPRV) is aiming for cm/s level spectroscopic measurements, equivalent to a ~$10^{-10}$ fractional wavelength measurement across wide wavelength ranges (hundreds of nanometers)[6]. This precision is orders of magnitudes smaller than the line widths in the stellar spectra; this goal is driven by the velocity signature of an Earth-like planet orbiting a Sun-like star, which imparts a ~9 cm/s ($3e^{-10}$) velocity signal. Achieving such precisions necessitates the use of novel calibration sources that provide a rich density of spectral features and are intrinsically very stable over many years.

These two requirements have driven the development of broadband ‘astro-comb’ laser frequency comb systems [7][8][9] (see Chapter 12 in the roadmap), which combine unparalleled frequency accuracy with a high density of stable emission features. The primary drawbacks to these astro-comb systems are their intrinsic complexity (leading to limited reliability), limited wavelength coverage in particular in the blue part of the spectrum, and overall cost. These systems require significant upfront investment (>$1M), and incur long-term costs due to maintenance and consumables [10].

Etalons, being intrinsically optically simple, are attractive for astronomical spectroscopy as they can produce a dense, comb-like spectrum similar to an LFC at a fraction of the cost and complexity. Additionally, the design flexibility of FPEs is high, with the free-spectral-range (FSR), i.e. spectral distance between adjacent peaks, finesse (ratio between the FSR and the full width at half maximum of the spectral peaks), and operating bandpass easily customizable for different spectrometers. These benefits have made etalons attractive calibration sources for the highest precision measurements, and most EPRV facilities now use etalons as part of their calibration strategy. Existing etalon systems include air-gap planar-planar cavities fed by single mode or multi-mode optical fibers [11][12][13][14][15][16], entirely-fiber-based interferometer systems [17][18], as well as microresonator-based systems [19]. Each system has distinct benefits and drawbacks, though all require technological improvements to reach the stability levels required for future EPRV instruments (see **Figure 1**).

| Etalon type | Cavity Material | Advancements needed |
|---|---|---|
| Fiber-fed air-gap Fabry-Perot 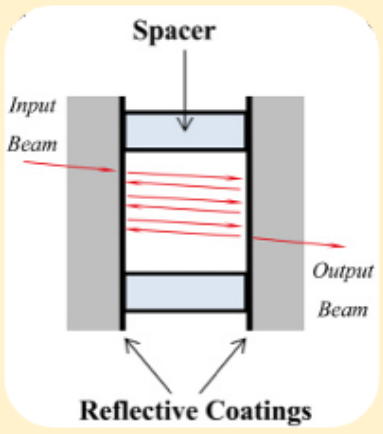  [23] | Air/Vacuum 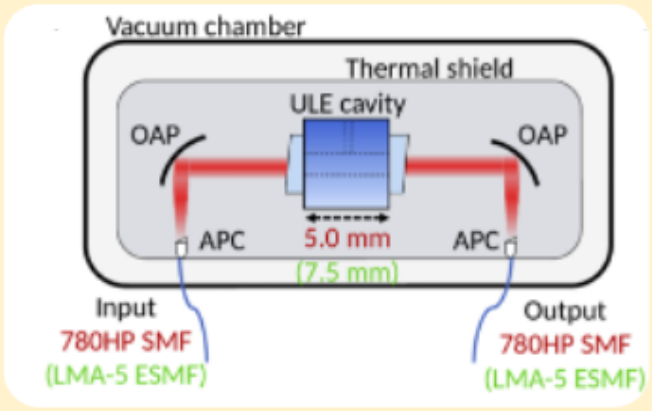  [15] | Improved mirror dispersion control, stability, independent absolute referencing, decreased cost and package size. |
| Fiber-based Fabry-Perot 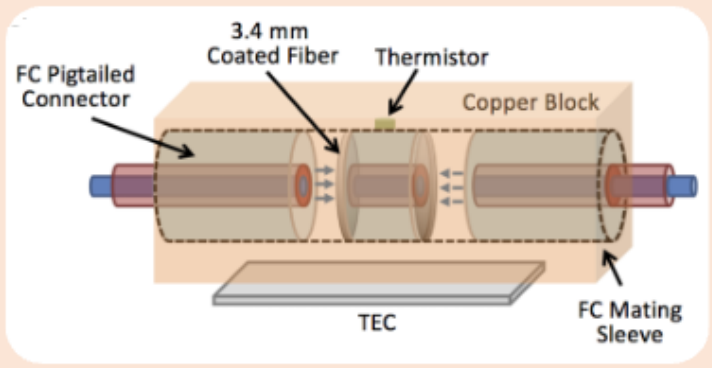  [16] | Fused silica fiber 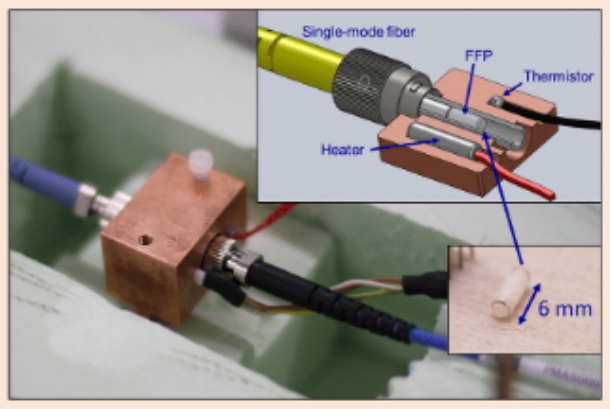  [17] | Polarization stabilization, cavities using endlessly-single-mode fiber architectures, stabilized chromatic dispersion. |
| Whispering gallery mode resonator 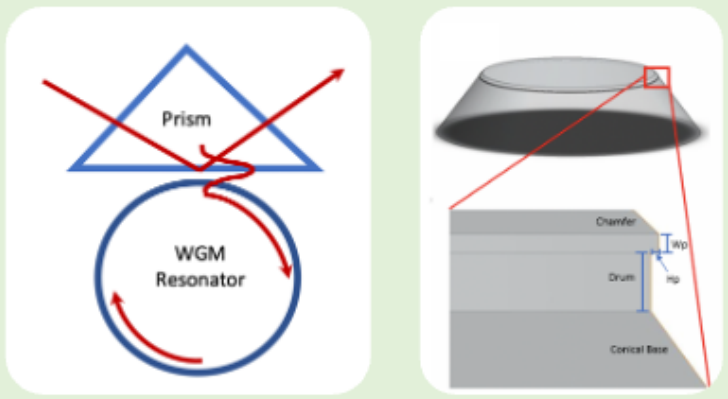  [18] | MgF2, CaF2 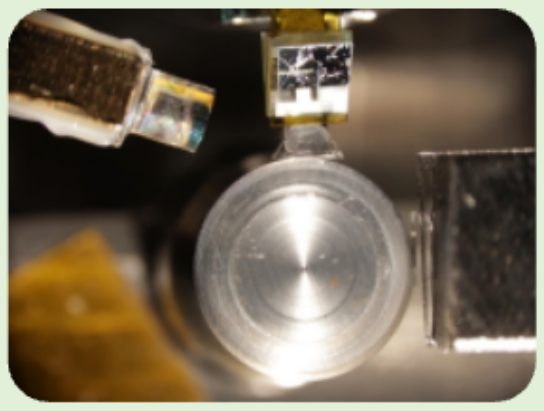 [18] | Fully-single-mode operation, stabilized injection, lower CTE and dn/dT waveguide materials, improved thermal control. |

***Figure 1** - Examples of Fabry-Perot-Etalon (FPEs) used in astronomical spectroscopy. FPEs are valuable tools for precise wavelength calibration of high resolution spectrometers. In all cases, the interferometric cavities act as sharp filters of a continuum spectrum, providing a dense series of lines for high precision calibration.*

## Current and Future Challenges

Etalon calibrators are typically intended as a simultaneous reference observed alongside with the stellar spectrum to track the instrumental drift during the exposure. To anchor the line positions of the etalon itself, an absolute reference like a Thorium-Argon (ThAr) hollow cathode lamp or an LFC can be used to calibrate the FPE periodically. Another solution is to concurrently monitor the wavelength of one or several peaks of the etalon to a very high precision by comparing it to an atomic frequency standard, e.g. a well-known optical transition; this technique tracks the drift of the etalon in real time [18]. In all cases, the etalon requires the relative peak position to be stable to a very high degree. As fundamentally passive optical elements (no innate frequency locking to an atomic standard), these 'astro-etalons' require excellent thermomechanical control to stabilize the output spectrum at the levels required for EPRV applications (~10e-10). While real-time monitoring of a single peak, as described above, can relax this requirement somewhat, this is unlikely to track any chromatic effects. This represents a technical challenge, especially for cavities designed around bulk materials such as $SiO_2$ (in the case of fiber-based etalons) or crystalline materials such as $CaF_2$ or $MgF_2$ (used in whispering-gallery-mode resonators) which have high coefficients of thermal expansion (CTE) and refractive index thermal coefficients (dn/dT). Even 'air-gap' etalon designs, which typically have spacers manufactured out of low-expansion materials such as Zerodur or ULE [20][21][22], still require thermal isolation and precise temperature control to maintain spectral stability (though not at the level of more standard glass materials). At the required level of control, even changes in the amount of injected light must be considered to avoid absorption-driven thermal variations.

Free-space or air-gap FPEs have been used extensively in RV applications [11][12][13][14][15][16] over the past ten years. These systems are optically simple, and revolve around a planar-planar cavity design fed by single-mode optical fibers. However, these systems require careful engineering to properly stabilize the cavity, including precise vacuum and temperature control. The structure of the etalon and the optical alignment require specialized techniques to maintain stability (optical contacting, athermal mount design, etc). Recently, it was shown that, in addition to an overall drift, bulk etalons show structured systematic drifts that vary as a function of wavelength (see Figure 2) [15][23]. The physical mechanism for this chromatic drift has yet to be identified, though it is likely associated with some variation in the FPE mirror stack properties. Careful investigations into the long term behavior of the optical dielectric coatings under realistic conditions (temperature stability, illumination, vacuum level) is needed to understand and mitigate these effects.

Fiber-based etalons have the benefit of being compact and easy to thermally stabilize, though birefringence issues can easily dominate the spectral stability noise floor. Unlike bulk etalons with a vacuum cavity, single-mode-fiber-based etalons exhibit birefringence due to inherent properties of the fiber that forms the cavity (see Figure 2). Crucially, fiber cavities based on non-polarization-maintaining fiber [17][18][24] produce two distinct resonance modes, one for each polarization. The spectral displacement between these two modes is set by the cavity birefringence, and the relative amplitude between transmitted polarization modes is determined by the polarization conditions of the light source [25]. While careful polarization control can mitigate these problems to a high degree, this instability still limits the utility of fiber-based FPEs when aiming for the highest precision, as typical astronomical spectrometers won't be able to easily resolve the separated transmission modes. The superposition of the two modes may manifest as a single peak in the spectrometer, but the effective line centroid will drift should the polarization conditions change.

Whispering Gallery Modes (WGM)-based etalons [18] show great promise as compact spectral filters that produce a wide bandwidth of sharp features (see Chapter 10 of the roadmap). WGM etalons are dielectric cavities with curved surfaces where light is trapped by total internal reflection from the dielectric boundary along which it travels; no optical coatings are required. Constructive interference occurs when the round-trip path length along the curved surface in the cavity is an integral of the wavelength. WGM resonators operate at any wavelength as long as the dielectric is transparent and the resonator surface is well-polished to reduce surface scattering losses.

Unlike ULE glass or Zerodur FP etalons, WGM etalons fabricated from optical crystals like $MgF_2$ and $CaF_2$ have significant thermal expansion that must be compensated to achieve a frequency stability suitable for PRV measurements. This has been attempted by engineering a composite WGM etalon that balances thermal expansion and thermo-refractive behavior [27].

**Advances in Science and Technology to Meet Challenges**

In the coming decade, the priority must be to develop stabilized systems that support broadband operation (100’s of nm), with extreme stability (<10e-10 fractional stability), and which are polarization insensitive. To meet these requirements, from the context of EPRV applications, multiple advancements must be made in a number of areas:

In the case of classical air-gap etalons, a more thorough understanding of interferometer mirror relaxation, dispersion, and coating stability are needed to better contextualize the observed behavior of current FPE systems (see Figure 2, [15], [23]). A deeper exploration of manufacturing techniques and athermal materials, with a keen eye on maximizing long-term (months-years) stability is also needed to improve upon the current state-of-the-art, if FPEs are ever to become stand-alone EPRV calibrators. Advancements in these styles of FPEs will hinge on improved dispersion stability in mirror coatings, and advanced methods for optically contacting mirror substrates to etalon spacers. In both the air-gap and fiber-based FPE systems, further investigation of ‘hard’, low-dispersion mirror coatings is also crucial for improving stability at the cm/s level chromatically.

In fiber-based systems, migrating towards using polarization maintaining, endlessly single-mode-fibers as the cavity material may offer a solution to the current limitations of these systems (polarization modes, narrow bandwidth operation). However, to establish a fully polarization maintaining system, the alignment of optical systems will have to be strictly controlled, especially during construction of the etalon cavity. Additional technical challenges are present if broadband performance is to be guaranteed. Air-gapped fiber etalons and multimode-fiber-based versions are worth exploring in the future. Fiber-based systems have the benefit of very small thermal mass, which makes them quick to respond to thermal tuning. This facilitates driving a servo loop to lock an etalon line on external laser sources.

Advancements in WGM-based designs will require exploration of waveguide materials that support hundreds of nm of bandwidth with pure single-mode operation, as is required for broadband spectroscopic applications. WGM etalons have yet to be demonstrated in an observatory environment, and will likely require a significant amount of engineering and testing to be implemented in a complete system ready for on-sky operation.

For any etalon architecture, a system that is referenced to an atomic standard (or standards) at one or even multiple wavelengths provides potentially much higher precision than a passive system. The number of wavelength 'anchors' required to fully characterize the etalon dispersion drift is unknown, though recent studies have shown that the chromatic drift can be highly structured [15][23]. This indicates that the chromatic drift due to slow changes in the mirror coatings must be calibrated regularly against an absolute, broad-band reference (like a ThAr lamp or LFC), while faster drifts (within a night) can be corrected with the frequency reference.

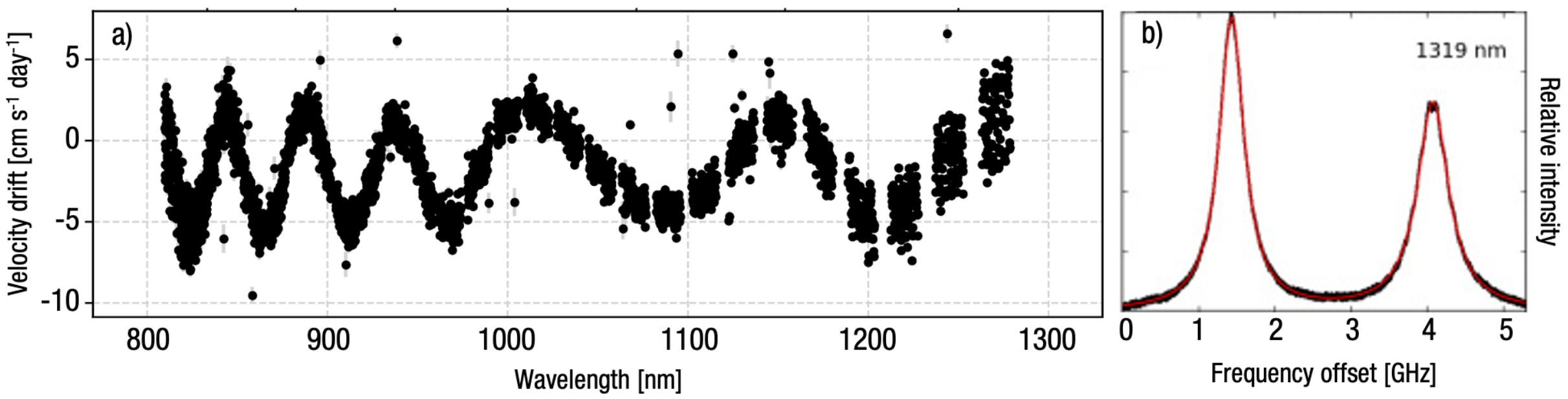

***Figure 2 -** <u>Left</u>: Chromatic drift of an air-gapped 'astro-etalon' system, showing systematic structure as a function of wavelength [18]. This is a key area of research for future etalon systems that aim for the most precise (10e-10 or lower) long term performance. <u>Right</u>: Observed birefringence in fiber-based etalon system, highlighting the two widely-separated polarization modes. These modes vary with incident polarization conditions, which are highly sensitive to input fiber geometry and thermomechanical drifts [23].*

## Concluding Remarks

The simplicity and rigidity of the FPEs sets them apart as practical systems for generating evenly spaced spectral peaks for calibration purposes. Calibrators based on passively-stabilized, bulk-optics, vacuum-gap FPEs have already demonstrated <10 cm/s stability performance, when used in combination with an absolute reference. To operate independently from any other calibration source, the absolute wavelengths of the FPEs' transmitted peaks have to be anchored. One possible way is to laser-lock the etalon to spectroscopic references. The ultimate achievable broadband accuracy of such devices remains unknown; in particular, the chromatic behavior of the locked etalons has to be studied more extensively.

Photonic Etalons, such as fiber-based FPEs and chip-based microresonators have much smaller footprints compared with their bulk counterparts, and can be thermally tuned easily. The compactness of these photonic etalons enables new calibration applications, such as in space-borne missions, where traditional vacuum-enclosed solutions are not viable due to dimension or weight limitations.

The key technical challenges going forward include material limitations, thermal stability, polarization control, and dispersion behavior.

# 15 | Future of Photonic Beam Combining Technologies for Interferometers

Karine Perraut[1], Guillermo Martin[1], Laurent Jocou[1], Elsa Huby[2] and Denis Mourard[3]
**[1] Université Grenoble Alpes, CNRS, IPAG, Grenoble, France**
**[2] LESIA, Observatoire de Paris, Université PSL, CNRS, Sorbonne Université, Université Paris Cité, Meudon, France**
**[3] Université Côte d'Azur, Observatoire de la Côte d'Azur, CNRS, Laboratoire Lagrange, France**

## Status

Optical long-baseline interferometry (OLBI) combines signals from separate telescopes and provides reconstructed images at the exquisite spatial resolution of the synthetic mirror whose aperture is sized by the separation between the telescopes. At visible and near-infrared wavelengths (e.g, in V, R, J, H, K bands; [0.5-2.5]µm), OLBI is routinely in operation at the Very Large Telescope Interferometer (VLTI, Chile), at the CHARA Array (California), and at the NPOI and LBTI arrays (Arizona), giving access to angular resolutions smaller than 1 millisecond of arc. Such an exquisite resolution opens the road for studying complex environments and determining accurate angular diameters of thousands of stars over the Hertzsprung-Russell diagram which is invaluable in the era of *Gaia* and space missions like PLATO or ARIEL [1].

While this technique has been for a long time hampered by its limited sensitivity (light is not collected by the giant synthetic mirror), GRAVITY at the VLTI [2] has revolutionized near-infrared interferometry by providing milli-arcsecond resolution imaging up to K~19th magnitude and 20-50 µarcsec astrometry, in addition to polarimetry and spectroscopy capabilities. GRAVITY has led to groundbreaking results covering the broad range of astrophysics which was targeted by OLBI long ago: the precision tests of Einstein's theory of general relativity and notably the strong experimental evidence that the compact mass in the Galactic Center is a Schwarzschild-Kerr hole [3, 4]; direct observations of the broad line regions surrounding quasars at sub-parsec scale [5]; high-resolution K-band spectra of several exoplanets [6]; the first resolved microlensed images [7]; and the resolution of several dozens of protoplanetary disks at a spatial scale where planets form [8]. For all these fields, zooming in with OLBI is of the utmost importance to reveal and constrain the physical phenomena at play in these complex and time-variable environments.

The GRAVITY breakthrough comes from a combination of fringe tracking, cooled optics, high-quality metrology, infrared adaptive optics, low-noise detectors, the upgrade of the VLTI infrastructure, and single-mode beam-combination. Focusing on the four-telescope (4T) beam-combiners at the heart of the instrument, following the pioneering integrated optics components in the H band installed at IOTA [9] and in the PIONIER instrument at the VLTI [10], the design of a pairwise static ABCD combiner has been optimized for the K band. The main challenges of the GRAVITY developments are the throughput in this specific spectral range where silica is no longer highly transmissive, operation in vacuum at a temperature as low as 200 K, and a back-illumination with a 1.908 µm metrology laser. Huge efforts were undertaken to optimize the technological process: doping process to increase the refractive index difference, annealing at 1000°C to remove the OH bonds present in the silica, manufacturing the 20 mm x 50 mm combiners in 3 stitched photolithographic reticles, as well as mastering the stress inside the component to limit the birefringence. All the challenges have been successfully tackled, leading to the most complex integrated optics beam combiner in this very specific spectral range [11].

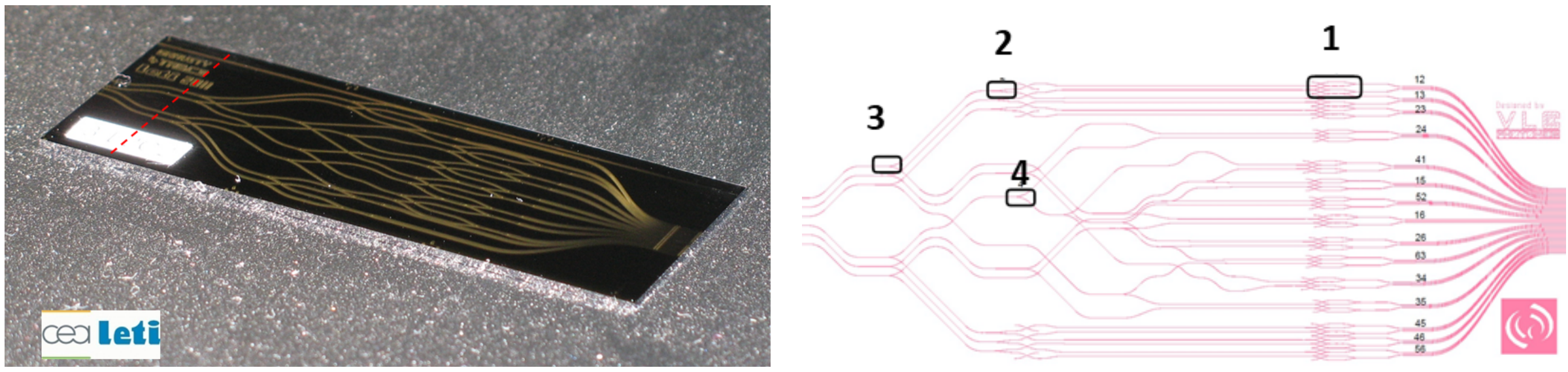

***Figure 1 -*** *Left. Four-telescope beam-combiner of GRAVITY operating in K-band, designed and manufactured by photolithography of doped-silica on-silicon waveguides by CEA-LETI [11]. Right. Design (by VLC Photonics-OCA based on [12]) of the SPICA six-telescope fringe-sensor operating in H band [13]. Each beam is divided in 5 equal parts thanks to a smart succession of 60/40 (3), 50/50 (2), and 66/33 (4) couplers. Each of the 15 possible pairs is combined inside an ABCD circuit (1), sampling the fringes with fixed phase shifts of 0, π/2, π, and 3π/2. This permits the instantaneous measurements of the phase-delay. Optical paths are equalized to avoid any bias in closure phase measurements.*

## Current and Future Challenges

**Increasing the number of beams.** Snapshot imaging of variable and complex environments makes the recombination of more than 4 telescopes a key driver for the next photonic developments, with challenges about the length, the throughput, and the complexity of the combiners. As a prerequisite, the fringe coding, i.e., the design of the circuit and the achieved signal-to-noise ratio, should be optimized. As an example, the 6T ABCD beam-combiner designed for the fringe-tracker of the visible SPICA instrument at the CHARA array optimizes the coding of the 15 fringe patterns. This 82-mm long chip exhibits a throughput higher than 50% for both polarizations over the whole H band [13]. Similar developments are ongoing for interferometers deployed on monolithic telescopes, such as pupil remapper instruments. Currently, a visible 9T-combiner is being optimized for the FIRST instrument [14], and a complementary infrared multi-beam nuller interferometer is being developed for the GLINT instrument [15] (see Chapter 18 for more details). Increasing the number of input beams to enhance the imaging capabilities of future instruments will come at the expense of complexity and a loss in signal-to-noise ratio. For more and more numerous input beams, within the context of limited wafer sizes, one future challenge would be to connect several chips together, which would make the interface, the optical path balancing, and the integration more complex.

**Improving the beam-combiner transmission.** Improving the OLBI sensitivity requires pushing the sensitivity of the beam-combiners by improving the intrinsic transmission of the materials, the coupling of the light into the waveguides, and the light collection at the output. Using larger refractive index contrast waveguides allows one to better confine the light, implement smaller radii bends and thus realize shorter optical functions, leading to more compact chips. This tackles the potential wafer size challenge. As a by-product, more compact chips could be more easily packaged and cooled down, notably for cryogenic and space applications. On top of that, more compact optical functions opens up the possibility to integrate more capabilities on a single-chip (detector and/or metrology laser hybridation [16], phase control, see chapter 22 for more details) assuming manufacturing technologies are compatible.

**Mastering spectral bandwidth and birefringence.** As several astrophysical diagnoses are based on spectral analysis, operating in different bandwidths (V, R in visible; J, H, K in near-infrared; L, M, N in mid-infrared [17]) is fundamental. As a consequence, efforts have to be made to optimize these technologies for a given band. Routine observations with precision astrometry and polarimetry requires mastering metrology propagation (not necessarily in the spectral range of the operation wavelength of the instrument), and birefringence as well as diattenuation inside the beam-combiner. As long as they are not time-variable at the observation timescale and calibratable through day-time procedures, none of these effects are insurmountable and it might even be possible to compensate for some at a system level (e.g., internal birefringence can be compensated with external $LiNbO_3$ waveplates).

## Advances in Science and Technology to Meet Challenges

**Towards more compact combiners.** For greater than 8 telescopes, planar lithographic chips require wafer sizes exceeding current technological limits when using low index difference waveguides (typ. 0.001). One option is to use high refractive index contrast waveguides (e.g., $SiN/SiO_2$ technologies with $\Delta n=0.5$), where high confinement enables sharper bends (radii of tens of microns), and therefore compact devices with multiple optical functions (splitting, directional coupling, wavelength multiplexing) [18]. The issue that arises is that high numerical aperture (NA) waveguides require tapers to maximize the coupling with input/output fibers. These tapers have a highly polarization-dependent transmission, which requires specific efforts like the development of polarization independent mode converters [19] achieving high NA to low NA adiabatic transitions.

**Enabling complex multi-channel and multi-band designs.** Thanks to the Ultrafast Laser Inscription (ULI, [20]) technique, the refractive index of a substrate can be locally modified, in particular at different depths into the substrate. Interestingly, this technique can be applied to a large assortment of materials, thus allowing for optimized devices for different wavelength bands. The versatility of ULI allows for 3D waveguide architecture, expanding the variety of possible designs, with a large number of inputs and avoiding in-plane crossings, thus reducing crosstalk and propagation losses. Applications include combining light with waveguide lattices (see Chapter 16 for more details), multimode to singlemode conversion devices (like photonic lanterns [21]), pupil remappers, converting a 2D input matrix into a 1D array (Dragonfly [22], GLINT [15]) and 3D directional couplers [23]. Challenges in ULI include the stabilization of the laser power during waveguide fabrication, and especially for writing waveguides deep inside the substrate, as power and focusing have to be finely tuned to ensure homogeneity of the refractive index modification.

**On-chip integration of injection, modulation, and detection.** The densification of the optical chip with new functions and many baselines is part of the efforts deployed to achieve fully integrated beam-combiners. Active functions based on electro-optic [24], thermo-optic [25] or piezo-optic [26] materials are being developed to allow for modifications of the waveguide's effective refractive index, and thus phase control, at rates reaching the MHz. This kind of hybrid device not only enables on-chip phase modulation but also fringe or null tracking [24, 27]. To increase the global throughput by optimal direct injection into the waveguide, microlenses can be directly integrated on the input/output facets of the chip thanks to 3D-printing [28]. In addition, novel approaches of detection consisting of growing single pixels directly over the waveguides increase readout time and output coupling efficiency [29]. This can be coupled to on-chip spectrometry using ULI (SWIFTS [30]) to obtain a fully integrated spectro-interferometer.

Different examples of the solutions discussed are shown in Figure 2.

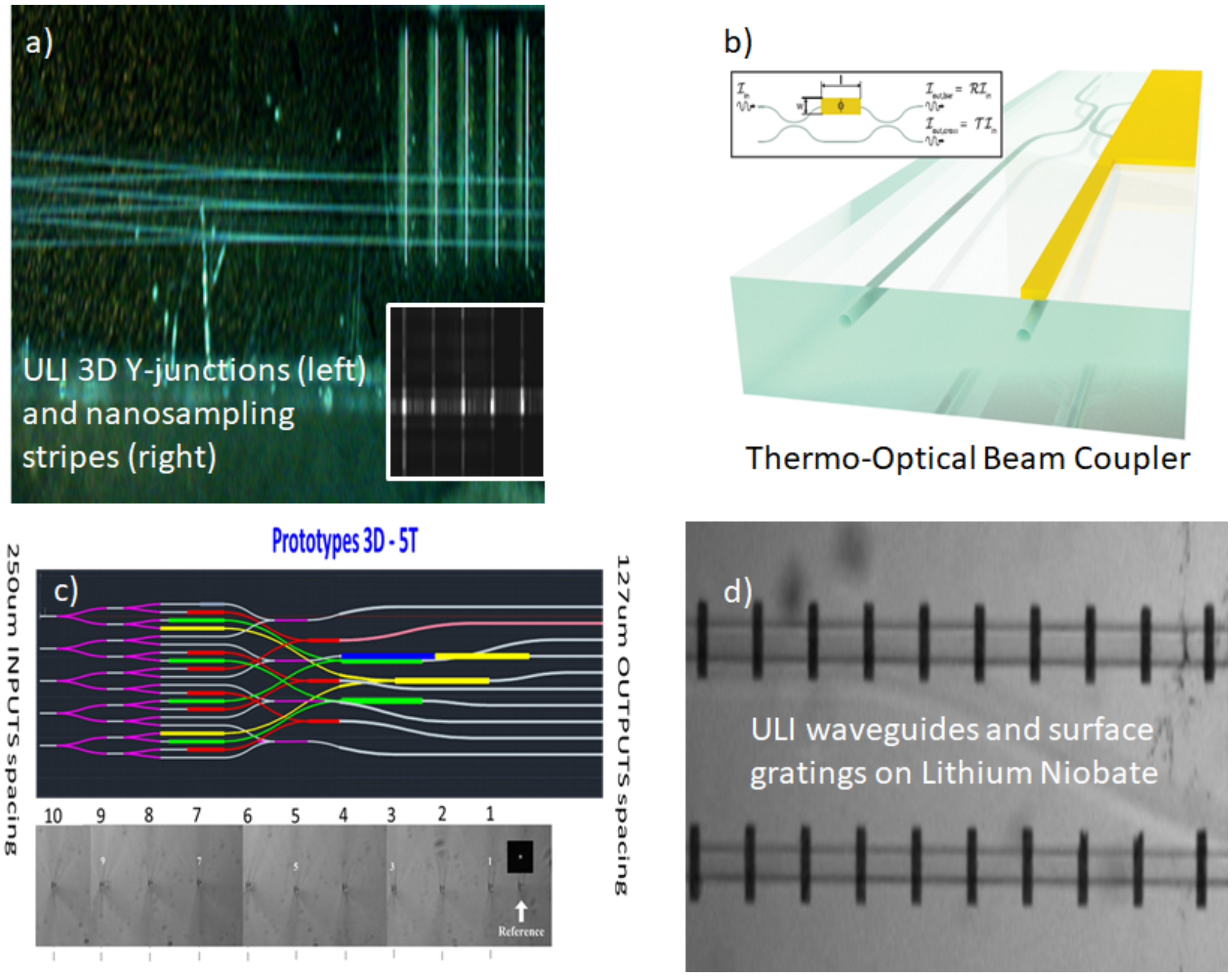

***Figure 2 -*** *Examples of some ULI beam combiners; a) Image of a 3D 3T beam combiner, with sampling grooves at the output waveguides [24], b) Scheme of a thermo-optic directional coupler [25]. Both examples a) and b) are singlemode in the near infrared. c) Scheme of a visible 5T 3D beam combiner for FIRST and white light image of the 10 outputs [31] and d) mid-infrared waveguides and surface gratings written in* $LiNbO_3$ *[32].*

## Concluding Remarks

In recent years, astronomy has greatly benefited from the coming of age of photonic technologies for beam combination, opening new opportunities for optical long-baseline interferometry. The initiation and strengthening of collaborations between research teams and technology centers now allows for the routine development of components with mature technologies, and for a dynamic synergy to tackle the development and/or optimization of specific circuits (especially for non-standard telecom spectral bands), including new functions. Notably, the advent of high density pupil-remapping projects (full 30 inputs of FIRST at SUBARU, SPIDER project with 518 sub-apertures [33]) requiring input facets with tens of waveguides in a mm-size 2D distribution will be challenging, as refractive index modification must be homogeneous whatever the depth on the substrate. The requirements for the next generation of combiners will necessitate the exploration of new technologies as the substrate size and the yield of planar photonic circuits become limiting. The ULI process is probably one of the most promising alternatives as it allows for 3D-mapping, it optimizes the circuit length, produces 3D tapers and waveguides that are in principle independent of polarization, and it is well suited for waveguide fabrication in active materials and materials having transparency windows where classical waveguide fabrication techniques are not functional.

## Acknowledgements

*GRAVITY has been developed in a collaboration by the Max Planck Institute for Extraterrestrial Physics, LESIA of the Paris Observatory and IPAG of Université Grenoble Alpes / CNRS, the Max Planck Institute for Astronomy, the University of Cologne, the Centro Multidisciplinar de Astrofisica Lisbon and Porto, and the European Southern Observatory. We acknowledge the funding support of CNRS/INSU, Agence Nationale de la Recherche contract #ANR-06-BLAN-0421, LabEx OSUG@2020 (Investissements d'avenir ANR10LABX56), LabEx FOCUS ANR-11-LABX-0013, Action Spécifique ASHRA of CNRS/INSU and CNES, and Observatoire des Sciences de l'Univers de Grenoble. SPICA-FT has been developed thanks to fundings from CNRS/INSU, OCA, UCA, Idex Jedi, and Région Sud. This work has been partially supported by the National Research Agency (ANR) through the French 'Recherche Technologique de Base' Programme. The FIRST project has received support from Action Spécifique Haute Résolution Angulaire (ASHRA) of CNRS/INSU co-funded by CNES and from the French National Research Agency (ANR-21-CE31-0005).*

# 16 | Future Prospects of Discrete Beam Combination Techniques

Abani Shankar Nayak[1], Stefano Minardi[2] and Stefan Kraus[3]
**[1] Institut für Angewandte Physik, Friedrich-Schiller-Universität Jena, Germany**
**[2] Ams-OSRAM, Jena, Germany**
**[3] Department of Physics and Astronomy, University of Exeter, UK**

## Status

Optical long-baseline interferometry (OLBIN) is an important tool of modern astrophysics, enabling high angular resolution imaging of stellar surfaces, tight binaries and proto-planetary/accretion disks. OLBIN facilities allow the interferometric combination of light collected by telescopes tens to hundreds of meters apart and are typically complex structures, where misalignment and vibration of the optical components can compromise the visibility of interference fringes. To counter the drawbacks of bulk optics at least in the beam combination instrument, Kern et al. [1] proposed the use of integrated optics (IO) components, which was successfully validated in the IONIC [2] and IONIC3 [3] instruments, and their successors, the PIONIER [4] and GRAVITY [5] instruments at the Very Large Telescope Interferometer (VLTI) facility that are currently delivering impressive high-angular observations of a multitude of objects [6, 7]. IO technologies (see Chapter 15 for more details) are able to miniaturize complex optical setups on a few centimeter long chip that consists of several photonic functions such as waveguides, Y-splitters, X-couplers, and phase-shifter. Besides the reduction of footprint, IO devices offer much higher thermo-mechanical stability and a reduction of the maintenance costs as their bulk optics counterparts. More relevant to science is the opportunity to integrate on a small chip, an instrument for the simultaneous combination of many telescopes and deliver precise measurements of visibility amplitudes and closure phases (CP), two quantities required for the interferometric synthesis of images. IO devices are usually manufactured on a variety of silicon-based materials by means of photolithographic processes, which constrain the waveguide-network to a plane. Due to the planar topology, multi-telescope beam combiners feature waveguide crossovers, which are sources of crosstalk [8] between the interferometric baselines and reduce the precision of the measured visibilities.

A possible solution to this limitation of planar photonic integrated circuits (PICs) is to exploit 3D manufacturing such as ultra-fast laser inscription (ULI - ultrashort laser pulses are used to locally modify the optical properties of a substrate material [9, 10, 11]) and avoid waveguide crossovers [12]. Therefore, a 3D-photonics approach to multi-telescope beam combination is the discrete beam combiner (DBC - see Fig. 1), a ”photonic lattice” consisting of a periodic array of parallel single-mode waveguides (WGs) that interact through the evanescent field of individual waveguide modes. The name DBC derives from the term ”discrete optics”, indicating optical systems where light is mainly bound to a discrete set of evanescently coupled WGs [13]. Initially conceived by Minardi & Pertsch [14] for OLBIN applications, the DBCs have been the focus of R&D studies aiming at measuring visibilities and CPs in multi-channel interferometers. These studies were not limited to numerical modeling [15], but included also the characterization of DBC prototypes operating at astronomical R- [16], J- [17], H- [18], L- [19]-bands and combining up to 3- [16], 4- [19, 20], and 6 [17]- telescopes. Recently, a 4-channel DBC was tested on-sky at the William Herschel Telescope during a pupil remapping experiment [18].

Compared to pairwise IO combiners [5], DBCs have a very simple design and employ only straight, relatively short single-mode waveguides, thus avoiding bending losses and have low propagation losses. Therefore, the main losses are derived from the coupling of starlight in the device, which could feature a pupil remapper [21] or a coherent reformatter [22] that has inherent bending or transition losses. Retaining high throughput or sensitivity is important for any interferometer, especially for facilities that combine many apertures for high-fidelity imaging, including proposed next-generation facilities such as the 'Planet Formation Imager' (PFI, [23, 24]) or the Magdalena Ridge Optical Interferometer (MROI [25]). ULI inherently allows low cost manufacturing of the components in a variety of materials (see Chapter 8 for details) with the transparency windows extending from the visible to the mid-infrared, the latter being crucial for the exploration of science cases related to dusty objects such as protoplanetary disks or active galactic nuclei. One important science objective, both for VLTI and CHARA, is to enable interferometry at very high resolving power > 20,000 to study molecular composition and kinematics from spectrally resolved lines (see Fig. 1). Other science cases, where DBCs could be potentially used are the characterization of exoplanet atmospheres [7], the study of the gas kinematics in the accretion/outflow launch region of young stars [26, 27, 28], and the measurement of spin-orbit alignment of exoplanet host stars [29]. Beyond OLBIN applications, DBCs coupled to a photonic lantern may enable wavefront sensing for astronomical adaptive optics systems (see Chapter 3 for details).

**Current and Future Challenges**

Science cases requiring high-fidelity interferometric imaging need a dense sampling of the coherence function (uv-plane coverage) of the starlight. Usually this is achieved by collecting visibility data across many different baselines by changing the location of the combined telescopes and/or exploiting earth rotation. This approach is possible only assuming the stability over time of the target itself. Fast varying astronomical targets (e.g. novae, variable stars, dynamical inner-disk environments, orbital motion of companions/planets) but also the truthful imaging of extended structures like protoplanetary disks requires the simultaneous combination of larger arrays of telescopes and enable a better coverage of the uv-plane within a snapshot. For instance, for a hypothetical planet at 1 au around a solar-mass star in the Taurus star-forming region, the orbital motion is up to 0.12 mas per day, which is comparable to the beam size in a PFI-like facility with maximal baselines of B = 1.2 km at observing wavelengths of $\lambda$ = 1 μm ($\lambda/2B$ = 0.09 mas). Therefore, such an array needs to achieve sufficient uv-coverage for imaging in a single night. The DBC offers a straightforward approach for scaling the beam combiner to larger arrays of telescopes and the resulting PIC is far simpler and less prone to fabrication errors than pairwise IO combiners that have ABCD structures [5]. However scaling up DBCs to larger telescope arrays will face challenges such as: 1) the factorial scaling of the algorithm currently used to select the configuration of the input sites of the DBC, 2) the photometric loss due to the quadratic scaling of the number of output waveguides (a general problem for all-in-one combiners), and 3) the decreased throughput due to the requirement to have longer devices. To date, a 6-input DBC has already been fabricated and tested in the lab [17], but the design of e.g. an 8-input DBC would require a smarter approach to meet these challenges.

Another key requirement for any OLBIN beam combiner are high instrumental fringe contrast, throughput, and broadband operation, which determines the device sensitivity. If we compare the performance metrics of a 4-input DBC (see Fig. 1) fabricated using ULI technology with its counterpart, the 4-input pairwise combiner [5] of the PIONIER instrument [4], we find that a DBC used with narrow-band light source (CW laser at 1600 nm) has an instrumental visibility (99%) and exhibit a closure phase accuracy (0.5°) comparable to that of the PIONIER combiner. However, astronomical applications need to combine light over a bandwidth comparable to an atmospheric transparency window (10% of carrier frequency). Because of the strong chromatic dependence of the evanescent mode coupling, the challenge of current generation DBC is to achieve high instrumental visibility with a broadband light source. Laboratory measurements and simulations have shown that the chromatic nature of the DBC's transfer function reduces the accuracy of the retrieved visibilities with increasing bandwidth [16, 20]. A solution to this problem is to disperse the output light with a resolving power > 50 and calibrate each individual color channel, as shown by Saviauk et al. [16].

A further requirement from astronomy is to improve the precision of the visibility measurements and achieve a better dynamic range of the reconstructed image. This is crucial for the observation of exoplanets and protoplanetary disks, where the contrast between the colder object and the bright central star is huge. The SNR of visibility amplitude and phase would require ratios >1000:1. Besides coupling the DBC to low noise detectors, the control of depolarization effects in the waveguides [30] and the improvement of the transfer matrix calibration procedure are further challenges to be addressed.

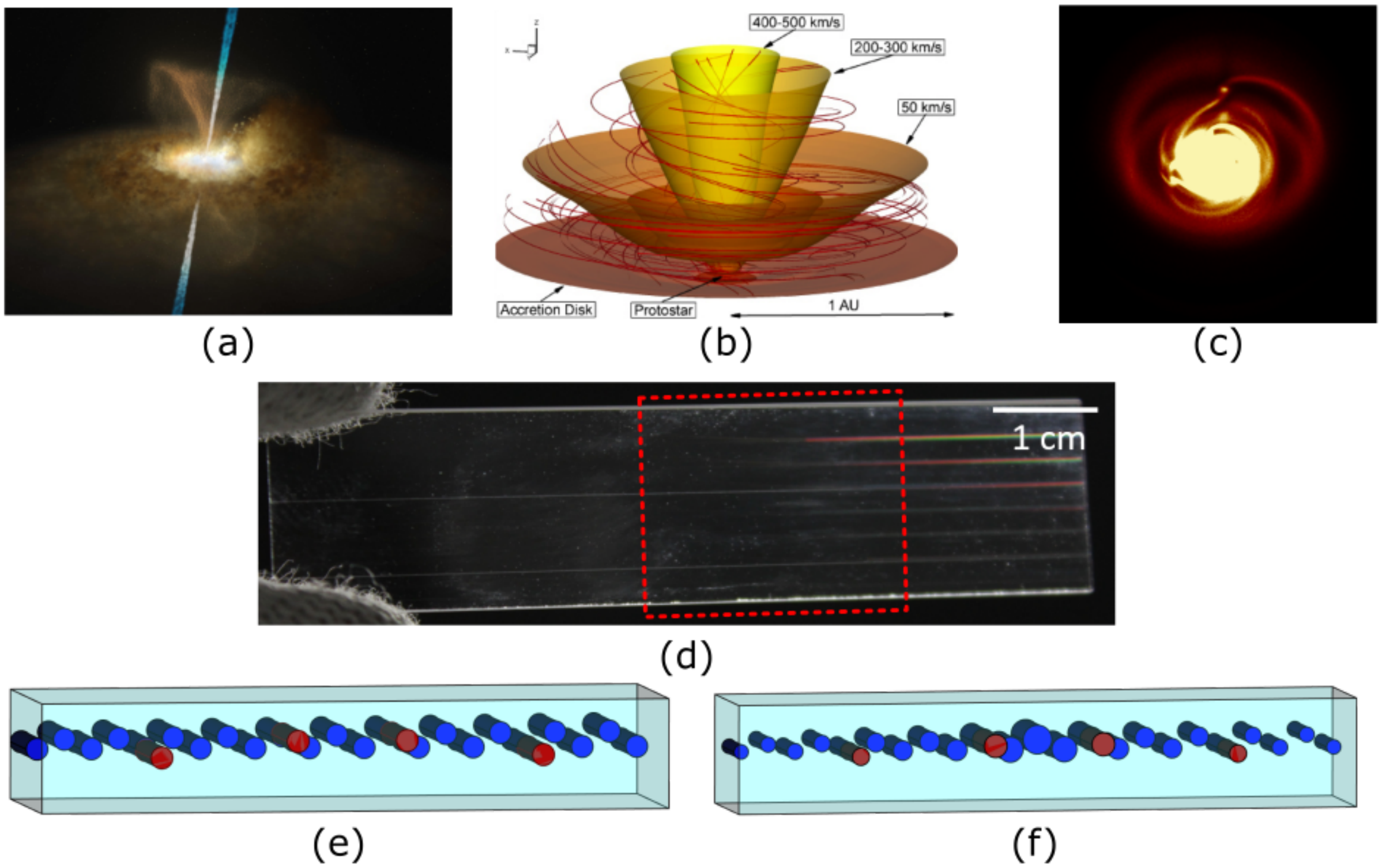

***Figure 1 -*** *The science cases potentially addressable by a DBC-based interferometric combiner are shown in (a), (b) and (c). (a) Artist illustration of the environment around Active Galactic Nuclei (credit: ESO/M. Kornmesser and L. Calcada). (b) A model showing the magnetic field topology and velocity structure in the accretion region around a T Tauri star (credit: Romanova & Owocki [31]). (c) Radiative transfer simulations of a protoplanetary disk with embedded protoplanets, computed for wavelength of 10 µm (credit: Dong, Whitney & Zhu [23]). (d) The IO chip used for on-sky tests at the William Herschel Telescope consists of pupil remappers, DBC (highlighted in red area) and coherent reformatters. Note that several DBC devices are contained in the red area (credit: AIP/A. Dinkelaker). (e) Design of current DBC with identical coupling coefficients across the width of the device. (f) Design of a possible non-uniform DBC that has apodized coupling coefficients to enable broadband light measurements and precise estimation of visibilities.*

## Advances in Science and Technology to Meet Challenges

*Broadband light extension:* Achieving high accuracy in the measurement of visibilities across a broad spectrum is still an issue for DBCs. A compromise between chromatic dispersion control, bend losses and coupling strength may be achieved by building arrays of periodically bent and/or apodized waveguides [32], which could increase the broadband operation of the DBC (see Fig. 1 for a schematic of a nonuniform DBC). Alternatively, DBC can be used in spectro-interferometric settings, e.g. by employing a hybrid integration approach by bonding the DBC chip to an IO-based arrayed waveguide grating (see Chapter 6 for details) or an Echelle grating [33].

*Polarization:* Depolarization effects in waveguide interferometers decrease the interferometric contrast and reduce the accuracy of the retrieved visibilities and CPs. While low birefringence has already been achieved in Eagleglass ULI waveguides operating at J- and H-band, this is not the case for ULI on other substrates (e.g. Infrasil [34], GLS [19], IG2 [35]) suitable for other astronomical bands. In principle it may be possible to optimize processes also on these glasses, e.g. by introducing appropriate annealing processes [36].

*Scalability:* An analytical approach has to be developed to solve the issue of factorial scaling of the current algorithm used to select the input sites of the DBC. This may be facilitated by relying on a supermode model rather than on the coupled mode approximation. However, from sensitivity considerations, all-in-one combiners perform worse when the number of combined telescope increases, but a cluster approach to combine subsets (3- or 4-) of the telescope array may be an effective strategy for imaging interferometry methods.

*Calibration:* An accurate calibration procedure of the DBC transfer matrix is essential to obtain good results in the on-sky observations [18]. Thus future implementations of DBC components would require special care in the development of a suitable calibration test-bench with low noise detectors. For instance, an adaptive optics system [37] could be included in the optical characterization setup for a more robust determination of the DBC's transfer matrix.

## Concluding Remarks

The above works have demonstrated the viability of DBC techniques and in particular, as compared to concurring IO beam combiner design, its potential for the scalability of the beam combination to large arrays of telescopes. However, the development of a full instrument based on DBC requires further technological research to address issues like the chromatic dispersion, the algorithm for the input site choice and, on a second level, the tackling of depolarization effects in the device and the improvement of the transfer matrix calibration procedures and test-benches. Research on these issues is currently ongoing and the future years may bring a DBC component to sky, triggering new instrument development for instance for VLTI, CHARA, NPOI, MROI, or aperture masking interferometry. Such devices could also help enable the next generation of interferometric facilities and promise to solve some of the major challenges facing modern observational astrophysics.

## Acknowledgements

*A.S.N acknowledges the combined support from Deutsche Forschungsgemeinschaft (326946494) and Bundesministerium für Bildung und Forschung (03Z22AN11). S.K. acknowledges support from an European Research Council (ERC) Consolidator Grant (Grant Agreement ID 101003096). The authors would like to thank Dr. Aline Dinkelaker for her valuable comments that improved the clarity of the paper.*

# 17 | Realizing Extremely Long-baseline Interferometers by Exploiting Photonic Technologies

Narsireddy Anugu[1], Gautam Vasisht[2] and Ludovic Grossard[3]
**[1] The CHARA Array of Georgia State University, Mount Wilson Observatory, Mount Wilson, CA 91203, USA**
**[2] Jet Propulsion Laboratory, California Institute of Technology, Pasadena, CA, USA**
**[3] XLIM Research Institute, University of Limoges, France**

### Status

The next leap in understanding important astrophysical phenomena e.g. protoplanetary disks and planet formation, young and mature exoplanets, environments of supermassive black-holes, requires optical/IR measurements on 10-100 microarcsecond (μas) angular scales; this is equivalent to an array with multiple km baselines in the near-IR (0.8 – 5 μm) [1-2]. For these reasons, km-scale fiber linked interferometers have been discussed since the 1980s for both ground- and space-based applications [3]. The OHANA project [4] in the 2000s was the first attempt to fiber-link the large telescopes on Mauna Kea, and succeeded in obtaining fringes between the two Kecks.

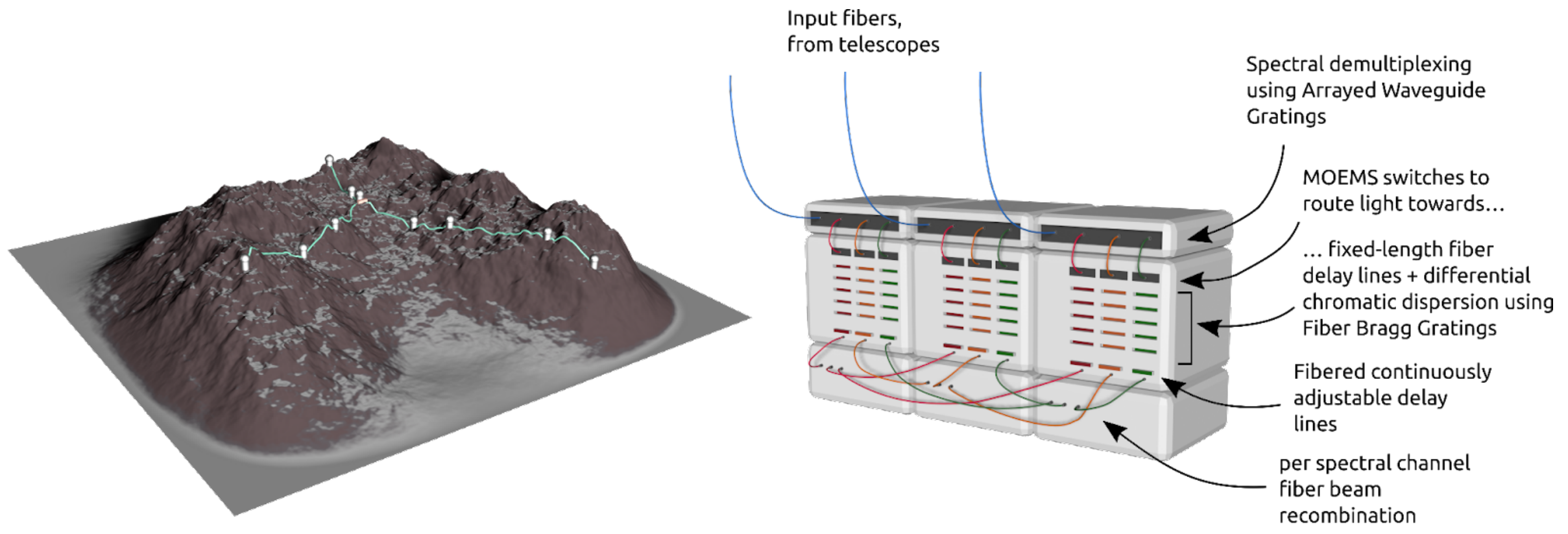

***Figure 1 -*** *(Left) A conceptual future extremely long baseline interferometer (E-LBI) that can be implemented with fiber beam transport on uneven terrain. (Right) A compact proposed fiber stabilization and delay line system.*

On km baselines fiber light transport becomes quite attractive relative to classical beam trains as one largely avoids: (i) higher infrastructural expense ( $\propto$ baseline $B$; larger optics, vacuum pipes, pumping systems, etc.) and large transport optics (diameter $\propto\sqrt{\lambda B}$) for minimizing diffraction losses (ii) site and environmental constraints due to the need for straight beam passage between telescopes and the laboratory (see Fig. 1). This constraint can block expansion of existing arrays, for example, both CHARA (B < 330m) and VLTI (B < 200m) are limited by site topography, and (iii) operational complexity, as it is harder to maintain the pupil and field from the telescope to the laboratory.

The alternative use of fibers was first discussed by Froehly in 1981 [3], when it became apparent that single-mode fibers (SMF) could offer significant advantages for both beam transport and combination. In a ground-based fiber interferometer, starlight collected by the telescope passes through a corrective adaptive optics system before injection into SMFs [4-9].

Fibers from each telescope transport light to the combiner laboratory, where path lengths are equalized using internal delays. Ideally all delay equalization would be in-fiber, however gross km-scale delays are not easily implemented in traditional fiber because of strong material dispersion and so effective dispersion management needs to be realized [10-15]. Fine delay equalization, however, can be done in fiber by stretching fiber segments [10-13] prior to beam combination [4,8]. Alternatively, as is done today, gross and fine delay equalization can be in free-space using mirror arrangements and moving delay line carts [4,8,11-12], but this breaks the realization of an all fiber, all photonic interferometer.

**Current and Future Challenges**

The main challenges in km-baseline fiber interferometer can be summarized upfront:

1. Transmission losses in fiber transport across various astronomical bands in the infrared are high. Currently only silica fibers in H-band can compete with mirror transport over km distances.
2. In-fiber delay compensation remains a major current and future challenge due to dispersion
3. Polarization and optical path difference (OPD) properties of the transport fibers could be actively monitored via laser or comb metrology. A challenge remains in demonstrating this over long distances.

Although fibers offer clear simplifications over distance, they add challenges due to their intrinsic properties (Table 1) – material absorption, chromatic dispersion [10-15], polarization birefringence, and thermal and vibration induced OPD [16-17]. Fibers are dispersive media, and dispersion due to fiber length mismatch, and fiber bending/stretching causes arbitrary phase delay as a function of the wavelength, which leads to a shift in the zero-order fringes and effectively loss of fringe visibility for broadband operation [10]. To minimize dispersion in the first order, all the static fiber lengths need to be equalized. Upstream air dispersion can be compensated with a double wedge glass system, translating one wedge with respect to the other to adjust thickness as implemented for CHARA [18]. Second-order differential chromatic dispersion can be compensated using a fiber stretcher or glass-based double wedge [10-15]. Unless dispersion can be minimized or controlled, in-fiber delay compensation is the major future challenge.

Due to their opto-elastic properties, the E-field vectors emergent from fibers are modified (unpredictably, chromatically) because of the bending and twisting, strongly reducing the fringe visibility. This can be remedied by splitting polarization states upstream, and using polarization maintaining (PM) fiber transport, with any upstream differential birefringence corrected by adding controlled amounts using $LiNbO_3$ plates [19-20]. Fibers also induce total optical path change due to thermal-mechanical stresses. Differential OPD is introduced by unequal longitudinal thermal gradients (OPD change of ~1 mm/km/K), or fast differential vibrations [16-17] which smears fringes on millisecond times or longer. While OPD errors can be reduced by careful passive insulation [8,16-17] and controlling the vibration environment, the OPD of fibers (and possibly the polarization state) can be monitored by using laser metrology and servo stabilized [16-17].

As part of the CHARA Michelson Array/ALOHA project [21], this passive and active OPD stabilization of fibers was achieved with an accuracy better than 4 nm RMS [16-17] for fiber lengths of 200 m. In this, a single metrology laser source operating at 1064 nm was split in two, injected into the same SMF used for the on-sky light transportation, and the signals from the two fibers were combined in the beam combiner laboratory and used to measured the differential OPDs by detecting fringe motion caused by the temperature fluctuations and vibrations. Next, these instrumental OPDs were corrected using fiber stretchers [16-17]. In 2022, on-sky fringes [priv comm] at 810 nm were achieved at the CHARA Array between the South1 and South2 telescopes with fiber beam transport and with a first active fiber OPD stabilization system. However, we note that the fringes were obtained using the free space delay lines of CHARA, at a narrow bandwidth of 4 nm centered around 810 nm, and using a single polarization. Using a fiber for each polarization and making fringes separately in each polarization is interesting as it also brings science opportunities, e.g., for studying dust around AGNs. Perhaps in-band optical frequency comb metrology could be conceived to monitor the state of the fiber. These on-sky results are very encouraging. However, several challenges have to be overcome in obtaining fringes at broadband wavelengths and without splitting the polarizations.

**Advances in science and technology to meet the challenges**

An extremely long baseline (>1km) interferometer (E-LBI) with multiple telescopes (>12) is conceivably the next "big project" after the completion of the extremely large telescopes. Realizing an all-fiber, all-photonic interferometer can make such an ambitious instrument quite cost-effective. We relist the challenges but herein offer potential solutions for development in the coming several years.

<u>Transmission:</u> A critical challenge is the efficiency and therefore sensitivity of the fiber interferometer, key to accessing faint primary targets (K >= 10) which in turn impacts sky coverage. Although the transmission of silica fibers in the J-H bands is good at ~0.3-0.5 dB/km, and that of ZBLAN fibers at K is tolerable, losses are much higher for other bands and are extreme in the mid-IR (5-13 μm) [Table 1 and see chapter 5]. Since mid-IR fibers have widespread uses, there is considerable research [22-23] towards improving transmission with different materials and variations such as hollow-core fibers. Heterodyne interferometry and frequency mixing techniques offer alternatives by shifting the thermal wavelengths into the radio and NIR wavelengths respectively [See Chapter 21]. Currently, these techniques are limited to a narrow spectral bandwidth and lower sensitivity but they remain interesting for the future.

<u>Wavelength coverage:</u> Another challenge is maximizing broadband wavelength transportation while keeping the SMF properties and maximum fiber-coupling. For shorter wavelengths, the waveguide can become multimoded. For broadband wavelengths, the input beam size may not match the mode field diameter of the SMF leading to non-optimal fiber beam coupling. A promising alternative is the use of polarization-maintaining photonic crystal fiber, which are single mode over the entire transparency range of silica (0.3-2 μm). The linear losses of this type of fiber are currently too high (> 10dB/km) to consider for the km scale baselines. Still this technology remains attractive in the future if losses and costs are reduced.

Fiber stabilization: Despite the availability of a state-of-the-art solution for the fiber stabilization for a two-fiber interferometer using a dedicated metrology laser [16-17], we need a solution for multiple telescopes, e.g., six telescopes of CHARA [5]. The main challenge would be after splitting the coherent metrology laser beam, how to send it to multiple telescopes separated on km scales and recombine the beams by building a multi-beam combiner to detect differential OPDs compared with a reference fiber.

Fiber delay lines: Existing interferometers use either free space air (VLTI/CHARA/NPOI) or vacuum-pipe delay lines (MROI) employing movable optomechanical trolleys. The delay lines are complex, and a low cost all-fiber solution would be attractive for a fiber interferometer. Only a few tens of cm of OPD [9-11] have been demonstrated by fiber stretching. However, fiber stretching can only come with side effects such as additional polarization, dispersion, and transmission issues, all of which depend on the stretch-state and wavelength [11-13]. Large fixed delays could be provided based on mechanical switches, e.g. Micro-optoelectromechanical (MOEM), that can allow the addition/removal of static fibers with a library of fixed lengths to the optical path similar to pipes of pan (POP) mirrors at CHARA [25] (see Fig. 2). Photonic dispersion via an arrayed waveguide grating spectrometer [see Chapter 6], would allow for an input beam delivered by the fiber delay lines to be dispersed into discrete bands of interest, with each spectral channel coupled to a unique output fiber. In this way, the corresponding spectral channels from different telescopes could be combined with channel specific beam combiners, rather than broadband combiners, allowing for more optimal devices to be used. In addition, fringe tracking, via phase shifters on the beam combiner or fiber stretchers, could be more effectively used to compensate over the narrower bands of each channel. The approaches and technologies described here would require a comprehensive maturation program of technology development and demonstration before they can benefit long baseline interferometry.

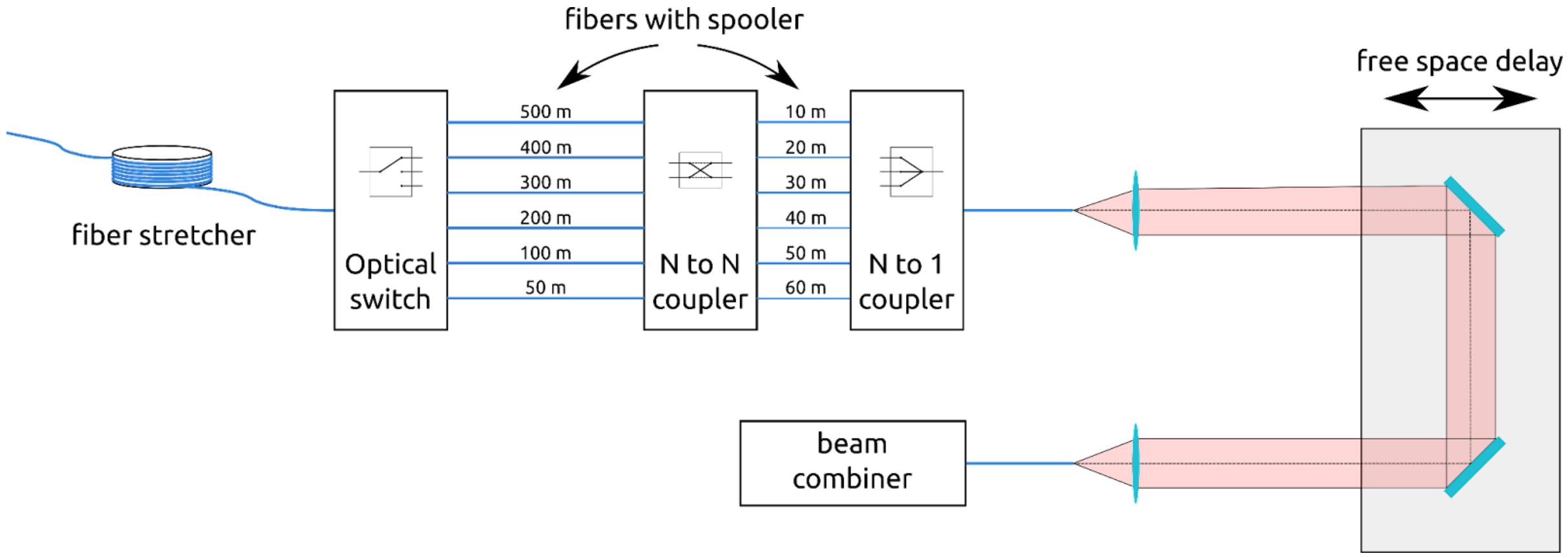

***Figure 2 -*** *A schematic diagram of three stage delay lines: fiber stretcher, fiber delay lines with mechanical switches and free space delay. Since the additional delay is in glass, we would have to correct for dispersion differences to match the vacuum path in other telescope arms. This dispersion correction can potentially be realized with a custom dispersion compensating Fiber Bragg Grating (FBG) for each static fiber length. A combination of (i) longer static optical delays in fibers with fiber switches and small delays using (ii) fiber stretching or (iii) even free space delays of a few meters that address fast fringe tracking correction could be a solution to the current limitation of entirely free space delay lines.*

## Concluding Remarks

For extreme long-baseline interferometry, beam transportation with SMF offers a unique advantage. Recent improvements in the transmission, wavelength coverage, and fiber control of the single-mode fibers may already provide partial solutions to some existing challenges. However, active technology development in this area is required to maximize the transmission, wavelength coverage, and usage of fiber delays. These breakthrough technology developments could deliver transformative science results by expanding the baseline size beyond the current < 1 km barrier.

## Acknowledgments:

*We wish to thank Robert Ligon and François Reynaud for fruitful discussions. We would like to acknowledge suggestions about dispersion compensating FBGs and AWGs made by Nemanja Jovanovic.*

***Table 1 -*** *A compilation of Fiber optic materials that may be used in optical/IR interferometry. Hollow core fibers (HCF) and hollow core photonic bandgap fibers (HC-PBF) are not explicitly listed but have been demonstrated in silica and chalcogenide glasses. They are an active area of research [23-24].*

| Fiber Type | Transmittance | Advantages | Disadvantages |
|---|---|---|---|
| Silica | High transmittance optical through H band<br><br>0.3 dB/km (1.2-2.0 μm)<br><br>5 dB/km (0.5-1.5 μm) | Non-toxic<br><br>Non-hygroscopic<br><br>Rugged, stable<br><br>Photonic crystals and PMF available<br><br>HC-PBF in rapid development | Brittle without proper coating |
| Fluoride:<br><br>ZrF4 (ZBLAN)<br><br>InF4 | High transmittance in K-L bands<br><br>1 dB/km at 2.5 µm<br><br>3-4 dB/km at 3.5 µm | Non-toxic<br><br>Stable with temperature<br><br>Low dispersion<br><br>Solid Core Photonic structures developed | Brittle without proper coating<br><br>ZBLAN hygroscopic |
| Chalcogenide:<br><br>As-S<br><br>Ge-As-Se-Te (GAST)<br><br>Ge-As-S-Se (GASS) | Transmittance in 0.7-10 µm<br><br>1 dB/m at 4.0-6.0 µm (As-S)<br><br>0.2 dB/m (8.0-11 µm GASS)<br><br>Theoretical expectation for HC-PBF is 1 dB/km | Non hygroscopic<br><br>HC-PBF in development<br><br>SC Photonic structures developed | High refractive index (n)<br>High dn/dT<br>Toxic<br>Fragile |
| Polycrystalline IR fiber (silver halide) | Transmittance in 3-17 µm<br><br>0.2 dB/m (9-13 µm) | Non-hygroscopic<br><br>Non-toxic | Long term photodarkening<br><br>SMF difficult to manufacture |

## 18 | Nulling Interferometry With On-chip Beam Combination

Marc-Antoine Martinod[1], Romain Laugier[1], Michael Ireland[2] and Jeffrey Jewell[3]
**[1] Katholieke Universiteit Leuven, Belgium**
**[2] The Australian National University, Canberra, Australia**
**[3] Jet Propulsion Laboratory, California Institute of Technology, Pasadena, CA, USA**

### Status

Exoplanetary science is amongst the most active fields in astronomy. While thousands of exoplanets have been discovered, few have been characterized, and even fewer have been detected from the habitable zone to the ice line. These regions are key to understanding planetary formation **[1]** and studying the exoplanets' features such as the surface **[2]**, the atmosphere **[3]** or spectral biosignatures **[4]**. Such studies require careful measurement of the orbital parameters as well as the spectrum of the exoplanets, particularly in the infrared, ranging from 1.5 µm for close-in or young giant planets, up to 18.5 µm for characterizing temperate planets from space **[5].** However, these observations are challenged by the small angular separation between the planet and its host star and the high contrast needed to discriminate the faint light of the planet from the overwhelming glare of the star. Although relevant planetary contrasts in the literature range from $10^3$ to $10^{10}$, the post-calibration contrast range for infrared nulling interferometry ranges from $10^4$ for 50 Myr giant planets or typical hot Jupiters **[6]** to $10^9$ for 4 µm thermal emission from an Earth-like planet around a solar-type star **[5]**.

Nulling interferometry produces interference so that the on-axis light is nulled out (and redirected to a separate path), while the faint, off-axis source's light can be gathered and sent to a sensor or spectrograph **[7]**. This technique provides all the capabilities to address these scientific questions and tackle the challenges: resolving and revealing the star's surroundings at high contrast, measuring the relative position of companions, and spectrally analyzing their light. To maximize the impact of the nulling, we need stable suppression and a diversity of spatial information to improve observing efficiency. Doing so in the mid-infrared enables compelling observations with less difficulty to meet the contrast requirement than at shorter wavelengths.

Nulling can equally be performed by combining the light from several telescopes as in long-baseline interferometry or by segmenting a single dish like aperture masking and using a photonic integrated-circuit (PIC). In all cases, the technical challenges for PIC developments remain similar. To date, the only attempt to observe on sky with an on-chip nulling beam combiner was by the GLINT instrument (Fig. 1), using integrated-optics to combine four sub-apertures from a monolithic telescope simultaneously with modal filtering **[8]**, and spectroscopy **[9]**. The upcoming ASGARD/NOTT instrument (formerly Hi-5) operating in the L-band (3.5-4.0 µm) will use a PIC to implement the latest developments of nulling with the long baselines and large apertures of the VLTI **[10]**. Although they are currently lossy, photonic technologies offer several assets for nulling including: modal filtering, flexibility in the circuit design, stability, compactness and low mass. The last two are also attractive for the prospect of space missions **[11]**. These assets make the PIC an ideal platform to combine at least 5 apertures simultaneously and provide a minimum contrast of $10^5$ (to get imaging capability). To achieve such a threshold, PICs will deliver most of the suppression but innovative data processing techniques can help enhance the contrast by an additional order of magnitude, such as numerical self-calibration **[12]** or machine learning techniques **[13]**. In the next sections, we will explore the challenges of using photonics for nulling and how we can address them over the next decade.

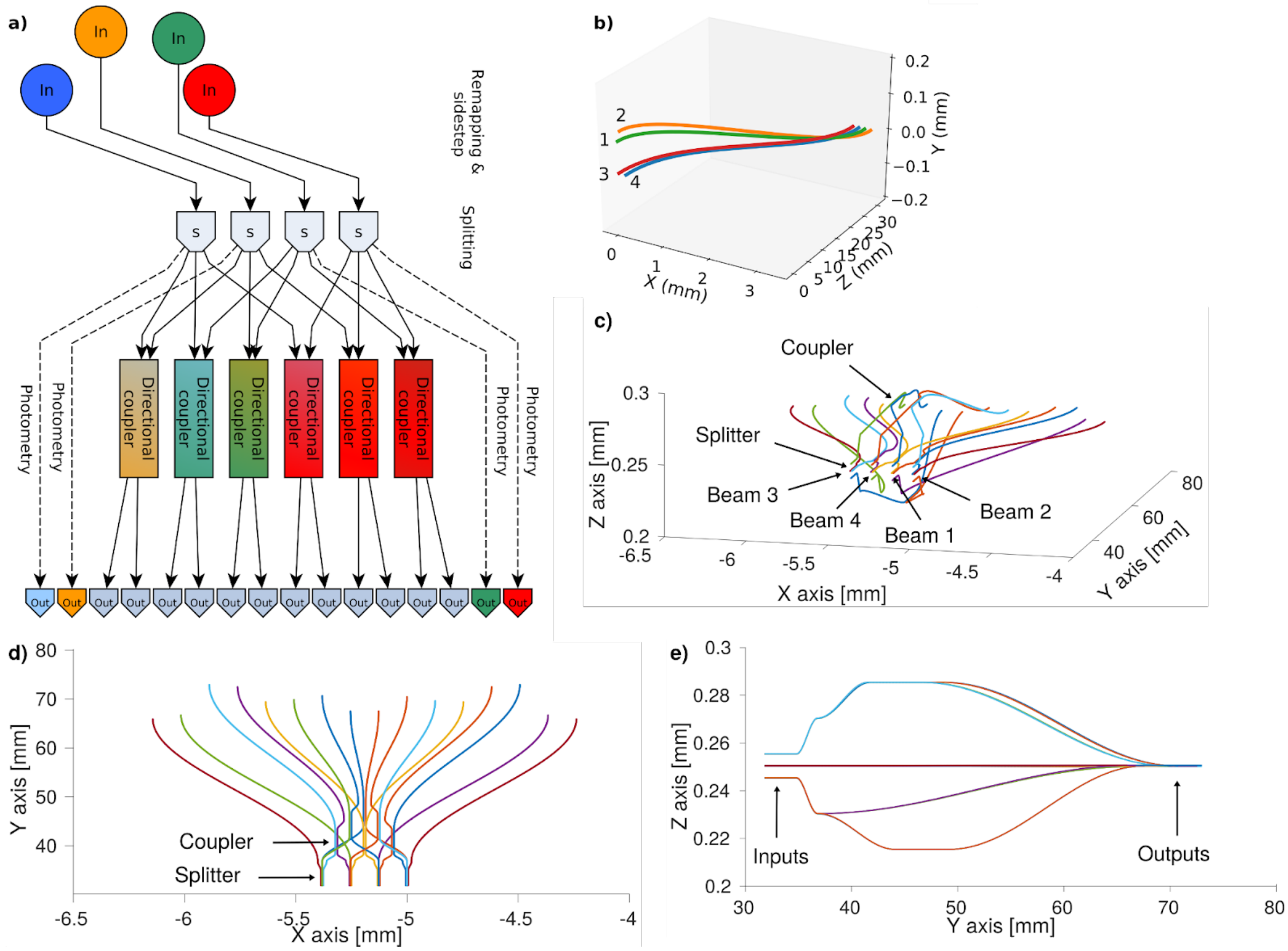

***Figure 1** - Schematics of GLINT's integrated-optics chip. **a)** Conceptual layout of GLINT showing the inputs (a pupil segmented into 4 sub-apertures), the sidestep, the remapping, the splitters, the directional couplers and the photometric paths, and the outputs. **b)** Schematics of the pupil remapper of the chip, coherently transforming the 2D configuration of the inputs (on the left) matching the desired pupil sampling pattern into a 1D configuration (on the right). The waveguide paths have been optimized to match their optical path lengths despite taking different routes. **c)** Perspective view of the beam combiner of the chip. **d)** Plan view in which light propagates from the 4 inputs at the bottom towards the top, encountering 4-way splitters and codirectional couplers. **e)** Right-side view of the chip showing the locations of the inputs and the outputs. The axis scale proportions in all the schematics differ for clarity in the drawing. Credit for b-e: **[9]**.*

## Current and Future Challenges

The most obvious challenge is the injection of the light into a single mode waveguide. A high quality wavefront and stable point-spread function, either from a space telescope or from an adaptive optics system (AO), is required to reach the best coupling **[Chapter 2]**. Any light failing to couple in the waveguide propagates through the bulk of the chip and can re-couple into the waveguides or contribute to a variable incoherent background. Using a sidestep on the input waveguides to get out of the stray light cone solves this issue **[14]** (Fig. 1b and 2b). Other solutions to investigate could be adding baffles inside the chip or modifying the surface of the chip to prevent total internal reflections.

Temporal fluctuations of the light that gets coupled into the waveguide leads to imbalance in the recombination, and additional light leakage. For this reason, input waveguides must feature photometric taps to sample the light of each beam **[15]** for real time monitoring and post-processing. However, they do not allow an active optimization of the injection against variable environmental conditions like vibrations or atmospheric turbulence **[next section]**.

Once in the chip, the light of the on-axis source can be nulled. However, any instrumental deviation from the correct coherencing of the combined inputs leads to a leakage of the starlight in the nulled output that is difficult to disentangle from the planet's signal. Therefore, it is critical to control the amplitude and the phase of the light being combined. An AO system and a fringe tracker are necessary to handle these but they are insufficient. Simple combiners, like the Bracewell configuration, are sensitive to such light leakage. Combiners that are less sensitive to such effects and solutions providing a balanced and broadband coupling of light are needed **[next section]**.

Additionally, the astrometry of the observed planet requires more diversity in the sampling of the UV plane; hence it is a driver for more complex nulling chips to simultaneously perform nulling on multiple baselines. GLINT pioneered it by using beamsplitters to attempt the simultaneous nulling of all six baselines **[9]**. However, differing optical paths inside the chip make nulling on more than two baselines simultaneously impossible. While such imperfections can be calibrated in an instrument like GRAVITY, they must be minimized for nulling devices to efficiently suppress the on-axis starlight.

Fabrication with the Ultrafast Laser Inscription (ULI) technique in borosilicate and Gallium Lanthanum Sulfide glass also face challenges. The low refractive index contrast delivered by this fabrication technique limits compactness (hence scalability and interest for space missions), especially to implement the aforementioned sidestep. Lithography provides higher index contrast but it does not offer the ability to realize circuits in three-dimensions like balanced 3D tricouplers **[next section]**. In addition, nulling the maximum numbers of baselines simultaneously requires an accurate match of the optical paths between one input and the fed baselines. To date, there is no such criterion during the fabrication.

While the fabrication and the choice of the material are mature for chips used at 1.5 µm, there is little experience at longer wavelengths. Some possible materials have been identified (Si, Al2O3, Ge, GLS) but lithography and ULI are not routinely used for nulling chips working at wavelengths >3 µm. Efforts are required in the fabrication process to deliver cost-effective couplers and splitters that meet the requirements.

## Advances in Science and Technology to Meet Challenges

All the challenges above can be overcome. The photonic lantern could be a promising solution to provide active control for injection **[Chapter 3]**. Such devices have been directly inscribed inside ULI-fabricated chips in the past **[24]** but their use for active injection has still to be demonstrated. In addition, photonic beam combiners are especially interesting for space-borne instruments but the injection of starlight in space-borne photonic devices is still to be demonstrated (e.g. the PICsat mission **[20]** that was lost). It would open the door to low-cost instrument concepts for space observatories.

In order to provide the possibility of building up to more complicated solutions utilizing photonics technologies, characterization should provide not just the coupling ratio but also the required wavelength-dependent relative input phase in order to achieve a deep null, and how polarized light propagates through the chip. Such characteristics are rarely available in the literature, except in the case of simulations.

Active control for on-chip tuning (e.g photoelectric **[17]**, thermo-optical **[18]**, piezo-electric systems **[23]**) can relieve the requirement on the optical path match between the waveguides upstream of the couplers by allowing for tuning of the input phase. Careful configuration of the baselines and sophisticated combiners could help to reduce the leakage due to the spatial coherence of the source or the atmospheric turbulence. The main illustrations of these configurations and combiners include the double-Bracewell **[19]** and the kernel-nulling **[16, Chapter 19**] variants. Another part of the difficulty in perfecting optical path control lies in the non-common path between the fringe tracker and the nuller. Three-dimensional tricouplers offer a combination of two bright outputs instead of one, allowing a measurement of the residual phase of the combination. Furthermore, symmetry in this 3D design is favorable to achieving a broadband null **[21]**.

However, the integration of the tricoupler leads to the challenge of the dynamic range of the spectrograph's detector. Different sensors, providing short integration time and low latency for the bright channels and longer integration time and low dark current for the dark outputs could be used; such setups have yet to be tested**,** and are also required for the use of photonics lanterns for assisting with the active injection. A broadband nuller also requires an achromatic phase shift that simple air-delay lines cannot provide. A promising concept of an achromatic phase shifter of 180° made in ULI and placed upstream of a tricoupler has been explored **[22].** Its association with a tricoupler enables the scalability of broadband nullers and is compatible with complex circuits. Last but not least, perfecting such phase shifters is crucial to enable nulling with new photonics platforms.

The push toward longer wavelengths demands operation at cryogenic temperatures. Temperatures below 100K are required to make the background emission of the chip negligible compared to the telescope's thermal emission at 3.5 µm **[16]**. This constraint is a challenge for devices using glues or other bonding techniques between multiple chips, V-grooves, or microlenses **[Chapters 23 and 24]** and may impact the refractive index contrast. The nulling instrument NOTT **[10]** will pioneer the development of photonics that can withstand these operating conditions. This nuller features a sidestep, an achromatic double-Bracewell combiner and photometric outputs embedded in a single chip with ULI technique (Fig. 2).

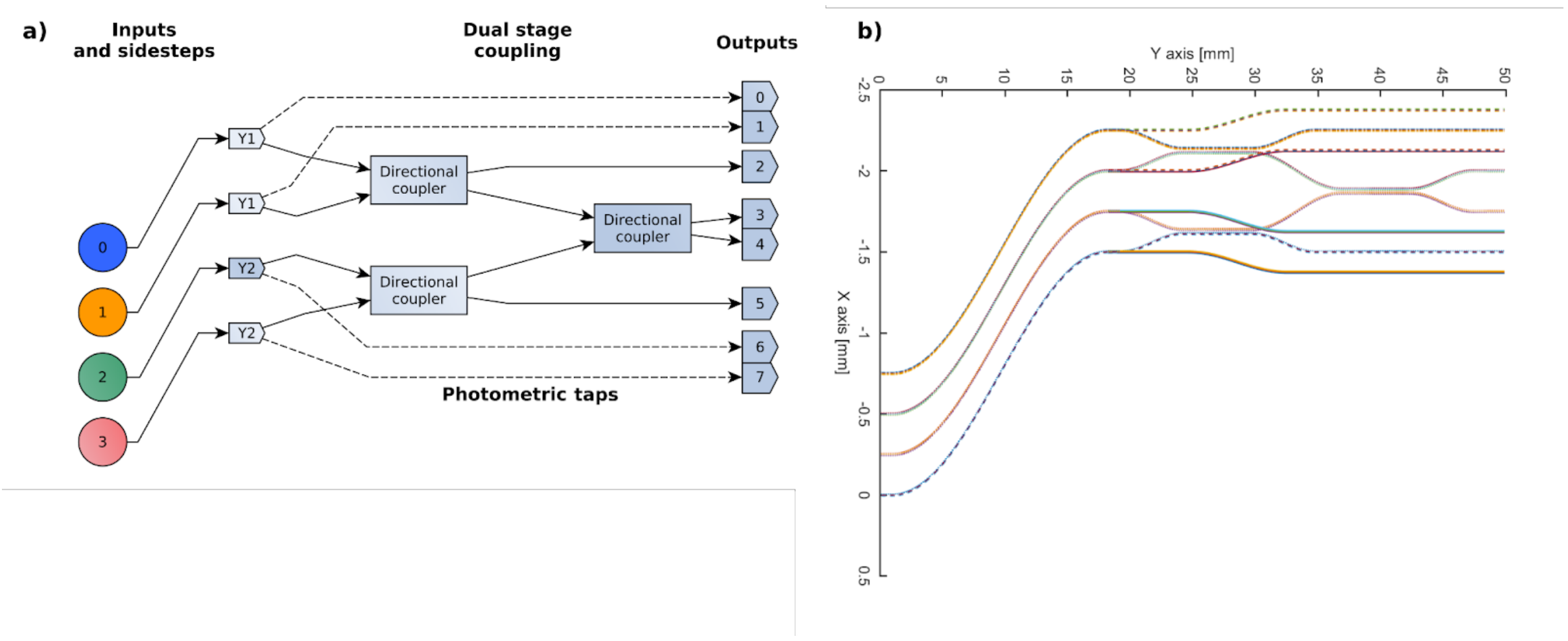

***Figure 2 -*** *Diagram of the photonic chip of Nott.* ***a)*** *Conceptual diagram showing the inputs, the sidesteps, the dual stage coupling (double-Bracewell) featuring tapered directional couplers to get achromatic coupling ratios, the photometric taps and the outputs.* ***b)*** *Top-view of schematics of the chip.*

## Concluding remarks

The prospect of nulling puts additional constraints on the requirement of interferometric beam combination. The need for precise matching of input phase and amplitude throughout the operation imposes a thorough characterization of the solutions in both phase and amplitude. While their host stars are bright, the planets are faint, leading to a need for throughput that encourages finding solutions to injection, achromatic coupling strategies and efficient on-axis light suppression.

Observations in the infrared region above 3.5 µm, require further development of circuitry with the mentioned requirements in a cryogenic environment for more ambitious ground and space projects, like the detection of different molecular signatures in the atmospheres of giant young planets. While the simple Bracewell combination is not expected to be sufficient for the more ambitious goals, the challenges mentioned here pave the way toward efficient and sensitive nulling interferometers. Synergies about contrast and angular resolution between a space nulling interferometer and space IR-O-UV observatories can also be explored.

## Acknowledgements

*This work has received funding from the European Research Council (ERC) under the European Union's Horizon 2020 research and innovation program (grant agreement CoG - 866070) and from the European Union's Horizon 2020 research and innovation programme under grant agreement No. 101004719.*

## 19 | Kernel Nulling Self-calibration

Nick Cvetojevic[1], Frantz Martinache[1] and Sylvestre Lacour[2]
**[1] Université Côte d'Azur, Observatoire de la Côte d'Azur, CNRS, Laboratoire Lagrange, France**
**[2] LESIA, l'Observatoire de Paris, Université PSL, CNRS, Sorbonne Université, Université Paris Cité, Meudon, France**

### Status

The use of interferometric nulling for the direct characterization of extrasolar planets is an exciting prospect, but one that faces many practical challenges when deployed on telescopes. Despite the success of nulling interferometry [1], most photonic implementations thus far have been of the pairwise Bracewell type, and face considerable difficulties when implemented on-sky. With the key requirement of nulling interferometry being the stable destructive interference of two or more incoming interferometer arms, an exceptionally high level of up-stream wavefront control and pathlength matching is necessary to achieve this. This is needed to stabilize the injection into the single-mode waveguides, and provide high-speed correction of the differential piston between the arms which directly impacts the photometric leakage term in the null channel. With the key scientific driver of this kind of interferometry being high-contrast detections close to the stellar host, a large and time-varying photometric leakage is undesirable.

Various numerical self-calibration methods [2-5] have been demonstrated to retrieve the astrophysical null in the presence of wavefront residuals, by exploiting the statistical distribution properties of the signal and various noise sources. However, these approaches have some limitations, with the model-fitting being fairly complex and scaling dramatically with more inputs, and the need for large datasets to obtain enough measurements to build accurate statistical models. Non-nulling interferometry typically makes extensive use of the production of self-calibrated observables, like closure phases [6,7], and their generalized form, kernel phases [8], to sidestep the limitations brought about by wavefront residuals. Bringing together the benefits of self-calibration of interferometric observables and the photon-noise suppression of nulling is a powerful and exciting tool as it opens a previously unexploited parameter space.

The concept of creating a directly self-calibrated nulling observable capable of removing the effect of residual wavefront errors has been proposed in the past, by using a double-Bracewell architecture [9-11], and by exploiting the measurement of fringes in the leaked light of the nulled channel [12]. However, a more versatile solution exists where specially designed beam combiners can produce nulled outputs whose linear combinations produce self-calibrated observables [13]. These observables, called kernel-nulls, are robust to upstream differential pistons or phase errors. In general, kernel-nulls can be created by subtracting the measured intensities of two nulled outputs, provided the electric field in the two outputs are complex conjugates, with one possible 3-input configuration shown in Figure 1. In practice, this is easily achieved in photonic interferometers by constructing a beam-combination architecture where input beams are recombined by mirroring phase offsets.

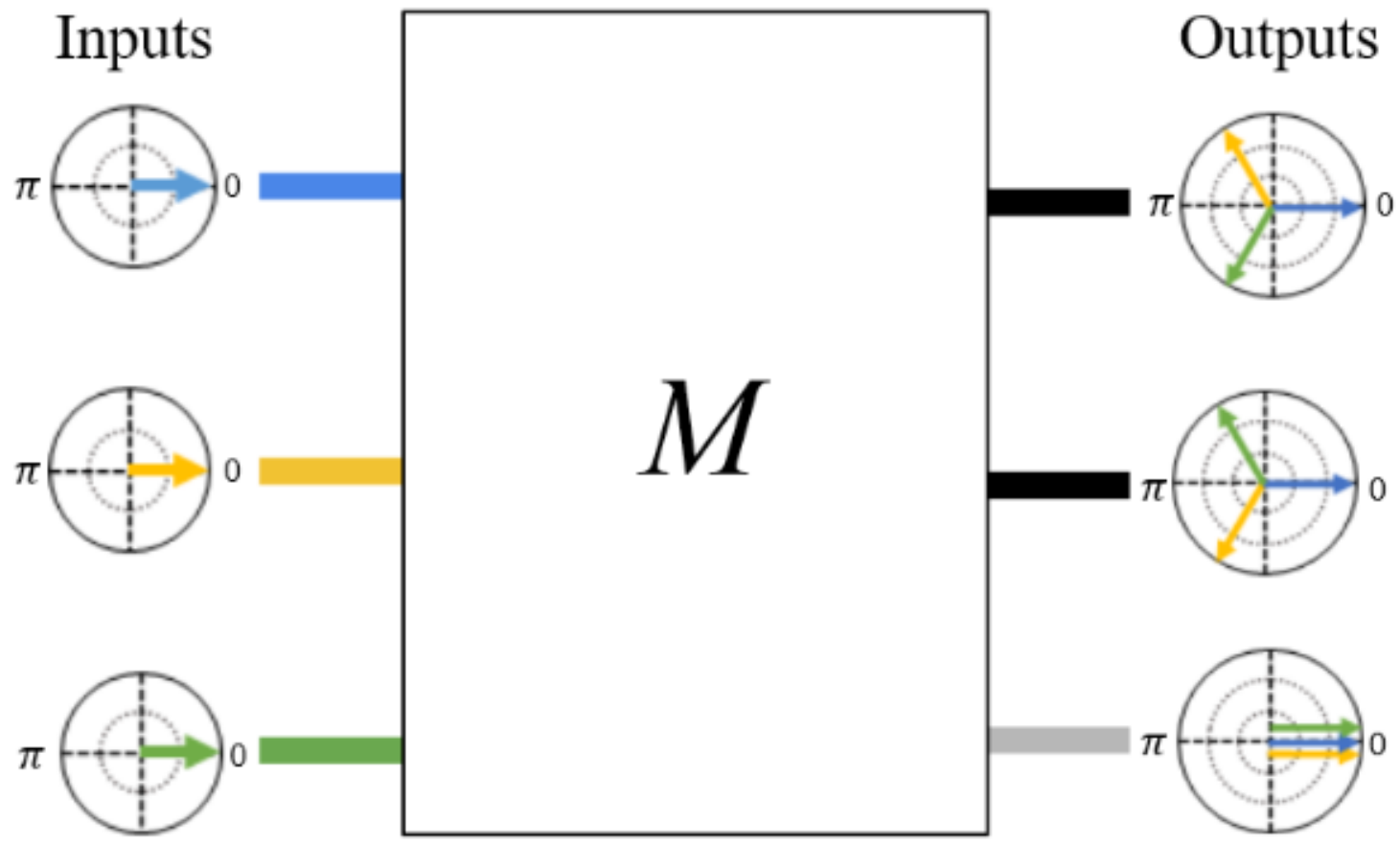

***Figure 1*** *- The electric field is represented by colored phasors on a complex polar plot, with the length of the vector being the amplitude and the angle showing the phase. For three in phase inputs, the Kernel Nuller creates three outputs. One bright channel where the input E-fields are in phase creating constructive interference, and two nulled channels where the input beams are given a ±2π/3 phase offset before being combined. In these channels the E-fields interfere destructively creating a null.*

Recent laboratory work has demonstrated the successful creation of kernel-nulls in 3-input [14] and 4-input [15] photonic beam combiners. These initial combiners avoided the use of classic pairwise combination using directional (evanescent field) couplers, and instead integrated the beam splitting and recombination into a single photonic component. This component is the multimode interference coupler (MMI), a planar multimode waveguide section (often rectangular) where multiple incoming waveguides are allowed to interfere spatially to form the required intensity profile at the MMI end face. By varying specific parameters (such as the length and width of the multimode waveguide section), the exact beam-combination required to achieve kernel-nulling can be directly designed into the MMI.

### Current and Future Challenges

In addition to the challenges shared with most astrophotonic interferometers, the unique beam-combination techniques required for kernel nulling raise additional technological challenges. The key requirement faced by any implementation of kernel nulling is the stable, accurate, and broadband pathlength control of the incoming beams prior to a beam combiner. While traditional pairwise Bracewell nulling also faces this challenge, kernel nulling often necessitates second, or third stage beam combiners, all of which must be accurately path-matched to an accuracy of a few degrees in phase. While bulk-optic discrete designs are being explored for kernel nulling, particularly in the context of space interferometers [16], the required stability and accuracy are challenging to achieve. Integrated photonic solutions are far more accommodating in this respect, but add complexity beyond what is seen in standard nullers. Because kernel nullers cannot function in a pairwise fashion (nulling one interferometric baseline at a time), and must interfere all incoming beams together in the null channels, any design using two-by-two beam combiners must be cascaded into multiple subsequent recombinations.

An alternative method to cascading 2-beam couplers is to instead collapse all the combinations into a single MMI. While utilizing MMIs as the primary beam combination component has distinct advantages (such as simplicity, robustness to manufacturing errors, and scalability), it does come with its own set of drawbacks. The primary challenge arises from the multimode nature of the interference inside the MMI, which is inherently wavelength-dependent. While more tolerant than its directional-coupler counterparts, this unwanted chromatic behavior can further be exacerbated by manufacturing errors, particularly in high refractive index contrast materials. While traditional directional couplers are also chromatic, the methods of creating broadband MMIs are different and need to be developed further in an astronomical context. Additionally, kernel-nulling imposes further constraints on the level of phase offset accuracy, and photometric balance in the splitting, as the self-calibration degrades the further they deviate from the ideal. Unlike traditional architectures, where each splitter and coupler are individual components on a chip and can be developed independently, utilizing an MMI for both tasks requires a more thorough design procedure to ensure it maintains performance across a wide wavelength band.

While a 4-input kernel nuller has been demonstrated in the laboratory using a single 4x4 MMI, this method does not produce all the possible kernel null observables. To make a complete kernel nuller for 4 input beams requires a second stage of beam-recombination after the initial nulling stage, shown in Figure 2. The practical impact of this is that any processes used upstream to phase up the interferometer arms (delay lines, deformable mirrors) can no longer be used without dephasing the first stage. Thus, either the pathlength-matching of the waveguides in the second stage must be made to very high tolerance, or the use of on-chip phase control is paramount. Thermo-optic phase actuators (TOPAs) integrated into the photonic chip are capable of correcting pathlength errors due to manufacturing or environmental drift, in a reproducible and stable manner. However, because the phase delay is created by a change in the effective refractive index of the waveguide, it is wavelength dependent. While the chromaticity is negligible for small TOPA strokes (less than 10 degrees of phase), correcting larger pathlengths of a few π becomes difficult to do achromatically. As these devices will be required to work over large wavelength bands the underlying waveguide structure inside the TOPAs need to also be made as achromatic as possible. This is especially true if correcting for OPD errors that arise before injection into the photonic chip, when attempting to stabilize an atmospherically perturbed wavefront for example.

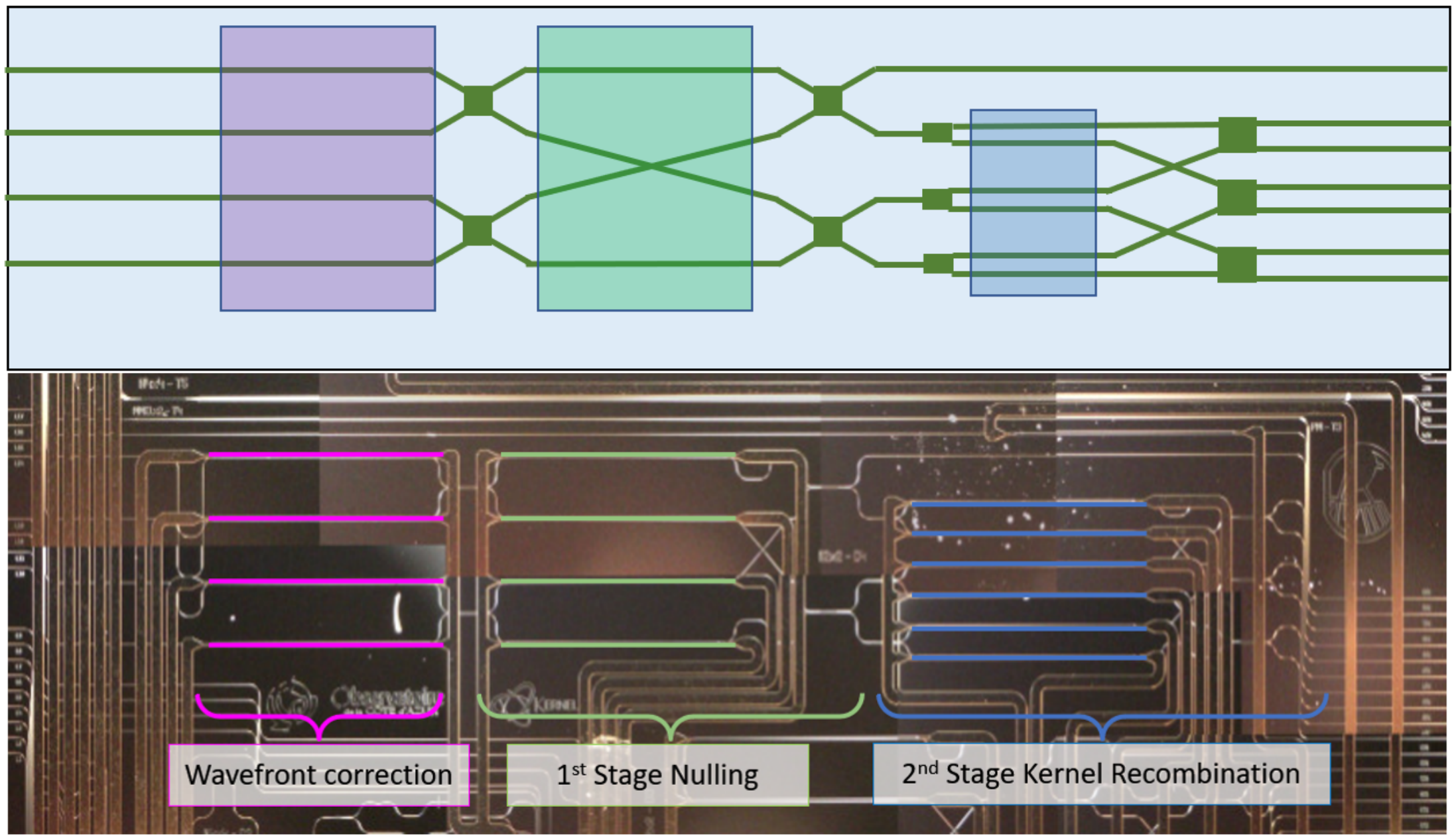

***Figure 2*** *- A chematic of a 4-input Kernel nuller device (top) achieved by the use of traditional waveguides, 2x2 directional couplers, and 1x2 splitters (green). TOPAs (colored) are used to control the internal pathlength matching of the interferometer arms inside the chip and correct for any error coming from either outside the chip (pink), inside the first stage nulling (green), and the beam-recombination that creates the kernel nulls. Microscope image (bottom) of the fabricated device, on a Silicon Nitride platform. Here the TOPAs are colored to their corresponding functio*

## Advances in Science and Technology to Meet Challenges

To do efficient astrophysical nulling one should sample as many baselines as the telescope arrangement allows, and create the complete set of kernel nulls for that number of inputs. With state of the art optical long baseline interferometers having four (VLTI) or six (CHARA) telescopes, one can create efficient Kernel nulling devices with a small number of inputs. This is similar for long baseline concepts in space, which would likely have 4 or 5 inputs. However, in the case of the ELTs a pupil segmentation regime would likely require many more inputs (30 or 40) to make full use of the aperture area. While the exact trade-offs regarding the ideal number of apertures to get the best sensitivity are being studied, it is likely that future on-sky devices will require efficient broadband operation over at least one wavelength band, and potentially handle many more inputs simultaneously.

To date, only 3-input and 4-input kernel nullers have been realized. However, it is possible to create a kernel nuller with an arbitrary number of inputs, which can probe a much greater UV coverage on-sky. The drawback is the waveguide routing becomes very complex so new design work needs to be undertaken to realize this in a low-loss manner. Ultra low-loss waveguide crossings, routing, and pathlength matching are critical to achieving this.

To make broadband MMIs, a more complex design is required than those typically used as standard for telecommunication applications. The internal wavelength-dependency can be compensated for by creating a non-uniform effective refractive index inside the multimode waveguide [17], either by changing the width of the MMI along the propagation direction, or by inserting more complex structures inside the cavity. The use of sub-wavelength grating structures inside the MMI has been shown to greatly increase the operating bandpass [18], and recent development in inverse design using deep neural networks makes the design of these complex structures far easier to derive [19]. To enable this, either high accuracy fabrication methods capable of achieving the required feature sizes need to be available for commercial foundries, or post processing using E-beam/Ion-beam lithography. Furthermore, robust modeling and design tools need to be developed to reliably and rapidly design prototypes for astronomical use. Similarly, the process of making broadband TOPAs requires modification of the waveguide core and thus has similar requirements. However, a more powerful technique to create true integrated dispersion compensation is the integration of low resolution wavelength dispersion, such as by arrayed waveguide gratings (AWGs, see chapter 7 for more details), in combination with traditional TOPAs. In this concept an AWG would disperse each input into a number of wavelength channels (few 10's of nanometers wide), with each one having a TOPA phase actuator to correct the phase individually, or a Mach-Zehnder configuration for amplitude control, before being recombined into a single waveguide by another AWG. This would enable direct control of the amplitude and phase at various wavelength channels over a broad wavelength band (see chapter 13 for details). To enable this, ultra low-loss waveguide crossings must be developed because of the extra routing required after the AWGs , along with active control systems for a much larger number of TOPAs.

**Concluding Remarks**

The self-calibration of nulling observables enabled by kernel nulling can greatly reduce the impact of upstream wavefront residuals on the performance of nulling interferometers. This relaxes the requirements for upstream optics such as delay-lines, fringe-trackers, and deformable mirrors, and enables the use of nulling interferometry in non-ideal conditions. While the fundamentals have been demonstrated at near-infrared wavelengths, to truly exploit this technology for future astronomical instrumentation the photonic components that underpin a kernel-nuller should be developed to work over a wide wavelength band and at different wavelength regimes. Furthermore, the use of active-photonics in the form of on-chip phase control, enables the practical use of more complex designs that have thus far been beyond the reach of astrophotonic interferometry.

**Acknowledgements**

*The authors acknowledge the funding from the European Research Council (ERC) under the European Union's Horizon 2020 research and innovation program (grant agreement CoG - 683029).*

# 20 | Nulling Interferometry with Optical Fibers and Photonic Lanterns

Yinzi Xin[1], Sergio Leon-Saval[2] and Eugene Serabyn[3]
**[1] Department of Astronomy, California Institute of Technology, Pasadena, CA, USA**
**[2] Sydney Institute for Astronomy (SIfA), School of Physics, The University of Sydney, Australia**
**[3] Jet Propulsion Laboratory, California Institute of Technology, Pasadena, CA, USA**

## Status

The combination of diffraction due to a telescope's finite pupil diameter and scattering due to imperfect wavefronts leads to bright halos of light surrounding typical stellar images. This severely limits a telescope's ability to detect faint sources near bright stars, such as that from exoplanets and circumstellar dust. As a result, optical techniques that enable high contrast observations at the smallest possible angular separations are of great interest. Interfering the light arriving at different parts of a telescope aperture (sub-apertures) can potentially yield improved resolution, and appropriately phasing the sub-apertures can interferometrically suppress (null) starlight. This technique of nulling interferometry [1] across a telescope's aperture can enable observations closer to a star than classical full-aperture coronagraphic imaging, providing access to separations where exoplanets are expected to be brighter and more numerous. Nulling interferometry is increasingly turning to photonic devices to reach deep stellar rejection. In this work, we focus on approaches that inject an interferometrically-combined beam into a photonic device (e.g. a single-mode fiber (SMF) or photonic lantern [2]), exploiting the interplay of the beam's phase with the spatial mode structure of the device, as well as leveraging the spatial filtering the device provides. These systems rely on appropriately phasing the different parts of the telescope pupil prior to beam combination, which results in a class of nulling interferometry instruments architecturally distinct from those employing pairwise beam-combination.

The first photonic nuller was the Palomar Fiber Nuller (PFN) [3], which combined the light from a pair of sub-apertures on the Hale Telescope by focusing them in anti-phase (π radians out of phase) onto the tip of an SMF. Anti-phasing provides an asymmetric stellar point-spread-function (PSF) that cannot propagate within a symmetric spatial mode, thus excluding the on-axis starlight [4,5]. At the same time, single-mode operation also serves to greatly reduce the sensitivity of the null to telescope aberrations [6]. While the PFN was based on multi-axial focal-plane beam-combination at an SMF, co-axial pupil-plane beam combination could also be used prior to SMF injection, either with a standard beam splitter, or by focusing a pair of sub-aperture beams from opposite sides of the pupil onto a focal-plane grating to superpose their +1 and -1 orders [7]. However, combining light only from small pupil sub-apertures implies an efficiency loss. This can be obviated (Fig. 1) by applying a relevant phase pattern to the entire pupil, such as a vortex-phase or simple π radian phase-step (i.e., "phase-knife") pattern [8,9,10], which yield respectively, circular and linear fringes. Such a Vortex Fiber Nuller (VFN), has recently been deployed at the Keck Observatory [11,12]. A Photonic Lantern Nuller (PLN) [13] leverages the same principles, but replaces the SMF with a mode-selective photonic lantern (MSPL) [14], which maps specific linearly polarized (LP) modes into individual single-mode outputs [15], resulting in multiple ports that null on-axis starlight but have different spatial sensitivities to off-axis light sources. While such techniques largely eliminate the need for free-space beam-combiner optics, one can go a step further, and carry out both beam combination and phase shifting within a photonic chip (see Chapters 15, 16, and 18). Additional layers of beam combination may also be useful for reducing residual stellar leakage in the null that result from device imperfections. Here we discuss the current and future challenges of fiber and photonic lantern-based approaches and the technology advances needed to meet the challenge of deep stellar nulling.

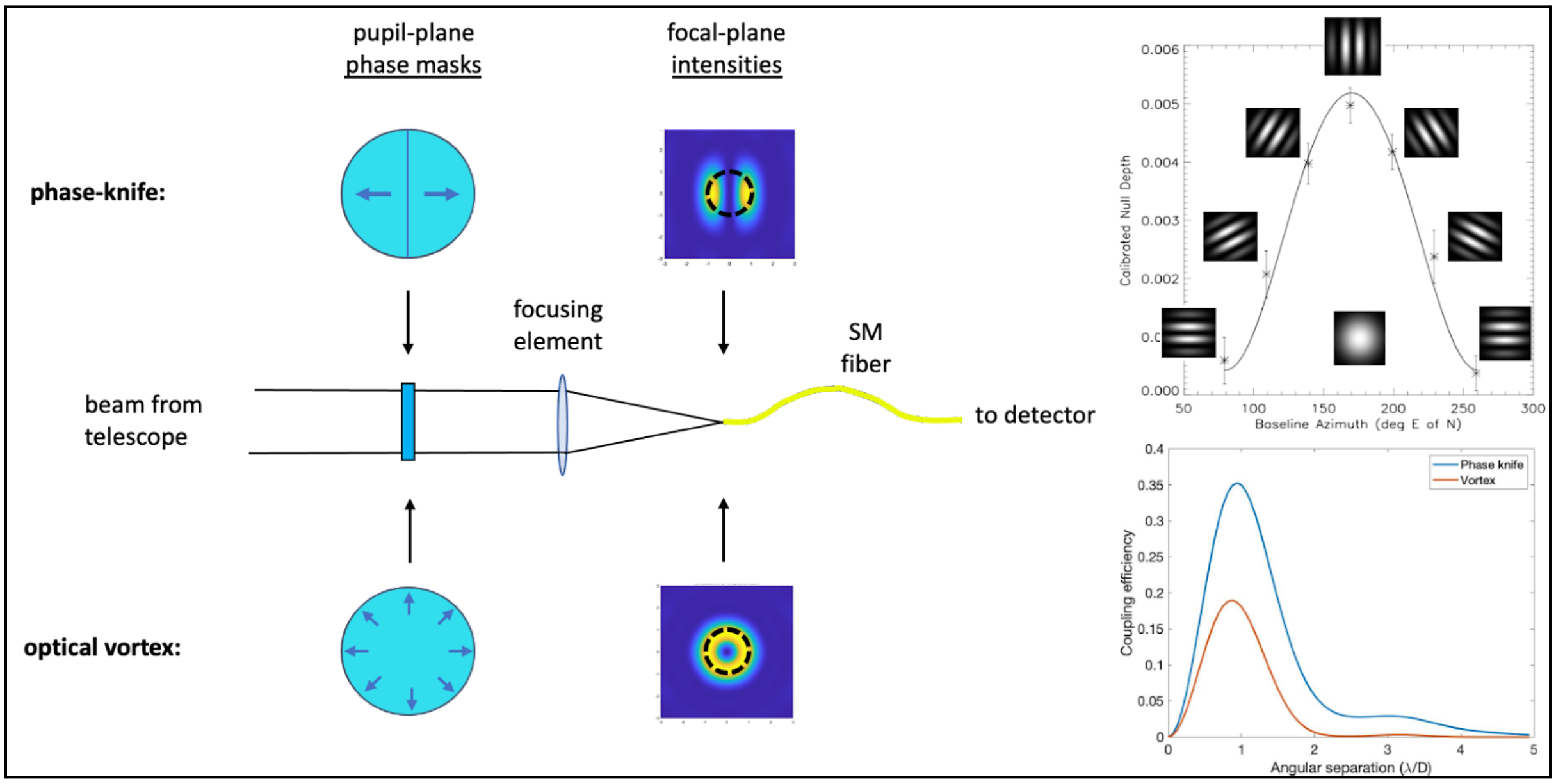

***Figure 1 -*** *Left: Schematic fiber nuller layout. A phase mask is first applied in the pupil plane. Here, a phase knife is depicted on top, and a vortex on the bottom. After the mask, the beam is focused and injected into a single-mode fiber that routes the light to a detector or spectrograph. The PSFs for the two cases are also shown with approximate single-mode acceptance angles indicated by the dashed circles. Top right: The null-depth rotation curve measured on the spectroscopic binary star eta Peg using the PFN [3], showing both the binary response and the few x 10-4 leakage due to the primary star's finite size. The inserts show the fringe orientations for each measurement (the binary orientation was roughly horizontal) and the PSF of the full telescope aperture near the bottom of that panel. Bottom right: Comparison of fiber-nuller coupling efficiencies [10] as a function of off-axis angle, in units of λ/D, where λ is the wavelength and D the telescope diameter. On-axis starlight is nulled, but throughput for off-axis sources peaks near λ/D.*

## Current and Future Challenges

Deep nulling requires an exquisite balancing of beam parameters, with any imbalance in amplitude or phase resulting in increased stellar leakage [1,16]. With commercial SMFs, multi-axial fiber-nullers have reached differing null depths as a function of wavelength and bandpass, with raw laboratory null depths ~ $10^{-6}$ with a 633 nm laser [5] and ~ $7 \times 10^{-5}$ for broadband near-infrared (~1.65 microns) light [17]. Meanwhile, laboratory tests of the VFN have achieved null depths of ~$6 \times 10^{-5}$ with a 635 nm laser [9]. This null depth is consistent with known static aberrations in the optical testbed, but the VFN will also eventually be limited by fiber-related performance limitations in systems with better wavefront quality.

These limitations can be attributed to a number of fiber imperfections that do not usually receive much attention, but will need to be understood for further progress, such as the deviation of the fiber beam profile from the theoretically symmetric single-mode shape (which can lead to a small coupling to asymmetric input modes), or light leaking into the fiber core from the bright starlight originally entering the fiber cladding (which can limit stellar rejection). At longer mid-infrared wavelengths (~10 microns), there are fewer materials available to make fibers, and fiber lengths are limited [18, 19], both potentially limiting the approach to true single-mode rejection.

Nulling with photonic lanterns will also be limited by practical limitations in photonic lantern design and manufacturing. The PLN relies on the symmetry properties of LP modes, as nulls are generated by the orthogonality of the stellar (on-axis) electric field with specific LP modes. Imperfections in mode shape may break these symmetries, which will cause on-axis light to leak into the nulled channels. Manufactured photonic lanterns will have some amount of cross-talk amongst modes, also causing some leakage from the non-nulled ports. The limit of cross-talk currently achievable in MSPLs is around ~1% [S. Leon-Saval, private communication], limiting achievable contrast ratios to the $10^{-2}$ level. Moreover, the modal structure of a photonic lantern varies with wavelength, deviating from perfect LP modes as the wavelength diverges from the design wavelength. Significant theoretical and experimental work is necessary to understand the fundamental limits of photonic lanterns, effectively characterize them in the lab, and implement them in a nulling configuration.

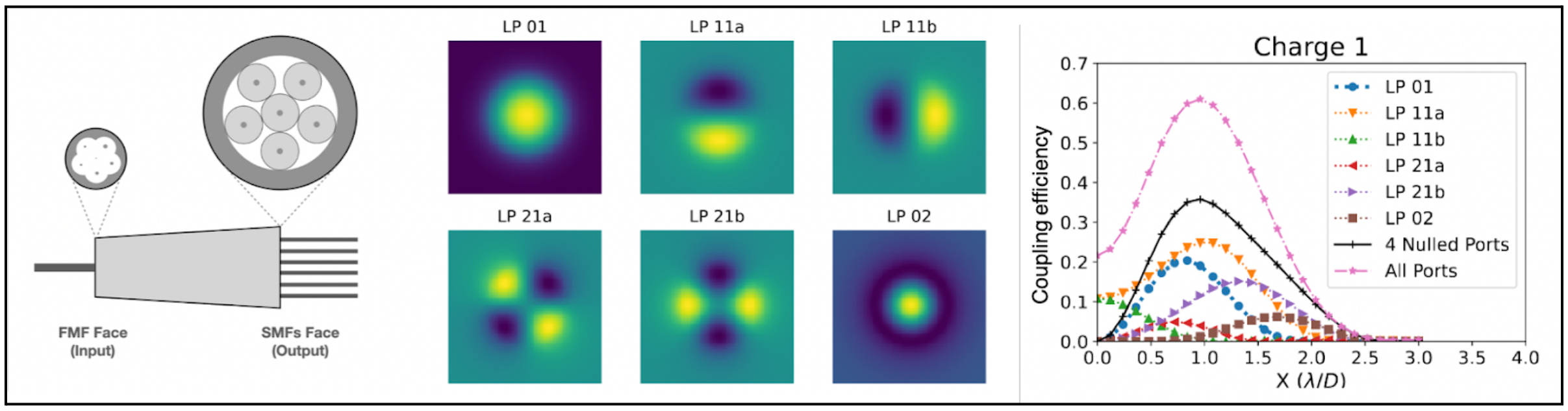

***Figure 2 -*** *Left: Schematics of a 6-port mode-selective photonic lantern spatial-multiplexer fiber system. Each LP mode at the few mode fiber (FMF) face is mapped to one of the six single-mode ports of the SMF face, such that light with an LP mode shape at the FMF side will result in flux in the corresponding SMF core. The device is bi-directional, so light injected into one of the SMF ports will propagate into the LP mode corresponding to that port at the FMF face. Middle: The field amplitudes of the first six LP modes, corresponding to the ideal modes of six-port MSPL. Right: Coupling efficiency for each port as a function of off-axis separation when used with a charge 1 vortex phase mask. Note that the coupling efficiency is not circularly symmetric for all ports, and that x-axis cross-sections are shown here.*

## Advances in Science and Technology to Meet Challenges

In the past, optical fiber-based nulling beam combiners have used a standard single-mode step-index fiber. However, the active optical fiber community has developed a myriad of new and improved optical fiber types, e.g. extremely low-loss (low scattering) graded index fibers [20], photonic crystal fibers (a.k.a. microstructured fibers), hollow-core fibers [21], and low-loss mid-IR (1-18 um) optical fibers [22]. We envision that exploring these new SMF technologies for fiber nulling could provide extended wavelength ranges (e.g. single-mode hollow core fibers in the infrared), improved null depths by increasing cladding mode suppression and reducing mismatch with the fundamental mode (e.g. cladding mode suppressed photosensitive fibers), and improved fundamental mode input selectivity (e.g. using single-mode graded index fibers with a larger effective index separation between the fundamental mode and the second mode, which leads to a large acceptance angle separation).

Some limitations of photonic lanterns have already been studied and improved in devices for other applications. For example, new telecom MSPL devices provide a broadband operation (more than 600 nm of bandwidth in the near-IR) and highly circularly symmetric mode structure outputs (resembling quasi-perfect LP modes) [23]. However, the need to push input modal cross-talk to its limit is unique to high contrast astronomy, since a cross-talk of 10-20 dB is considered sufficient in telecommunications applications [24]. Significant work via both simulations and fabrication and testing is needed to understand the intrinsic limitations of broadband mode selectivity and cross-talk in MSPLs, as well as the steps needed to push designs towards those limits. Additionally, robust methods for experimentally measuring cross-talk have yet to be fully established. Current characterization techniques include measuring modal purity through near and far field mode profiles [25], or leveraging refractive index differences to disperse the modes [20]. However, these approaches do not comprehensively measure the photonic lantern's structure. Other techniques, such as interferometry, phase diversity, or neural network methods need to be explored in order to fully characterize a photonic lantern's complex modal structure.

New geometric structures and waveguide properties may further improve the multi-mode-to-single-mode fiber mapping, hence increasing on-axis starlight rejection and exoplanet spatial information. In the meantime, the VFN and the PLN are relatively new instrument designs, and new instrument structures may also open different approaches. For instance, the use of hybrid MSPL [26] behind a vortex could potentially bring the best of both worlds: SMF nulling, wavefront sensing, and coupling optimization on a single device without introducing non-common paths. Techniques to remove residual stellar leakage, such as wavefront control or back-end beam-combination, will need to be explored. Criteria that future photonic lanterns must meet for viable broadband PLN usage still need to be established, and the performance of PLN designs with currently manufacturable photonic lanterns still need to be experimentally characterized. In parallel, however, many of the known and predicted limitations can be addressed by advances in photonics technology.

**Concluding Remarks**

The injection of an interferometrically-combined beam into an optical fiber or a photonic lantern is a promising approach to nulling interferometry, encompassing instrument configurations such as the Palomar Fiber Nuller, Vortex Fiber Nuller, the Photonic Lantern Nuller, and the recently proposed Phase-knife Nuller. Besides the fiber itself, several of these approaches also rely on specialized pupil-plane phase masks, which can generally be manufactured with existing technologies such as spatially variant liquid crystal polymers or microstructures. Real-time configurable phase masks would be a further step forward. However, the stellar null can be degraded by fiber and photonic lantern imperfections, such as modal shape impurity, cross-talk, and cladding leakage, which will need to be mitigated in the future. Recent advances in the design and fabrication of fibers and lanterns may already provide solutions to some of these limitations, but further research will be necessary to understand them and push the limits of these techniques for higher contrast astronomy applications.

**Acknowledgements**

This work was supported by the National Science Foundation under Grant No. 2109232, and by the National Science Foundation Graduate Research Fellowship under Grant No. 1122374. Part of this work was carried out at the Jet Propulsion Laboratory, California Institute of Technology, under contract with NASA (80NM0018D0004).

# 21 | Heterodyne Interferometry and Frequency Mixing Techniques

Jean-Philippe Berger[1], Azzurra Bigioli[2], Guillaume Bourdarot[3], Ludovic Grossard[4], John Monnier[5] and François Reynaud[4]
**[1] Institut de Planétologie et d'Astrophysique, UGA, CNRS, France**
**[2] Katholieke Universiteit Leuven, Belgium**
**[3] Max Planck Institute for Extraterrestrial Physics, Garching, Germany**
**[4] XLIM Research Institute, University of Limoges, France**
**[5] University of Michigan, Ann Arbor, MI, USA**

**Status**

Several important fields of astronomy such as planet formation or black-hole environments studies request observations at angular resolutions that lie far beyond the performances of current facilities. Optical interferometers offer the potential of reaching resolutions of a few tens of micro-arcseconds. Yet, the technologies at the heart of existing interferometers cannot be simply extrapolated to future arrays with kilometric baselines and tens of telescopes since they are costly, complex and require high maintenance (see e.g [1]). One of the major simplifications of such infrastructures consists in replacing the free-space optical trains by photonic links. In the visible and near infrared, fiber links can be envisioned (see Chapter 17). Unfortunately, such technologies are much less mature in the mid infrared domain (e.g. from 3 to 20 µm) where several key astrophysical processes such as planet formation, or dust formation and evolution, leave an important radiative imprint (see chapter 5 for early but interesting developments of specialty fibers and chapter 8 for new photonics technologies). Newly developed heterodyne and frequency mixing techniques have the potential to expand to thermal wavelengths the concept of optical links, i.e. systems that avoid most of the free-space propagation from the telescopes to the central interferometric correlation unit,. Moreover, the advent of quantum optics coupled with heterodyne technologies allows potentially revolutionary concepts to be explored.

**Heterodyne** interferometry uses the interference beat between the celestial signal and a local oscillator (LO) synchronized in phase between all telescopes to generate radio-frequency signals whose correlation will allow image of the brightness distribution of the object to be reconstructed (see figure 1). This approach allows the multiple splitting of the signal requested to correlate the electromagnetic field of one telescope with all the others to be compensated by amplification without loss of signal. The Berkeley Infrared Spatial Interferometer was a complete mid-infrared (10 µm) infrastructure that demonstrated, through valuable astronomical observations, the feasibility of astronomical heterodyne interferometry albeit with some sensitivity limitations [2]. It used two $CO_2$ lasers synchronized in phase as local oscillators that were propagated to the telescopes and interfered with the celestial signal on ~3GHz HgCdTe detectors.

**Frequency mixing techniques** use the non-linear sum frequency generation effect to shift the spectrum of the collected light from the thermal infrared to the near infrared or visible as close as possible to the telescope focus where the signal is extracted from the surrounding thermal noise. It thus allows the coherent propagation of the converted mid-infrared light with silica optical fibers or guided optics (figure 2). The nonlinear interaction between the astronomical signal and a highly coherent laser beam occurs in a second order nonlinear waveguide. In 2015, first interference fringes on the sky were obtained using Periodically Poled Lithium Niobate (PPLN) nonlinear ridge waveguides with a 1550 nm to 630 nm frequency up-conversion interferometer installed on the CHARA telescope array [3]. On-sky tests were also performed at 3.5 µm on a single interferometric arm. Starlight up to magnitude Lmag=2.8 at 3.5 µm was successfully detected after conversion to 820 nm in a PPLN ridge waveguide [4].

**Heterodyne interferometry with entangled photons** has been proposed to mitigate the local oscillator photon shot noise that adds in quadrature at each output. This background acts as a substantial noise floor for wavelengths shorter than ~50 µm. The use of N=1 quantum states or heralded single photons via entanglement (e.g., from spontaneous parametric down-conversion) promises to suppress shot noise and was explored for the case of heterodyne interferometry by Gottesman et al [5]. When such a Quantum Local Oscillator (QLO) is split, sent to two telescopes, then mixed with starlight, the photon counts can be scrutinized for coincidence at the combiners. Two detected photons at the same time in different telescopes means the QLO and a stellar photon mixed and non-local interference can be recovered. A preliminary laboratory proof of principle was presented in [6]. There is still a 50% theoretical loss in efficiency compared to “direct detection” as sometimes the two photons show up at the same telescope, but the improvement over shot-noise limited operation can be dramatic.

**Current and Future Challenges**

Both heterodyne and frequency mixing techniques suffer from significant sensitivity limitations with respect to “classical”, direct interferometry. Both have intrinsic narrow spectral bandpasses caused respectively by the detector bandpass and the non-linear up-conversion process with a narrow-band laser.

In **heterodyne interferometry** the beating of a highly coherent local oscillator with celestial thermal radiation adds a corresponding shot noise penalty that severely limits its practical use in astronomy in the mid-infrared. Improving the sensitivity of the measurement chain requires increasing the overall spectral bandpass. This is achieved by two means: augmenting the detector temporal bandwidth (and thus, spectral bandpass) and using several spectral channels. First, detector bandwidths of the order of 30 GHz would be highly desirable (from current ~3 GHz commercial devices). This characteristic, at 30 THz (~10 µm), corresponds to a 10 nm spectral bandwidth. Secondly, in order to span a significant fraction of the astronomical N [8-13 µm] and Q [15-27 µm] bands, one should aim at developing focal planes combining a high number of such detectors. This would allow adjacent spectral bins to be sampled by dispersing the celestial light onto the detectors and having it interfere with a mid-infrared laser frequency comb in which each line is tuned on the desired pixel central wavelength. This would require mid-infrared frequency combs with repetition rates of the order of several tens of GHz capable of producing of the order of 100 lines in order to provide several hundred nanometers of spectral coverage (see Chapter 4 for details). Moreover, local oscillators at each telescope have to be synchronized in phase to allow the correlation. Therefore, a dedicated synchronization scheme between each oscillator is required. Finally, since classical digital radio correlation techniques with such high bandwidth signals request prohibitive digitization and computing power [7], technologies capable of correlating high bandwidth signals have to be demonstrated.

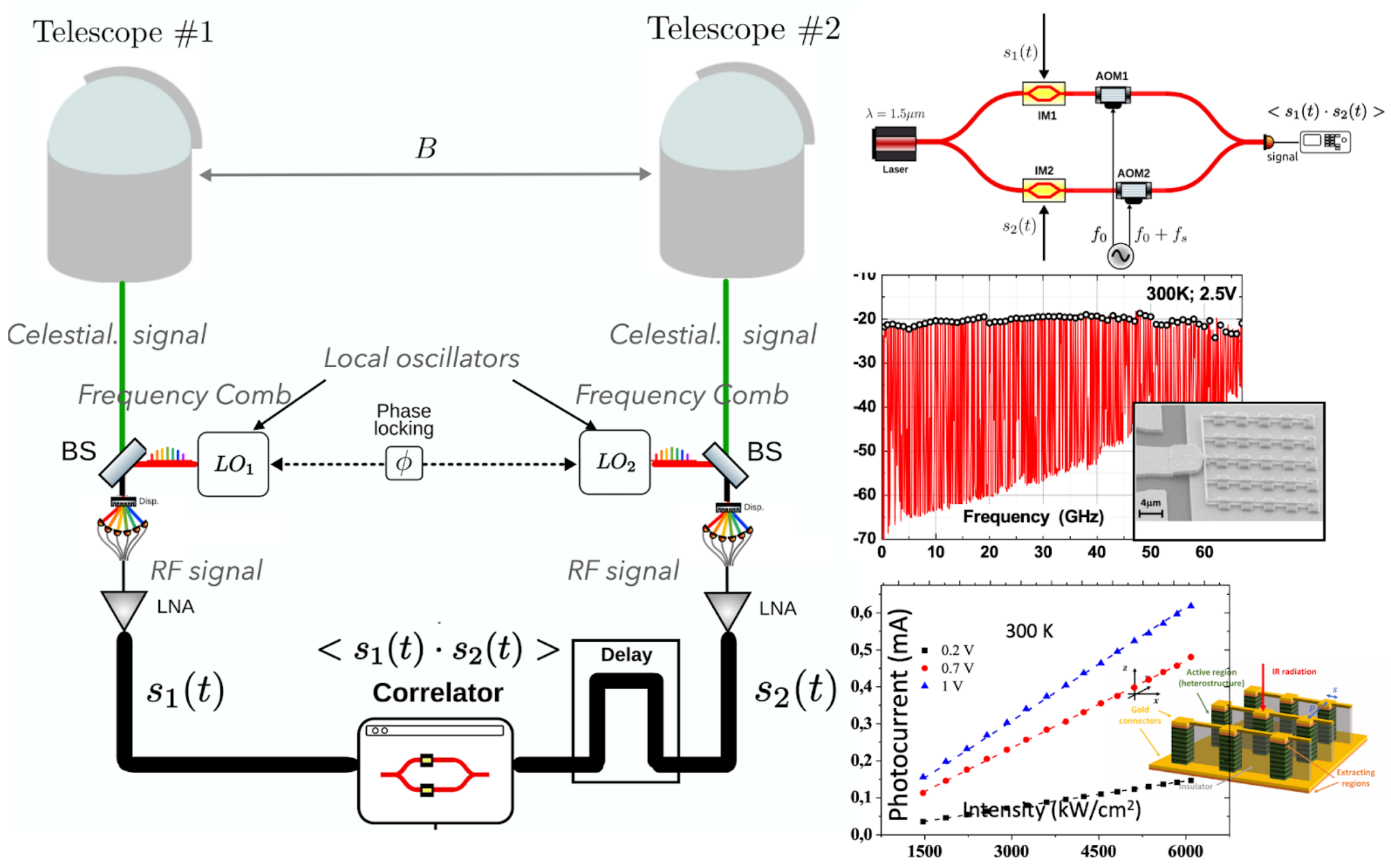

*Left: Schematic principle of a heterodyne two telescope interferometer. Right: Top: Schematic of a photonics two way correlator (see e.g. [22]). Right: Middle: Heterodyne beat spectrum on a quantum well photodetector at 10µm + SEM photograph of a multiple quantum well structure [10]. Right: Bottom: QWIP photocurrent curves as a function of incident intensity showing the linearity. Schematic of the gold patch embedding the quantum wells and waveguide structure.*

**Frequency conversion**'s limited bandpass has to be tackled as well. For example, the current bandwidth with a single line pump laser at 3.5 micron is only 37 nm for a 2-cm long piece of PPLN. Two possible strategies can be envisioned: 1) injecting several pumps, for example using a comb, in the same crystal and demultiplexing through a multichannel spectral mode [8] or 2) multiplexing different signals on different crystals (with different periodicity to ensure wavelength sampling) which presents the advantage of allowing the conversion of large bandwidth signals with only one monochromatic pump laser. A second important challenge related to sensitivity is to improve the overall efficiency of the conversion chain which requires better mode overlap between the astronomical and pump fields in the nonlinear ridge waveguide, coupling efficiency at the input and output of the waveguide, conversion efficiency and possibly noise-free detection in order to compensate for the losses.

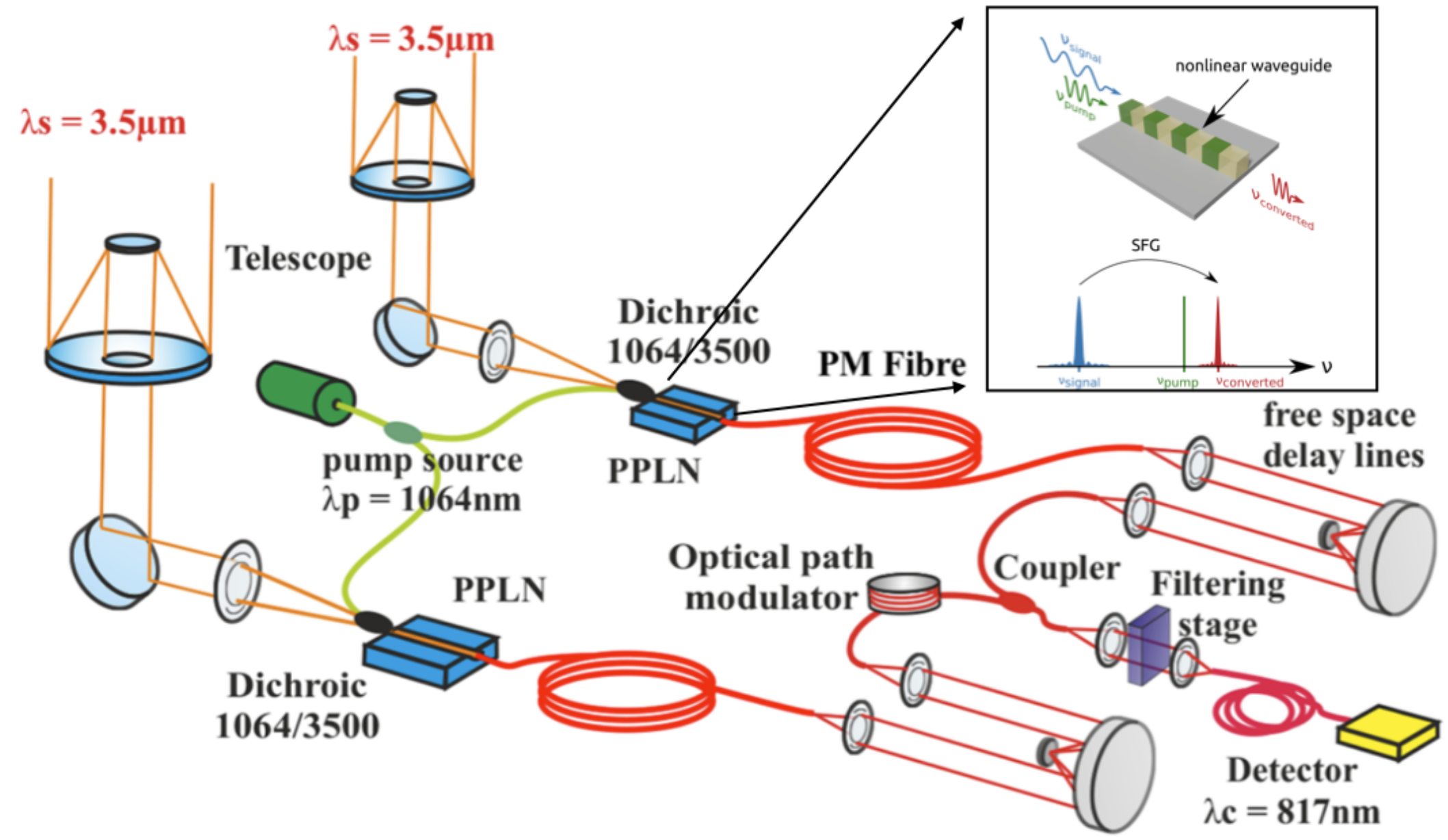

Schematic of a two telescope interferometer with coherent upconversion of the celestial signal through a nonlinear crystal (ALOHA project).

The **heterodyne interferometry with entangled photons** technique offers the unique promise of operating at all wavelengths. Yet, it is difficult to implement as the entangled photons still need to be coherently distributed in real time with near unity transmission, though future advances in quantum communications may make this practical [9]. Similarly, as for heterodyne and frequency up-conversion schemes, there is a similar limitation in bandwidth. Therefore, a dense spectral-multiplexing should be envisioned to cover a broad astronomical bandpass.

**Advances in Science and Technology to Meet Challenges**

*High bandwidth detectors:*

N-band heterodyne interferometry requires high dynamic range, fast and sensitive single-pixel detectors. To overcome the limited speed (<3 GHz) of the HgCdTe detector, the most advanced alternative for the 10 µm spectral range is provided by Quantum Well Infrared Photodetectors (QWIPs), and Quantum Cascade Detectors (QCDs) in unbiased mode. Thanks to their unipolar electronic transport (no hole mobility is involved), they can achieve flat responses up to more than 70 GHz [10] and saturation intensity up to $kW/cm^2$ [11]. A 20 GHz-3dB QCD for 4.6 µm radiation has entered the market as of 2021. Following the integration of a QWIP into a metamaterial acting as a micro-antenna and a cavity [12], the research on photonic-enhanced QWIPs and QCDs is increasingly promising [13]. An optimized photonic architecture could combine speed with a Quantum Efficiency of at least 40%, allowing the electromagnetic field confinement in the material to be maximized without increasing the doping density and thus the dark current [14]. Other fast technologies have been proposed, namely using graphene [15]. However, their short dynamic range, with a response saturation for power

< 10 μW, prohibits the regime where the laser shot noise dominates over the detector noise contributions.

*(Synchronized) Mid-infrared Laser Frequency Combs:*

In order to cover a sufficiently large spectral bandpass of the astronomical mid-infrared filters, laser frequency combs (LFC) are needed to act as local oscillators. They should span several 100 GHz with repetition rates or comb line spacings corresponding to the spectral bandwidth of each detector (e.g several 10's of GHz). This can be achieved by difference frequency generation in quadratic nonlinear media [16]. Quantum Cascade Lasers are another promising technology since they were demonstrated to operate as frequency combs [17]. Further research could consider QCL chips that include several emitting heads tuned to study specific lines and continuum. Ultrafast optical parametric oscillators (OPO) with several GHz repetition rates have been demonstrated in the near infrared and [18], this could be one way forward in the mid-infrared. Beyond that, local oscillators need to be synchronized in-phase across the array. One could find inspiration for this in the demonstration of the stabilization of a mid-infrared (10.6 μm) laser to an atomic clock through a 43 km fiber link. A near infrared comb is locked onto a primary frequency reference and propagated on a phase-stabilized link to each telescope station. At the telescope level it is upconverted to the mid-infrared to provide the local oscillator locking signal [19].

*High bandwidth correlation.*

The correlation of high bandwidth signals requested by heterodyning could be done numerically at the price of huge computing power [7]. However, analog photonic correlation schemes using amplitude modulators have proven to be a robust way of propagating and correlating very large bandwidth (up to 100 GHz) heterodyne signals over telecom fibers (see e.g. [18]). They should be investigated further.

*Reliable nonlinear crystals*

The deployment of parametric conversion schemes requires the manufacturing of reliable nonlinear crystals. This in turns begs for robust industrial processes that allow a proper control of the spectral acceptance, waveguide structure and the management and reduction of the noise due to spontaneous parametric down conversion.

*Miscellaneous developments*

Both heterodyne and frequency upconversion techniques would dramatically benefit from compact high transmission discrete or continuous kilometric delay lines. The first one because it needs an external co-phasing instrument capable of measuring atmospheric phase fluctuations, the later because it is intrinsically a direct interferometry technique that requires optical path compensation. Finally single-photon local oscillators are required as well as fast time-stamped photon-counting detectors in the search for the optimal quantum protocol to achieve the best possible visible and near-infrared telescope resolution of faint astronomical objects. More generally, the availability of true quantum networks, developed in other industrial contexts, would represent a major advance for this application

## Concluding Remarks

The expansion of optical astronomical interferometric arrays requires an increase in their imaging capabilities, angular resolution and sensitivity. This cannot be simply extrapolated from current technologies. Technological breakthroughs in the field of photonics and quantum technologies present new opportunities to envision radically different array architectures. Heterodyne interferometry, both classical and with single-photon local oscillators, as well as frequency up-conversion offer the perspective of arrays with infrastructure free links thus dramatically reducing their complexity.

# 22 | Efficient Hybrid Photonic and Electronic Integration for Realizing Multi-functional Astrophotonic Instruments

Sherif Soliman[1], Katarzyna Ławniczuk[2], Nemanja Jovanovic[3], Jeroen Duis[1] and Ronald Broeke[2]
**[1] PHIX Photonics Assembly, Enschede, Netherlands**
**[2] Bright Photonics BV, Eindhoven, The Netherlands**
**[3] Department of Astronomy, California Institute of Technology, Pasadena, CA, USA**

## Status

The ultimate goal for astrophotonics is to be able to realize multi-functional and efficient integrated photonic instruments. Given that various functions, such as dispersion, spectral filtering, switching, etc, have been developed and optimized over a range of material platforms (SiN, SOI, InP, etc), it will be necessary to exploit hybrid integration - the integration of disparate technologies, materials and chiplets - to be able to assemble these and realize multi-functional instruments.

Four of the most critical aspects which are currently preventing the mainstream adoption of astrophotonic technologies are 1) low throughputs (fiber to chip, propagation and bend losses, device losses, grating/vertical couplers, etc), 2) difficulties with scaling to large channel count devices needed for large bandwidths and high resolutions, 3) efficient integration with detectors and 4) issues with feeding large-port-count fiber arrays to planar photonic chips. Astrophotonic technologies can also be used to more generally support observations by enabling high bandwidth data links driven by the extremely high data rates of the square-kilometer array (SKA) (0.1-1 TB/s), for example [1], or for free space laser communications systems, critical to future astrophysics space missions. In addition, numerous astronomy applications require wide optical bandwidths, which imposes selection criteria on transparency and absorption in different photonic material platforms. For these reasons, developing efficient hybrid integration processes is key to the success of astrophotonic technologies.

To address issues related to coupling to high index contrast photonic integrated circuits (PICs) from low index contrast fiber arrays, interposer devices can be used. These consist of tapered waveguides that provide better mode matching between technologies which can also augment the pitch between the waveguides simultaneously. For example, the Teem photonics interposer solution, referred to as the waveguide array to fiber transposers (WAFTs), allows for efficient coupling (<0.7 dB) between Si based PICs and silica-based optical fiber arrays [2,3]. However, they are now also being used with SiN and InP PICs as well. These WAFTs consist of ion exchange waveguides imprinted into glass-based PICs. The company provides three versions of the WAFT, including 1) an edge coupling version which combines fiber spacing concentrators (FSC) with spot size converters/tapers (SSCs) 2) a top coupling variant which adds a mirror at the tip of the structure and combines it with FSCs that provides wafer level and grating coupler compatibility (see in Fig. 1(a)) and 3) an evanescent coupling version [2].

In a similar vein, being able to couple the light efficiently from 2D fiber arrays, (e.g. a multi-core fiber - MCF) and a PIC is also crucial. Two possible solutions exist for this currently including using a 2D array of grating couplers on the chip to inject the light from above [4] (see

Fig. 1(c)) and/or utilizing fan-out structures that remap a 2D array of waveguides to a linear array [5]. However, neither option is efficient currently, with the former suffering from losses associated with the grating couplers due to manufacturing imperfections, their narrow bands of operation and the stringent requirement of the angle of incidence of the input beam for efficient coupling and the latter limited by realizing mode-matched waveguides and tapers with the ultrafast laser inscription process. These technologies both require further development.

Integrating detectors, be it 2D or 1D arrays or single pixel devices with PICs is also critical to eliminating losses, minimizing cross-talk and realizing integrated instruments. We refer the reader to a full discussion in chapter 23.

Hybrid lasers based on III-V materials like GaAs or InP can also be integrated onto a Si or SOI chip [6] and could be used for either calibration or metrology via say soliton microcombs [7].

To drive active components, including lasers, detectors, phase shifters, etc, electronic integration is necessary. Typical electronic integration involves mounting PICs onto PCBs and using micro-wire bonding. But this does not scale when hundreds or thousands of active components are needed. Innovative solutions are being explored where the electronic integrated circuits (EIC) are etched into the PIC for transceiver applications. Different technologies are being proposed to meet the PIC and EIC integration requirements like advanced packaging substrates and new fiber-to-chip connectivity solutions [8]. Regardless of how the electronics are integrated, astronomical applications will demand low noise characteristics.

Photonic instruments may also require micro-optics, either micro-transfer printed [9, 10], 3D printed [11, 12] or metasurfaces etched into the PIC [13 14] to feed flip-chipped PICs (PICs mounted above the primary chip) and/or detectors [15].

While hybrid integration refers to integrating two or more PICs/devices into a single package, heterogeneous integration involves combining two or more material technologies into a single PIC chip, for instance, III-V materials for lasers or detectors onto Si/SiN PICs [16-18]. Some of the key approaches to achieving heterogeneous integration include flip-chip assembly, heterogenous bonding, micro-transfer printing, and epitaxial growth. Each technology is at a different level of maturity and has different alignment accuracy requirements (summarized in Table 1 in [10]). These technologies offer a more compact alternative to ‘hybrid integration’, but have their own limitations in integrating material platforms such as high alignment accuracy (for all), complex back-end process flow (for heterogeneous bonding), and lattice matching (for epitaxial growth) [10].

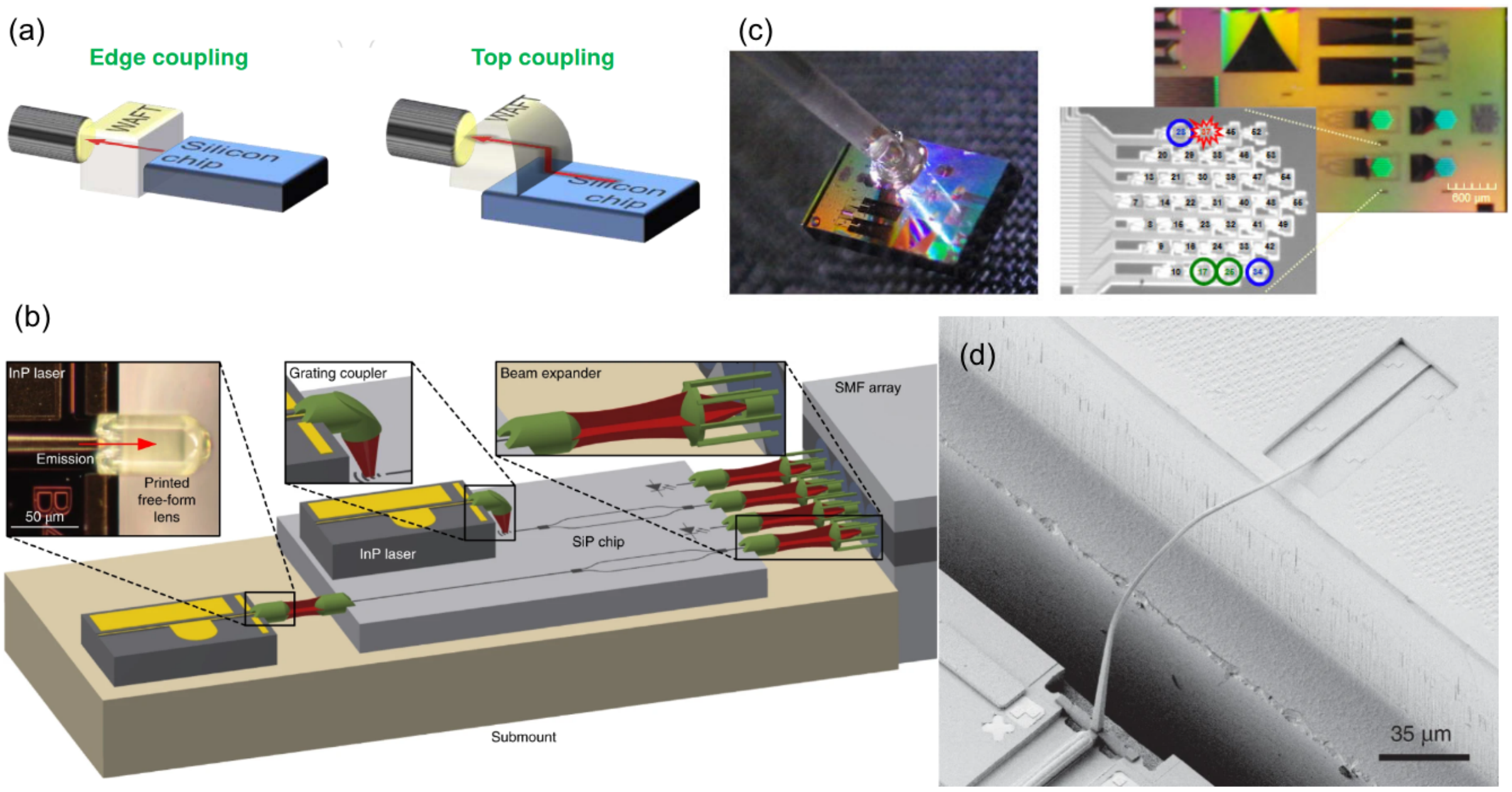

***Figure 1 -*** *(a) Shows two different WAFT interposer variants [2]. (b) Conceptual as well as microscope images of 3D printed micro-lenses and micro-mirrors connecting PICs and fiber arrays with 88% efficiency [12]. (c) An example of integrating a MCF with a PIC [4]. (d) An SEM of a photonic wire bond [32].*

## Current and Future Challenges

Optical mode matching is critical to efficiently couple the light between different photonic components realized in different material platforms. As light may be delivered to the PIC instrument via an optical fiber or may need to pass to other optical components located on another chip, mode matching is key to ensure maximum throughput. This is achieved with spot size converters (SSCs) that relax the tolerances on in- and out- coupling requirements. SSCs are often realized as a taper, including lateral and inverted tapers, segmented and 3D tapers. Inverted tapers are prevalent but are typically highly polarizing (birefringence and diattenuation) and when large exit beams are needed (i.e. 10 microns for example), require a very small and precise exit taper width, making precise mode size selection challenging. Segmented tapers are more tolerant, but require extremely small feature sizes, i.e. sub-wavelength to prevent diffractive effects. This requires special processing (e.g. e-beam lithography), which can be very challenging for mass production. Finally, 3D tapers can mitigate the polarization sensitivities and deliver a circular output beam, but are harder to implement and are only offered by a handful of foundries (e.g. LioniX [19], Fraunhofer [20]) as part of their multi-project wafer (MPW) or custom runs.

Due to the limitations in SSC performance, SSCs can be combined with interposer devices, like WAFTs. The WAFTs outlined above can generate circularly symmetric modes that can be tapered from 3 to 10 microns, with high efficiency, and insertion losses of < 1.5 dB [2]. This reduces the requirement on the more complex and limited on-chip SSCs, and by combining the two, efficient transitions could be realized. Despite the existence of these technologies, optimizing each to maximize overall performance, especially over broadbands still remains a challenge.

Another loss that comes about from hybrid integration is that of Fresnel reflection brought about by the different indices of the materials used. In some applications, index matching glues are used between devices. To minimize the residual Fresnel reflections between the chips and the glue, anti-reflection coatings are placed on the end faces of the two chips. AR coatings range from single wavelength operation (e.g. for narrowband lasers) to coatings functioning over very broad spectral bands of > 100 nm wide. Challenges remain with respect to finding suitable bonding adhesives with high transmission across all spectral regions of interest to astronomy applications (<380 nm up to >5 microns), as well as implementing low loss coatings across various material platforms and broader bandwidths.

There are multiple challenges and approaches towards integrating laser sources onto PICs [10]. Among the existing methods are: monolithic integration of lasers using InP-based material platforms, hetero-epitaxial growth of lasers on Si substrates, as well as hybrid integration approaches via micro-transfer-printing, flip-chip, vertical coupling, and lensed coupling, as well as die-to-wafer and wafer-to-wafer bonding, with some approaches schematically shown in Fig. 2(a) [21]. The challenges for hybrid integration are high alignment accuracy required in case of e.g. flip-chip, vertical coupling, and lensed coupling; as well as throughput, reliability, scalability, yield, and process and test complexity in case of e.g. die-to-wafer and wafer-to-wafer bonding. The main challenge in case of monolithic integration, in the context of astrophotonics instruments, is limited material transparency and optical gain bandwidth, typically covering telecommunication and data-communication C- and O-bands, as well as insufficient throughput (high losses).

When many active devices are needed on the chip, be it photodetectors, lasers, amplitude or phase shifters, it’s necessary to also miniaturize the electronics. Etching the EIC into the PIC is a solution that allows scaling to hundreds or thousands of active devices and is been demonstrated [22] and is being offered commercially [23], but is challenging due to the mismatch between EIC and PIC lithographic technologies and not offered on a large range of material platforms. Further, even if this was solved, the number of devices that can be placed on a chip will eventually be limited by the finite size of the photonic components, rather than the electronics.

Hybrid integration of different photonic components is challenging since the components are typically more fragile than electrical components and have extremely tight optical alignment tolerances,  often <1 um to minimize losses between devices/chips [24]. This is exacerbated by coefficient of thermal expansion (CTE) mismatches as low noise detection in astronomy is typically conducted at 70K for near infrared observations. In the same vein, more general space qualification (e.g. radiation hardening and vibration/shock) testing for hybrid integrated devices needs to be undertaken to prepare them for any future missions.

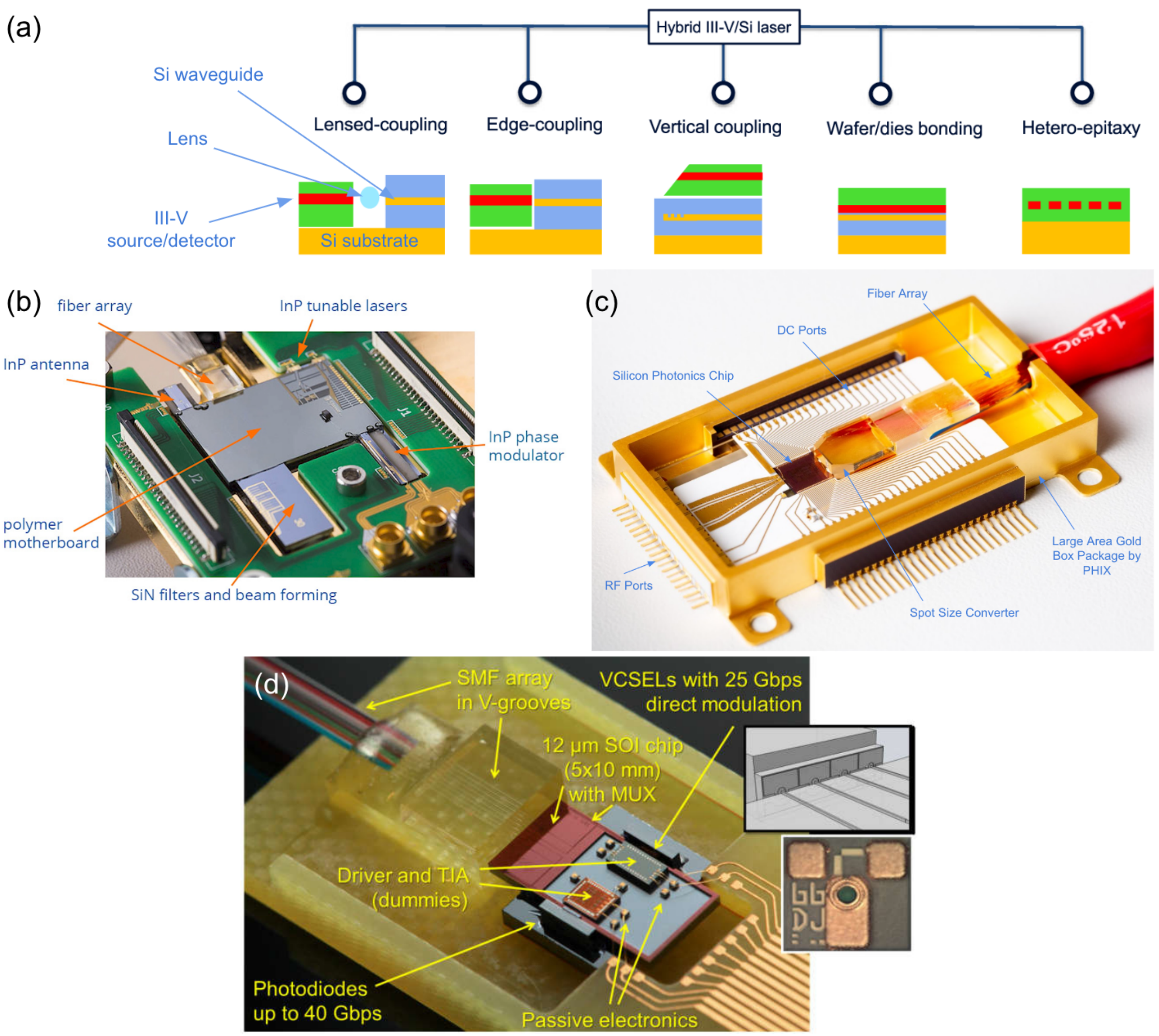

***Figure 2*** *- (a) A schematic diagram of several different processes by which lasers or detectors can be integrated with PICs [2]. (b-d) show examples of hybrid packages. (b) A photo of a millimeter wave transceiver for the Teraway project [25, image courtesy of Phix Photonics Assembly]. (c) A silicon PIC with a fiber array and a WAFT assembled in a box using Phix's standard packaging [26, image courtesy of Phix Photonics Assembly]. (d) Mechanical test assembly that illustrates a transceiver concept based on 12 µm SOI waveguides and directly modulated VCSELs. The VCSEL-SOI integration concept is also illustrated schematically. A microscope image of a VCSEL is shown on the right [27].*

## Advances in Science and Technology to Meet Challenges

The goal is to reach fiber-to-waveguide and chip-to-chip coupling efficiencies of >95% per facet (< 0.4 dB). This has been demonstrated in fiber-to-SiN waveguides over a broad band (1450-1650 nm) using inverted tapers [28]. However, this required lithographic processing with very small feature sizes, which is not offered as part of standard contact/stepper lithography fabrication and hence is not currently scalable. To enable this, more foundries need to offer new and improved processing techniques which allow for smaller feature sizes (necessary for inverted and segmented tapers) as well as 3D tapers as part of their standard services. Design tools also need to be advanced to optimize the SSCs given these new limitations for broadband applications.

The ion exchange technology for WAFTs could potentially be used to realize 3D circuits in future. Fanout devices that not only route the output of a MCF to a PIC, but also provide optimal mode matching between the disparate technologies could be realized in this way for the first time.

Alternative technologies such as SU8 polymer interposers or tapered waveguides (eg: bilayered inverse taper in lithium niobate [29]) could also be used in place of WAFTs. In addition, there are a host of other fiber tapers including capped adiabatic tapered fibers, lensed fibers [30], or cascaded tapers and cantilever tapers [31]. Interconnects are also critical to connect different photonic devices. One example is shown in Fig. 1(d), where a photonic wire bond is used to transmit light from a laser to a waveguide [32]. Polymer interconnects such as those using ZPU12 polymer additionally provide the required flexibility for interconnects [33]. The optical losses of PolyBoard PICs were reported to be ~$0.7\,dB/cm$ at 1550 nm. The fabrication process of the PolyBoard platform provides the added advantage of achieving multilayer waveguide structures by using alternating waveguide and cladding layers and a semi-automated assembly process has been demonstrated using commercially available equipment which eases the scaling from prototypes to production. A completely different approach utilizes 3D printed micro-optics to achieve mode-matching between PICs (Fig. 1(b)), beam shaping, and PIC-to-fiber-array efficiencies as high as 88% [12], albeit without the flexibility of interconnect approaches.

Finally, advances in heterogeneous integration are critical to minimize the losses, attain high stability and scalability, and thereby, maximize the integration advantage in astrophotonics. The key challenges of heterogeneous integration approaches are the necessity of high alignment accuracy (< 0.5 µm) and the high pressures (200 mg/bump) that may be needed for bonding [34, 24]. However, new approaches such as mass-reflow of copper bumps have demonstrated a final self-alignment with ~0.5 µm accuracy in flip-chip assembly of SOI grating couplers and a microlens die, with an initial misalignment tolerance of 10 µm with only 5.5 mg/bump of applied pressure [24]. To achieve the highest possible coupling efficiencies with small waveguide modes (~3 µm) in high-contrast materials, alignment accuracies of ~0.1 µm will be needed. Further advances are also required in making such precision assembly processes scalable for various other material platform combinations.

**Concluding Remarks**

Processes such as heterogeneous integration and hybrid integration are critical to realizing compact all-photonic instruments for astronomy. Developments in regards to advancing SSCs, developing 3D interposers, scaling processes to integrate many detectors with chips, enabling efficient and cross-talk free waveguide crossings, improving processes to integrate micro-optics, and advancing the design and performance of vertical and grating couplers are all critical to realizing efficient astrophotonic instruments.

**Acknowledgements**

*We would like to acknowledge informative technical discussions with Pradip Gatkine.*

# 23 | Semiconductor Detectors and Integrated Photonics

Donald F. Figer
**Center for Detectors, Rochester Institute of Technology, Rochester, NY, USA**

## Status

Semiconductor detectors are central to astronomical discovery and have been the principal tool used in the vast majority of high impact discoveries in the field over the past 40 years, such as the discovery of dark energy, dark matter, exoplanets, and the identifications of black holes in the centers of galaxies, including our own. These detectors convert photons of sufficient energy into charge carriers that they then integrate in a storage well that is often formed as a pn junction. They are the detector of choice for a large range of wavelengths spanning x-rays through mid-infrared wavebands (~0.1 to 30,000 nm). Their application includes imaging and spectroscopy, and they are the primary detectors in the Hubble Space Telescope, James Webb Space Telescope, and the vast majority of telescopes on the ground. In general, they are integrated at the focal planes of bulk optical systems. They are almost always implemented as two-dimensional arrays of photodetectors (PDs).

The most commonly-used arrays are in charge-coupled devices (CCDs) and CMOS active pixel sensors. CCDs move integrated charge to a readout circuit. CMOS devices convert charge to voltage within a pixel and use a readout integrated circuit (ROIC) to transfer the voltage to the output. Simple ROICs consist of a source-follower per pixel, row/clock-select circuits, and a buffer that connects to external electronics. Complex ROICs perform operations such as multiple sample correlated reads. Very low noise semiconductor detectors can now count single photons and resolve photon numbers at optical [1] and infrared [2] wavelengths. Some other single photon detectors use Geiger-mode avalanche gain, such as SPADs (single-photon avalanche diodes) [3], but those generally have significantly lower fill factor and cannot count photon number. Single-element detectors (PDs) are rarely used in astronomy applications, but they are useful in photonic circuits [8]. A number of groups have implemented PDs using a variety of fabrication processes and material systems]. For example, [12] present a recent PD architecture using InP waveguides with monolithic p-i-n PDs integrated in a two-step organometallic vapor phase epitaxy (OMVPE) process. Also, [13] present a Ge PD integrated on an SIO based waveguide platform.

A new class of CMOS detectors enable applications where it is important to detect, and sometimes count, individual photons. These devices have very low read noise, typically less than 0.5 $e^-$. Examples include devices from Gigajot Technology [1], Hamamatsu [10], and BAE Systems [11]. In all cases, the small capacitance of the sense node produces a very large conversion gain, on the order of hundreds of microvolts per electron. This technique will work for individual detector elements integrated in a photonic integrated circuit (PIC), but it requires a PIC process that provides a small design node and integration of complex electronics, two things that are not common at the present time. A short-term solution could be to use these imaging arrays with ad hoc coupling of photonic elements, such as fibers, or to use micro-lenses. Other types of low noise detectors, such as EMCCDs, MKIDs, SNSPDs, etc. could have some utility, and are discussed elsewhere in this volume.

The ultra-low read noise of emerging CMOS array detectors presents enormous opportunities for photonic circuit applications. As an example, one can implement a circuit which splits light in a multi-mode fiber into single-mode fibers and detect that light without degrading the signal-to-noise ratio as would be the case with a detector that cannot detect single photons. Similarly, noiseless rebinning provides flexibility when pixel sizes do not perfectly match waveguides and the beam is oversampled.

**Current and Future Challenges**

While semiconductor detectors for astronomy have evolved primarily into two-dimensional (2D) formats, such as for imaging applications, integrated photonic implementations may benefit from other formats because the detection no longer needs to be made at a focal plane. In fact, it is challenging to integrate 2D detector formats into planar photonic circuits, making it difficult to take advantage of their near-gigapixel array formats. One-dimensional (1D) arrays are more easily integrated into planar architectures through edge-coupling, but they also present challenges when attempting to connect many optical channels from the chip to the array because of routing complexities to avoid waveguide crossings, which introduce loss and cross-talk. Finally, the most flexible format is single PDs that can have arbitrary physical relationships to each other.

Regardless of dimensional format, there are several challenges to realize the potential of integrated photonics when used with semiconductor detectors, i.e.: 1) materials, 2) routing and coupling, 3) form factor, 4) fabrication and 5) temperature control. In several important ways, dimensional format dictates the nature of the challenges in the aforementioned categories. For instance, coupling for 2D arrays requires some way for the PIC to emit light orthogonal to its surface, whereas the light can be in-plane for 1D and PD formats.

Detector materials are chosen for their wavelengths of detection, and their compatibility with PIC materials, so they can be either deposited or pick-and-placed on the PIC (Si, SOI, SiN, InP). The detector materials detect light with wavelengths shorter than their bandgap wavelengths, which for silicon is ~1.1 um, allowing visible applications. Ge, InP, InGaAs, HgCdTe, and InSb have suitable band gaps to detect infrared light.  Lattice mismatch presents serious challenges, limiting the combinations of detector and PIC materials that can be used. Coupling and fabrication techniques vary widely for each combination of materials.

Routing light inside a PIC to a detector pixel needs to be carefully considered to prevent excess loss and cross-talk. This becomes extremely challenging to achieve when thousands of pixels are needed. In this case, ejecting the light orthogonal to the plane of the chip as soon as it's ready, reduces the complexity of maintaining many  low loss, cross-talk free channels inside the chip. It also provides a natural way to use a 2D array or a second PIC with PDs with a planar PIC device. This vertical ejection could be achieved with grating [23]. Both of these devices are extremely challenging to realize efficiently currently and are not offered as part of the cost-effective multi-project-wafer (MPW) fabrication runs offered by foundries. Other challenges of this approach include mode-matching, which may require micro-optics and anti-reflection coatings to minimize losses. To exploit 1D edge-coupled arrays, low loss, cross-talk free routings become the primary challenge. Directly integrated PDs on the PIC can help circumvent the routing issue but are still susceptible to Fresnel reflection losses if not carefully engineered.

The form factor of PDs is one limitation when considering direct integration onto the PIC. For example, PDs can be integrated via direct growth or bonding after picking-and-placing the PD and can be coupled to in a number of ways, including via evanescent coupling (see Fig. 1 for details and also [31]) and butt-coupling. The architecture in the figure for example needs relatively long pathlengths for the evanescent coupling to be efficient. It is also labor intensive to produce, and thus relatively expensive. On the other hand, coupling efficiencies can approach ~100%.The PDs footprint will depend on the type of optical coupling and the materials used. Fitting a large number of PDs directly on the chip while avoiding excessive crosstalk from waveguides is challenging.

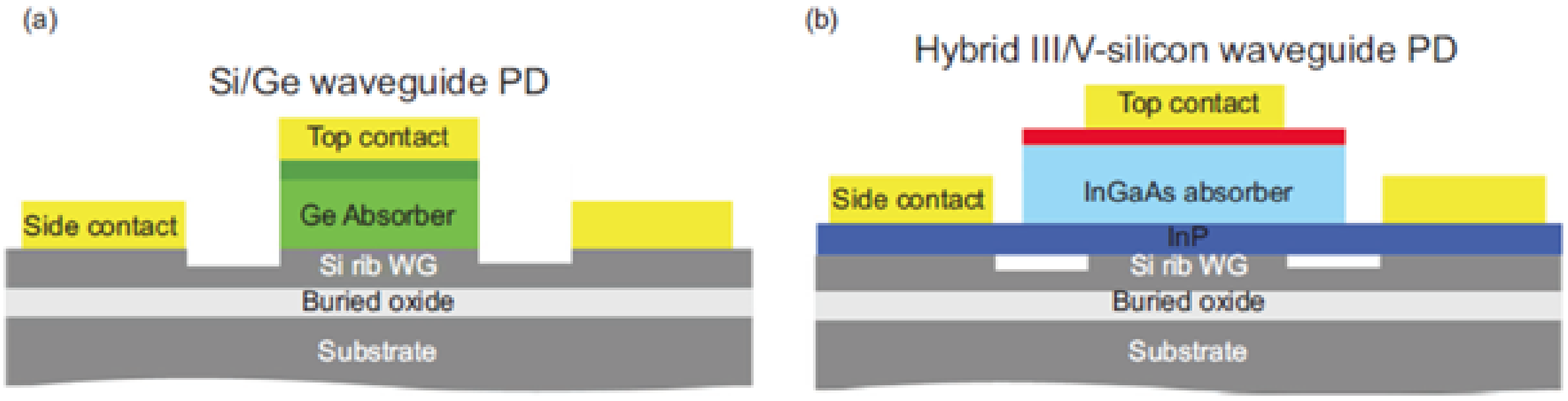

***Figure 1 -*** *The figure shows an evanescently-coupled waveguide (WG) below absorbers containing two different materials, Ge (a) and InGaAs (b) [8].*

Fabrication might be the principal future challenge in realizing the promise of PICs with detectors. Ideally, the fabrication process would provide a menu of elements and materials that are manufacturable and integrable in a foundry setting, such as described in a process design kit. A central challenge is the cost required for all the specialized tooling that is needed for growth and integration. Typically, the tools cannot be used for multiple material systems due to risk of contamination.

Temperature can be a relevant environmental challenge in several different ways. One, some semiconductor detectors require low temperatures, e.g., T~70 K, in order to function with low dark current. This requirement implies an architecture that is very bulky and resource-intensive, requiring cryogenic dewars and complex readout electronics. Another way in which temperature is important is in architectures where elements need very stable temperatures. One example is the fine tuning of frequency that is commonly implemented for ring resonators (whispering gallery modes). If these elements are near a detector, and the detector generates transient heating, particularly through clocking lines, then there will be a coupling between the two elements that will require careful thermal management. Another example is for signal modulation [9].

**Advances in Science and Technology to Meet Challenges**

Although 2D detector arrays offer a large number of pixels, they can cost 5-20x the cost of a custom PIC in the IR. Despite the fact that PICs can be replicated to reduce cost, highly replicable astrophotonic instruments could be limited by the cost of IR detectors. There is a strong demand for low-cost, high sensitivity IR arrays to address this limitation and enable massively replicated instrument designs.

The materials mismatch challenge can be reduced by developing custom anti-reflection coatings and choosing materials with better index-matching to minimize losses due to coupling. On-die integration is a common solution to reduce these loses and also minimize path-length loses in waveguides/fiber, but that then makes fabrication and integration more complicated for non-silicon-based devices that require significant processing outside the standard CMOS foundry flow and more complex heterogeneous integration, such as flip-chip processing [7, 30]. See also [31] for monolithic heterogeneous integration. Another method to reduce the effects of materials mismatch is to use intermediate buffer layers. This reduces defect growth and dark current [14]. Foundries would ideally offer some of these newer processes and the photonic design kit (PDK) must accommodate mismatched materials in the design; very few combinations are now supported.

Several advances are needed to eject light vertically out of the PIC to PDs or detector arrays efficiently, [30]. Firstly, vertical and grating coupler design and performance needs to be improved to realize the efficient ejection of a high quality beam over a broad wavelength range. Secondly, foundries need to offer these specialized components as part of MPW fabrication runs and provide them in standard PDKs. Cross-talk free waveguide crossings are needed to exploit butt-coupling of the edge of a PIC with a 1D or 2D detector array with low loss. Multilayer integration schemes that rely on evanescent coupling between layers have been demonstrated [15, 20], but they need further development to extend the bandwidth. Gratings are also used to eject light out of the circuit plane [22]. An example uses ring resonators to inject resonant photons [20] into a two-dimensional detector array. Another example is the use of a dielectric antenna [21]. One could use these kinds of devices to obtain a spectrum at approximately a thousand spatial locations for a megapixel array. This might be considered to be an “intermediate” solution while fully integrated architectures mature. In addition to losses in couples, it is often the case that the optical elements are larger than the pixel spacing, which in turn introduces another inefficiency in the system. New foundry capabilities are addressing the fabrication challenge through technology transfer from academic and government labs, such as in AIM Photonics which can accommodate simple PDs within PICs in both Si and Ge. Integrating other materials is more of a challenge and outside of the current process. Europe has foundry capability for integrating InP and Si [7].

Another geometry that is difficult to accommodate with a flat array of detectors is a curved focal plane, for example as would be produced by the Rowland circle of a photonic Echelle grating or an arrayed waveguide grating. In these cases, it may be more convenient to use PDs arranged along an arc.

Finally, progress is needed in the arena of single photon counting and photon number resolving detectors in PIC format. Some photonic techniques use small beams, ~1 micron, which implies small pixel size for the optimum coupling. This is fortuitous because the majority of semiconductor foundry capacity is optimized for the small pixels in smart phone cameras. Indeed, the Gigajot devices have pixels of roughly the same size and single photon sensitivity.

## Concluding Remarks

Semiconductor detectors are the workhorse for astrophysics applications over a broad range of wavelengths, and they are likely to become key elements in PIC implementations for those applications in the future. They have already achieved single photon sensing and photon number resolving capability at room temperature, making them an attractive option for integration into PICs. While there are challenges to using them, several research groups have made progress in implementing them, especially in single-detector (PD) formats. To realize much higher detector numbers, one must decide whether it is more cost-effective to implement currently-available high pixel count 2D and 1D arrays, such as in formats having many pixels in arrays, or to use lower dimensional form factors.

## 24 | Integrated Superconducting Detectors for Optical and IR Astrophotonics

Benjamin A. Mazin[1], Alexander B. Walter[2] and Chang-Ling Zou[3]
**[1] Department of Physics, University of California, Santa Barbara, CA, USA**
**[2] Jet Propulsion Laboratory, California Institute of Technology, Pasadena, CA, USA**
**[3] CAS Key Laboratory of Quantum Information, University of Science and Technology of China, Hefei, Anhui, China**

### Status

Astrophotonics harnesses the power of photonic integrated circuits (PICs) and other guided light technologies to make components or even entire instruments for Astronomy. For all astrophotonic applications a detector is needed for the final conversion of photons into electrical signals. For example, the detectors can be designed by putting the photoelectric materials on the top of the waveguide, such that the propagating photons can be efficiently absorbed through the evanescent field of the waveguide [Sprengers et al. 2011, Pernice et al. 2012]. Below we discuss two different superconducting detector technologies that are currently being pursued for integration with astrophotonics: Superconducting Nanowire Single Photon Detectors (SNSPDs) [Natarajan, et al. 2012, Wollman, et al. 2021, Zadeh et al. 2021] and Microwave Kinetic Inductance Detectors (MKIDs) [Day et al. 2003, Szypyrt et al. 2017, Walter et al. 2020]. There is also ongoing work using optical Transition Edge Sensors (TESs) [Nagler et al. 2021, Lita et al. 2022]. Superconducting detectors are naturally compatible with single-mode waveguide spectrometers due to the sub-micron dimensions of an SNSPD hairpin or an MKID inductor (see Fig. 1b and d). While superconducting detectors can add complexity, the sensitivity of these technologies can enable new capabilities, reduce the size, weight, and power (SWaP) and cost of instruments, and improve performance.

Superconducting detectors like SNSPDs or MKIDs detect individual photons that are absorbed in a superconducting film. They are known for high speed and low noise. The critical advantage of superconducting detectors over conventional technologies is the lack of read noise of a true photon counting detector without after-pulsing and their negligible dark counts ($<10^{-5}$ cps for SNSPDs and $<10^{-3}$ cps for MKIDs). This makes them ideal detectors for low-flux applications involving a single-mode spectrometer, such as characterization of exoplanet atmospheres or observation of faint extragalactic sources.

SNSPDs were first fabricated on a GaAs waveguide over a decade ago [Sprengers et al. 2011]. Soon after, Pernice et al. [2012] directly coupled SNSPDs to a Si waveguide on-chip photonic network and demonstrated 91% detection efficiency. Since then, SNSPDs have been integrated on many waveguide platforms including Si, SiN, AlN, GaAs, LiNbO3, and diamond, as well as coupled to Arrayed Waveguide Gratings (AWGs), on-chip echelle gratings, Rowland circle spectrometer, quantum dots, on-chip photonic crystal cavities, MEMs switches, and other photonic components. See Ferrari et al. [2018] for a review. The relatively simple structure of SNSPDs and their exquisite timing performance with UV to Mid-IR demonstrations of <5 ps jitter and up to $10^7$ Hz maximum count rate make them favored detectors for many science and commercial applications of integrated quantum photonics systems [Moody et al. 2022]. While MKIDs lack this history in photonics, their similarly high performance and their natural multiplexability into large arrays make them interesting for astronomical applications.

The complex, high speed, power-hungry, and often custom readout electronics required for superconducting detectors, as well as their cryogenic cooling requirements, can make them undesirable if their high performance is not required. SNSPDs are notoriously difficult to multiplex into the large arrays required for astronomical imaging or spectroscopy [Zhao et al. 2017, Wollman et al. 2019]. MKIDs on the other hand, are naturally frequency multiplexable and the 20,440 pixel MEC instrument at the Subaru Telescope combined with the SCExAO instrument is the largest superconducting camera in the world [Walter et al. 2020]. Both technologies require expensive cryogenic cooling with NbN, NbTiN, WSi, or MoSi SNSPDs operating at ~<4K while TiN or PtSi MKIDs operate even colder at ~0.10 K. Recently, efficient SNSPDs working at 7K have been reported [Gourgues+2019], opening up several applications with a low-cost cryostat for ground- and space-based telescopes.

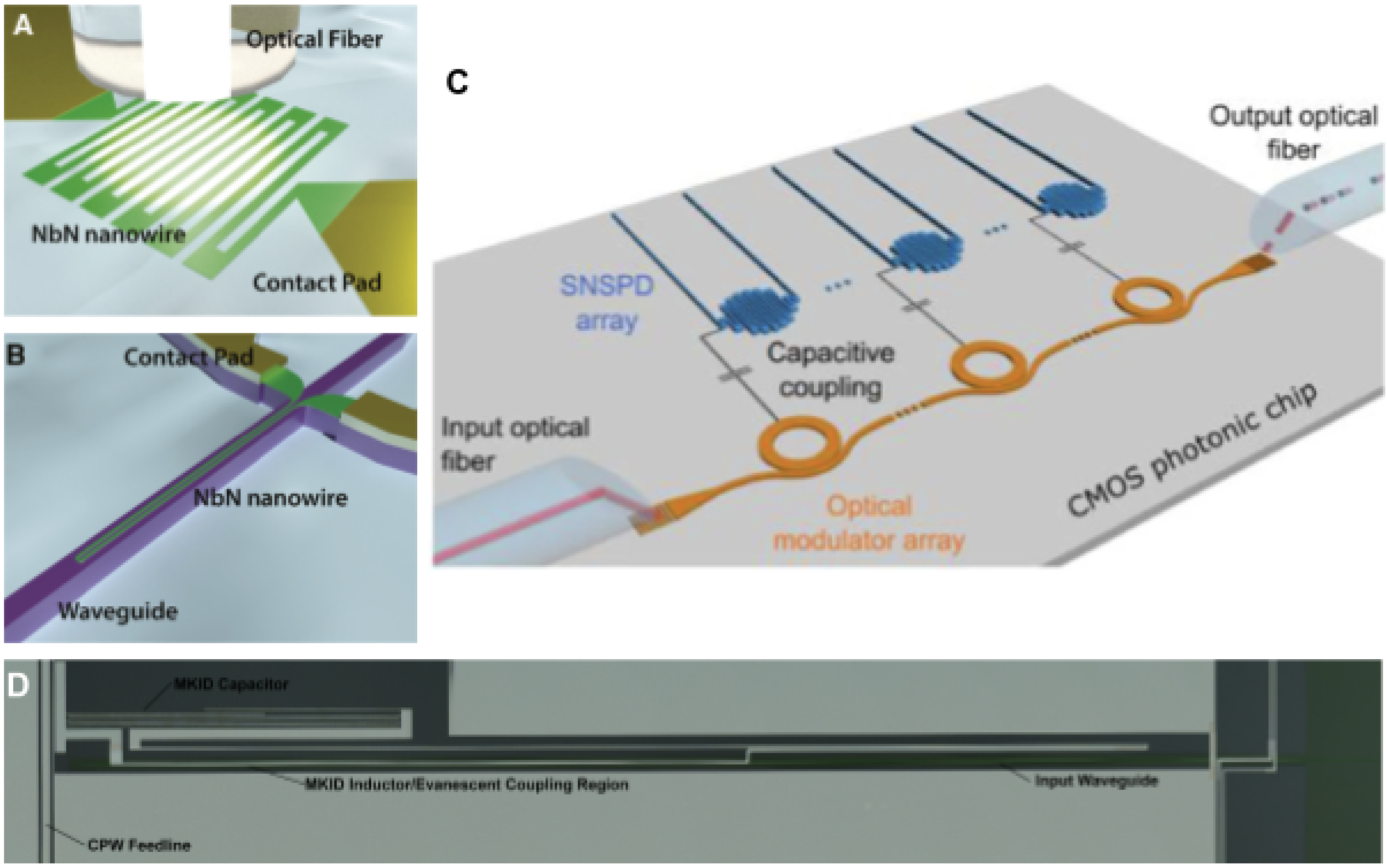

***Figure 1*** *- A and B) (from Ferrari 2018) A fiber coupled SNSPD and waveguide integrated SNSPD respectively are depicted. The fiber coupled SNSPD will typically employ anti-reflection coatings and be embedded into an optical cavity to raise the coupling efficiency, whereas, the waveguide integrated SNSPD can take advantage of evanescent coupling to reach near unity efficiencies. C) (from de Cea 2020) A photonic readout scheme transforms the electrical signal from an SNSPD into an optical signal on a fiber using micro-ring resonators. By tuning the resonators to different wavelengths the pixels can be wavelength division multiplexed. D) A microscope image of a waveguide integrated MKID. The waveguide is the dark green strip running horizontally across the center of the image. The MKID inductor, made of Niobium and Hafnium, partially overlaps the waveguide for evanescent coupling.*

**Current and Future Challenges**

The largest SNSPD array to-date, a 32x32 array of WSi nanowires, was demonstrated by Wollman et al. [2019] using a row-column multiplexing technique. Conventionally, SNSPDs are DC biased near their critical current such that when a photon is absorbed a local hotspot forms. The highly resistive hotspot temporarily redirects the bias current into the readout circuit thus generating an AC pulse signal. In the row-column multiplexing scheme this signal is split with a resistive-inductive network such that half is propagated along the row and column feedlines respectively. Since photon events are recorded by searching for coincidence events in the rows and columns this readout scheme is susceptible to confusion during multi-photon events. Furthermore, the splitting of the signal as well as parasitics in the resistive network reduce the signal-to-noise ratio presenting a major challenge for scaling to megapixel arrays. Other multiplexing schemes exist including thermal row-column [Allmaras et al. 2020], frequency multiplexing [Doerner et al. 2017], and SFQ readouts [Yabuno et al. 2020] but these are less mature and have increasingly complex fabrication steps that generally include multiple lithographic layers with complicated electrical wiring. Coupling light directly on-chip with integrated photonics adds to this complexity due to the experimentally challenging growth or transfer of high-quality functional material, including semiconductors, two-dimensional material layers, and also superconducting nanofilms, to dielectric photonic materials. These functional materials also introduce complexities in the nanofabrication processes.

MKIDs consist of lithographed microwave resonators whose resonant frequency temporarily changes when a photon interacts with the superconducting metal of the detector. The magnitude of this frequency change, usually measured as the phase shift of a microwave probe tone, encodes both the energy (R=35 at 400 nm) [Zobrist et al. 2022] and arrival time (1 μs) of the photon. Since an MKID resonator has high quality factor Q and unity transmission off-resonance, many resonators can be read out over a single microwave feedline allowing for easy fabrication into large arrays. MKIDs require a very clean and uniform substrate to deposit their superconducting film on in order to attain the very high internal Q's (>2e5) and low two level system (TLS noise, see [Gao et al 2008]) required for optimal performance. One challenge with realizing waveguide-integrated MKIDs is to use dielectric materials which simultaneously have low optical loss and low TLS density, and to mitigate the effects of TLS noise from amorphous dielectrics such as $SiO_2$. Maintaining high cleanliness and inter-material compatibility at interfaces represent the primary challenges when simultaneously fabricating photonic components.

A cryogenic operating temperature creates additional complications. First, the optical properties of PICs can change slightly from room temperature, adding another variable to fabrication. This is not well explored in literature. Second, efficient self-aligned coupling from a single-mode optical fiber to a PIC at sub-Kelvin temperatures has remained a challenge for the community, as the maintenance of sub-micron alignment of the optical fiber is necessary to maintain efficient coupling. A space-based astrophotonics instrument must have an alignment that survives the vibrations of the launch, and then the autonomous cooling en-route to the final orbit. Thirdly, increased cost, complexity, and SWaP is needed to cool these superconducting detectors. This may be prohibitive for optical astronomy but useful for mid-IR astronomy where cryogenic cooling is always required for reduced thermal backgrounds.

## Advances in Science and Technology to Meet Challenges

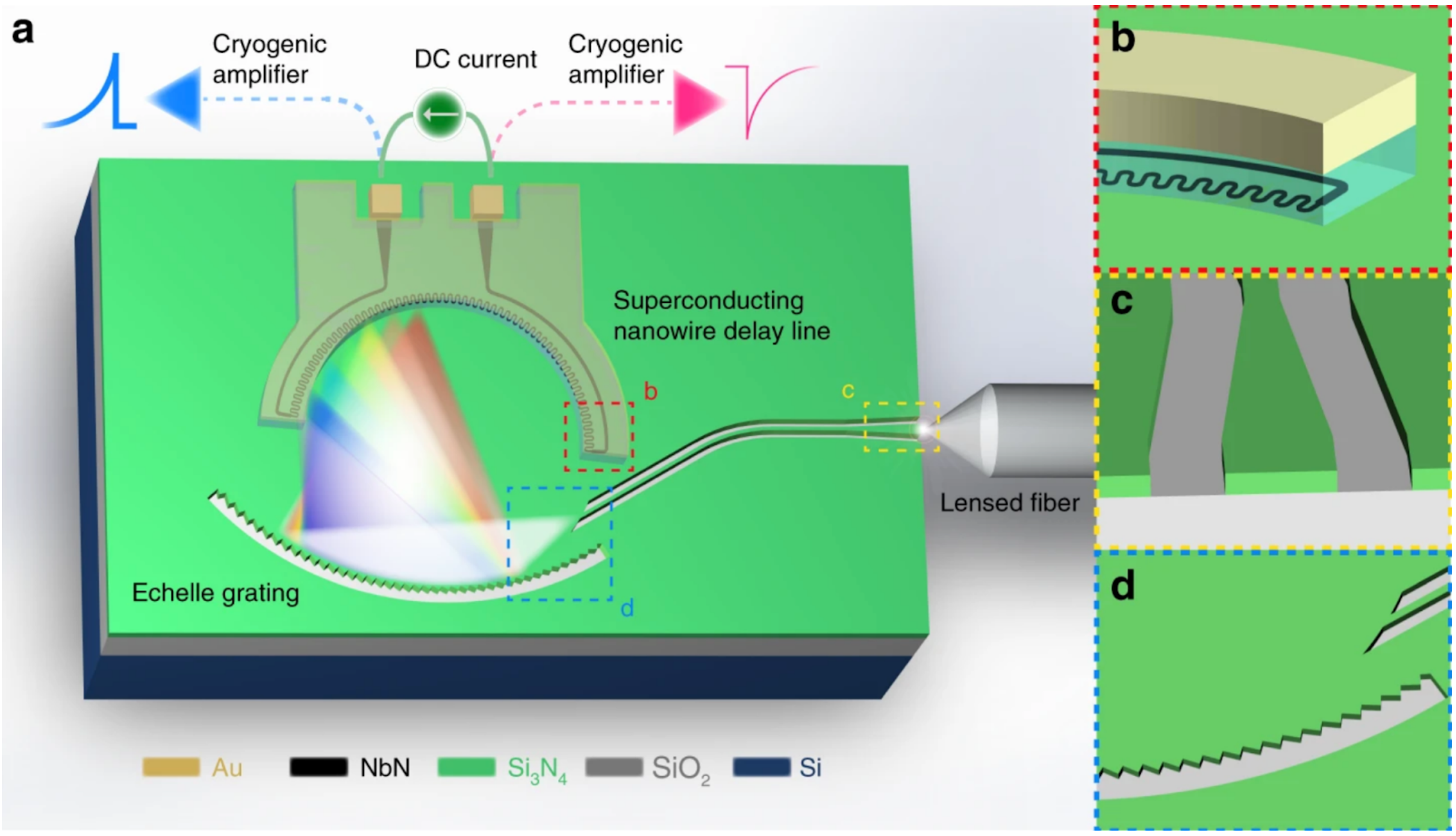

***Figure 2*** *- Reproduced from Cheng et al. 2019. Light from a waveguide is dispersed with a photonic Echelle grating onto a superconducting nanowire implementing an on-chip spectrometer.*

A novel application of integrated photonics could implement on-chip waveguides into the readout circuit of an imaging array of SNSPDs. In de Cea et al. [2020] they demonstrate a UV sensitive MoSi SNSPD pixel electrically coupled to a photonic microring resonator as depicted in Fig 1c. The voltage pulse from a photon event in the detector is used to forward bias the resonator influencing its refractive index and imprinting the signal on an optical carrier propagated along a Si waveguide. The signal is then coupled to a single-mode fiber through a vertical grating coupler and readout at room temperature. This approach allows for wavelength division multiplexing of SNSPDs which avoids the system tradeoffs of other multiplexing techniques by preserving high sensitivity, high efficiency, high maximum count rates/dynamic range, and a robustness to crosstalk or pickup noise. Future developments will need to demonstrate many pixels multiplexed to a single optical fiber and mitigate the thermal power dissipation of the microring resonators. Ultimately, an imaging array of SNSPDs suitable for astronomy may utilize several different multiplexing techniques which could be tailor fit to the application.  Megapixel arrays of SNSPDs could emerge by the end of the decade, but a more dedicated research effort would produce them considerably sooner. Apart from light intensity measurements, the single-photon detections by SNSPDs might also benefit the ‘intensity correlation measurement’ for astronomy applications [Barbieri+06]. For example, the second-order correlation and even high-order correlations of photons have been demonstrated within a single photonic chip recently [Najafi+15, Cheng+2022].

A highly desired photonic technology for astronomy is the monolithic integration of spectrographs with detectors that offer a compact solution and high efficiency by avoiding losses when imaging the guided light to on-chip detectors. They would consume significantly less SWaP, enjoy higher stability, be cheaper to produce, and be easier to use. For the faint sources targeted by astronomers, photon counting, low dark rates, low read noise, broadband sensitive detectors like SNSPDs or MKIDs are ideal. A moderate resolution (2.2 nm) but very small bandwidth (24 nm at 1550 nm wavelengths) has already been demonstrated with SNSPDs coupled to narrowband nanophotonic circuits [Kahl et al. 2017]. More recently, Cheng et al. [2019] demonstrated the first broadband on-chip spectrometer with 200 effective spectral channels between 600 and 2000 nm wavelengths. Light manipulated by a waveguide is dispersed by a lithographically defined Si3N4 Echelle grating onto a NbN SNSPD which is read out with time-of-flight multiplexing (Figure 2). The design allows for easy scalability to higher spectral resolution. For this proof of concept the system detection efficiency was ~<0.1% at 1550 nm although future implementations could reach 50% with some significant optimizations of the 8-9 dB nanowire absorption loss and 7-8 dB diffraction efficiency loss in the grating. Dedicated efforts are also needed in scaling the number of on-chip detectors integrated with the PICs (eg. several thousands for spectrographs).

Work is on-going at UCSB to build waveguide integrated MKIDs with the goal of integrating these detectors with an AWG spectrograph, as shown in Figure 1d. This PIC spectrometer uses the intrinsic spectral resolution (R~35 at 400 nm now, up to R~100 in the future) of the MKIDs to determine what order a photon came from [O'Brien 2020], allowing an octave or greater bandwidth spectrometer with significantly higher spectral resolution than the MKIDs alone can deliver (R=1-100k). This on-chip spectrometer could be extremely compact, allowing tens to hundreds of them to reside in a single dewar.

## Concluding Remarks

Combining photonics with on-chip superconducting detectors allows instrument designers to bring extremely powerful light processing tools together with the most powerful photon counting detectors available. We are at the very start of this field, but the promise is considerable. Potential instruments include high-resolution fiber-fed integral field spectrographs for exoplanets direct imaging using high dispersion coronagraphy, inexpensive time-resolved high-resolution spectrometers, mid-IR space-based exoplanet transit spectrometers, wavefront sensors [Norris 2020], on-chip interferometric instruments [Martinod 2021], and significantly less expensive and more stable extreme precision radial velocity (EPRV) instruments.

## Acknowledgements

*Thanks to Matt Shaw, Miguel Daal, and Majid Mohammed for comments and advice.*